\documentclass{pasa}

\usepackage{graphicx}
\usepackage{aas_macros}
\usepackage[breaklinks=true]{hyperref}
\usepackage{color}
\hypersetup{colorlinks,citecolor=blue,linkcolor=blue,urlcolor=blue}
\usepackage{multicol}

\newcommand{\heii}{He\,{II}}
\newcommand{\civ}{C\,{IV}}

\newcommand{\oiii}{[O\,{III}]}
\newcommand{\ciii}{C\,{III}]}

\newcommand{\bpass}{{BPASS}}

\title[BPASS Version 2.1]{Binary Population and
  Spectral Synthesis Version 2.1: construction, observational verification and new results}
\author[Eldridge, Stanway et al.]{J.J. Eldridge$^1$\thanks{email:j.eldridge@auckland.ac.nz}, E.~R.~Stanway$^2$\thanks{email:e.r.stanway@warwick.ac.uk}, L. Xiao$^1$, L.A.S. McClelland$^1$, G. Taylor$^1$, M. Ng$^1$, S.M.L. Greis$^2$, J.C. Bray$^1$ 
\affil{$^1$Department of Physics, University of Auckland, New Zealand}
\affil{$^2$Department of Physics, University of Warwick, Gibbet Hill Road, Coventry, CV4 7AL}
}

\jid{PASA}
\doi{10.1017/pas.\the\year.xxx}
\jyear{\the\year}

\begin{document}

\begin{frontmatter}
\maketitle

\begin{abstract}
 The Binary Population and Spectral Synthesis (\bpass) suite of binary stellar evolution models and synthetic stellar populations provides a framework for the physically motivated analysis of both the integrated light from distant stellar populations and the detailed properties of those nearby. We present a new version 2.1 data release of these models, detailing the methodology by which \bpass\ incorporates binary mass transfer and its effect on stellar evolution pathways, as well as the construction of simple stellar populations. We demonstrate key tests of the latest \bpass\ model suite demonstrating its ability to reproduce the colours and derived properties of resolved stellar populations, including well-constrained eclipsing binaries. We consider observational constraints on the ratio of massive star types and the distribution of stellar remnant masses. We describe the identification of supernova progenitors in our models, and demonstrate a good agreement to the properties of observed progenitors. We also test our models against photometric and spectroscopic observations of unresolved stellar populations, both in the local and distant Universe, finding that binary models provide a self-consistent explanation for observed galaxy properties across a broad redshift range. Finally, we carefully describe the limitations of our models, and areas where we expect to see significant improvement in future versions.

\end{abstract}

\end{frontmatter}


\section{INTRODUCTION }

The counterplay between observation and theory, the constant iterative process by which models interpret data and data in turn enhances and improves models, is central to the exercise of modern astrophysics. Observations by their very nature are never entirely complete, nor infinitely precise; it is impossible to fully understand the interior of a star or the properties of an unresolved stellar population from observational data alone. Instead, observations are compared to theory and a prescription developed which best fits both the data and our understanding of the physical laws and processes governing the system under observation. In making the comparison, theoretical or numerical models may be discounted or modified, while they, in turn, suggest further observations that may help distinguish between competing interpretations.

Increasingly important for interpreting observations of populations (rather than individual instances) of astrophysical objects is the concept of stellar population and spectral synthesis.  Stellar population synthesis combines either theoretical or empirical models of individual stars according to some mass and age distribution, to predict the ratios of different stellar subtypes, the frequencies of transient events, or the probability of recovering an unusual object as a function of luminosity, density or evolutionary state.  Spectral synthesis combines these stellar population models with stellar atmosphere models, and sometimes with non-stellar components such as dust and nebular gas, to predict the colours, line strengths and other observational properties of an observed population \citep[see][for a recent review]{2013ARA&A..51..393C}. 

It is important not to underestimate the assumptions that go into these models and the uncertainties associated with them. Population synthesis models have been in existence for several decades, with popular examples including Starburst99 \citep{1999ApJS..123....3L} and, for galaxy evolution studies, the GALAXEV models of \citet{2003MNRAS.344.1000B}. These codes are still in active development and continue to add to their initial model grids, with v7.0 of Starburst99 recently supporting stellar models from the Geneva group which incorporate rotational mixing effects \citep{2014ApJS..212...14L}. However work in recent years has demonstrated that there is still much to improve in population synthesis \citep[][and references therein]{2012IAUS..284....2L,2013ARA&A..51..393C}, refining models to address a number of physical processes that are currently not included. Stellar atmosphere models are being revisited to better match the spectra of observed stars across a broad range of metallicities, and evolutionary states. The underlying stellar evolution models are also undergoing an unprecedented increase in accuracy \citep[see e.g.][]{2012ARA&A..50..107L}.

Perhaps most importantly, there is an increasing recognition in the community that it is impossible to ignore the effects of stellar multiplicity when modelling the evolution and observable properties of young stellar populations, whether or not these are resolved.  \citet{2012Sci...337..444S} estimated that 70\% of massive stars will exchange mass with a binary companion, influencing their structure and evolution. While the exact fraction likely varies with environment, stellar type, mass ratio and metallicity, recent estimates of the binary or multiple fraction from both the Milky Way \citep{2013ARA&A..51..269D,2015ApJ...799..135Y,2016MNRAS.457.1028S} and the Magellanic Clouds \citep[e.g.][]{2013MNRAS.436.1497L,2014ApJS..215...15S} suggest that field F,\,G,\,K stars may have an observed binary fraction of 20-40\,percent, while that for O and B type stars may reach 100\,percent. 

There are several codes and groups that study and account for interacting binaries in population synthesis \citep[e.g.][]{2008ApJS..174..223B,2004NewAR..48..861D,2002MNRAS.329..897H,2009A&A...508.1359I,2009ARep...53..915L,2013A&A...557A..87T,1996MNRAS.280.1035T,2004A&A...419.1057W}. However \textit{spectral synthesis} including interacting binary stars has not received the same amount of attention. For a considerable time only two groups have worked on predicting the spectral appearance of populations of massive stars when interacting binaries are included. These are the Brussels group \citep[e.g.][]{1998A&A...334...21V,1999NewA....4..173V,2000A&A...358..462V,2003A&A...400...63V,2003A&A...400..429B} and the Yunnan group \citep[e.g.][]{2005MNRAS.364..503Z,2008ApJ...685..225L,2007MNRAS.380.1098H,2014ApJS..215....2H,2015MNRAS.447L..21Z}. The predictions of these codes are similar; binary interactions lead to a `bluer' stellar population and provide pathways for stars to lose their hydrogen envelope for stars where stellar winds are not strong enough to remove it. However both codes, as far as we are aware, do not make their results freely and easily available. For a full and details description on the importance of interacting binaries and the groups undertaking research in this field we recommend the review of \citet{2017PASA...34....1D}.

The Binary Population and Spectral Synthesis, \bpass, code\footnote{ \texttt{http://bpass.auckland.ac.nz}} was initially  established explicitly to explore the effects of massive star duplicity on the observed spectra arising from young stellar populations, both at Solar and sub-Solar metallicities \citep{2009MNRAS.400.1019E}. In particular it was initially focused on interpreting the spectra of high redshift galaxies, in which stellar population ages of $<100$\,Myr and metallicites a few tenths of Solar dominate the observed properties \citep{2012MNRAS.419..479E}. We have also endeavored to make the results of the code easily available to all astronomers and astrophysicists who wish to use them.

\bpass\footnote{Note that the term \bpass\ can be used interchangeably to refer to the stellar evolution code, the spectral population synthesis code, the resulting models or the collaborative project exploring and exploiting these models.} is based on a custom stellar evolution model code, first discussed in \citet{2008MNRAS.384.1109E}, which was originally based in turn on the long-established Cambridge STARS stellar evolution code \citep{1971MNRAS.151..351E,1995MNRAS.274..964P,2004MNRAS.353...87E}. The structure, temperature and luminosity of both individual stars and interacting binaries are followed through their evolutionary history, carefully accounting for the effects of mass and angular momentum transfer.  The original \bpass\ prescription for 
spectral synthesis of stellar populations from individual stellar
models was described in \citet{2009MNRAS.400.1019E,2012MNRAS.419..479E}, while a study of the effect of supernova kicks on runaways stars and supernova populations
was described in \citet{2011MNRAS.414.3501E}. In the years since this initial work, a large number of additions and modifications have been made to the \bpass\ model set, resulting in a version 2.0 data release in 2015 which is briefly detailed in \citet{2016MNRAS.456..485S} and \citet{2016MNRAS.462.3302E}. It has been widely used by the stellar \citep[e.g.][]{2017ApJ...834..107B,2016arXiv160103850W} and extragalactic \citep[e.g.][]{2016MNRAS.459.3614M,2016ApJ...826..159S} communities but has not been formally described.

In this paper, we provide a full description of the inputs and prescriptions incorporated in a new version 2.1 data release of \bpass, as well as demonstrating the general applicability of the models through verification tests and comparisons with observational data across a broad range of environments and redshifts. We also describe all the available data products that have been made available both through the \bpass\ website and through the PASA journal's datastore. The more important new features and results included in this release of BPASS include: calculation of a significantly larger number of stellar models, release of a wider variety of predictions of stellar populations, inclusion of new stellar atmosphere models, a wider range of metallicities, binary models are now included for the full stellar mass range and a broader range of compact remnant masses are now calculated in the secondary models. 

In section \ref{sec:method} we present the numerical method underlying the \bpass\ models, discussing stellar evolution models (section \ref{sec:method_evol}) and population synthesis (section \ref{sec:method_popsynth}) as well as the combination of these results with stellar atmosphere models to produce synthetic spectra (section \ref{sec:method_atmos}).  In section \ref{sec:tests_resolved} we present tests verifying that our models successfully reproduce the properties of resolved stellar populations, and in section \ref{sec:tests_sne} we consider verification tests arising from stellar death in the form of supernova and transient data. In section \ref{sec:tests_unresolved} we consider a comparison between our spectral synthesis models and observed data for well-constrained coeval systems of stars (both clusters and eclipsing binaries). In section \ref{sec:complex} we consider the more complex star formation histories and nebular emission more commonly observed in the integrated light of galaxies across cosmic time. Finally, in section \ref{sec:future} we discuss aspects which remain a priority for future developments of the \bpass\ code, before summarising our conclusions in section \ref{sec:conclusions}.

We typically report object and model photometry (i.e. Figures \ref{FigJJ8a}-\ref{fig:mag_WR_inst}) in Vega magnitudes, with the exception of galaxy photometry drawn from the Sloan Digital Sky Survey (i.e. Figures \ref{fig:m33_clusters}, \ref{fig:mag_sfh}-\ref{fig:mag_sdss2}) which is shown in the survey's native AB magnitude system. Where required we use a standard flat $\Lambda$CDM cosmology with $H_0=70$\,km\,s$^{-1}$\,Mpc$^{-1}$,  $\Omega_\Lambda=0.7$, $\Omega_M=0.3$ and $\Omega_k=0$.


\begin{figure}
\begin{center}
\includegraphics[width=\columnwidth]{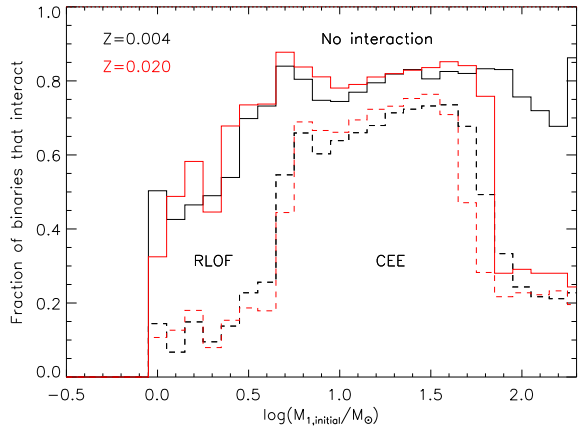}
\caption{The fraction of primary stars in our fiducial population that experience a binary interaction at two metallicities within the age of the Universe versus the initial mass of the primary star. The solid line represents the total fraction that experience Roche lobe overflow (RLOF). While the dashed line represents the number when RLOF progresses to common envelope evolution (CEE).}\label{Fig0a}
\end{center}
\end{figure}

\begin{figure}
\begin{center}
\includegraphics[width=\columnwidth]{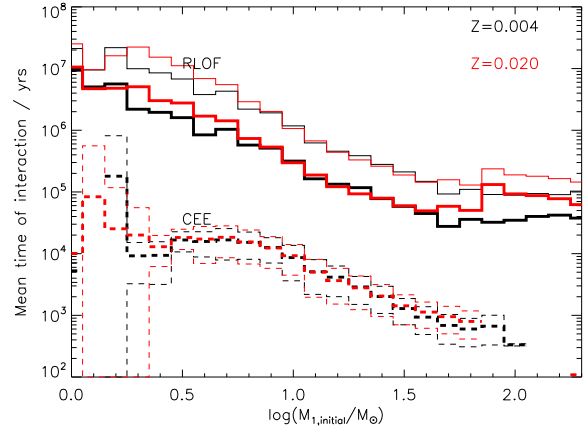}
\caption{The mean times that binary interactions last for in our fiducial simulation versus the initial mass of the primary star. The solid lines are for RLOF and the dashed lines for CEE. The thick lines are for the mean, the thin lines are at $\pm1\sigma$. \textit{As discussed below, the CEE time are significantly overestimated due to our method of including CEE in our detailed evolution models}.}\label{Fig0b}
\end{center}
\end{figure}

\section{NUMERICAL METHOD}\label{sec:method}

Complexity arises in the construction of synthetic stellar populations due to the various layers that have to be created and combined to obtain the final result. Most broadly these are the models of stellar evolution, the method to combine these into a population and finally predictions as to the appearance of this population in observational data. Below we detail each of these in turn and summarize numerical input values and parameter ranges in Table \ref{tab:inputs}.


\begin{figure}
\begin{center}
\includegraphics[width=\columnwidth]{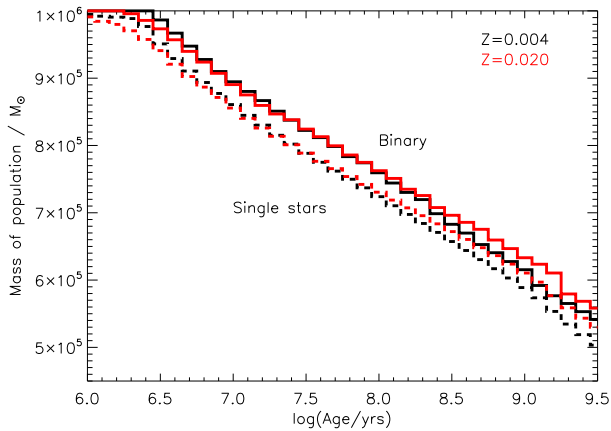}
\caption{The evolution of the fiducial stellar population mass (the mass contained in stars only) with time. We assume formation of an initial population with total mass $10^6$\,M$_\odot$ within the first Myr. Solid lines are for binary populations and dashed lines are for the single-star populations, and results are shown at two metallicities.}\label{Fig0c}
\end{center}
\end{figure}

\begin{figure}
\begin{center}
\includegraphics[width=\columnwidth]{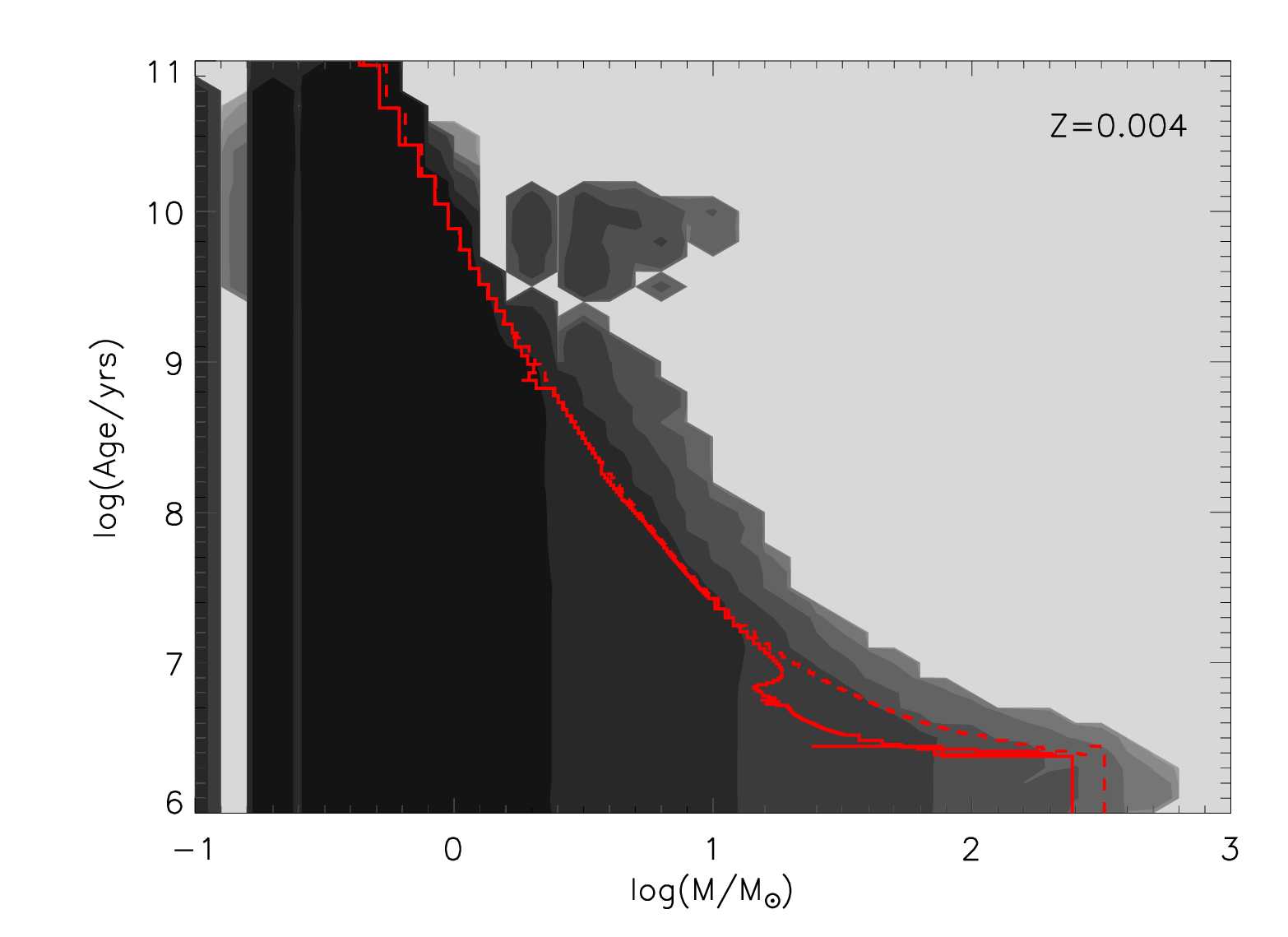}
\includegraphics[width=\columnwidth]{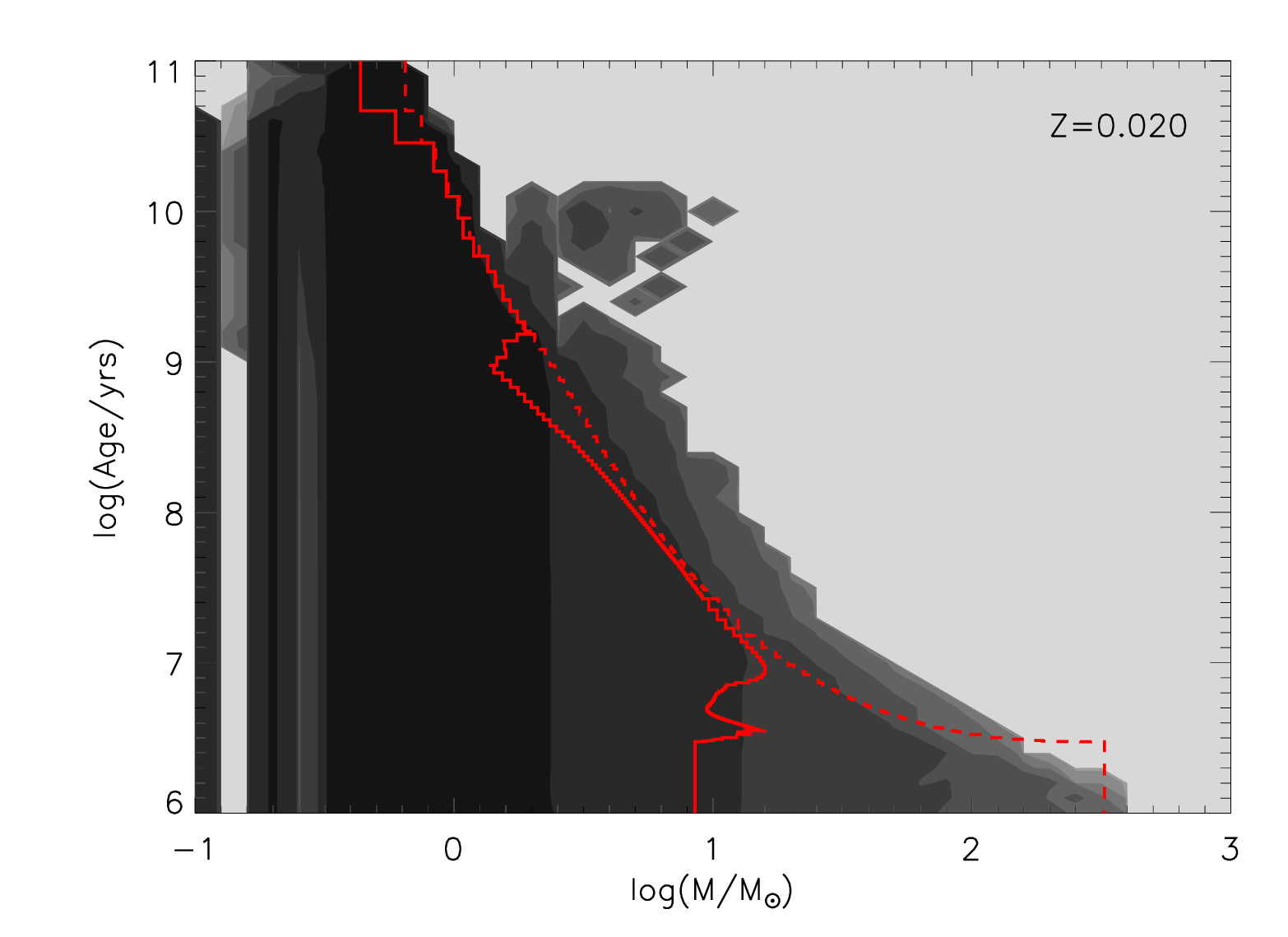}
\caption{Contour plots showing how the stellar mass function evolves with age. The greyscale contours represent a binary population with each contour being a change in number by an order of magnitude. The lines represent the maximum mass at any given time, the solid line for the initial mass of the star and the dashed line the final mass of the star with that lifetime.}\label{Fig0d}
\end{center}
\end{figure}

\subsection{The Stellar Evolution Models}\label{sec:method_evol}

\subsubsection{BPASS Single star models}

The stellar evolution used in \bpass\ is a derivative of the Cambridge STARS code that was first developed by \citet{1971MNRAS.151..351E}. It has since been updated by many authors but the most recent thorough description for the variant we employ was given in \citet{2008MNRAS.384.1109E}.
It is a \citet{1964ApJ...139..306H} type evolution code that solves for the detailed stellar structure using an adaptive numerical mesh and time step to track the stellar evolution. 
For our single star evolution code the models have initial masses ranging from 0.1 to 300\,M$_{\odot}$. We calculate every decimal mass between 0.1 and 10\,M$_{\odot}$, every integer mass between 10 and 100\,M$_{\odot}$ and beyond that increase the stellar mass by 25\,M$_{\odot}$ between models. We attempt to compute all our stellar models from the zero-age main-sequence to the end of evolution in a single run. We first calculate the models with a resolution of 499 meshpoints. For a small number of models numerical difficulties are encountered and we recalculate these model at a resolution of 199 meshpoints to overcome such problems. 

We assume an initial uniform composition of the stars such that the initial compositions of the hydrogen and helium abundances by mass fraction are given by $X=0.75-2.5Z$ and $Y=0.25+1.5Z$ respectively. Here $Z$ is the initial metallicity mass fraction which scales all elements from their Solar abundances as given in \citet{1993oee..conf...15G}. This is selected to match the composition of the opacity tables included in the models. The opacity tables are described in \citet{2004MNRAS.348..201E} and are based on the OPAL \citep{1996ApJ...464..943I} and \citet{2005ApJ...623..585F} opacities. 

There is no strong consensus in the literature regarding the definition of Solar metallicity. \citet{2014ApJ...787...13V}, for example, suggest the metal fraction
in the Sun is rather higher than the $Z = 0.020$ usually assumed, while some authors \citep{2002ApJ...573L.137A,2005ARA&A..43..481A} suggest that Solar metal abundances should be revised downwards to closer to $Z = 0.014$ \citep[also appropriate for massive stars within 500\,pc of the Sun,][]{2012A&A...539A.143N}. We retain Z$_{\odot}=0.020$ for consistency with our empirical mass-loss rates which were originally scaled from this value. We note that at the lowest metallicities of our models the small uncertainty in where we scale the mass-loss rates will cause similarly small changes in the mass-loss rates due to stellar winds. At the lowest metallicities mass-loss is primarily driven by binary interactions. The metallicities we calculate have $Z=10^{-5}$, $10^{-4}$, 0.001, 0.002, 0.003, 0.004, 0.006, 0.008, 0.010, 0.014, 0.020, 0.030 and 0.040 (i.e. 0.05\% Solar to twice Solar).

Convective mixing is modelled by standard mixing-length theory and the Schwarzchild criterion. We include convective overshooting with $\delta_{\rm OV}=0.12$. This value for the amount of overshooting was calibrated against observed double-lined eclipsing binaries. The calibrations involved assuming the area of both components was the same and overshooting was varied until the surface parameters of the stars were reproduced \citep{1995MNRAS.274..964P,1997MNRAS.285..696S,1997MNRAS.289..869P,2015A&A...575A.117S}. 

We do not follow rotational mixing in detail but rather adopt a simple paramaterization in which we assume a star is either not mixed or fully mixed depending on binary mass transfer (see next section). If the mass transfer from a companion exceeds 5\% of a star's initial mass, it is assumed to lead to either rejuvenation and/or quasi-chemically homogeneous evolution (QHE).  Rejuvenation is the consequence of rapid mixing of new material into the stellar core, resetting its evolution to the zero-age main-sequence, thereafter the star evolves normally. QHE follows if the
rotational mixing continues after mass transfer ends, disrupting the formation of shell burning and other internal processes. We have calculated simple QHE models by assuming certain stars are fully mixed throughout their main-sequence lifetimes. They therefore evolve to higher temperatures during this time, going `the wrong way' on the Hertzsprung-Russell diagram. We assume this evolution is only possible at $Z\le0.004$ and for masses above $20$\,M$_{\odot}$ \citep{2006A&A...460..199Y}.

The final detail of our single star models is the mass-loss scheme incorporated for stellar winds. There are two aspects of this scheme: first, the mass-loss rate to apply to a star during a certain phase of its evolution, and second, how the mass-loss rates scale with metallicity. In all our models we apply the mass-loss rates of \citet{1988A&AS...72..259D}, unless the star is an OB star where we use the mass loss rates of \citet{2001A&A...369..574V}. These are theoretical rates that do not include clumping but do match observed mass-loss rates \citep[e.g.][]{2015ApJ...809..135S} and similar calculations including clumping do show that for the luminous O star the rates are unchanged \citep{2012A&A...537A..37M}. For Wolf-Rayet stars, when the surface hydrogen abundance is less than 40\% and the surface temperature is above $10^4$K we use the mass-loss rates of \citet{2000A&A...360..227N}. At different metallicities we scale these mass-loss rates by $\dot{M}(Z)=\dot{M}(Z_{\odot})(Z/Z_{\odot})^{\alpha}$ and typically use $\alpha=0.5$, except in the case of OB stars where $\alpha = 0.69$ \citep{2001A&A...369..574V}. There is some evidence that the value should also be greater for Wolf-Rayet stars \citep{2015A&A...581A..21H} but this is currently not considered.

All our models evolve from the zero-age main-sequence up to the end of core carbon burning, or neon ignition for our most massive star models. Less massive models either form carbon-oxygen or oxygen-neon cores and evolve up the Asymptotic Giant Branch (AGB). We do not model AGB thermal pulses in detail nor incorporate specific AGB mass-loss rates at the current time. To do so is difficult with the STARS code and requires care to overcome numerical difficulties \citep{2004MNRAS.352..984S}. At the current time we limit the spatial and temporal evolution of our AGB models and so we do not observe thermal pulses. We find that the cores then grow up to the Chandrasekhar mass and fail when carbon is ignited in the CO core. This is unphysical and leads to overly luminous AGB stars in our populations. In future we will recalculate these models with realistic AGB mass-loss rates to terminate the evolution earlier or add a routine with in the population synthesis to use rapid synthetic model of AGB evolution \citep[e.g.][]{2004MNRAS.350..407I}. The latter option would allow us to investigate the importance of AGB stars in more detail.

Our lowest mass models (below approximately 0.8\,M$_{\odot}$) never ignite helium and end their evolution as helium white-dwarfs. Models above this mass and up to approximately 2\,M$_{\odot}$ fail at the onset of core-helium burning due to the core being degenerate at this time. Due to the rapid increase of luminosity in degenerate material with the STARS code we are unable to evolve through this event and the models fail. Therefore to model the further evolution of these stars we follow the method of \citet{2005MNRAS.356L...1S} and create more massive models where helium ignites in only a slightly degenerate core, we then decrease the mass of the star and allow the helium core to grow out until our model matches the parameters of the last failed helium-flash model. We  combine these tracks to make the evolution track complete the possible evolutionary paths of these intermediate mass stars. This is a reasonable approximation as recent more detailed calculations indicate that the helium flash does not explode the star nor cause significant structural changes \citep{2008A&A...490..265M}.

\begin{figure}
\begin{center}
\includegraphics[width=\columnwidth]{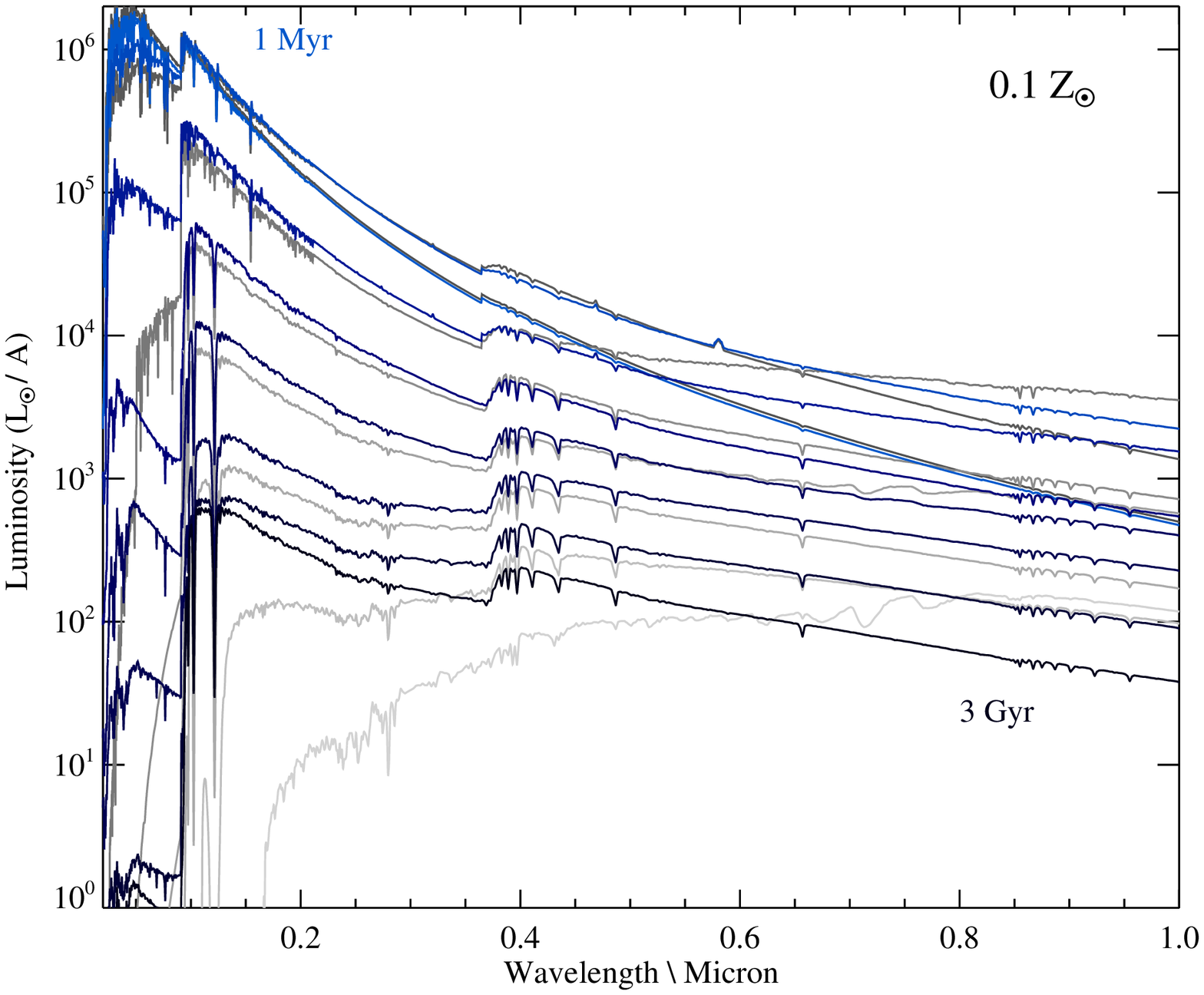}
\includegraphics[width=\columnwidth]{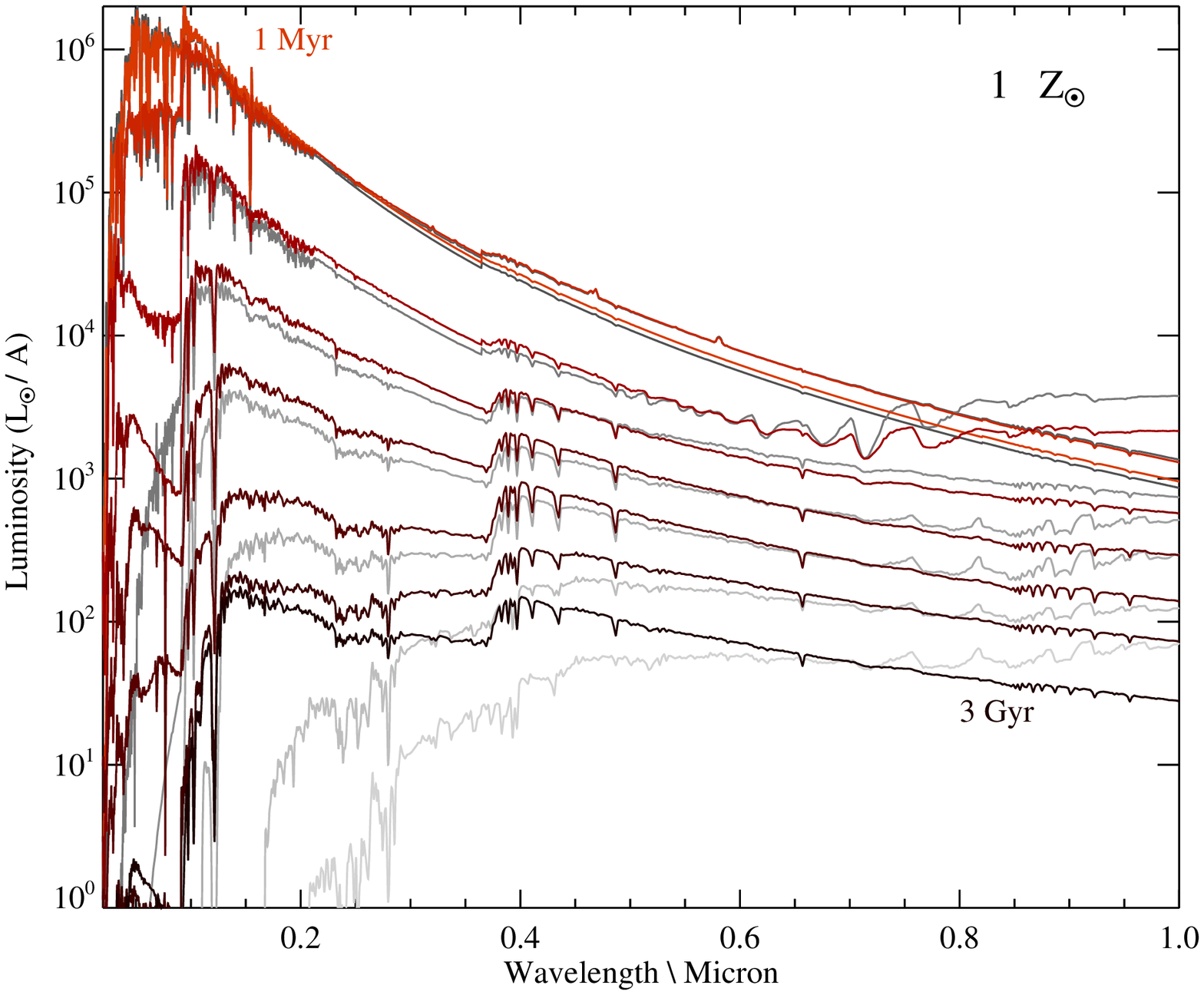}
\includegraphics[width=\columnwidth]{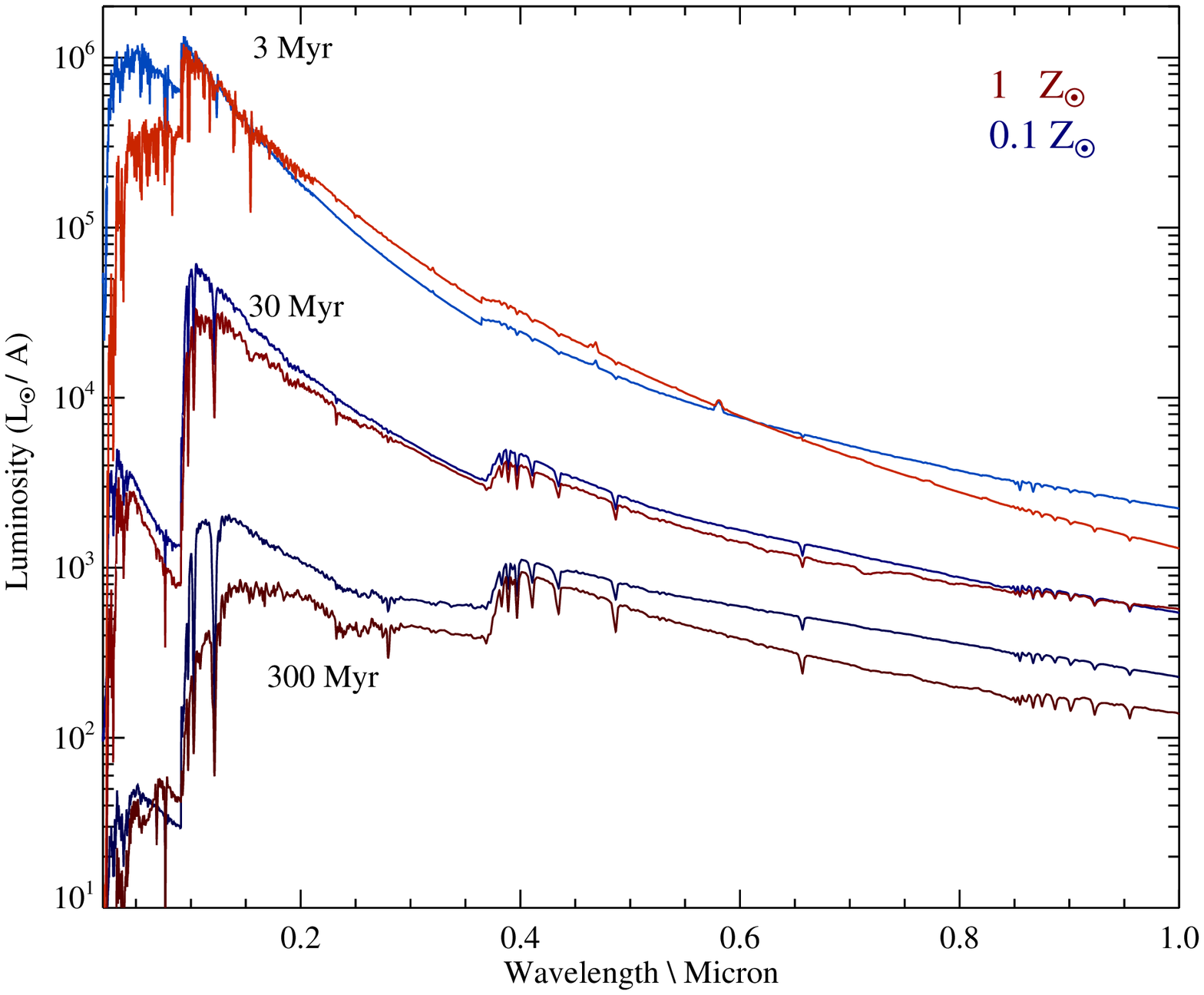}
\end{center}
\caption{The synthetic spectra produced for a co-eval population (i.e. instantaneous starburst) at times of 1, 3, 10, 30, 100, 300, 1000 and 3000\,Myr after star formation. Spectra are shown for binary populations (bold, coloured lines) and single stars (pale, greyscale line), and at metallicities of Z=0.002 (top) and Z=0.020 (centre). In the bottom panel we compare \bpass\ binary models at the two metallicities directly, at ages of 3, 30 and 300\,Myr.}
\label{fig:seds}
\end{figure}

\begin{figure*}
\begin{center}
\includegraphics[width=\columnwidth]{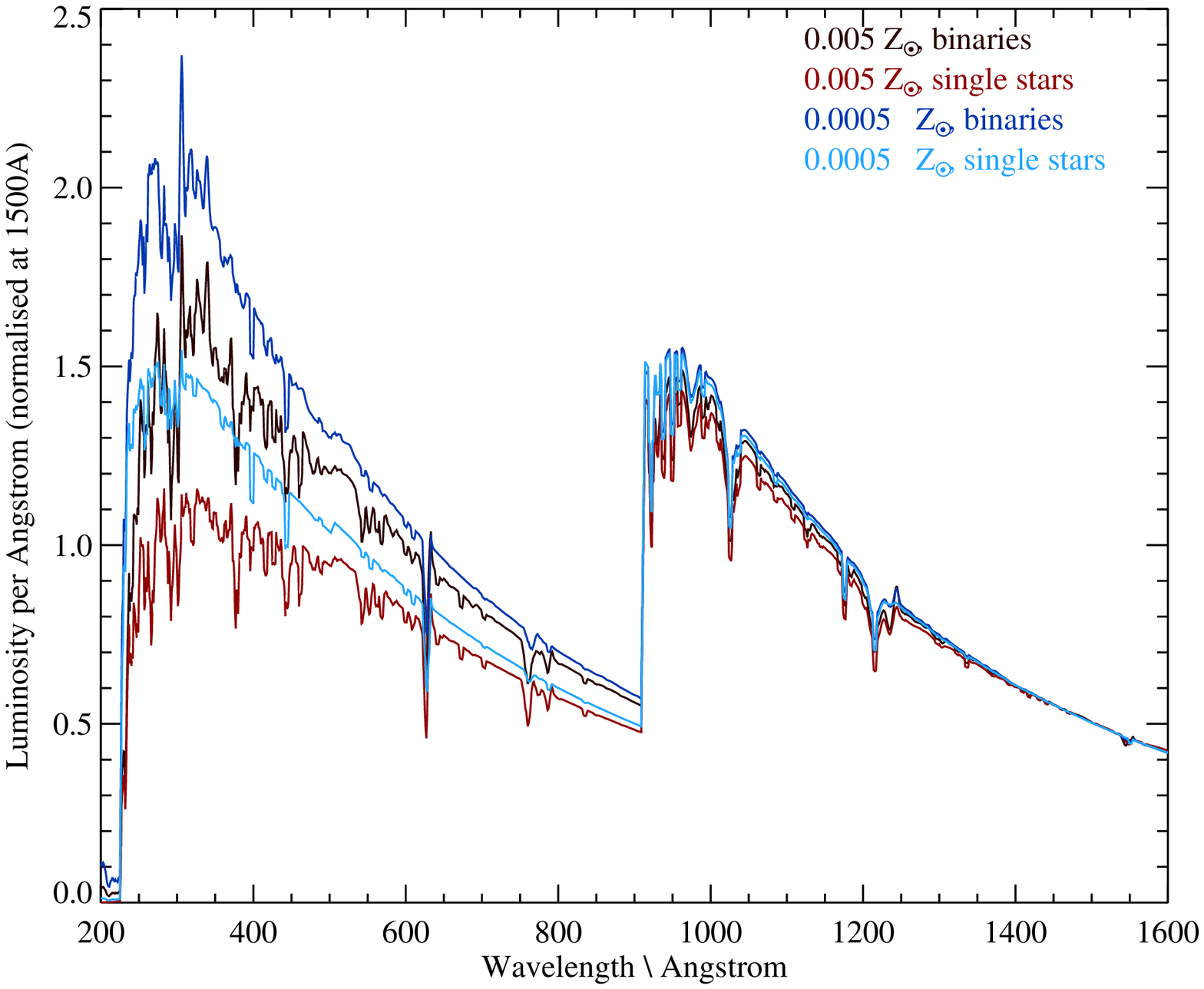}
\includegraphics[width=\columnwidth]{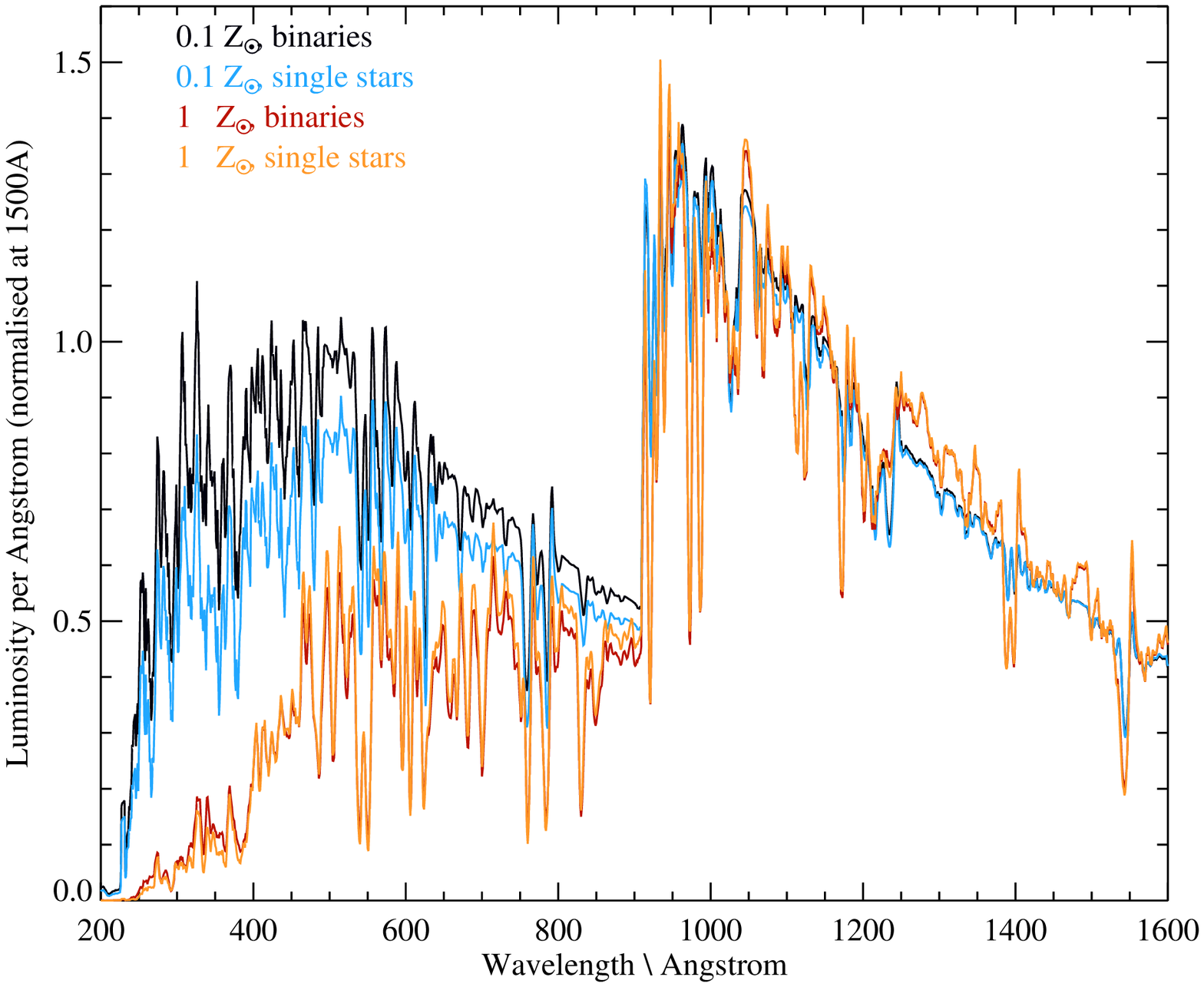}
\caption{The extreme ultraviolet (200-1600\AA) spectral region in synthetic spectra produced for a co-eval population (i.e. instantaneous starburst) at a time 30\,Myr after star formation. The two panels show the spectra as a function of metallicity (0.05, 0.5\% $Z_\odot$ in the left-hand panel, 10 and 100\% $Z_\odot$ in the right-hand panel) and \bpass\ single star vs binary evolution for each metallicity. The effect of binary evolution is to increase the hardness of the spectrum, and hence the ionizing photon output (total flux emerging short of 912\AA), particularly at very low metallicities. Synthetic spectra have been scaled to a common luminosity at 1500\AA.}
\label{fig:euv}
\end{center}
\end{figure*}

\begin{figure*}
\begin{center}
\includegraphics[width=\columnwidth]{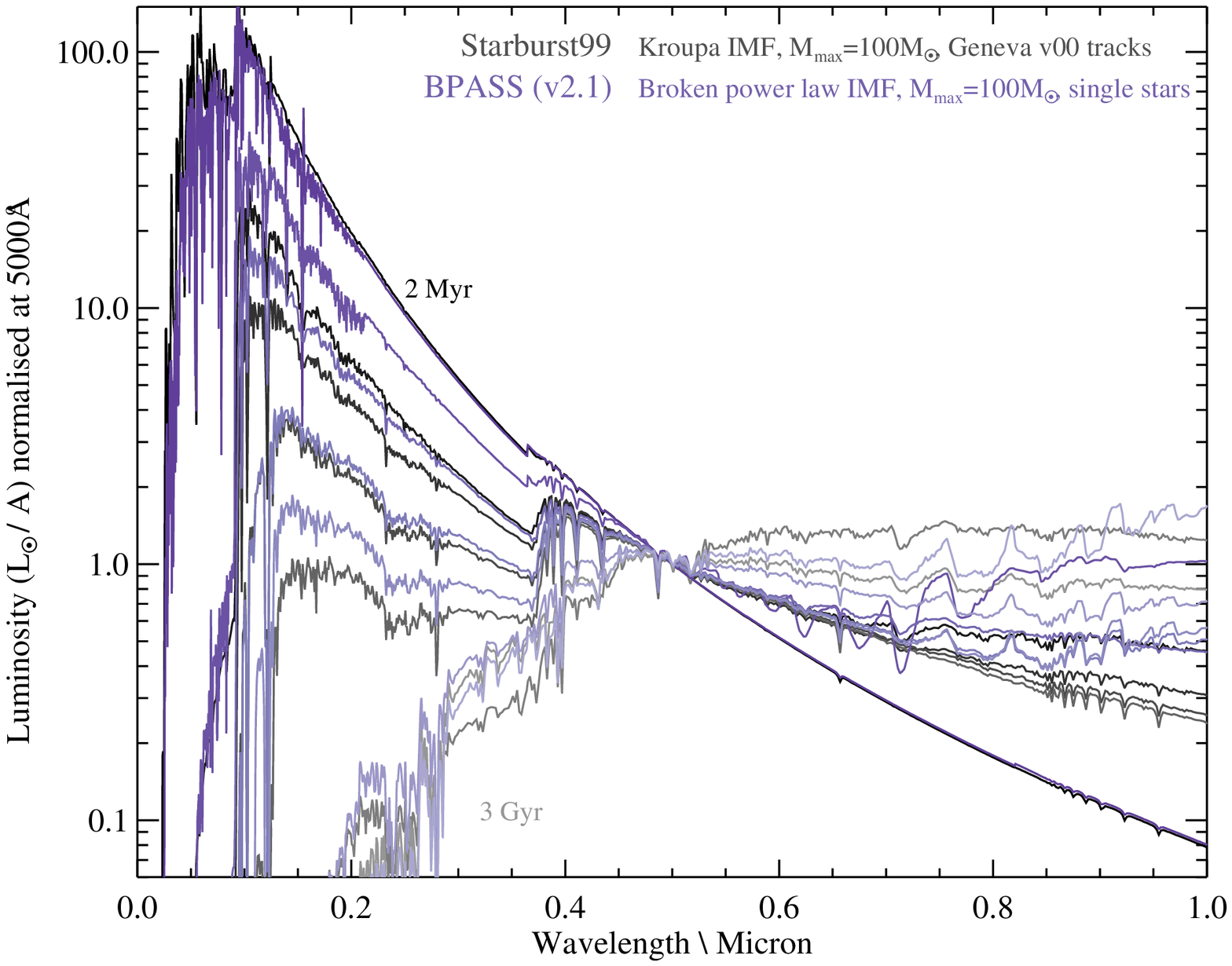}
\includegraphics[width=\columnwidth]{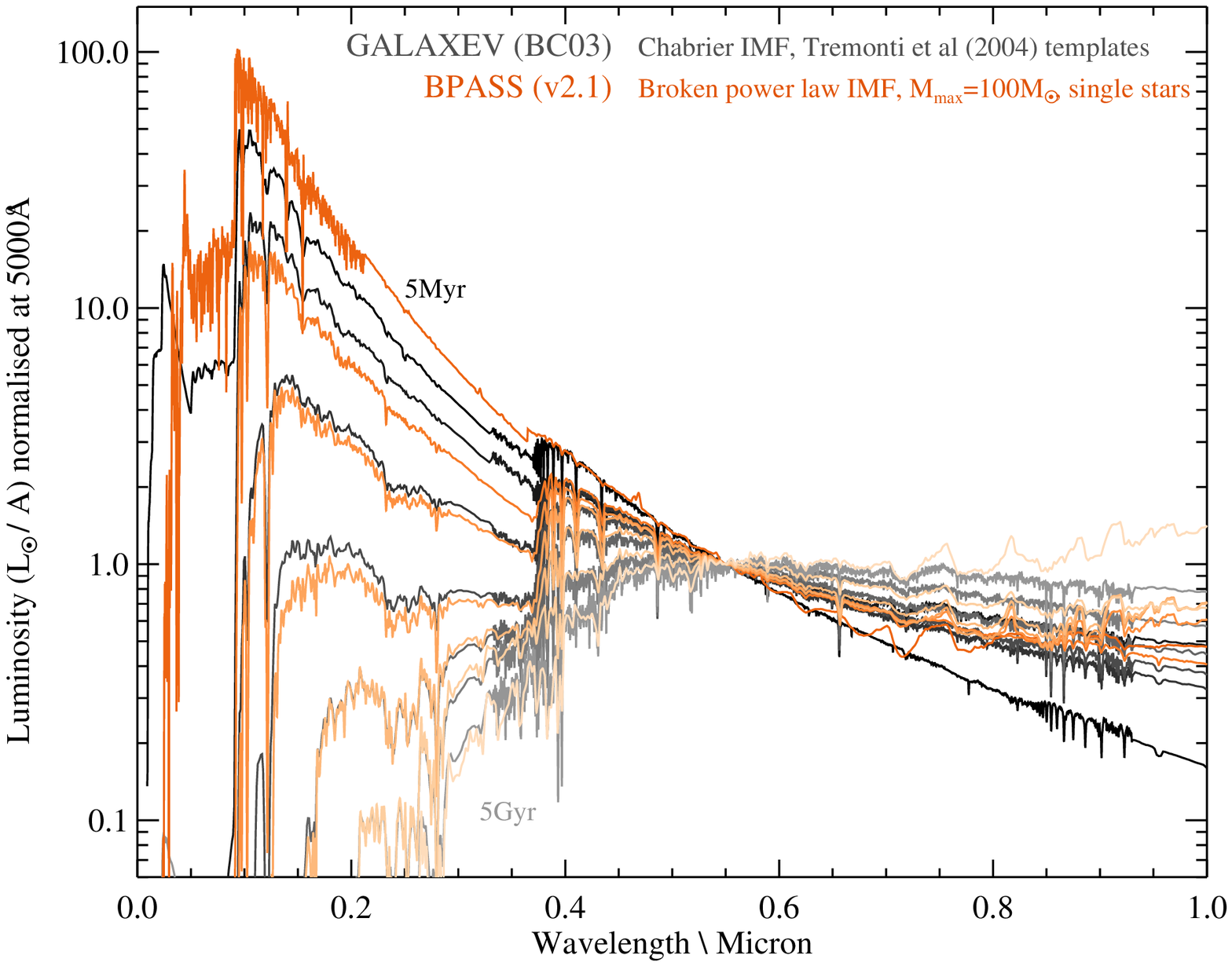}
\includegraphics[width=\columnwidth]{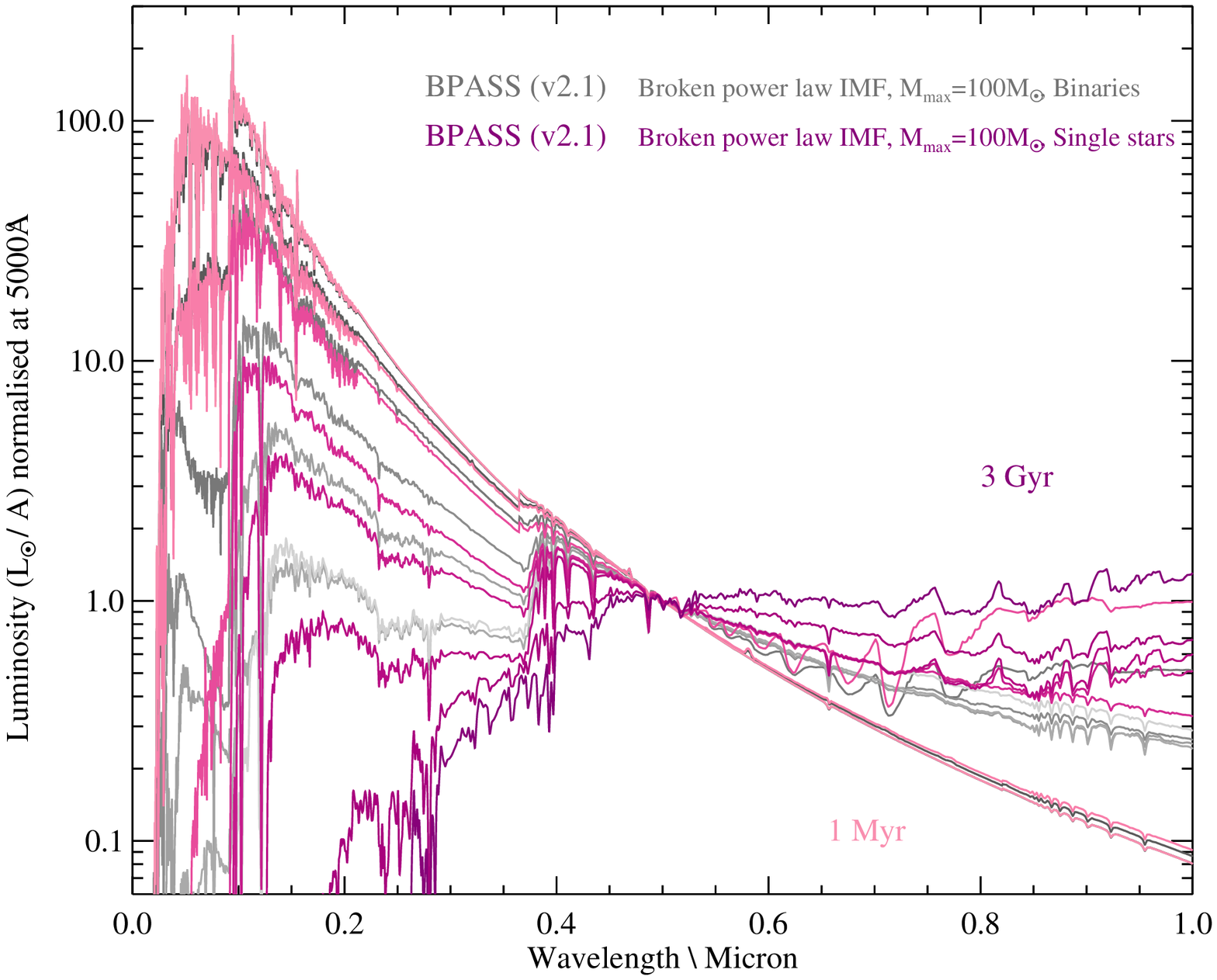}
\includegraphics[width=\columnwidth]{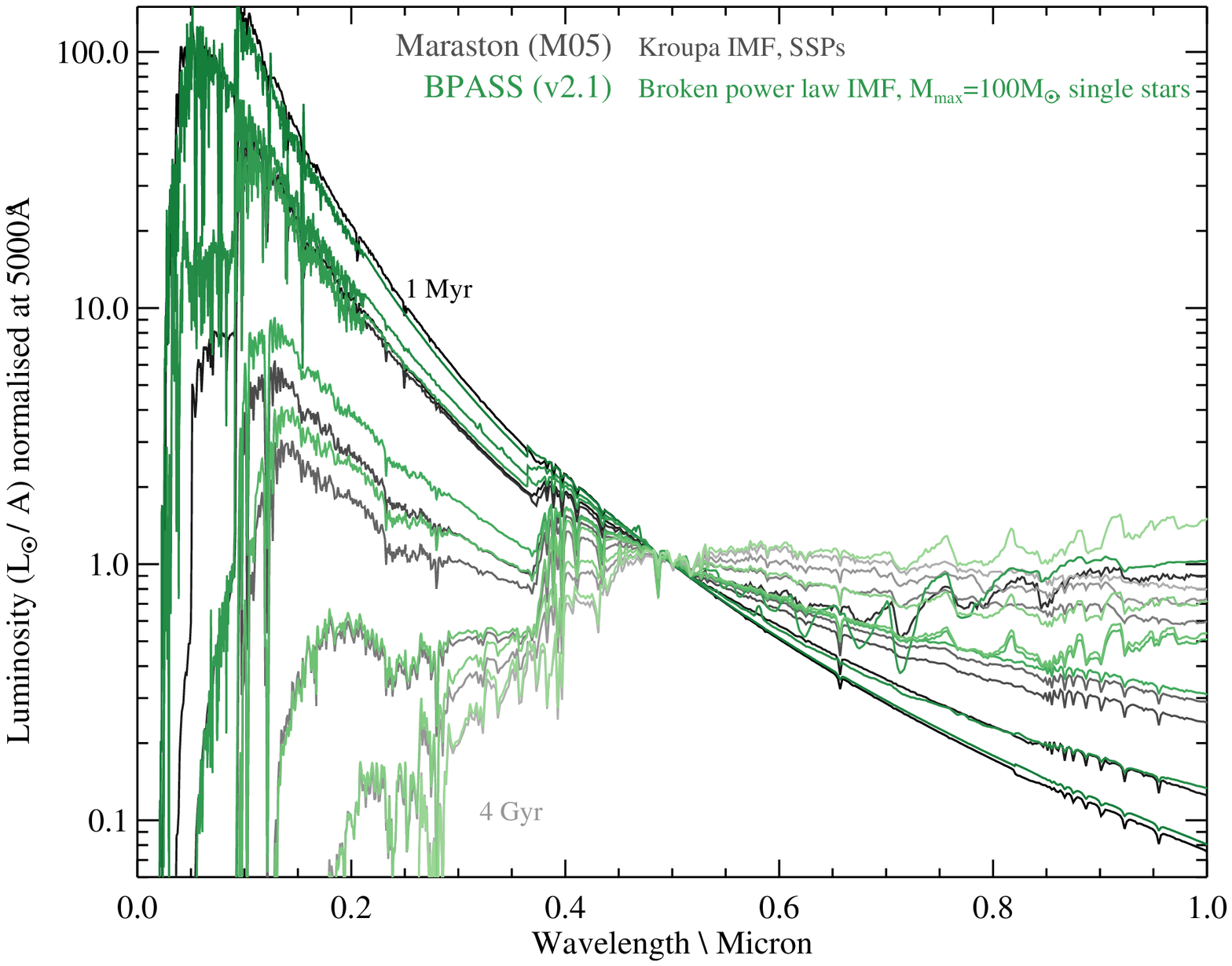}
\caption{A comparison between the simple stellar population models (i.e. instantaneous starbursts) produced by different stellar population synthesis codes at Solar metallicity. In three panels, \bpass\ single star models are compared to the closest available match in metallicity, age and initial mass function from publicly available synthesis codes. These are: (top right) the GALAXEV models of Bruzual \& Charlot (2003), specifically the galaxy templates used to classify galaxies in the Sloan Digital Sky Survey by Tremonti et al (2004), which use a Chabrier IMF;  (bottom right) the Maraston (2005) models, shown here using the Kroupa IMF; (top left) Starburst99 models (Leitherer et al), generated with the non-rotating Geneva model set and an IMF matching our power law slopes, and (bottom left) BPASS binary star models. Synthetic spectra have been scaled to a common luminosity at 5000\AA.}
\label{fig:sed_comparison}
\end{center}
\end{figure*}

\subsubsection{BPASS binary star models} 

Our binary models are identical to our single star models in nearly every respect but we additionally allow for extra mass-loss or gain via binary interactions. We assume the orbits of the binary are circular and thus described by Kepler's 3rd Law such that $(M_1+M_2)/$M$_{\odot} \propto (a/215R_{\odot})^3/(P/{\rm yr})^2$, where $a$ is the orbital separation and $P$ is the orbital period. We assume mass lost in stellar winds removes orbital angular momentum in a spherically symmetric shell around the star losing the mass. Thus the orbits widen over the evolution of the star. We only follow one star with detailed calculations during the evolution. This avoids wasting computational effort on calculating the evolution of a 1\,M$_{\odot}$ secondary star at the same time as a more rapidly-evolving 10\,M$_{\odot}$ primary. During the evolution of the primary star models we use the single star rapid evolution equations of \citet{2002MNRAS.329..897H} to approximate the secondary's evolution.  When the end state of the primary's evolution has been established, we then substitute the secondary's evolution with a detailed model, either using a single star model or calculating a new detailed binary model with a compact remnant, depending on whether the binary is bound or unbound. 

The grid of masses we use for primary evolution also ranges from 0.1 to 300\,M$_{\odot}$. We model every decimal mass from 0.1 to 2.1\,M$_{\odot}$, then 2.3, 2.5, 2.7, 3, 3.2, 3.5, 3.7\,M$_{\odot}$, every half M$_{\odot}$ from 4 to 10\,M$_{\odot}$, then every integer mass to 25\,M$_{\odot}$, then every 5\,M$_{\odot}$ to 40\,M$_{\odot}$ then 50, 60, 70, 80, 100, 120, 150, 200 and 300\,M$_{\odot}$. The grid of mass ratios, $q=M_2/M_1$, ranges uniformly from 0.1 to 0.9 in steps of 0.1. We establish initial binary periods with $\log(P/{\rm days})$ from 0 to 4 in steps of 0.2.

For the secondary models which have a compact companion, the grid of models we calculate is determined from the results of the population synthesis after the first supernova. The grid uses the same array of possible periods as for the primaries, however the secondary masses are  reduced to $M_2=0.1$, 0.2, 0.3, 0.4, 0.5, 0.6, 0.8\,M$_{\odot}$, every integer mass from 1 to 25\,M$_{\odot}$, 25, 30, 35, 40, 50, 60, 70, 80, 100, 120, 150, 200, 300, 400, 500\,M$_{\odot}$. The compact remnant masses are arranged in a grid of $\log(M_{\rm rem,1}/$M$_{\odot})$ from -1 to 2 in steps of 0.1.

Our numerical method to account for binary evolution is inspired by the method of \citet{2002MNRAS.329..897H} and was first described in detail in \citet{2008MNRAS.384.1109E}. However we have had to modify some of the aspects as they are difficult to implement into a detailed stellar evolution code. The key point in our models when binary evolution becomes very different to that of our single star models is when the radius of the primary star increases and so it fills its Roche Lobe. This is the equipotential surface beyond which any material is no longer most strongly gravitationally attracted to the primary star, and Roche lobe overflow (RLOF) can occur. We allow for this in our models by using the effective Roche lobe radius defined by \citet{1983ApJ...268..368E}, where a star will have an equivalent volume to that of its Roche lobe. At radii beyond this it will begin to overflow mass towards its companion star. The Roche lobe radius is given by,
\begin{equation}
\frac{R_{\rm L1}}{a} = \frac{0.49 q_1^{2/3}}{0.6 q_1^{2/3} + \ln (1+ q_1^{1/3})},
\end{equation}
where $q_1 = M_1/M_2$. When the star fills its Roche lobe we determine a mass-loss rate due to overflow of,
\begin{equation}\label{eqn:overflowmassloss}
\dot{M}_{R1}=F(M_1) \left[\ln \left(\frac{R_1}{R_{L1}}\right) \right]^3
\end{equation}
where
\begin{equation}
F(M_1)=3\times 10^{-6} [\min (M_1, 5.0)]^2.
\end{equation}
The value here was chosen by \citet{2002MNRAS.329..897H} to ensure mass transfer is stable within their code. We also found this to be the case in applying their rate. There has been significant amount of work on the exact details of RLOF \citep[e.g.][]{1988A&A...202...93R,1990A&A...236..385K,2007ApJ...660.1624S} and our model is rather simple. However upon the onset of RLOF the stellar model will continue to expand until either CEE occurs or the mass-transfer becomes stable. In future we plan to calibrate the RLOF against observed systems undergoing mass transfer.

This mass is assumed to be transferred to the secondary star or lost from the system. We limit the accretion rate onto the secondary star by assuming it can only accept mass at a rate determined by its thermal timescale, such that $\dot{M_2} \le M_2/\tau_{\rm KH}$. The rest of the material is assumed to be lost from the system, taking with it angular momentum from its orbit. In the case where the companion is a compact remnant with an initial mass less than $3M_{\odot}$ the accretion rate is limited instead by the Eddington luminosity. Above this mass we assume that super-Eddington accretion can occur due to the compact remnant being a black hole, where super-Eddington luminosities have been observed. In the Eddington-limited case any excess material and its angular momentum is lost from the system. These limits are also approximate and a more detailed accretion model will be required in future.

We do follow the rotation rate of the stars in our code but this has no direct physical consequence in the code beyond QHE due to spin-up in mass-transfer in binaries (See Section 2.2.2). We follow the rotation rate to keep track of the angular momentum held by the stars, since as stars approach CEE the Darwin mechanism can cause CEE to occur \citep{1880RSPS...31..322D}. While we attempted to implement tidal models such as those in \citet{2002MNRAS.329..897H} based on \citet{1981A&A....99..126H} we found that the most stable method was to only assume tidal synchronization once a star fills its Roche lobe. Assuming that the stars synchronize their rotation with the orbit, transferring angular momentum from the stars to the orbit or vice-versa. In future we will add in a more general model of tidal forces and also allow the rotation to induce extra mixing in the stellar interior.

If RLOF does not stop the increase of the donor star's radius, or if too much angular momentum is required to spin-up the donor star, then CEE occurs. In our code CEE is taken to occur in our models when the radius of the primary star is the same as the binary separation. At this point we switch the orbital evolution to a formalism based on the typical population synthesis model where the result of CEE is determined from comparing the binding energy of the envelope with the orbital energy \citep{2013A&ARv..21...59I}. We continue to determine the mass-loss rate from Equation \ref{eqn:overflowmassloss} but set an upper mass-loss rate limit of 0.1\,M$_{\odot} {\, \rm yr^{-1}}$ to avoid numerical problems of having such a high mass-loss rate. This unfortunately means we overestimate the CEE timescale in our models. To compare to the typical CEE mechanism it is assumed that the change in the orbital energy of the binary is equal to the binding energy of the envelope of the star that is lost in the interaction,
\begin{equation}
E_{\rm bind}=\Delta E_{\rm orb}.
\end{equation}
This can be expressed as,
\begin{equation}
-\frac{Gm_1 m_{\rm 1, envelope}}{\lambda R_1} = \alpha_{\rm CE} \left(-\frac{G m_1 m_2}{2a_i} +\frac{G m_{\rm 1, core} m_{2}}{2 a_{f}} \right)
\end{equation}
Where $m_1$ and $R_1$ are the mass and radius of the primary, $a_i$ and $a_f$ are the initial and final orbital separation, $\alpha_{\rm CE}$ is a const representing the efficiency of the conversion from binding energy to orbital energy and $\lambda$ is a constant representing how the structure of the star affects its binding energy \citep{1990ApJ...358..189D,2000A&A...360.1043D}.

In a detailed stellar model it is difficult to remove the entire envelope in a single timestep. Therefore we instead allow the mass-loss rate to increase as high as possible and relate the binding energy of the material lost to the change in the orbital energy. We do this by equating the binding energy of the lost material to the change in the orbital energy such that,
\begin{equation}
\frac{-G(M_1+M_2) \delta M}{R_1}  = \frac{G M_1 M_2}{a^2} \delta a
\end{equation}
which can be rearranged to give,
\begin{equation}
\delta a = \frac{a^2}{R_1}\frac{M_1+M_2}{M_2}\frac{\delta M_1}{M_1},
\end{equation}
where $\delta a$ is the change in orbital separation and $\delta M_1$ is the mass loss of the surface. The advantage of this method is that the structure of the star is naturally taken account of as we integrate over the removal of the envelope. The disadvantage is that because our time for CEE is a few orders of magnitude too large other sources of energy may limit how efficient CEE is. However the efficiency of CEE is uncertain.

Our CEE mechanism is meant to reproduce the envelope ejection and the in-spiral simultaneously. Eventually we find that CEE either results in removal of the stellar envelope or a merger. If the primary star eventually shrinks within the Roche lobe the CEE is taken to have ended. Mergers are assumed to occur when the companion star begins to fill its own Roche lobe. At this point the total mass of the secondary is added to the primary star. We typically find the stars that experience CEE on, or just after, the main-sequence tend to merge, while post-main sequence stars tend to only remove the hydrogen envelope. Also even in systems that do merge, mass loss prior to the event of the merger means that the final stellar mass is less than the total pre-CEE mass of the binary.

In Figure \ref{Fig0a} we show for our fiducial population (as described in Table \ref{tab:inputs}) the fraction of stars that interact, splitting this into those that experience RLOF or RLOF and CEE. We see that at masses of about $5<M<40$\,M$_{\odot}$, 80\% of binaries interact with around 90\% of those interactions leading to CEE. We note that for smaller mass ratios 100\% of interactions lead to CEE, while for our largest mass ratios only 80\% lead to CEE. Outside this mass range RLOF dominates the binary interactions. At high masses the stellar wind mass loss is already significant and so once RLOF starts only a little extra mass loss is required to avoid CEE, especially at higher metallicity where stellar winds reduce the number of systems that interact to 30\%. Below 5M$_{\odot}$ down to approximately 1M$_{\odot}$ the fraction of interacting systems decreases. This is due to the maximum radius of the stars decreasing so fewer stars fill their Roche lobe and interact. At the lowest masses, below 0.7M$_{\odot}$, the interaction fraction decreases to zero due to the stars being smaller than the orbital separation for their evolution witin the age of the Universe. This is likely due to our minimum period for our binary models of 1 day, as well simple binary model that does not include processes that would allow for periods to shorten and drive interactions. As a result, with the exception of those in the closest binaries, such sources do not overfill their Roche lobes and so never interact. Stars below $\sim$1\,M$_{\odot}$ never interact in our default model set as they remain compact during their main sequence lifetime, which is long compared to the age of the Universe.

We show the mean duration over which RLOF and CEE episodes occur in our binary models in Figure \ref{Fig0b}. More massive stars interact on shorter timescales than low mass stars, with RLOF ranging from $10^5$ up to $10^7$~years while CEE has much shorter timescales of $100$ to $10^4$ years. These times for our CEE events are much longer compared to the very short timescale of days or years that are predicted by dynamical simulations \citep[see][]{2013A&ARv..21...59I} as well as those found from observed events. However they are significantly shorter than the thermal and nuclear timescales of the stars so they should have little impact on the eventual predictions of binary evolution.

\subsubsection{Supernovae and Remnants}\label{sec:snrs}

At the end of the evolution of our models we check to see what remnant may be produced. Currently we assume that either a white dwarf, neutron star, black hole or no remnant will be formed. We assume that a core-collapse supernova will occur if central carbon burning has taken place and the CO core mass is greater than 1.38\,M$_{\odot}$ and the total stellar mass is greater than 1.5\,M$_{\odot}$. If this condition is not met then the remnant left will be a white dwarf with the mass of the helium core at the end of the stellar model; the secondary star will continue to evolve with a white dwarf in orbit around it. 

In the case that a supernova occurs  the remnant mass is determined by calculating how much material can be ejected from the star given an energy input of $10^{51}$~ergs, a typical supernova energy. This method was first outlined in \citet{2004MNRAS.353...87E} which provides more details. The mass that is not unbound goes into the remnant, thus providing a way to estimate the mass of any resultant black hole. This method is similar to others \citep[e.g.][]{2003ApJ...591..288H,2012ApJ...757...69U,2015arXiv151004643S,2015MNRAS.451.4086S}. For simplicity neutron star remnants are assumed to have a constant mass of 1.4\,M$_{\odot}$, while remnants more massive than 3\,M$_{\odot}$ are considered to form black holes. Finally if the helium core mass at the end of evolution is between 64 and 133\,M\,$_{\odot}$ a pair-instability supernova is assumed to occur which completely disrupts the star and leaves no remnant \citep{2002ApJ...567..532H}.

For any system with a core-collapse supernova we estimate the supernova category as type IIP, II-other, Ib and Ic. We identify the supernova types as described in \citet{2011MNRAS.414.3501E} and \citet{2013MNRAS.436..774E}. If the mass of hydrogen in the star is greater than $10^{-3}$\,M$_{\odot}$ then we have a type II supernova. If the total mass of hydrogen is greater than 1.5\,M$_{\odot}$ and the ratio of hydrogen mass to helium mass in the progenitor is greater than 1.05 we have a type IIP. When the hydrogen mass is less than $10^{-3}$\,M$_{\odot}$ a type Ib/c supernova occurs. If more than 10\% of the ejecta mass is helium then the supernova is assumed to be a type Ib, otherwise it is Ic. This leads to roughly twice as many Ib as Ic supernovae which is in agreement with the most recent attempt to accurately determine this ratio \citep{2017PASP..129e4201S}.

We also determine whether an event is likely to generate a Gamma-ray Burst (GRB) or Pair-Instability Supernova (PISN). For a long-GRB the star must have experienced QHE and have a remnant mass greater than 3\,M$_{\odot}$. Note that this is only one possible pathway for GRBs and limits their presence in our rate estimates to lower metallicities. Some type Ic supernovae arising from non-QHE systems also likely generate GRBs but we do not account for these. For PISN we apply mass limits from \citet{2002ApJ...567..532H} as detailed above. Finally any progenitor with a final mass between 1.5 and 2\,M$_{\odot}$ is identified as a low-mass progenitor with an uncertain outcome. These events may be faint and rapidly evolving, as described by \citet{2016MNRAS.461.2155M}.

We also estimate the type Ia thermonuclear supernova rate by two channels, first selecting out white dwarfs that begin with a mass below the Chandrasekhar limit but then reach or exceed this mass due to binary mass transfer. Second, we count double white dwarf binaries that merge due to gravitational radiation with the merger time calculated assuming circular orbits and the analytic expressions of \citet{1964PhRv..136.1224P}. Our resultant rates are highly approximate and we do not consider them rigorous or precise but include them for completeness. We caution that these should be used with care, and in future we may refine our type Ia models further for comparison with other predictions and observational data.

\begin{figure}
\begin{center}
\includegraphics[width=\columnwidth]{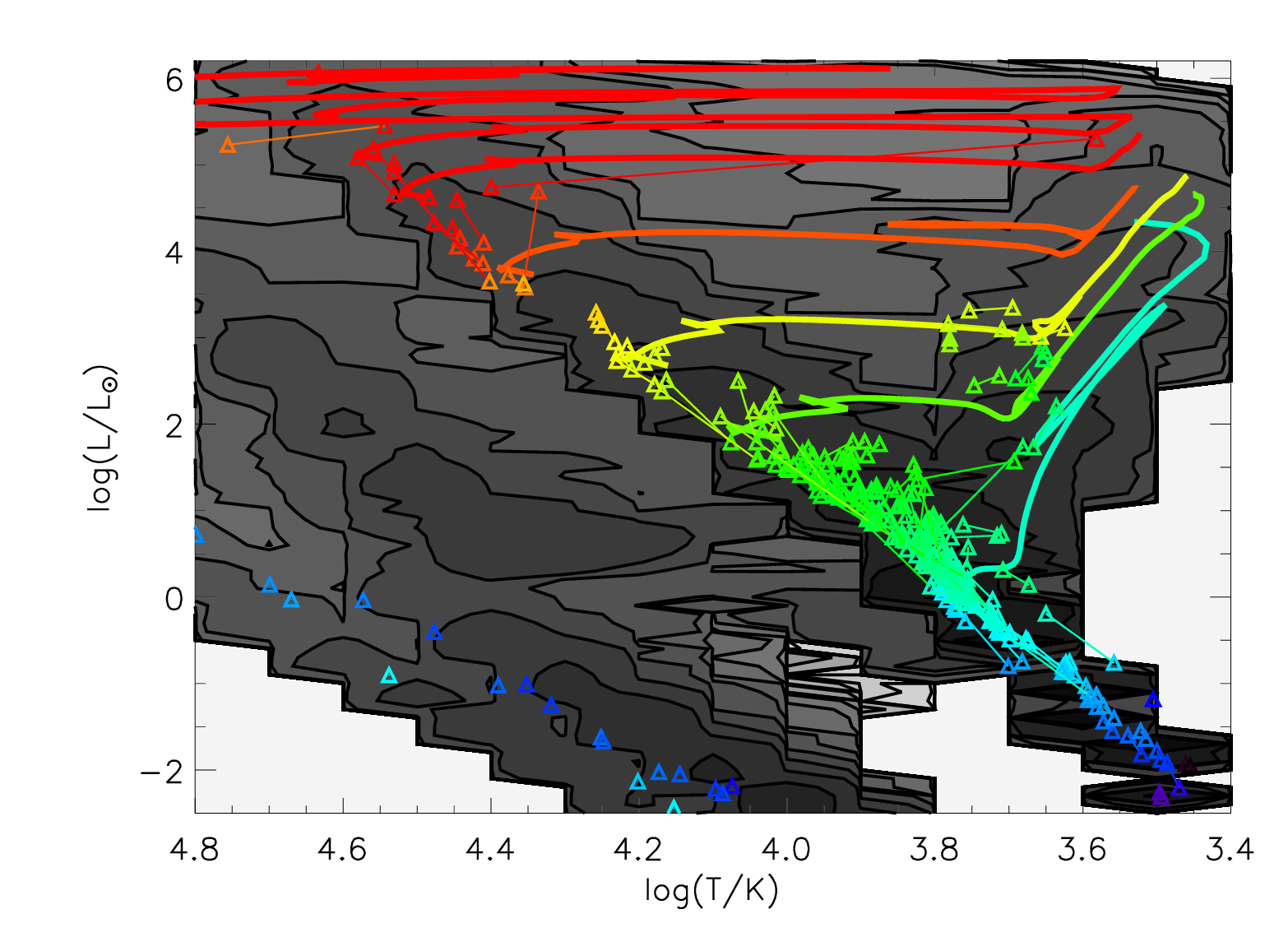}
\includegraphics[width=\columnwidth]{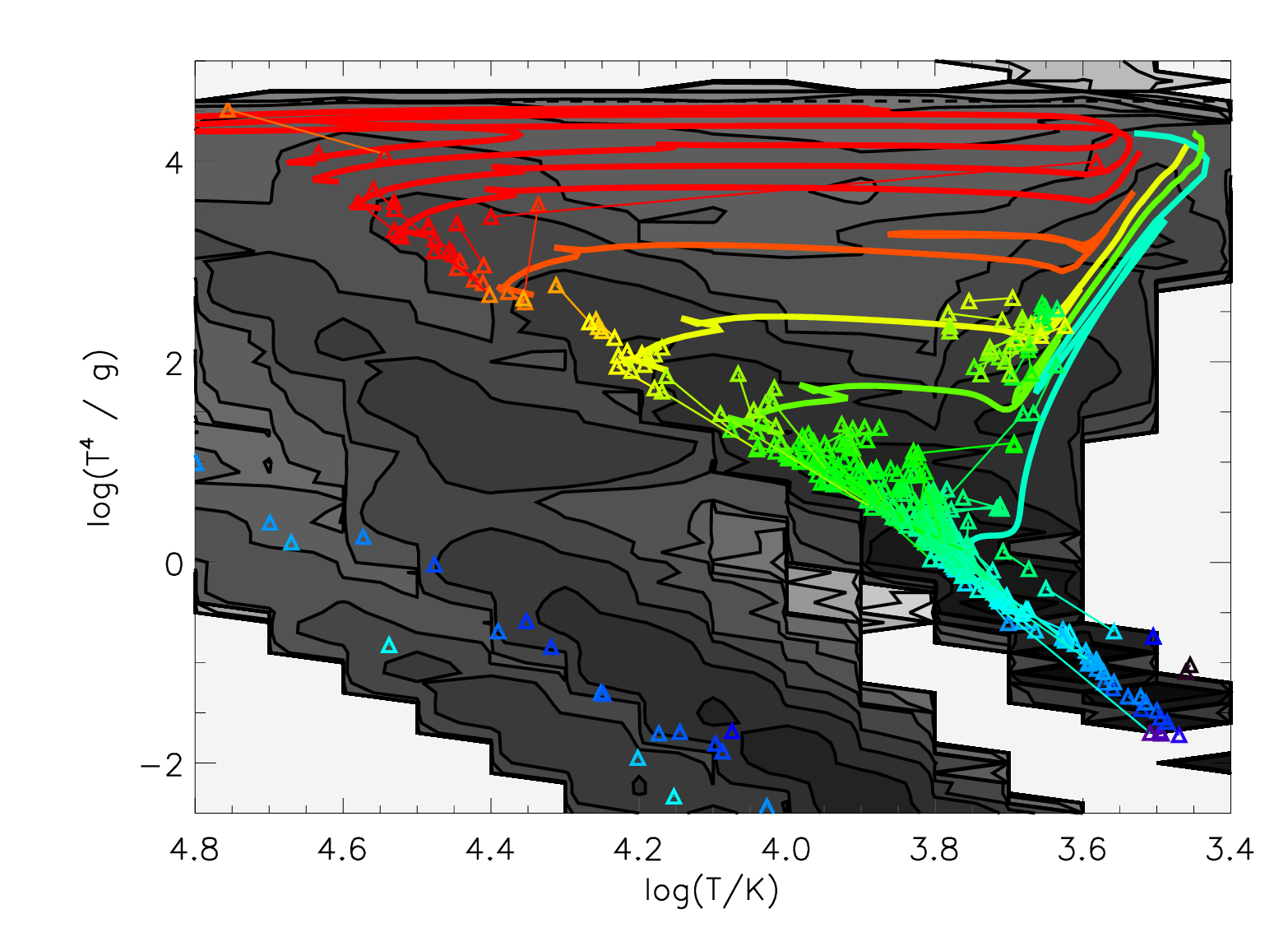}
\includegraphics[width=\columnwidth]{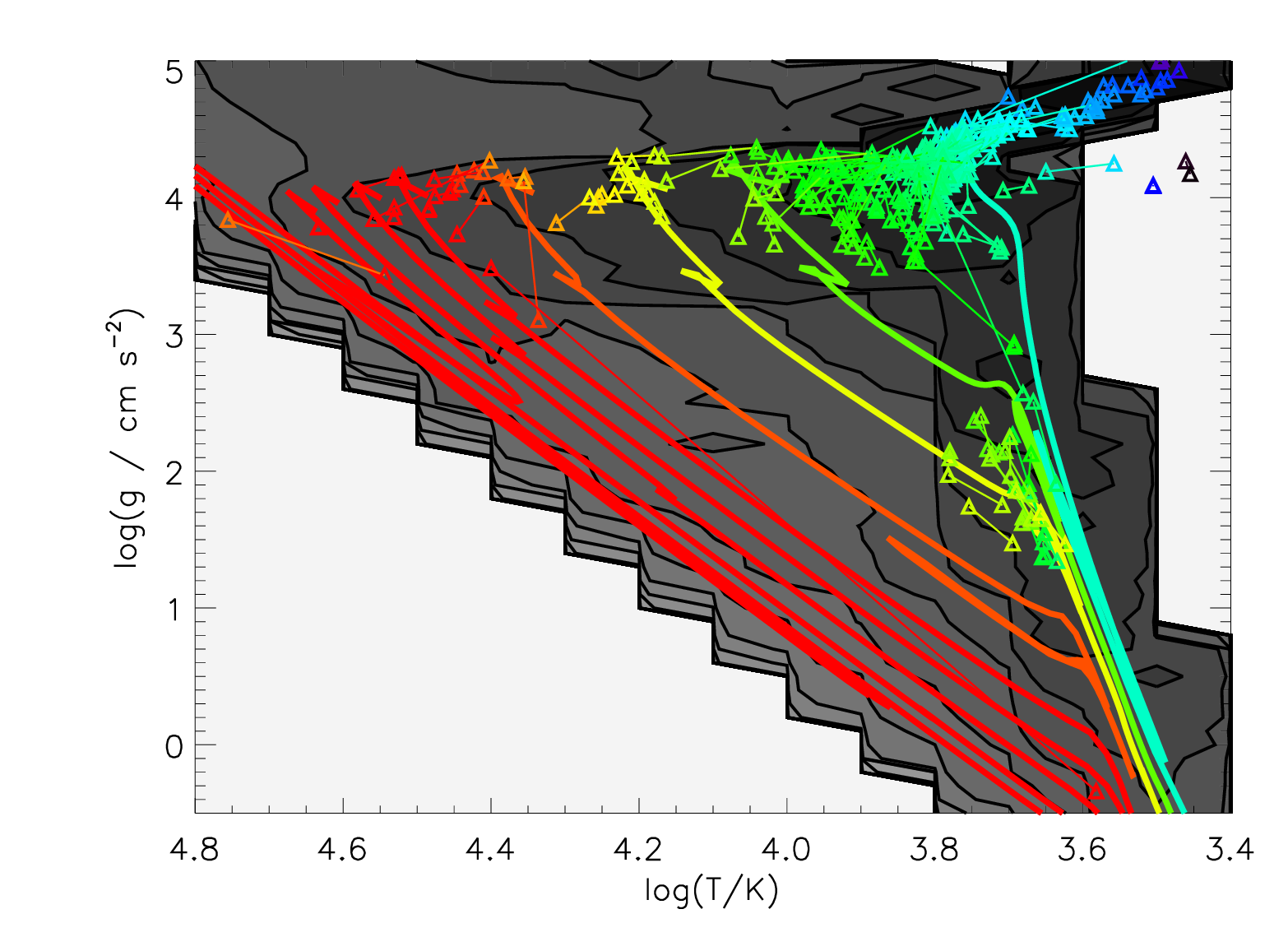}
\caption{Eclipsing binaries overplotted on evolution tracks for our binary models (greyscale, indicating the range of outcomes for a given initial mass). The observed population of eclipsing binaries are overplotted as small symbols, with lines linking binary companions. We show representative tracks for the photometric Herzsprung-Russell diagram and its spectroscopic counterpart, together with the log g-log T plane. The representative tracks have masses of 1, 3, 5, 10, 20, 30, 50 and 80\,M$_{\odot}$ and both they and the observed data are colour coded by mass. The observational data is mainly taken from \citet{2015ASPC..496..164S} and supplemented by VV Cephei, WR20a and $\gamma$-Velorum, as well as the white dwarf sample of \citet{2017arXiv170605016P} as detailed in the text.}\label{Fig1}

\end{center}
\end{figure}

\subsection{The Population Synthesis}\label{sec:method_popsynth}

\begin{table*}
\caption{BPASS v2.1 input parameter ranges for binary populations \label{tab:inputs}}
\begin{tabular}{lp{1.4\columnwidth}}
Parameter  &  Permitted Values\\
\hline\hline 
Primary Masses & 68 values from M$_1=$0.1, 0.2, 0.3, 0.4, 0.5, 0.6, 0.7, 0.8, 0.9, 1, 1.1, 1.2, 1.3, 1.4, 1.5, 1.6, 1.7, 1.8, 1.9, 2, 2.1, 2.3, 2.5, 2.7, 3, 3.2, 3.5, 3.7, 4, 4.5, 5, 5.5, 6, 6.5, 7, 7.5, 8, 8.5, 9, 9.5 10, 11, 12, 13, 14, 15, 16, 17, 18, 19, 20, 21, 22, 23, 24, 25, 30, 35, 40, 50, 60, 70, 80, 100, 120, 150, 200 and 300\,M$_{\odot}.$\\
Primary Mass Ratios & 9 values with M$_2$/M$_1$ = 0.1, 0.2, 0.3, 0.4, 0.5, 0.6, 0.7, 0.8, 0.9\\
Secondary Masses & 46 values from M$_2=0.1$, 0.2, 0.3, 0.4, 0.5, 0.6, 0.8, 1, 2, 3, 4, 5, 6, 7, 8, 9, 10, 11, 12, 13, 14, 15, 16, 17, 18, 19, 20, 21, 22, 23, 24, 25, 30, 35, 40, 50, 60, 70, 80, 100, 120, 150, 200, 300, 400 and 500\,M$_{\odot}$.\\
Compact remnant masses & 31 values from $\log(M_{\rm rem,1}/$M$_{\odot})$ = -1, -0.9, -0.8, -0.7, -0.6, -0.5, -0.4, -0.3, -0.2, -0.1, 0, 0.1, 0.2, 0.3, 0.4 ,0.5, 0.6, 0.7, 0.8, 0.9, 1, 1.1, 1.2, 1.3, 1.4, 1.5, 1.6, 1.7, 1.8, 1.9 and 2.\\
Period Distribution & 21 initial periods from $\log(P/{\rm days}) =$ 0, 0.2, 0.4, 0.6, 0.8, 1, 1.2, 1.4, 1.6, 1.8, 2, 2.2, 2.4, 2.6, 2.8, 3, 3.2, 3.4, 3.6, 3.8, 4.0\\
Stellar Ages & Output ages from $\log({\rm Age / yrs}) = 6.0$ to 11.0 (1\,Myr to 100\,Gyr)\\
Metallicity Mass Fractions & $Z = 10^{-5}$, $10^{-4}$, 0.001, 0.002, 0.003, 0.004, 0.005, 0.006, 0.008, 
0.010, 0.014, 0.020\ (Z$_\odot$), 0.030, 0.040\\
Initial Mass Functions & 7 different mass functions - fiducial version: has an IMF slope of -1.30 from 0.1 to 0.5M$_{\odot}$, and a slope of -2.35 from 0.5 to 300M$_{\odot}$. We have variations with other upper mass slopes of -2.00 and -2.70 and also versions with the upper mass of 100M$_{\odot}$. Our final alternative has a slope of -2.35 from 0.1 to 100M$_{\rm odot}$\\
Stellar Atmospheres & BASELv3.1 ($\log g= -1$ to 5.5, $\log(T/K)=2000$ to 50000 and $\log($Z/Z$_{\odot})=-3$ to 0.5. For metallicities below this minium we use v2.2 that extends to $\log($Z/Z$_{\odot})=-3.5$.)\\
					& PoWR (We use the SMC and LMC, WN grids with $X_{\rm surface}=0.4$, 0.2 and 0. For the Galaxy we use the WN grids with $X_{\rm surface}=0.5$, 0.2 and 0. At all metallicities we use the Galactic WC grid.)\\
                    & WMBASIC (new model grid for O stars, see section \ref{sec:method_atmos_wmbasic} and Table \ref {tab:wmbasic})
\end{tabular}
\end{table*}

\subsubsection{Binary distribution parameters}

In the population synthesis the individual stellar models are combined together to make a synthetic stellar population which can be observed and compared to observations of the real Universe. Together with the models described above we need to know the distribution of initial parameters required to create the population. The three main parameters are the initial mass function (IMF), the initial period distribution and the initial mass-ratio distribution. Each population is generated at a single initial metallicity. For our fiducial models we base our IMF on \citet{1993MNRAS.262..545K} with a power-law slope from 0.1 to 0.5\,M\,$_{\odot}$ of -1.30 which increases to -2.35 above this. The power law slope extends to a maximum initial stellar mass of 300\,M$_{\odot}$, although we note that due to mergers stars more massive than this can form in our binary populations. Our most massive stars have masses up to 570\,M$_{\odot}$ although these are very rare systems. In addition to this standard model we also calculate models with two different upper IMF slopes of -2.00 and -2.70, as well as varying the upper maximum initial stellar mass to 100\,M$_{\odot}$. Finally we calculate a model set with a constant IMF slope of -2.35 from 0.1 to 100\,M$_{\odot}$ (the standard \citet{1955ApJ...121..161S} IMF), for comparison with previous work.  

For the binary populations, we assume a flat distribution in initial mass ratio and log-period in all our models. The result of this is that while all of our stars are technically part of a binary population, in many cases the stars evolve as isolated individuals and are never close enough to interact. The actual interacting binary fraction will depend on the physical size of stars at different ages relative to their Roche lobes, and thus on mass and metallicity. Our upper period limit of 10000 days has been chosen so that about 80\% of massive stars ($M\gtrsim5$\,M$_\odot$) interact, as shown in Figure \ref{Fig0a}. From 5M$_\odot$ to 0.7M$_\odot$ this fraction decreases to about 50\%. At lower masses the separation of the orbits with a 1 day minimum is too wide for stars to interact within the age of the Universe. At Solar metallicity, many of the most massive ($M\sim100$\,M$_\odot$) stars also avoid interactions due to strong mass-loss in stellar winds. 

Observational constraints on these numbers are uncertain but our estimates of massive star interaction are comparable to the 70\% estimated for O stars by \citet{2012Sci...337..444S}. On the other hand, the flat distributions are probably over-simplistic. As discussed by \citet{2017PASA...34....1D} binary parameters (including the binary fraction) vary with initial mass. While nearly every massive star is in a binary, the binary fraction drops to 20\% at lower initial masses below a Solar mass, but the precise binary fraction as a function of stellar mass remains poorly constrained in the literature. In the Milky Way, \citet{2012Sci...337..444S} found that the observed period distribution is slightly steeper than flat, with a bias towards more close binary systems for O stars. However \citet{2012ApJ...751....4K} found a flatter period distribution in the Cygnus OB2 association that is consistent with \citet{1924PTarO..25f...1O}'s law, although their results did also suggest a slight preference for short period systems as well.  The most recent study by \citet{2017ApJS..230...15M} has quantified the period and mass ratio distribution in  an extensive compilation of previous works, and finds a slightly different period and mass ratio distribution to that we use.  We intend to modify our input distributions accordingly in a future version of \bpass.

The uncertainty in assumed period distribution is degenerate with uncertainties in the assumed model to handle Roche lobe overflow, common envelope evolution, tides and other binary specific processes. 
Hence we adopt this simple approach based on the well-studied massive star population \citep{2012Sci...337..444S}. If required, it is possible to simulate the effect of more complex populations and binary distributions by varying the mix between our single star and binary populations with age (since stars of different initial masses dominate in different time bins). 

We note that, because we assume orbits are circular, our closest observational comparison will be with the semi-latus rectum distribution not the period distribution. As shown by \citet{2002MNRAS.329..897H}, the outcome of the interactions of systems with the same semilatus rectum is almost independent of eccentricity. This is equivalent to assuming that systems are circularised by tidal forces before interactions occur. We only include tides in our evolution models when a star fills its Roche lobe as this can move the system into CEE if there is not enough angular momentum in the orbit to spin up the primary. We assume tidal forces are strong and the star's rotation quickly synchronises with the orbit. This is of course an approximation and there are recent studies have begun to explore how mass transfer may be different in eccentric systems \citep[e.g.][]{2007ApJ...660.1624S,2015mgm..conf.1820B,2016ApJ...825...70D,2016ApJ...825...71D}. However we note that the \bpass\ stellar models have been successfully tested to see if they can reproduce an observed binary system with a slight eccentricity even after mass transfer \citep{2009MNRAS.400L..20E}.

\begin{figure}
\begin{center}
\includegraphics[width=\columnwidth]{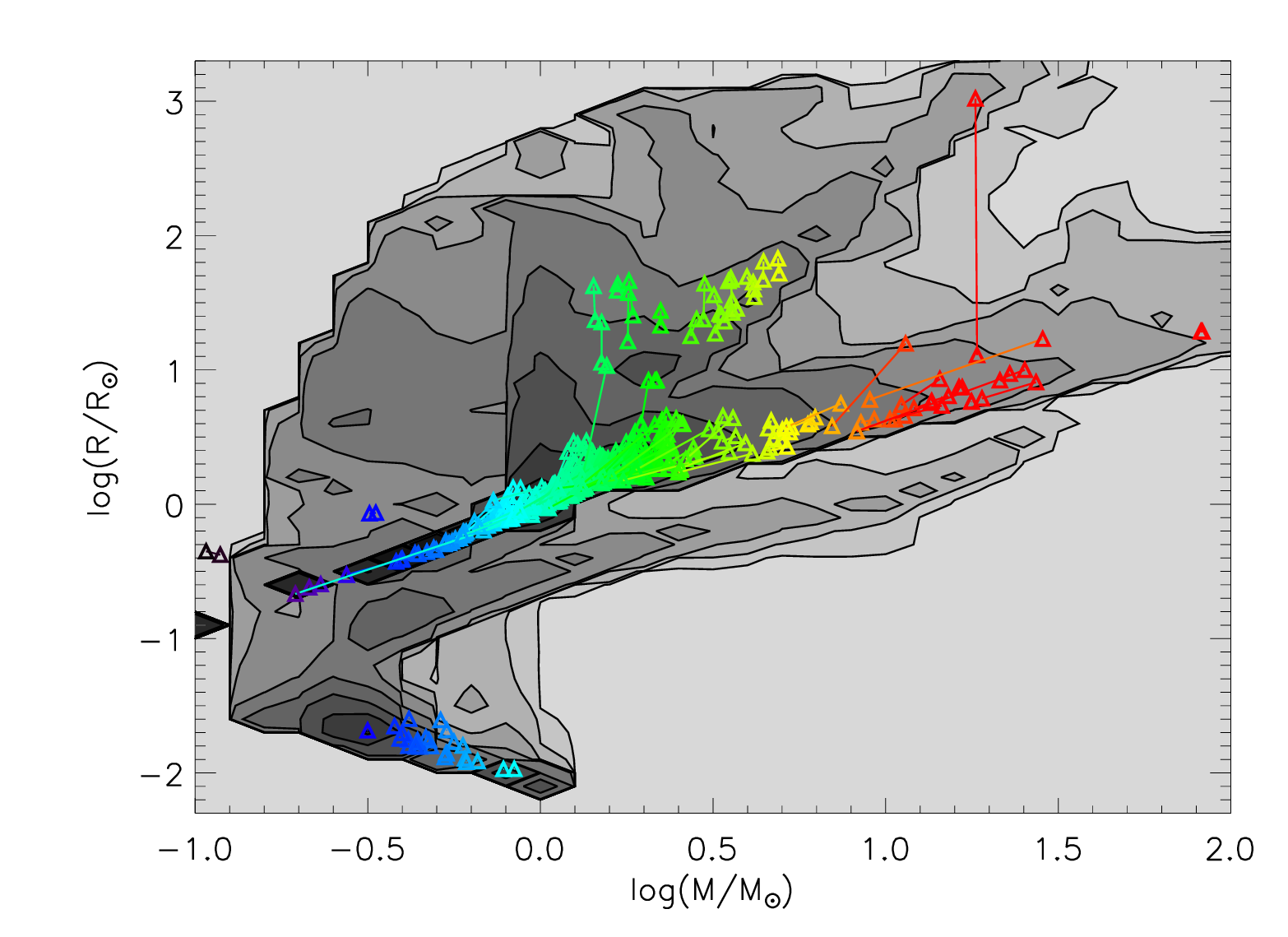}
\includegraphics[width=\columnwidth]{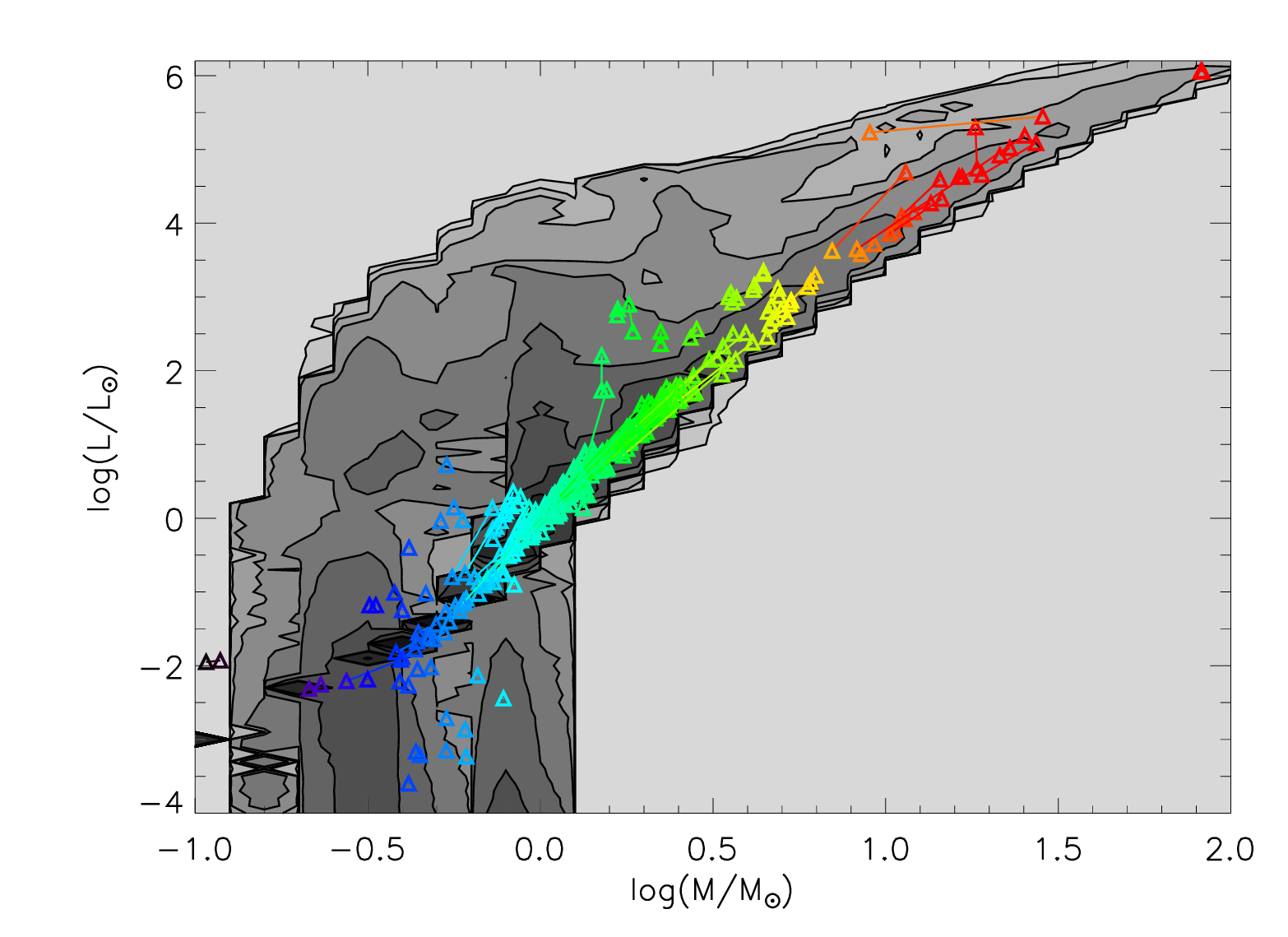}
\includegraphics[width=\columnwidth]{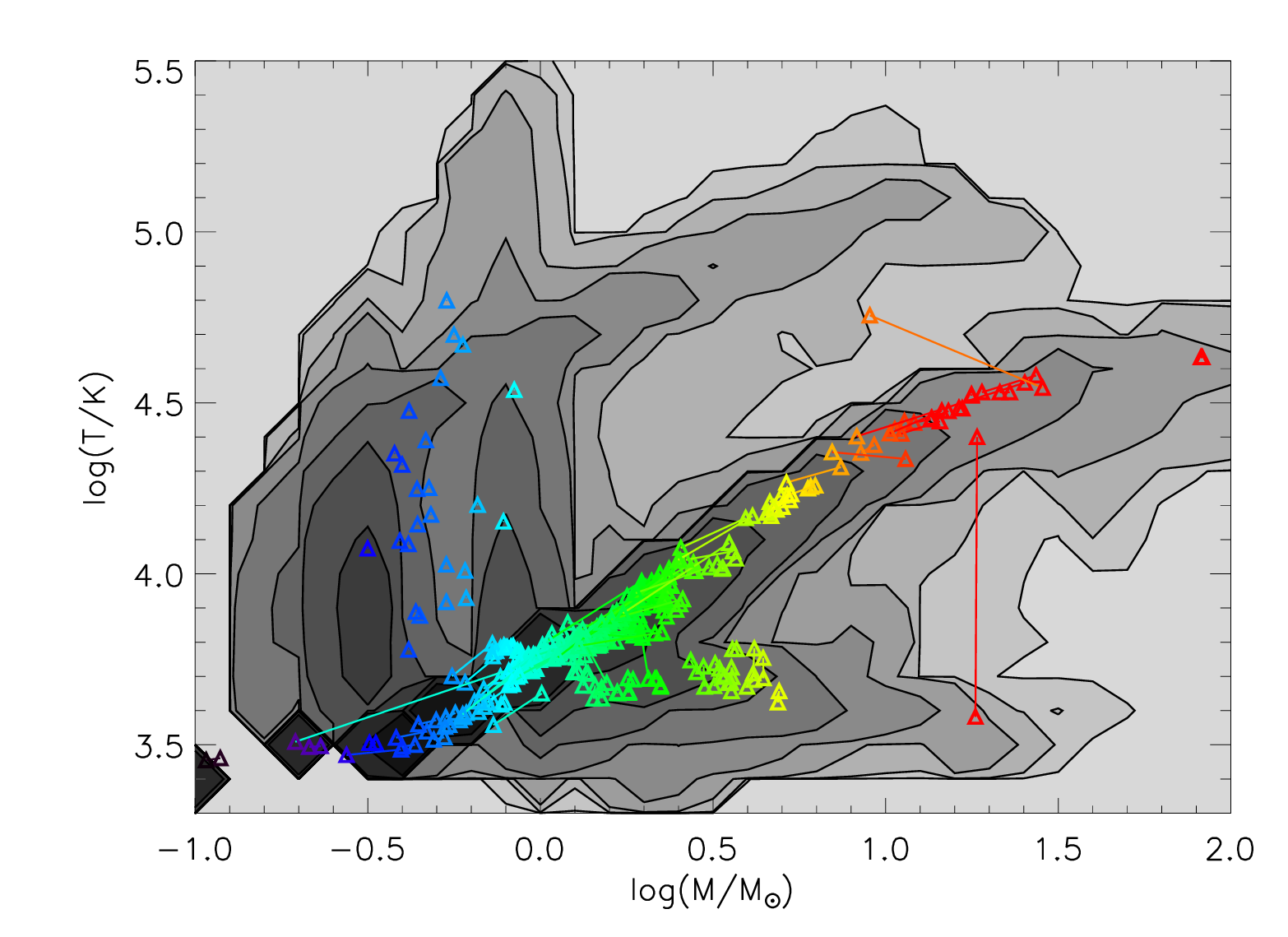}
\end{center}
\caption{Eclipsing binaries overplotted on evolution tracks for our binary models (greyscale, indicating the range of outcomes for a given initial mass). The observed population of eclipsing binaries are overplotted as small symbols, with lines linking binary companions, coloured by metallicity. We show distributions for key parameters. The observational data is mainly taken from \citet{2015ASPC..496..164S} and supplemented by VV Cephei, WR20a and $\gamma$-Velorum, as well as the white dwarf sample of \citet{2017arXiv170605016P} as detailed in the text.}\label{Fig2}
\end{figure}

\subsubsection{Binary evolution treatment}
We combine all the primary star models together according to these distributions. We then must consider how to account for the secondary stars. We pick out the stars that explode in supernovae by 
and estimate the remnant mass as described in Section \ref{sec:snrs}. For remnants less than 3\,M$_{\odot}$ we assume that the remnant receives a kick taken at random from a Maxwell-Boltzmann distribution with $\sigma=265$\,km\,s$^{-1}$ \citep[see][]{2005MNRAS.360..974H}. For remnants more massive than 3\,M$_{\odot}$ we assume the star has a kick that is reduced by a factor of the remnant mass divided by 1.4\,M$_{\odot}$, assuming that the kicks track a momentum distribution rather than a velocity distribution. There is some observational evidence that black-hole kicks should be weaker than for neutron stars \citep{2016MNRAS.456..578M}. We note that we have also investigated other models of neutron-star kicks but have not yet included these in the released BPASS models \citep[see][]{2016MNRAS.461.3747B}.

If the binary is disrupted then the companion star is modelled as a single star in its further evolution. Conversely if it remains bound then the post-supernova orbit is calculated, circularized and a secondary model of a star orbiting a compact remnant is used to represent its future evolution. Eventually if a second supernova occurs the same process is repeated so that the parameters of double-compact remnant binaries can be estimated. Note that the secondary may still undergo supernova, even if the primary does not: this possibility is followed and these later events are also considered.

We also check in our models for rejuvenation of the secondary stars. If a secondary star accretes more than 5\% of its initial mass and it is more massive than 2\,M$_{\odot}$ then the star is assumed to be spun-up to critical rotation and is rejuvenated due to strong rotational mixing. That is, it evolves from the time of mass transfer as a more massive, zero-age main sequence star \citep[this is similar to the method used by][]{1998A&ARv...9...63V}. At high metallicities we assume that the star quickly spins down by losing angular momentum in its wind so that there are no further consequences for its evolution. However with weaker winds at lower metallicity, $Z\le0.004$, we assume that the spin down occurs less rapidly and so the star remains fully mixed throughout the main-sequence and burns all its hydrogen to helium. We assume the star must have an effective initial mass after accretion $>20$\,M$_{\odot}$ for this evolution for occur, based on the work of \citet{2006A&A...460..199Y}. We note that we have discussed the importance of these quasi-homogeneously evolving (QHE) stars in greater detail in \citet{2011MNRAS.414.3501E,2012MNRAS.419..479E} and \citet{2016MNRAS.456..485S} showing there is strong observational evidence that they exist. We only consider QHE that arises from mass transfer, however, and do not yet include it in tidal interactions as done by \citet{2016MNRAS.458.2634M} and \citet{2016A&A...588A..50M}. In such a scenario, stars can be spun up by tidal forces acting between two stars in orbital periods of the order of a day, leading to both stars experiencing QHE; this is not yet implemented in \bpass.

\subsubsection{Binary Mass Function as a function of age}

We present some fundamental results from our populations in Figures \ref{Fig0c} and \ref{Fig0d} which concern the mass in the stars at different different stellar population ages. Figure \ref{Fig0c} shows that from an initial $10^6$\,M$_{\odot}$, the total mass of stars in a co-eval binary population slowly decreases with time as stars die and lose mass in stellar winds. We see that the single star populations tend to decrease faster than the binary populations initially, while the binary populations retain more of their mass until the first SN occurs. This leads to an offset between the binary and single star populations which persists until late times. Higher metallicity populations retain more mass at late times. This is because the higher metallicity stars have longer lifetimes due to being less compact with lower central temperatures so burning through their nuclear fuel takes a longer time. Although at early times the far stronger winds of high metallicity stars means more mass is lost at early times for the higher metallicity population. After this, the stars will evolve along the lower-mass pathways at an appropriate rate. The net effect of this is to lead to longer total stellar lifetimes and so a delay in the death of the stars and formation of remnants. By the age of 1\,Gyr, a Solar metallicity starburst will have lost 37\% (39\%) of its initial mass in the binary (single) star evolution case. This is towards the upper end of the  ``return fractions'' of material to the ISM  of $R\sim0.4$ for a \citet{2003PASP..115..763C} IMF and $R\sim0.3$ for a \citet{1955ApJ...121..161S} IMF found by \citet{2014ARA&A..52..415M}. At 20\% of Solar metallicity, both binary and single star return fractions increase by $\sim$2\%.

Figure \ref{Fig0d} illustrates this effect in greater detail, showing how the stellar mass functions vary with age at the same two metallicities. In both cases, due to rapid mergers of the most massive members of a stellar population, binary populations can have more massive stars than were initially assumed and these can retain mass for longer. At late times the most massive star is always beyond that expected from single-star populations: the phenomenon that leads to the observation of ``blue stragglers'' in globular cluster populations. The mass of the most massive star is greater at late times in high metallicity populations, even though at early times the mass of individual stars rapidly decreases due to the stronger mass-loss. 

We note here that beyond approximately 3 Gyrs our current models predict a population of more massive stars than would be expected. These are more than twice the maximum mass for a single star population at this age. These are systems where the primary evolved to a white dwarf after rejuvenating it's companion star. Our simple method of estimating the rejuvenation age currently uses the age at the end of the model rather than the time of the mass transfer. While this is adequate if the primary explodes in a supernova it is not for the case where our models include the time on the white dwarf cooling track. 


\subsection{Stellar Atmospheres and Spectral Synthesis}\label{sec:method_atmos}

\begin{table*}
\caption{WM-Basic input parameter ranges for new O star grids} \label{tab:wmbasic}
\begin{tabular}{llp{1.5\columnwidth}}
Luminosity Class & Parameter  &  Calculated Values\\
\hline\hline
Dwarfs & $\log g$  & 4.0 and 4.5 \\
 & $T$/kK &  50.0, 45.7, 42.6, 40.0, 37.2, 34.6, 32.3, 30.2, 28.1, 26.3 and 25.0\\
 \hline
 Supergiants & $\log g$ & 3.88, 3.73, 3.67, 3.51, 3.40, 3.29, 3.23, 3.14, 3.08, 2.99 and 2.95 \\
 & $T$/kK &  51.4, 45.7, 42.6, 40.0, 37.2, 34.6, 32.3, 30.2, 28.1, 26.3 and 25.0\\\end{tabular}
\end{table*}

While the stellar models described above contain a large amount of detail about the structure, physical conditions (e.g. temperature, surface gravity) and evolution of stars in a population they provide relatively limited information about their appearance. This is because the physics of stellar atmospheres comprises a second modelling challenge that can be as complex and difficult to calculate as a stellar evolution model. The computational time to calculate a model atmosphere for one set of surface conditions can be the same as the time to calculate the entire evolution of a single stellar model. Therefore some compromise must be found in connecting the evolution and atmosphere models. Some authors \citep[e.g.][]{2004ApJ...615...98R,2014A&A...564A..30G,2015ApJ...800...97T} have calculated sets of atmosphere models for specific phases of evolution of their stellar models and interpolate between these. Our solution has been to use pre-calculated grids of atmospheres that we then match to our evolution models to predict their luminous output spectrum. The exception is for extreme cases where no appropriate atmosphere grid is extant: we have created a new set of atmosphere spectra for hot OB stars from existing tools. We detail these grids in section \ref{sec:method_atmos_wmbasic} below.

\subsubsection{Existing Model Grids}\label{sec:method_atmos_grids}

Our base atmosphere models are those of the BaSeL v3.1 library \citep{2002A&A...381..524W}. This is a set of atmosphere models that extend over a large range of $\log g$, $\log T_{\rm eff}$ and metallicity. We interpolate between models to predict the spectrum of each star generated by our evolution code. This version only extends over metallicities from [Fe/H]=-2.0 to 0.5. We therefore at our lowest metallcities use the slightly older v2.2 grid of \citet{1998A&AS..130...65L} that extends down to [Fe/H]=-3.5.

We supplement these with Wolf-Rayet (WR) stellar atmosphere models from the Potsdam PoWR group\footnote{www.astro.physik.uni-potsdam.de/$\sim$wrh/PoWR/powrgrid1.html} \citep{2003A&A...410..993H,2015A&A...577A..13S}. We only use WR spectra once the surface hydrogen abundance drops below 40\% and the surface temperature is above $\log(T/{\rm K})=4.45$. A key change implemented in \bpass v2.0 and v2.1 is that there are now different metallicity nitrogen-dominated (WN) Wolf-Rayet atmospheres available \citep{2015A&A...579A..75T}. In our previous work \citep[such as][]{2009MNRAS.400.1019E}, only Solar metallicity were available and we artificially reduced the line luminosities of these models at lower metallicites. We now use the closest metallicity WN models to the stellar models being considered. When $Z\ge 0.0095$ we use the Solar PoWR models, below this limit we use their LMC models and when $Z < 0.0035$ we use their SMC models. For WC (i.e. carbon-dominated) Wolf-Rayet stars there are still only Solar metallicity models available, but we have not altered these spectra in any way in our current models, using the Solar models at all metallicities. This may lead to slight overestimates in the strength of carbon WC features in low metallicity spectra but the abundances in the stellar atmospheres during the WR phase are relatively insensitive to the initial metallicity (which has a larger effect on mass-loss rate). This will be modified as as further PoWR models become available. We switch to the WC models when there is no surface hydrogen and the helium mass fraction drops below 70\%.

As in \citet{2009MNRAS.400.1019E} we add in an excess He\,{  II} flux for massive Of stars as suggested by \citet{2008MNRAS.385..769B} since these are under-represented in atmosphere models. For non-WR stars with a surface temperature above 33000K and when $\log g \le 3.676\log(T/K)-13.253$, we increase the He\,{  II} stellar wind lines at 1640\AA\ and 4686\AA\ by 520\,L$_{\odot}$ and 33.8\,L$_{\odot}$ respectively. Due to the rarity of these cases, we find this has a minimal effect on the predicted population line fluxes.  Interestingly, \citet{2017arXiv170203108C} have studied He\,II emission from massive stars in the Tarantula Nebula and report a very high equivalent width for a young stellar population. We note that with our current prescription as described above, our v2.1 models produce a comparably high equivalent width at slightly older (but similar ages), see Section \ref{sec:wr_obs} and Figure \ref{fig:WR_bumps} below.

\subsubsection{New O star models}\label{sec:method_atmos_wmbasic}

In previous distributions of \bpass\ models, we supplemented the above with a set of O star models generated using the WM Basic code by \citet{2002MNRAS.337.1309S}. We have now recalculated this grid, and extended it, as summarised in Table \ref{tab:wmbasic}\footnote{These new models are the first of the two main differences between v2.0 and v2.1.}
 
For \bpass\ v2.1 models we have created new WM-Basic \citep{1998ASPC..131..258P} models based upon the grid determined by \citet{2002MNRAS.337.1309S}. The key differences are that we have extended the metallicity coverage and also added a new sequence of models with a  higher surface gravity to better match the evolution code predictions. We have kept the range of temperatures the same (25-50\,kK) but now added a grid of \textit{high gravity} stars with $\log g=4.5$ (c.f. 4.0 in the earlier grid). This is because we noticed that on the zero-age main-sequence many of our O stars had gravities higher than 4.0 which has a significant effect on the strength of some stellar lines. At low metallicities our QHE models also increase their surface gravity as they evolve with the stars shrinking as helium is mixed to the surface and the mean-molecular weight increases. Therefore these new models are important for predicting the correct spectrum.

We also calculate models at every metallicity we consider in \bpass\ rather than simply interpolating between fixed points as before. The key benefit is that we now have spectra at the lowest metallicities when $Z=10^{-5}$ and $10^{-4}$, where the influence of these massive stars on the synthetic spectrum is strongest. These are beyond the range for which observational validation data is available for stellar atmospheres and so constitute a theoretical extrapolation from better constrained atmospheres at higher metallicity. However incorporating these gives greater predicting power for stellar populations at these  extreme metallicities.

We have made these models available on the \bpass\ website and they have already been used in other published works \citep[e.g.][]{2017ApJ...840...44B}. We note that only the inclusion of the higher gravity models leads to any significant differences in model outcomes. We find that the ionizing flux predictions at low metallicites decrease by about 0.1\,dex, but that the far-ultraviolet luminosity of a population increase by a similar amount (see figure \ref{fig:ionize_comp} later). This is because the QHE evolution models have more correct atmospheres. These stars evolve to hotter temperatures as they burn hydrogen to helium. These models also allow us to predict O\,{  V} 1039\AA\ lines at very young ages that are comparable to those found in observations (A. Runnholm, private communication).

\subsection{BPASS model outputs}
\subsubsection{Spectral Energy Distributions}\label{sec:method_seds}

The basic output of \bpass\ is a set of moderately high resolution model spectra, gridded from 1 to 100,000\AA\ in 1\AA\ bins and measuring total flux in each wavelength bin in units of L$_{\odot}$ (i.e. producing a spectrum of $F_\lambda$ in L$_{\odot}$/\AA\ spanning from the extreme-ultraviolet to far-infrared wavelengths). These are generated for stellar population ages of 1\,Myr to 100\,Gyr\footnote{The increase of the upper age limit to 100Gyrs from 10Gyrs is the second of the two main differences between v2.0 and v2.1.} in increments of $\Delta\,\log({\rm age/yr})=0.1$ where this age is taken relative to the onset of star formation, assuming a co-eval stellar population of total mass $10^6$\,M$_\odot$. Examples are shown in Figure \ref{fig:seds} and \ref{fig:euv}, illustrating the optical-infrared spectral energy distributions as a function of age and the extreme ultraviolet as a function of metallicity and binary evolution respectively. 

In Figure \ref{fig:sed_comparison} we show a direct comparison between the simple stellar population models (i.e. co-eval starbursts) produced by different stellar population synthesis codes at Solar metallicity and allowed to evolve over time. In three panels, \bpass\ single star models are compared to the closest available matching models in metallicity, age and initial mass function from publicly available synthesis codes. Perhaps the most direct comparison is between \bpass\ and the Starburst99 models \citep[][hereafter S99]{1999ApJS..123....3L,2014ApJS..212...14L}, both of which were originally designed and optimised for young starbursts. The S99 models shown here were  generated using the non-rotating Geneva 2012/2013 stellar model set and a broken power law IMF with slopes matching our default IMF but with a maximum stellar mass of 100\,M$_\odot$. 

We also compare against two SPS codes designed primarily for modelling galaxies. 
The GALAXEV models of \citet[hereafter BC03]{2003MNRAS.344.1000B} are widely used for modelling mature galaxies and specifically we compare against the galaxy templates used to classify galaxies in the Sloan Digital Sky Survey by \citet{2004ApJ...613..898T}. These use a \citet{2003PASP..115..763C} IMF and are built on the Geneva and Padova single star stellar evolution models, combined with a large library of semi-empirical template spectra and additional atmosphere models calculated using the BaSeL code \citep{2002A&A...381..524W}. 
The \citet[hereafter M05]{2005MNRAS.362..799M} models were designed to confront the thermally pulsating AGB stage in more detail than previous models and use a prescription for fuel consumption to determine the contribution of post-main sequence stars and their role in determining the near-infrared colours of galaxies. They employ the Cassisi stellar evolution tracks \citep{1997MNRAS.285..593C,2000MNRAS.315..679C}, BaSeL atmospheres \citep{1998A&AS..130...65L} and a \citet{2001MNRAS.322..231K} broken power law IMF similar to ours (with an upper power law slope of -2.30 where we use -2.35, and a maximum mass of 100\,M$_\odot$).

As the figure demonstrates, the \bpass\ single star models are very similar to all three other synthesis models at ages of $<$1\,Gyr and in the ultraviolet, where the models are dominated by massive stars. The BC03 models are slight outliers, appearing redder at very young ages ($<$10\,Myr) than any of the other model sets. Our single star models remain similar to the output of other codes at a matching age through the blue optical bands, with subtle differences in the IMF and evolutionary timescales assumed for the most massive stars by different evolution codes leading to our single star models being slightly bluer than S99 and M05 models but redder than BC03 models at ages of $\sim$10-100\,Myr. 

\bpass\ models begin to diverge significantly from the other model sets in the red-optical and longwards at stellar population ages of more than a few times 100\,Myr. At these ages, the light from our single star populations are dominated by evolved massive stars, primarily AGB stars, whose molecular line features are clearly visible in the composite SED. Perhaps unsurprisingly our models are most similar in the infrared to the M05 models, which are also heavily influenced by the AGB phase, but we see a stronger effect still. Realistically, this comparison suggests that AGB stars are over-represented in our co-eval single star populations (either in terms of number or luminosity) at ages $>1$\,Gyr. In our favoured binary models (shown in the fourth panel of Figure \ref{fig:sed_comparison}), the number of AGB stars in a population is suppressed by the effect of binary interactions, but these interactions also keep the spectra rather bluer than those seen in other stellar population synthesis models at late times. {\it We note that extreme caution should be applied when using this version of \bpass\ in the red at ages $>1$\,Gyr as a result}; we return to this point in section \ref{sec:oldpops}.

\subsubsection{Released Data Products}\label{sec:method_atmos_outputs}

The spectral energy distributions described above for co-eval simple stellar populations form the core of the \bpass\ v2.1 data release.
Composite (i.e. non-co-eval) stellar populations are not provided as a part of our standard release but can be constructed by assuming a star formation history (see section \ref{sec:sfh}). The simplest case is a population forming stars continuously at a constant rate. Here the only potential difficulty lies in dealing with logarithmically spaced time intervals in the models since the total number of stars to be added to the composite spectrum depends on the width of the interval. Stars contributing to the first time bin (at log(age/years)=6.0) include all members of the population up to age $10^{6.05}$, the second bin stars in the age range $10^{6.05}$ to $10^{6.15}$ years and so forth.

We  calculate magnitudes for individual stellar models and for our synthetic populations in the standard $UBVRIJHK$ filters, as well as $ugriz$ SDSS filters, $FUV$(1566\AA), $NUV$(2267\AA) fluxes and several HST optical filters at each step. Additionally we calculate the production rate of ionizing photons based on the extreme ultraviolet flux since this is strongly affected by both binary evolution and metallicity (see Figure \ref{fig:euv} and \citet[][]{2016MNRAS.456..485S}). 

Given that our population synthesis tracks all stages of stellar evolution, we are also able to produce corresponding rates of supernovae (separated by type) and information on the compact remnant population and stellar class distribution at each time step.

\begin{enumerate}
\item Stellar Model Outputs:
\begin{enumerate}
\item Binary stellar models with photometric colours
\item New OB stars atmosphere models
\end{enumerate}

\item Stellar Population Outputs (all versus age):
\begin{enumerate}
\item Massive star type number counts
\item Core collapse supernova rates
\item Yields, energy output from winds and supernovae and ejected yields of $X$, $Y$ and $Z$.
\item Stellar population mass remaining
\item HR diagram  (isochronal contours)
\end{enumerate}

\item Spectral Synthesis Outputs (all versus age):
\begin{enumerate}
\item Spectral Energy Distributions
\item Ionizing flux predictions
\item Broadband colours
\item Colour-Magnitude Diagram (CMD) making code
\end{enumerate}

\item Available on request due to unverified status:
\begin{enumerate}
\item Approximate accretion luminosities from X-ray binaries
\item A limited set of nebula emission models
\end{enumerate}
\end{enumerate}

All outputs of the current \bpass\ v2.1 data release can be found at {\tt http://bpass.auckland.ac.nz}, and are mirrored at {\tt http://warwick.ac.uk/bpass}.


\section{OBSERVATIONAL VALIDATION ON RESOLVED STARS AND STELLAR POPULATIONS}\label{sec:tests_resolved}

It is vital to test the accuracy and predictive power of our models as far as we can. Because the most basic component of \bpass\ are the stellar models these are the first element we consider. In this section we explore a number of observational constraints on individual stars and resolved stellar populations, and use these to validate our models, commenting on successes and limitations of the \bpass\ stellar models.

\subsection{Main Sequence Stars and Eclipsing Binaries}

 A powerful validation test is the comparison of our models to double-lined spectroscopic eclipsing binaries. From these systems it is possible to accurately determine the radius, luminosity, surface temperature, mass and distance of the stars,  making them amongst the best understood individual stars other than the Sun. We use the list compiled by \citet{2015ASPC..496..164S}\footnote{www.astro.keele.ac.uk/jkt/debcat, accessed late 2016} that builds upon that of \citet{1991A&ARv...3...91A} which listed systems with observational uncertainties less than 3\%. 

The scope of the \citet{2015ASPC..496..164S} catalogue is explicitly limited to stars which do not show clear evidence for evolution affected by mass transfer. In this paper we supplement this sample with data on 3 additional (non-eclipsing) binary systems: VV Cephei \citep{1977JRASC..71..152W,1991A&AS...90..175B,1992A&AS...95..589H,2004ASPC..318..222B,2008AJ....136.1312H,2010ASPC..425..181B}, WR20a \citep{2005A&A...432..985R} and $\gamma$-Velorum \citep{2007MNRAS.377..415N}. We include these as they contain very masive stars, a Wolf-Rayet star and a red supergiant. Since we are concerned with the effects of very massive stars it is important to include these in our model tests. We have also supplemented the data with recent observations of white dwarfs in eclipsing binaries from \citet{2017arXiv170605016P} to test the validity of modes in this evolutionary state.

In Figure \ref{Fig1} we present contour plots of the density of stars expected from ongoing constant star-formation for 10\,Gyr at Solar metallicity for the Hertzsprung-Russell diagram, the spectroscopic HR diagram and the $\log g$--$\log(T/$K$)$ diagram. Contours are defined as timescale and IMF weighted probabilities of a star occurring at a given point in parameter space.  Each contour represents an order of magnitude difference in stellar number density. In each of these plots, representative stellar evolution tracks are shown, as well as the observed eclipsing binaries. Pairs of stars in a binary system are plotted linked by a thin line. Both tracks and observational data are colour-coded by stellar mass. In general the agreement by eye is good. Most stars occur in the regions expected for main sequence evolution, with a few objects consistent with identification as helium-burning giants.  As expected, the white dwarf sample of \citet{2017arXiv170605016P} populates the white dwarf cooling track in our models.

In Figure \ref{Fig2} we take the comparison slightly further and plot each derived parameter (radius, luminosity and temperature) versus mass. Observational data shows a clear trend that relates mass to each of these parameters: these are derivable from stellar structure theory by homology assumptions. Again qualitative agreement with the \bpass\ stellar models is generally excellent. The poorest agreement between models and data is for the WR binary star $\gamma$-Velorum. This is not an eclipsing binary and thus the radius estimate is dependent on calculation from a model atmosphere so over-interpretation of this discrepancy may be unwise. The radii of WR star models is uncertain due to the effect of clumping in the outer envelope. Models with this effect included increase the radii and decrease the temperature of WR stars, improving agreement \citep{2016MNRAS.459.1505M}. By contrast both the highest mass star, WR20a, and the red supergiant, VV Cephei, show observational properties consistent with the predicted properties of these stars in our models. 

Besides eclipsing binary stars, the observational samples with well-constrained physical properties suitable for testing our models become less clearly defined. For most stars we do not have a precise estimate of distance; this of course will change with time as Gaia continues its mission. Instead we can use the work of \citet{2014A&A...564A..52L} who defined the spectroscopic HR diagram; we have already shown an example of this in Figure \ref{Fig1}. This diagnostic involves a gravity-weighted luminosity. Combining the Stefan-Boltzmann law, $L=4\pi R^2 \sigma T_{\rm eff}^4$, with the surface gravity, $g=GM/R^2$, it can be shown that: 
\begin{equation*} 
\log\left(\frac{T^4_{\rm eff}}{g}\right)=\log\left(\frac{T^4_{\rm eff} R^2}{GM}\right) = \log\left(\frac{L}{4\pi \sigma M}\right).
\end{equation*}

\begin{figure*}
\begin{center}
\includegraphics[width=\columnwidth]{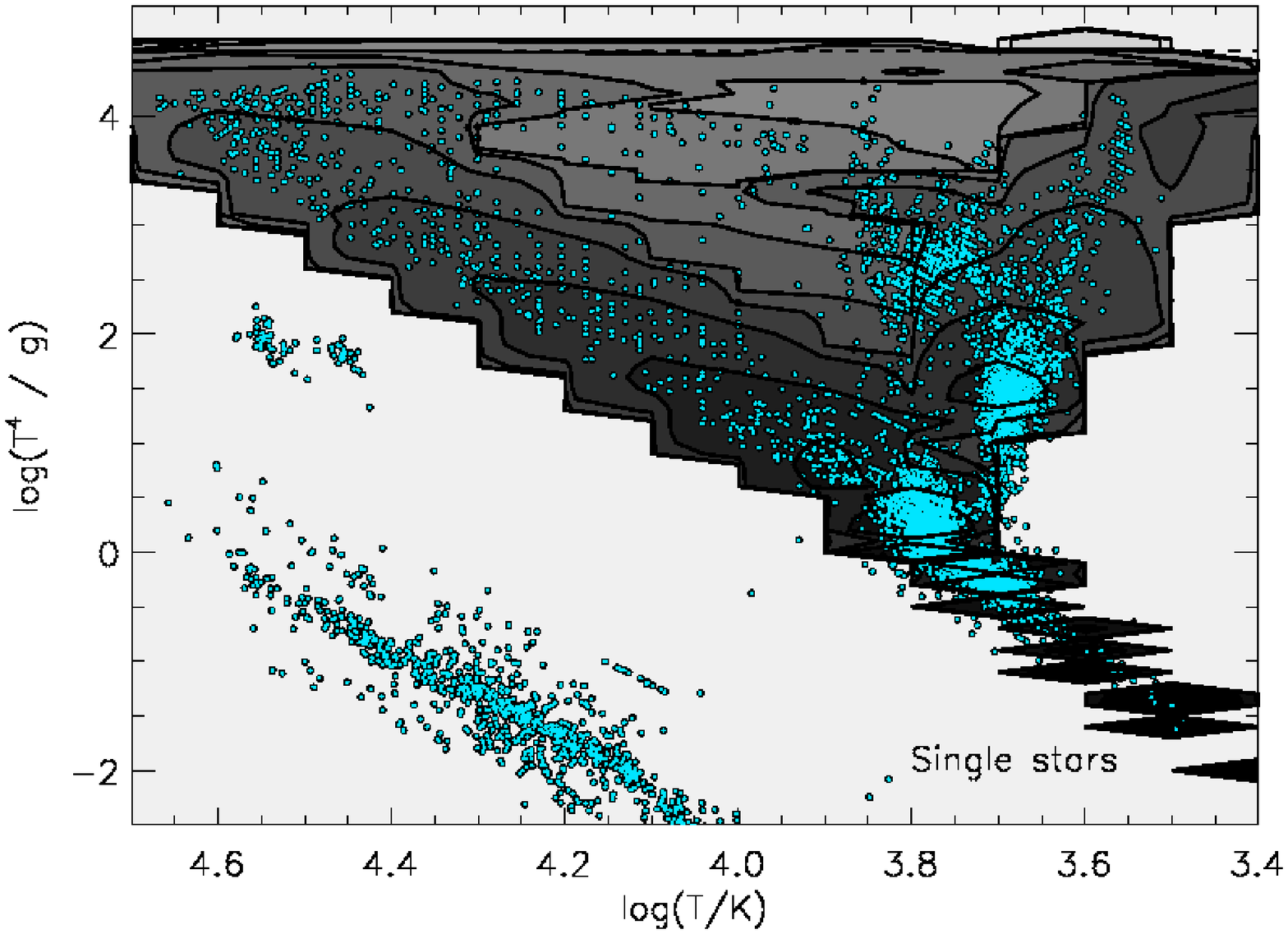}
\includegraphics[width=\columnwidth]{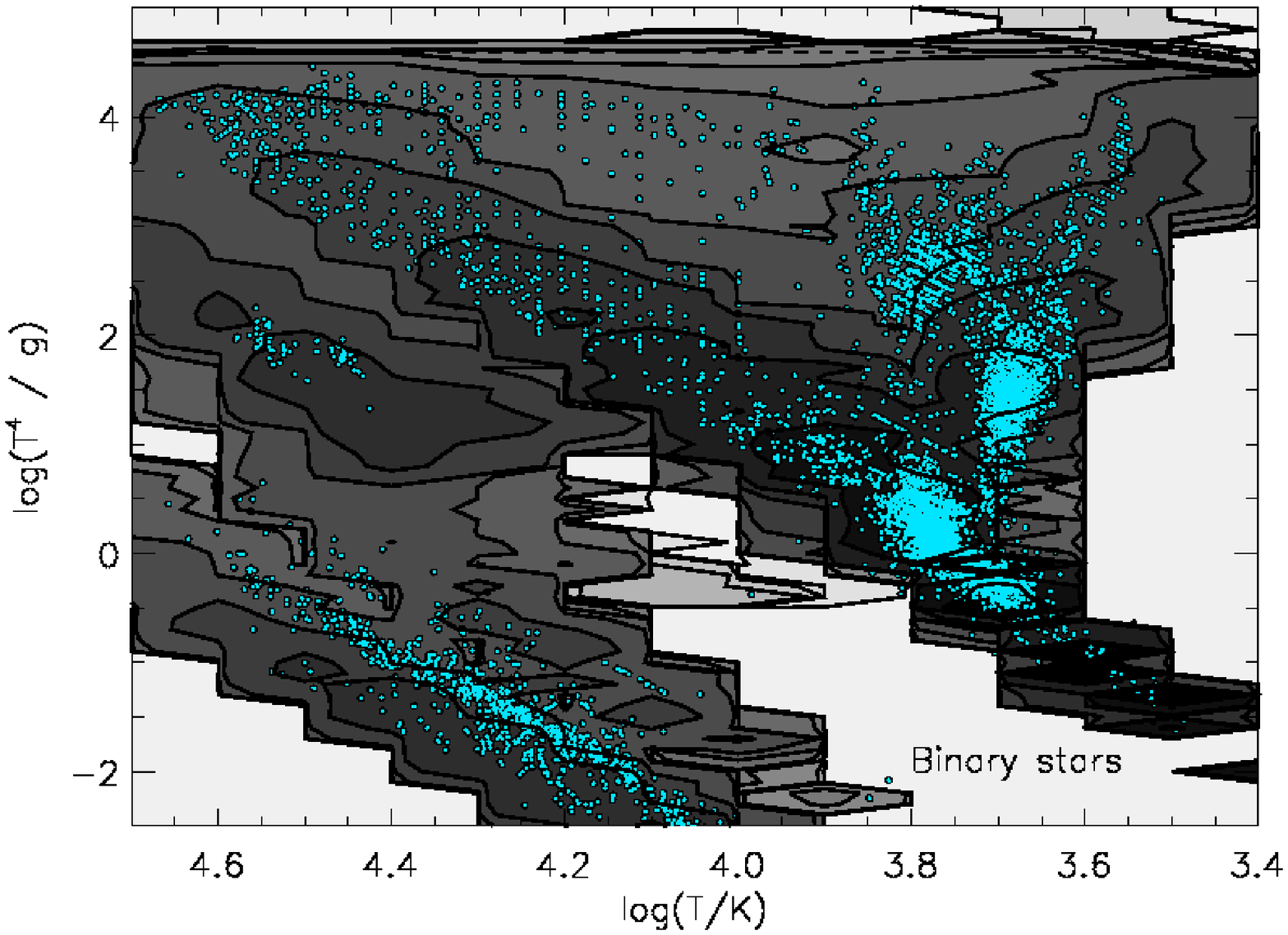}
\caption{Spectroscopic HR diagrams for single (left) and binary (right) stellar populations. This diagram replaces bolometric luminosity with the gravity weighted flux as described by \citet{2014A&A...564A..52L}. Here we overplot  the data from \citet{2014A&A...570L..13C} for Galactic stars near Solar metallicity as small blue points. Each contour represents an order of magnitude in number density of objects.}\label{FigJJ4}
\end{center}
\end{figure*}

This is a luminosity-related variable that can be derived from the spectrum of a star alone, without needing to determine the precise distance to the object, and thus significantly expands the sample of objects against which models can be tested. Here we consider the Galactic stellar catalogue of \citet{2014A&A...570L..13C} for massive stars with half-Solar metallicity and above as well as those that do not have metallicities. In Figure \ref{FigJJ4} we compare both \bpass\ single and binary star predictions to the distribution of observed stars in this parameter space. In general the agreement is good. In the binary populations, at Solar metallicity, we see only small differences in the hydrogen burning main sequence and the red giant branch relative to the single star models. The binary models predict more post-main sequence stars, especially blue and yellow supergiants. We are additionally able to predict the number of helium burning sdB/O stars, as well as the position of the white dwarf cooling track. The slight offset on the HR diagram between our white dwarf models and some of the observations is likely due to unknown white dwarf metallicities and perhaps gravitational settling that will affect the surface appearance of these stars and is not included in our models.

\begin{figure*}
\begin{center}
\includegraphics[width=2\columnwidth]{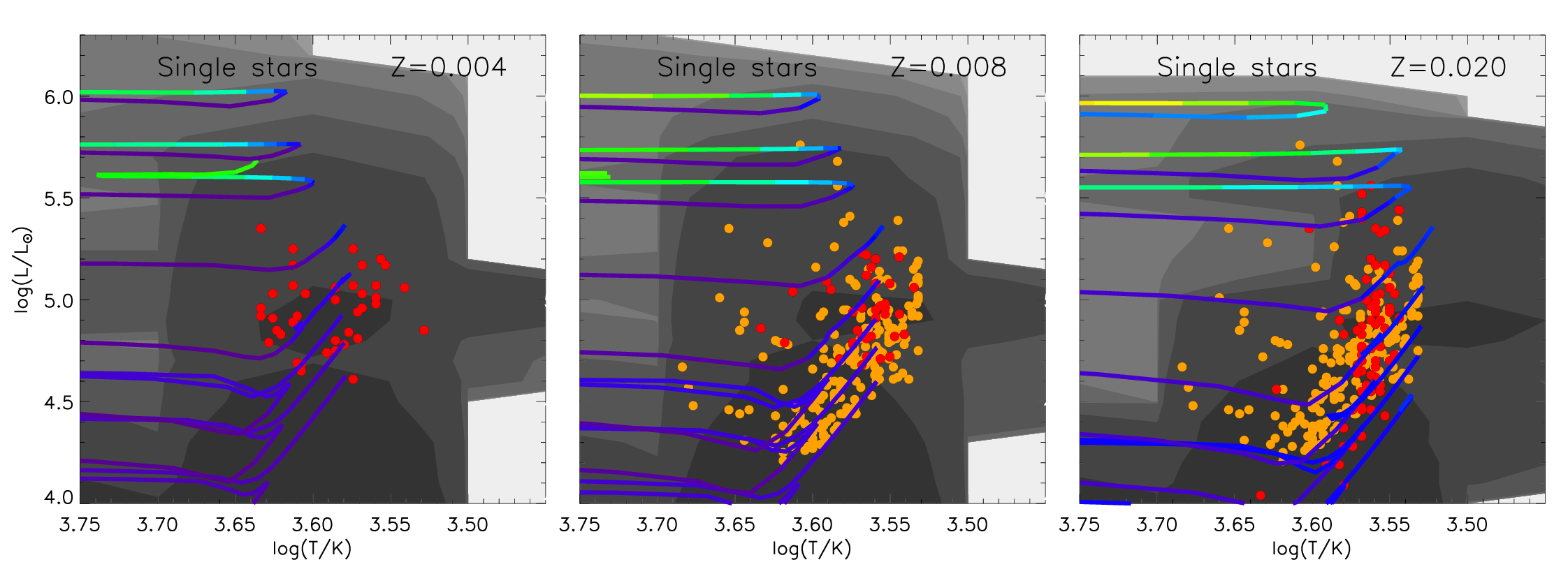}
\includegraphics[width=2\columnwidth]{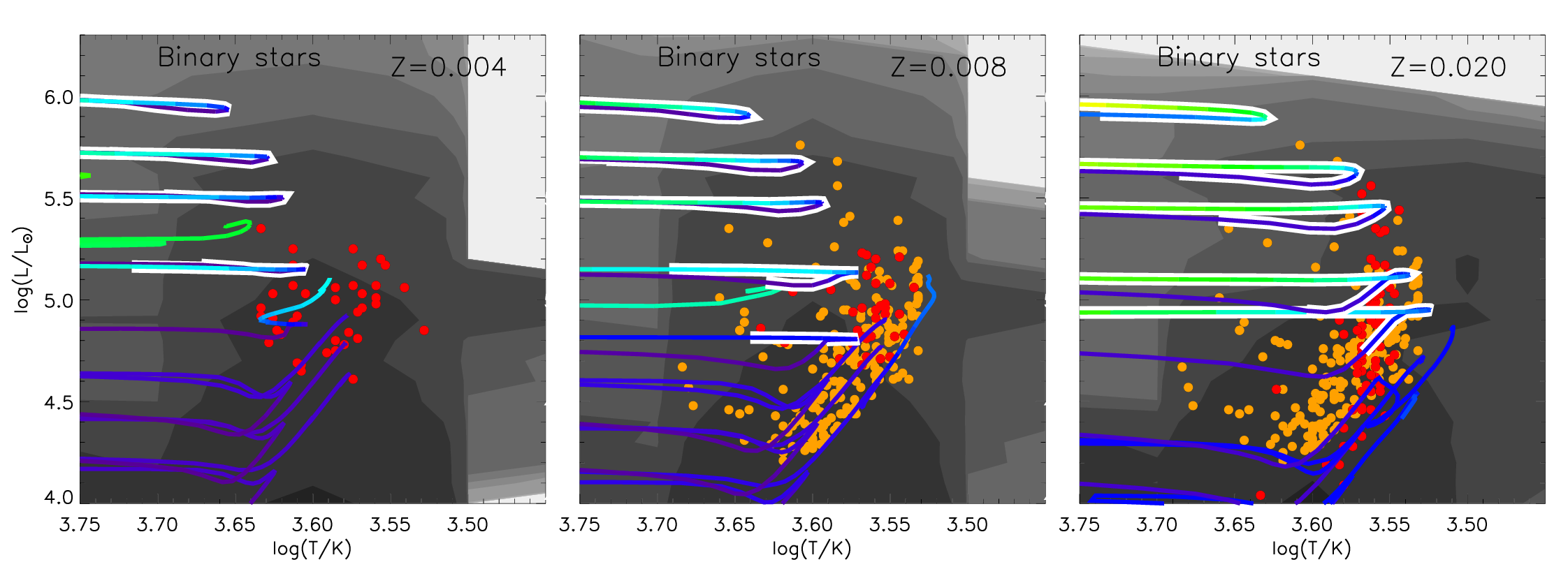}
\caption{HR diagrams focused on the red supergiant branch at three metallicities for single and binary stars. The observational data is from the samples taken from \citet{2005ApJ...628..973L,2009ApJ...703..420M,2012ApJ...749..177N,2012ApJ...750...97D,2016ApJ...826..224M}. For $Z=0.004$ we use RSGs from the SMC, for $Z=0.008$ we use those in the LMC and M33, for $Z=0.020$ we use those from the Galaxy and M31. The tracks and observed WR stars are colour coded to show the surface hydrogen mass fraction. The tracks have masses of 8, 10, 12, 15, 20, 30, 40 and 60\,M$_{\odot}$, the binary star tracks shown have initial periods of 1000~days and an initial mass ratio of 0.5, the section outlined in white indicates when a binary interaction is taking place. Each greyscale contour represents an order of magnitude in number density of objects.}\label{FigJJ5}
\end{center}
\end{figure*}

\begin{figure*}
\begin{center}
\includegraphics[width=2\columnwidth]{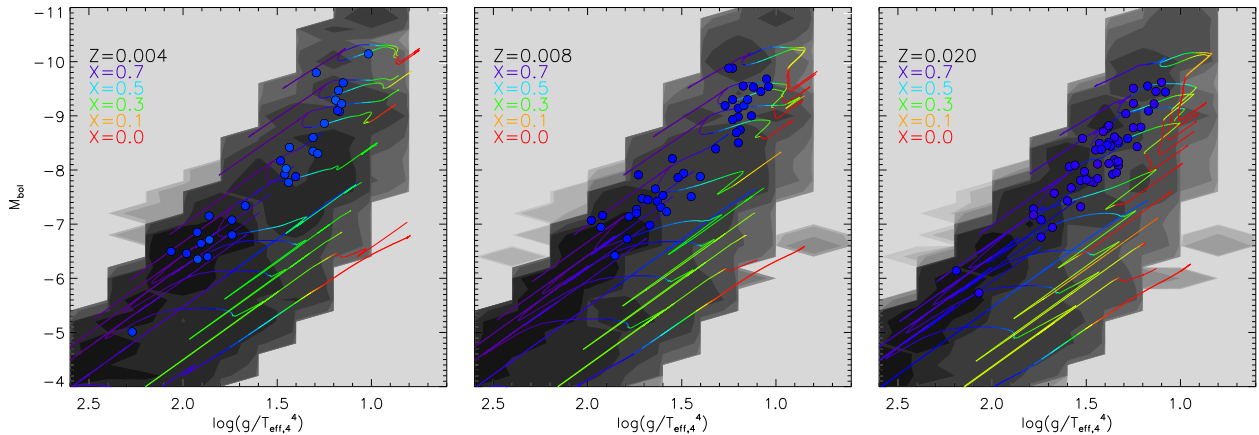}
\caption{Flux-weighted gravity versus bolometric luminosity for blue supergiants. Observational data taken from the compilation of  \citet{2015A&A...581A..36M}. Left-hand panel is for $Z=0.004$, central panel for $Z=0.008$ and right-hand panel for $Z=0.020$. The tracks have masses of 3, 5, 8, 10, 12, 15, 20, 30, 40 and 60\,M$_{\odot}$ and an initial binary period of 100\,days.}\label{FigJJ6}
\end{center}
\end{figure*}

\begin{figure*}
\begin{center}
\includegraphics[width=1.8\columnwidth]{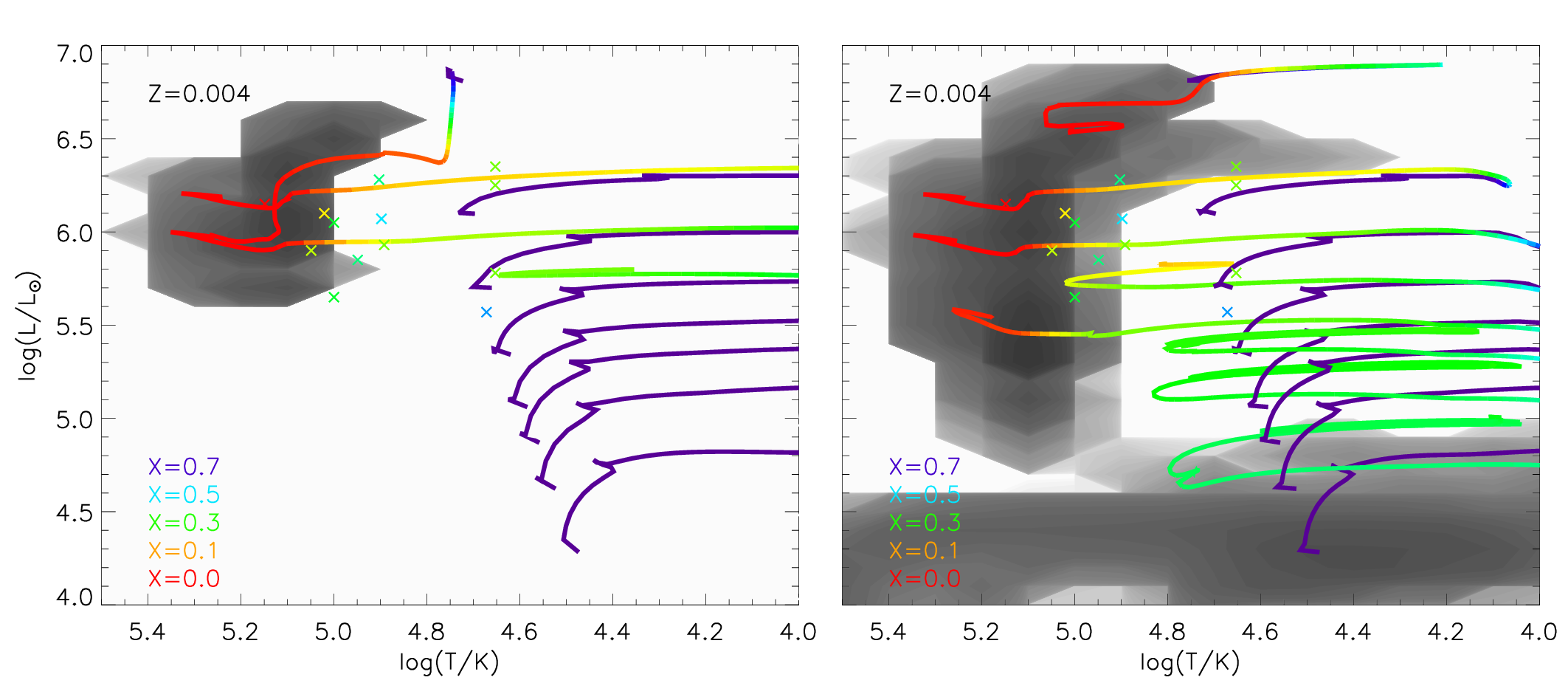}\\
\includegraphics[width=1.8\columnwidth]{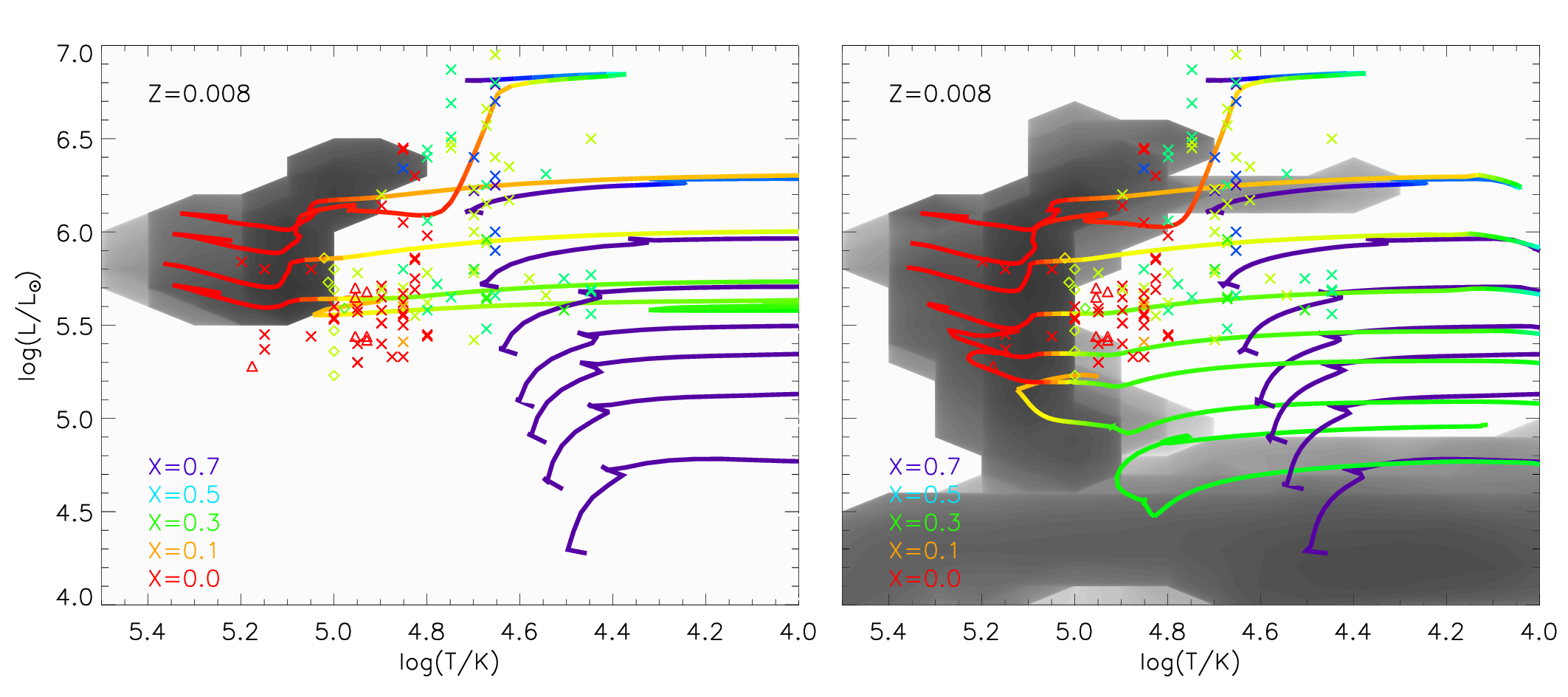}\\
\includegraphics[width=1.8\columnwidth]{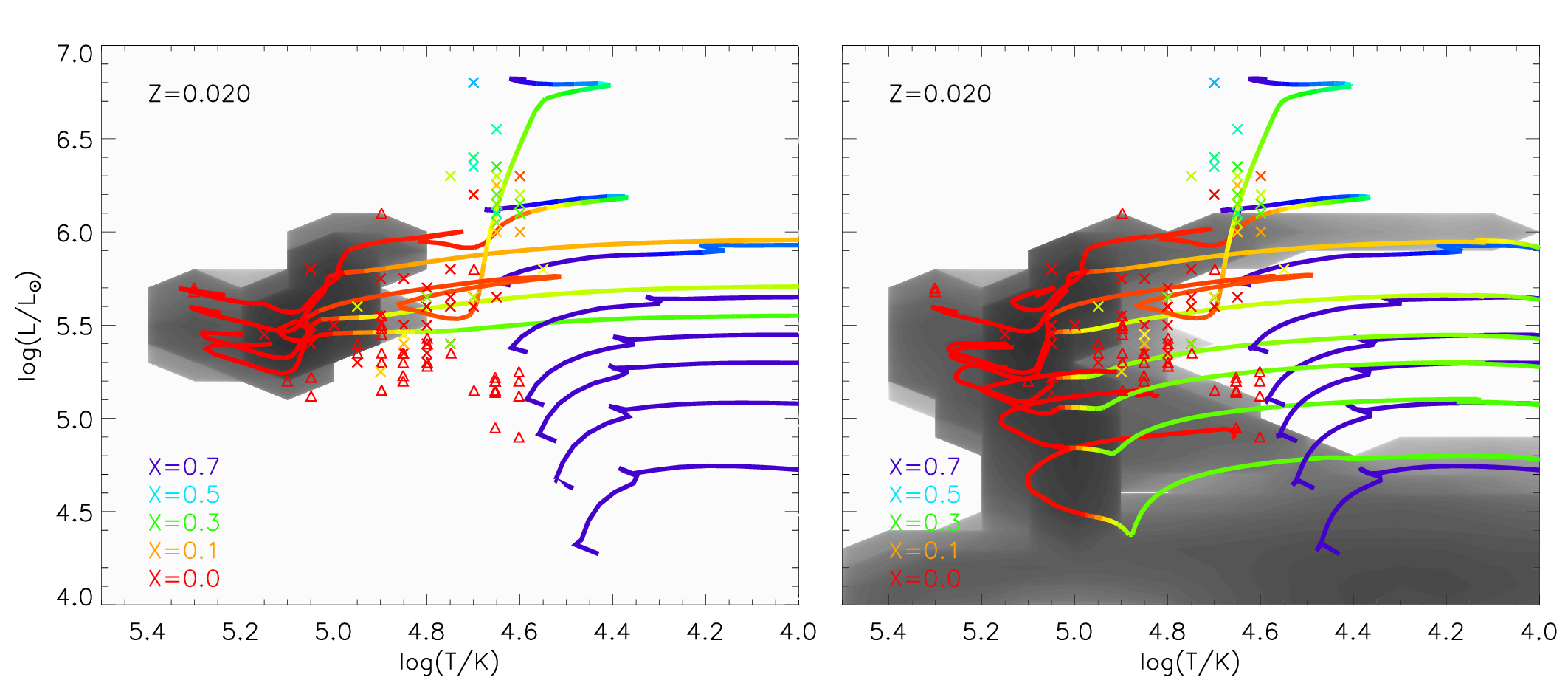}
\caption{HR diagrams with contours plots showing the expected location of hydrogen-free Wolf-Rayet stars in the SMC (top), LMC (middle) and Galaxy (bottom) panels. The left-hand panels are for the single star populations and the right-hand panels for binary populations. The tracks and observed WR stars are colour coded to show the surface hydrogen mass fraction. The observations are taken from: \citet{2003A&A...410..993H,2012A&A...540A.144S,2015A&A...581A..21H,2017A&A...598A..85S}. Triangles indicate WC/WO stars, crossed WN stars, diamonds indicate the WN3/O3 stars from \citet{2017ApJ...841...20N}. The tracks have masses of 15, 20, 25, 30, 40, 60, 100 and 300\,M$_{\odot}$ and an initial binary period of 100\,days.}\label{FigJJ7a}
\end{center}
\end{figure*}

\begin{figure}
\begin{center}
\includegraphics[width=\columnwidth]{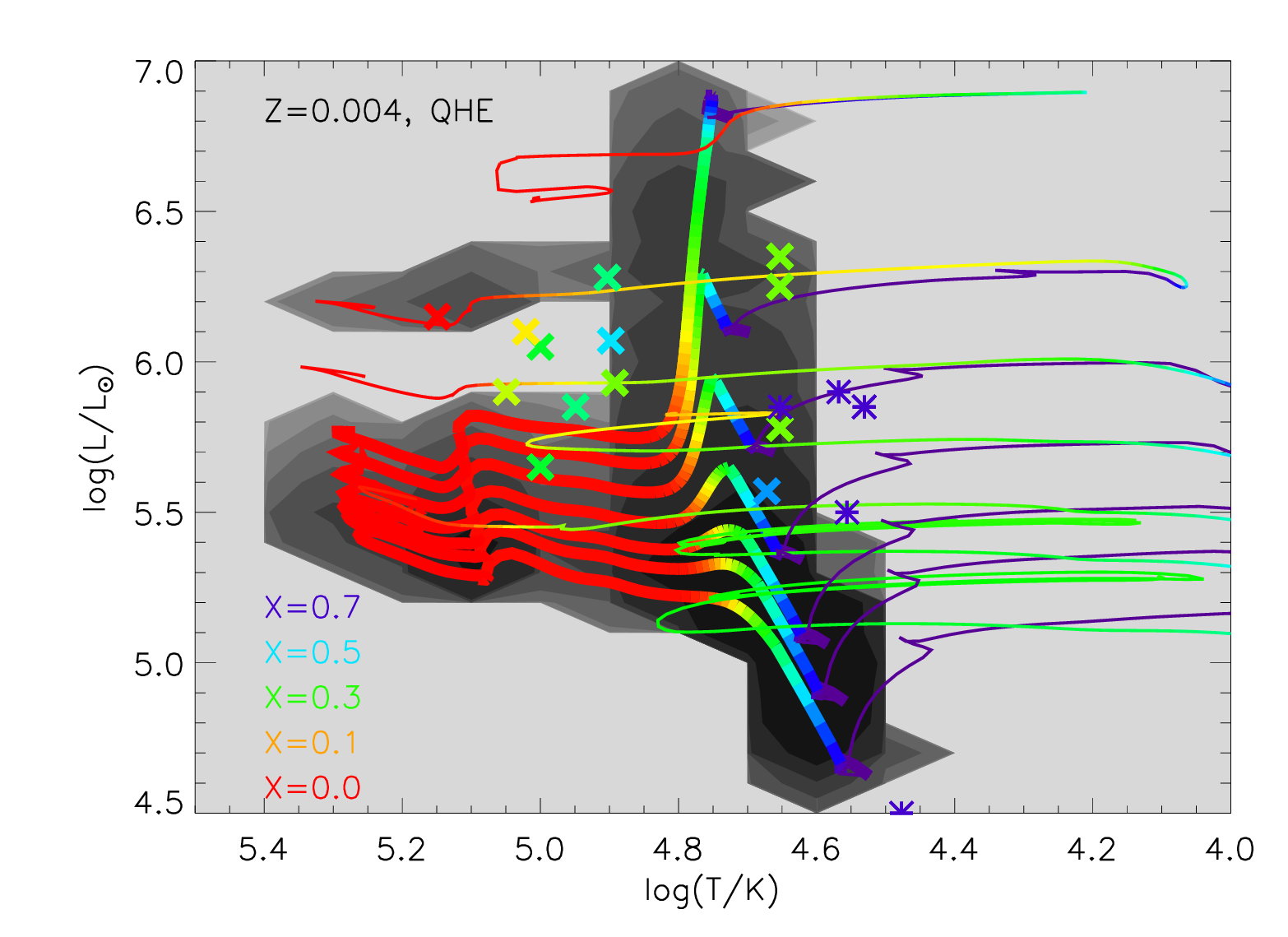}
\caption{HR diagram showing the evolution of QHE models at $Z=0.004$ compared to the properties of WR stars in the SMC. Crosses indicate WR stars, asterisks are the O star companions from \citet{2017A&A...598A..85S}. Colour coding is the surface hydrogen abundance. The thicker tracks are for QHE stars while thinner tracks are standard binary models. The tracks have masses of 20, 25, 30, 40, 60, 100, 300\,M$_{\odot}$ and an initial binary period of 100\,days.}\label{FigJJ7b}
\end{center}
\end{figure}

\subsection{Post main-sequence, giant and evolved stars}
While stars spend most of their time on the main sequence, the evolution after this phase can have significant implications for the observational characteristics of a stellar population, particularly if the population is unresolved. In this subsection we  discuss how well we reproduce observed properties of red supergiants (RSGs), blue supergiants (BSGs) and Wolf-Rayet (WR) stars.

The figures used in this section are made in a similar way to the figures in the previous section but focus on  different sub regions of the HR diagram for comparison with different observational samples. We limit the upper age of the population to 100\,Myrs for red supergiants, and 10\,Gyr otherwise.

\subsubsection{Red and Blue supergiants}
In Figure \ref{FigJJ5} we compare population density contour plots and 1000-day period stellar tracks to observed RSGs in the SMC, LMC, M31 and M33 taken from \citet{2005ApJ...628..973L,2009ApJ...703..420M,2012ApJ...749..177N,2012ApJ...750...97D,2016ApJ...826..224M}. The fit is qualitatively good. Based on the comparison between data and model contours, we infer that red supergiants typically have masses $\le 25\,M_{\odot}$. At higher masses, the number of observed supergiants decreases, especially in the SMC and LMC, which are at lower typical metallicities than the Milky Way. At all metallicities, the presence of relatively long period ($\sim$1000\,day) binaries appears to prevent red supergiants from getting too cool. The details of the RLOF and CEE experienced by these stars is described in Section 2.1.2. Binary interactions tend to only occur for the higher mass tracks at this orbital period, with RSGs below 12\,M$_{\odot}$  largely unaffected by being in such a wide binary. The reason being that while the RSGs have similar temperatures, the more massive and luminous RSGs are larger in radius and thus more likely to experience a binary interaction.

The next stars to consider are BSGs. These are post-main sequence objects that lie just off the main sequence at high luminosities as the stars travel through the Hertzsprung-gap. \citet{2015A&A...581A..36M} investigated these stars using rotating models; here we investigate similar evolutionary tracks but for binary models. We demonstrate in Figure \ref{FigJJ6} that we reproduce the  trend in bolometric luminosity versus flux weighted gravity for these sources. The BSG models contributing to contours are selected from  stellar models that are initially more massive than 8\,M$_{\odot}$, have $M_{\rm bol}\le -4$, $4.4 \ge \log(T_{\rm eff}/{\rm K})\ge 3.9$, $X_{\rm surf}\ge 0.4$, $\log g \le 3.5$ and have completed core hydrogen burning. The observed BSG sample  \citep{2008ApJ...681..269K,2008ApJ...684..118U,2009ApJ...704.1120U,2012ApJ...747...15K,2013ApJ...779L..20K,2014ApJ...785..151H,2014ApJ...788...56K} is split between the three plots at $\log(Z$/Z$_{\odot})\le -0.55$ for $Z=0.004$ and $\log(Z$/Z$_{\odot})\ge -0.2$ for Z=0.020 and in between these limits for $Z=0.008$. In general the observed BSGs lie over the contours of greatest probability in our models. At the highest bolometric luminosities and at $Z=0.002$ the data deviates to  slightly higher flux-weighted gravity than our models predict. At $Z=0.020$, this trend reverses, with the data suggesting slightly lower flux-weighted gravities.  This may result from a small discrepancy in mass, or perhaps metallicity between models and data.

\subsubsection{Wolf-Rayet stars}
Further comparisons can be made to WR stars, especially those that are hydrogen-free. We show the HR diagrams for these stars in the panels of Figure \ref{FigJJ7a} at three different metallicities and for single stars and binary populations. Here the contours represent where we expect to observe completely hydrogen-free stars which corresponds to when the tracks are red. The binary tracks in the right-hand panels now have an initial period of 100 days with a mass ratio of 0.5. These are compared against observed WR stars \citep{2002A&A...392..653C,2006A&A...457.1015H,2012A&A...540A.144S,2013A&A...559A..72T,2014A&A...565A..27H,2015A&A...581A..21H,2017A&A...598A..85S}.  Figure \ref{FigJJ7a} illustrates a known outstanding problem in that many WC stars and WN stars have cooler surface temperatures than predicted by stellar evolution models \citep{2003A&A...410..993H,2012A&A...540A.144S}.  One possible explanation is that clumping in the outer convective zone of the star may inflate the envelope \citep{2012A&A...538A..40G,2016MNRAS.459.1505M}. When we look at the comparison between our stellar models and observed WR stars, especially for hydrogen-free objects, in Figure \ref{FigJJ7a} we see the luminosity distribution is reproduced, but we see the expected temperature offset at LMC and Galactic metallicities, especially for the hydrogen-free WC stars (shown as triangles). Interestingly we do not see so large an offset in the SMC where the abundances of the observed sources are similar to those of nearby stellar tracks. 

\citet{2016MNRAS.459.1505M} found  that inflation can decrease the surface temperature of a WR stars by about 0.2\,dex. As can be seen, this is roughly the temperature decrease required to secure better agreement between the models and the observed WR stars. We do not include this effect in our current \bpass\ models as the physics is uncertain and we are still unclear whether the fitting method of the atmosphere models is entirely correct. If inflation effects are metallicity dependent this may explain why the offset is less pronounced at LMC metallicities. The relatively minor difference in temperature has little effect on the integrated spectrum of a stellar population, but we note that caution is needed when evaluating the properties of an individual WR star through model fitting.

Recently \citet{2017ApJ...841...20N} discovered a new class of WR stars, WN3/O3 stars. We have included them in the LMC panels in Figure \ref{FigJJ7a} as diamond symbols. We see that the models agree well with the observed luminosity, temperature and surface composition of both binary and single star stellar tracks. This suggests therefore that these are typical WR stars, however their reported mass-loss rates indicate  that the rates used in our models may be too high for such stars. We note that it has been suggested by \citet{2017arXiv170403516S} that these may be less massive helium stars rather than Wolf-Rayets, which would also be consistent with our binary models. However the projected distance from O stars they find of a few hundred parsecs could be understood if these objects are also runaway stars. The results of \citet{2011MNRAS.414.3501E} for type Ib/c SN progenitors suggest that 10 to 20\% of Wolf-Rayet stars could travel 100 parsecs, and further, away from their birth places (see their Figure A3).

\subsubsection{Quasi-chemically homogenous evolution}%
As discussed in section \ref{sec:method_evol}, a small subset of evolved stars are expected to undergo quasi-homogenous evolution and these models too require observational verification. QHE stars for the most part should look identical to more normal WR stars so will be difficult to separate out, and indeed there are no clearly unambiguous examples known. However their chemical composition may aid identification. Most typical WR stars form helium cores and then lose their hydrogen envelopes. By comparison QHE stars, when hydrogen rich, are still core-hydrogen burning so will exist for a longer period of time and also have different compositions at locations on the HR diagram close to the traditional hydrogen-burning main sequence. 

In Figure \ref{FigJJ7b} we compare the QHE tracks (shown with bold lines and evolving towards higher temperatures) to standard binary evolution tracks (thin lines, evolving first towards lower temperatures). Contours give the probability distribution of all models in our population synthesis which experience QHE. We see that in general the WR stars are in better agreement with the standard evolution tracks than with QHE models. As noted by \citet{2015A&A...581A..21H} and \citet{2017A&A...598A..85S} it appears that mass-loss rates may have been slightly overestimated in current models. We see however that the lowest luminosity WR star in this sample is much cooler and more hydrogen rich than the other stars. Its location matches quite closely that of our 40$M_{\odot}$ QHE track. This star, AB2, is perhaps the best candidate for a QHE (by mass transfer) star in the nearby Universe and deserves further study \citep{2009A&A...495..257M,2015A&A...581A..21H}.

\begin{figure}
\begin{center}
\includegraphics[width=\columnwidth]{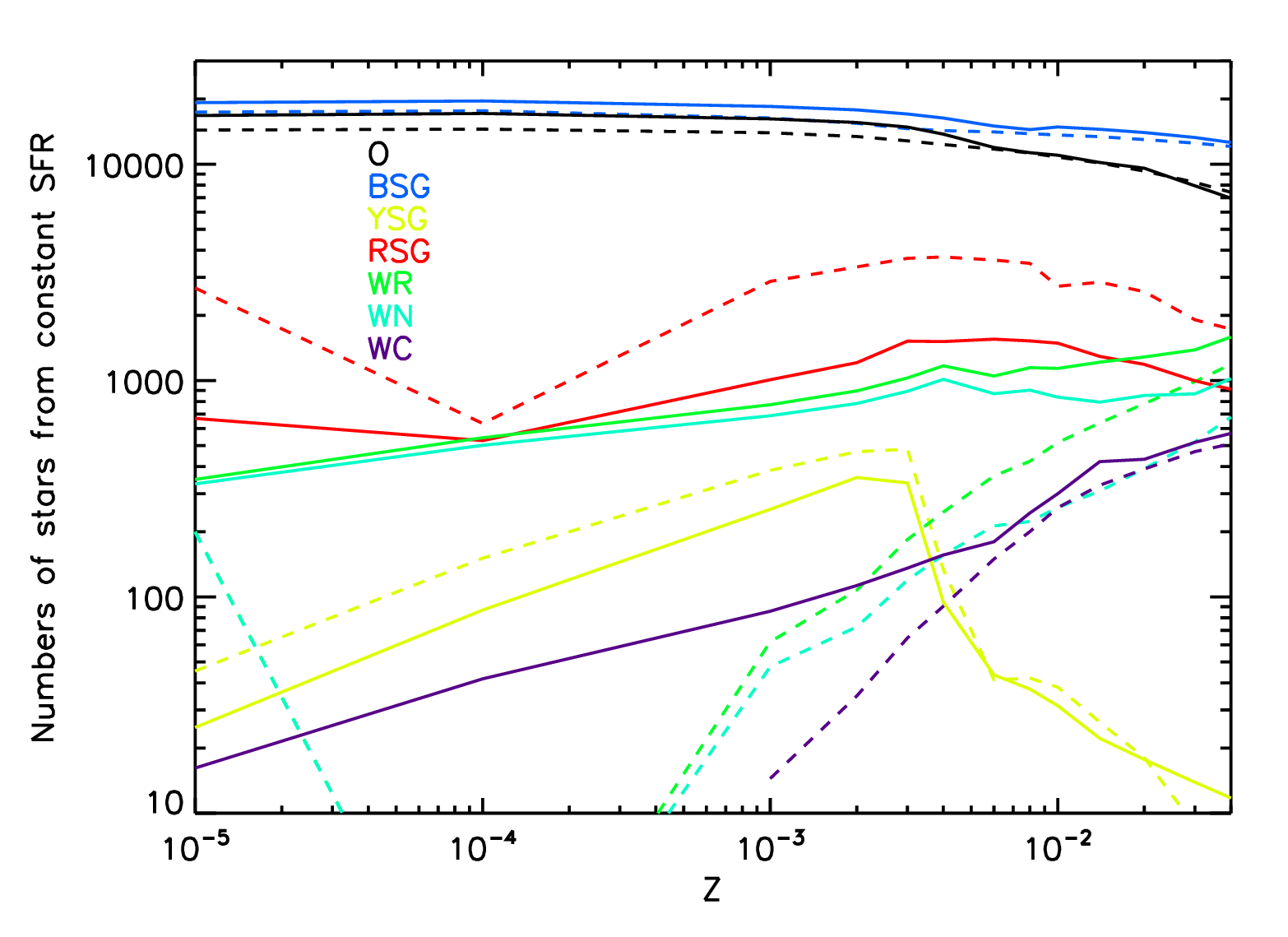}
\includegraphics[width=\columnwidth]{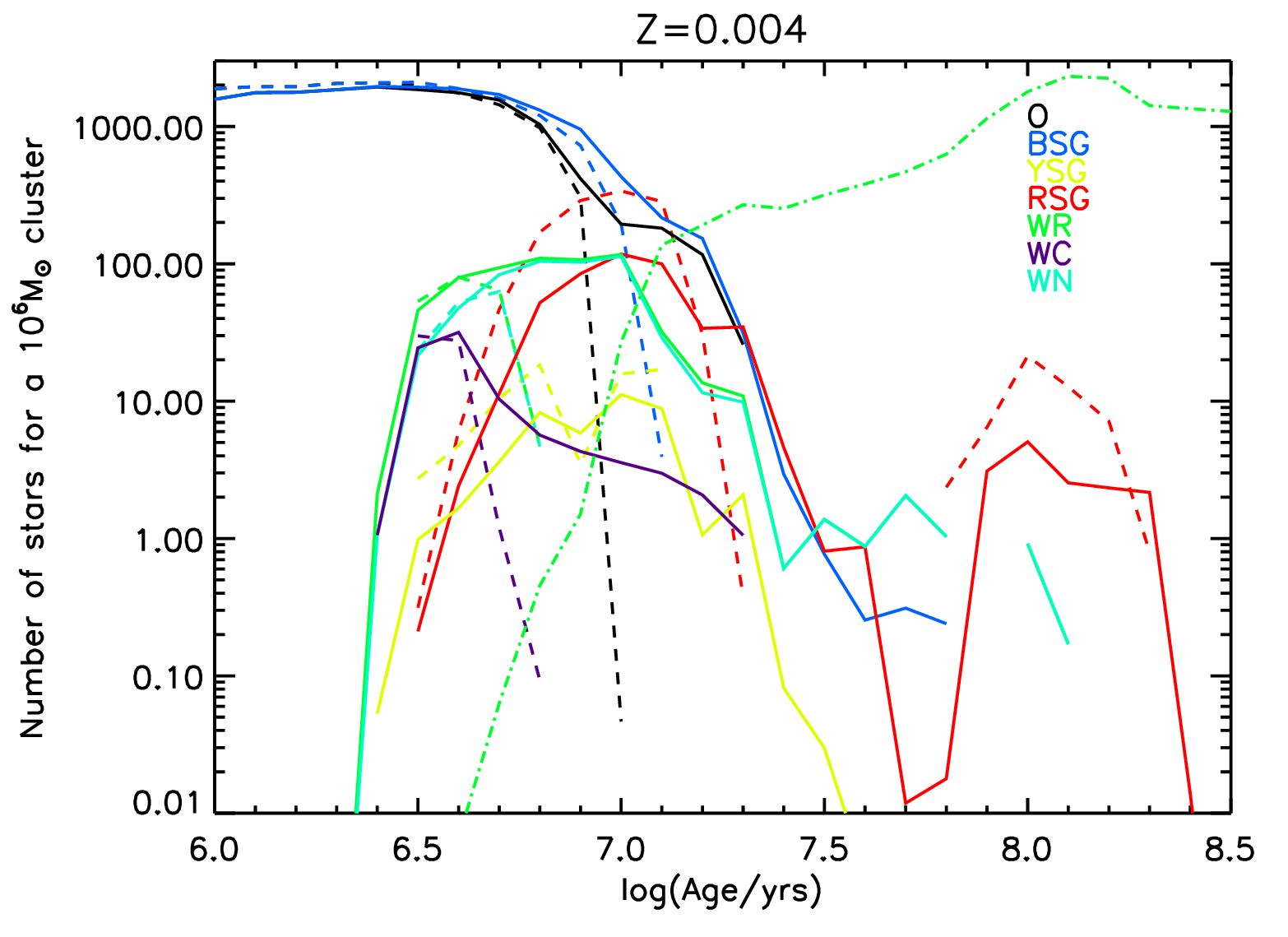}
\includegraphics[width=\columnwidth]{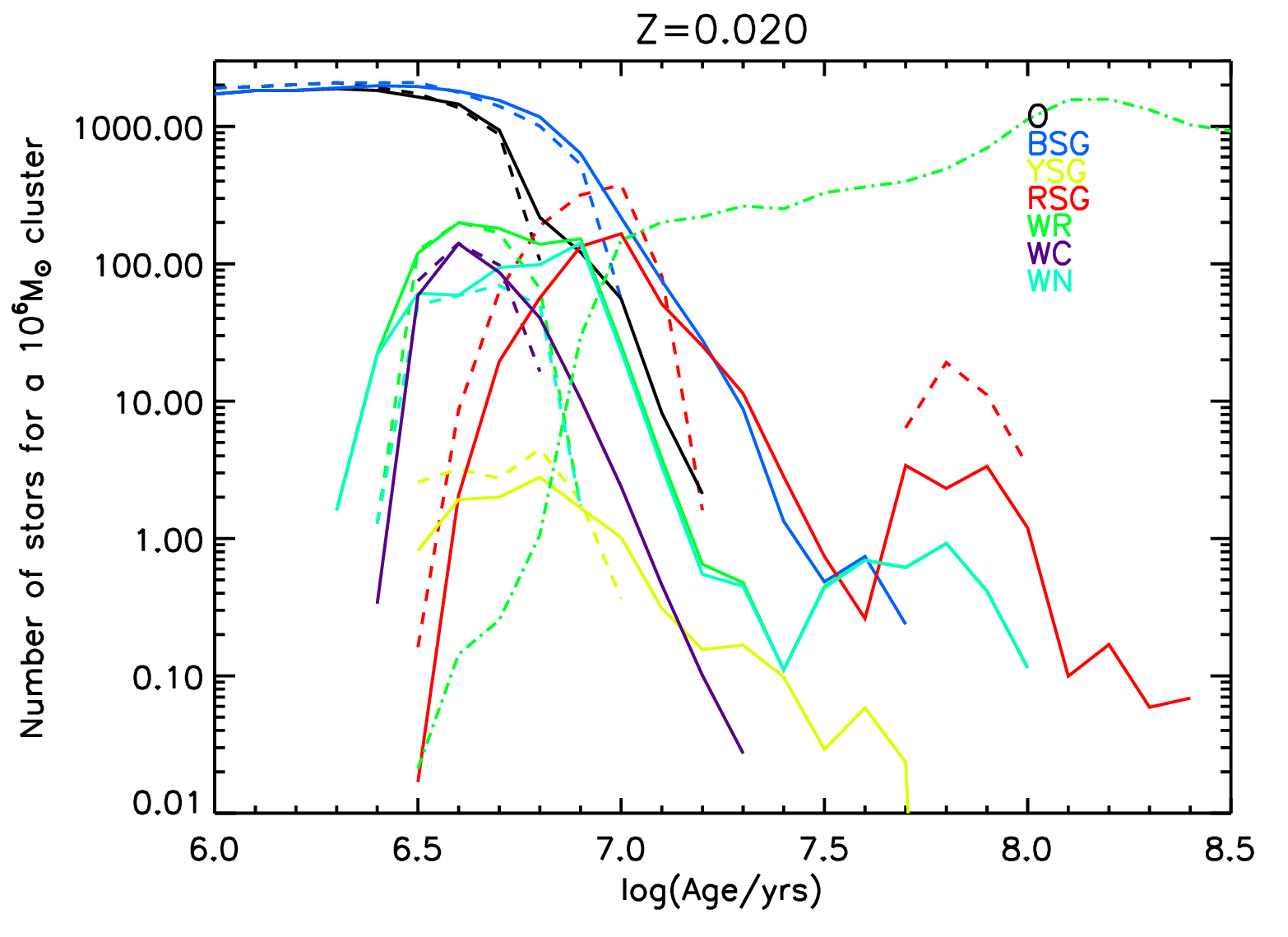}
\caption{Number of massive stars with log (L/L$_\odot$)$>4.9$ of different stellar types from our fiducial populations. Dashed lines are for single stars, solid binary populations. The dot-dashed line for the WR population includes lower luminosity, lower mass stars which otherwise satisfy the temperature and surface abundance criteria for WR stars. Top panel shows the metallicity variation of stellar type ratios assuming a population undergoing continuous star formation at a rate of 1\,M$_{\odot}{\rm \, yr^{-1}}$. Lower panels show time evolution of an instantaneous burst (i.e. a single-aged stellar population) at two different metallicities.  Where lines are interrupted, zero stars in that type category remain in the (finite-sized) population in a given time bin.}\label{FigJJ9a}
\end{center}
\end{figure}

\begin{figure*}
\begin{center}
\includegraphics[width=\columnwidth]{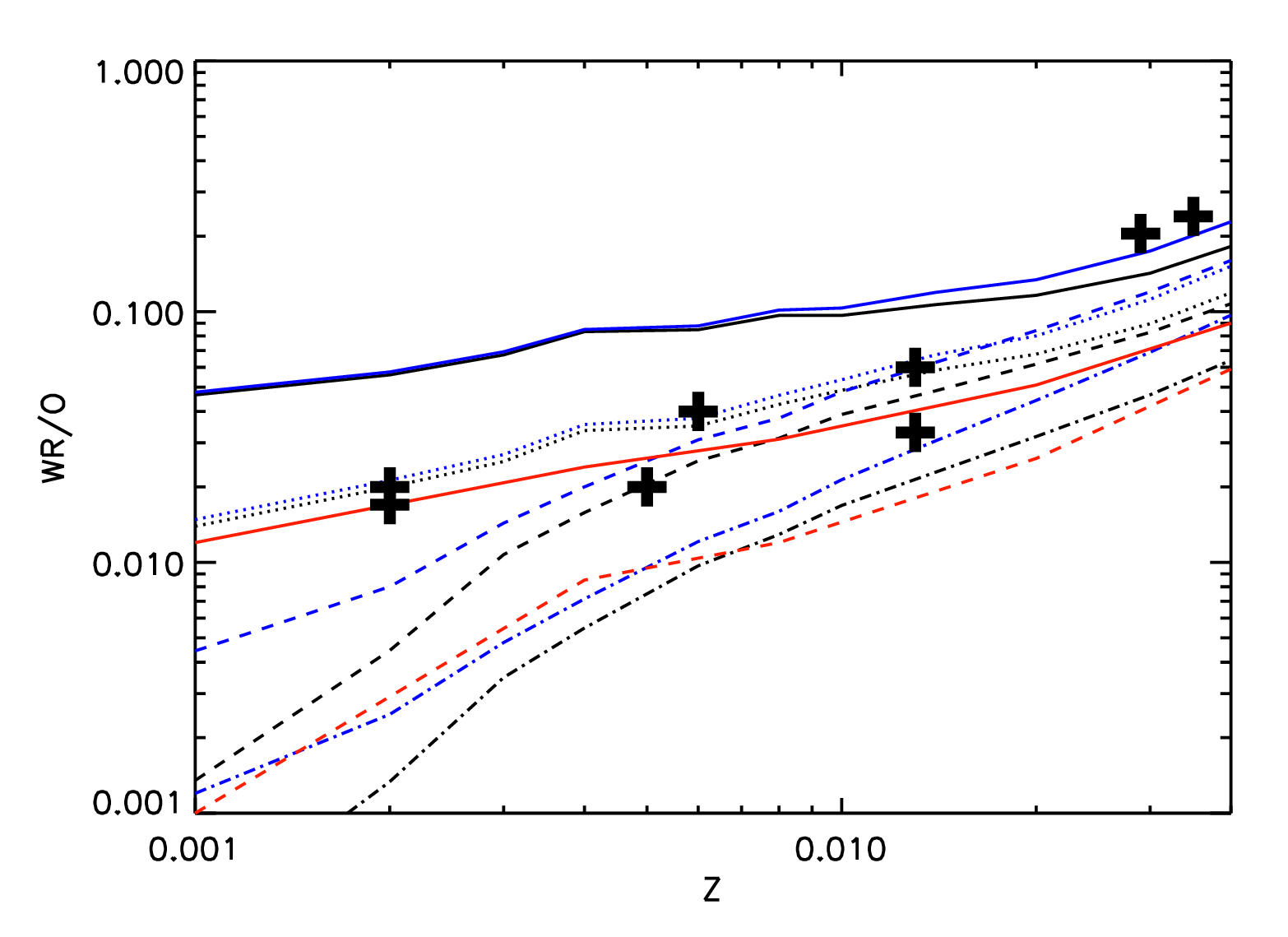}
\includegraphics[width=\columnwidth]{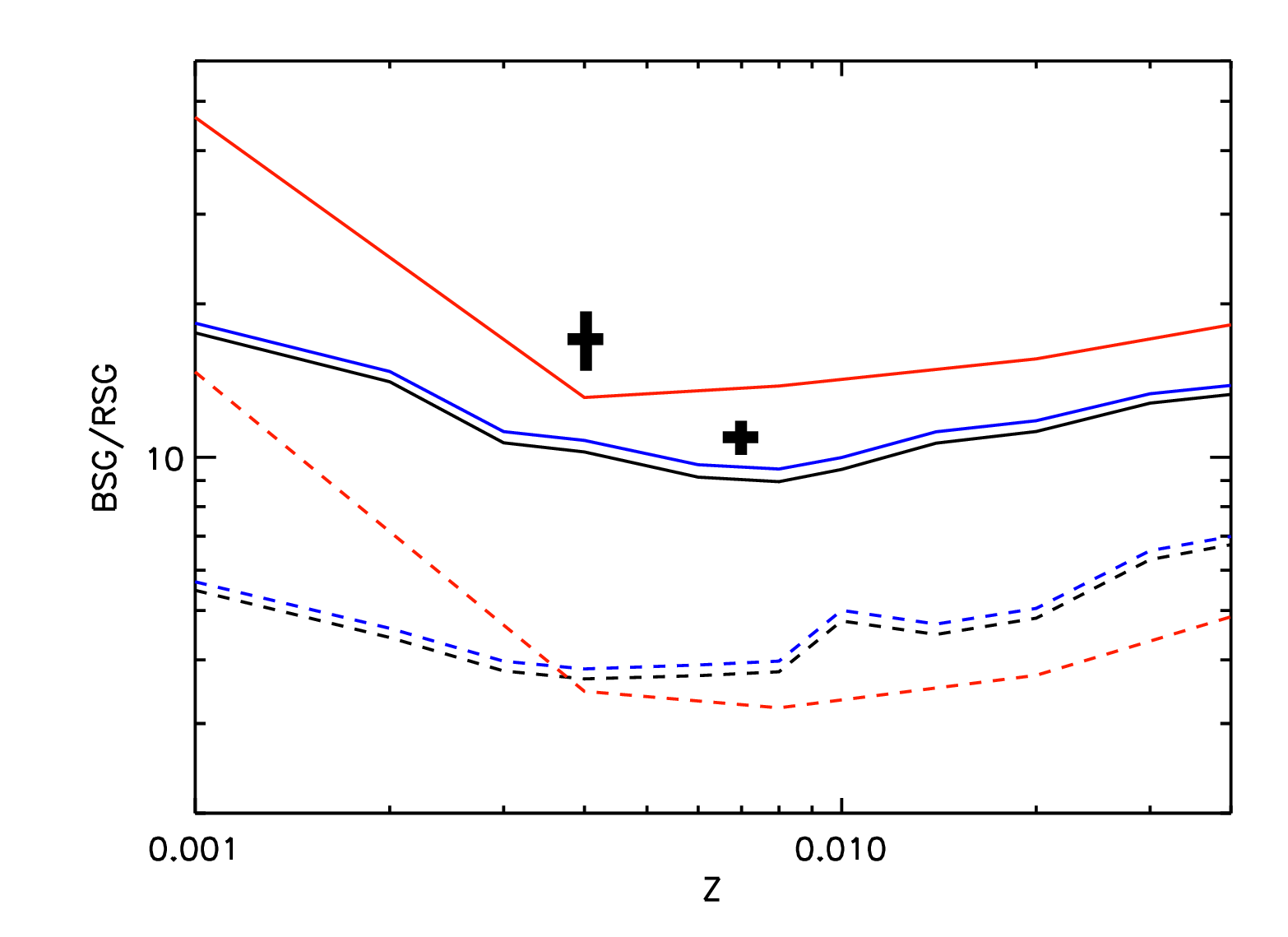}
\includegraphics[width=\columnwidth]{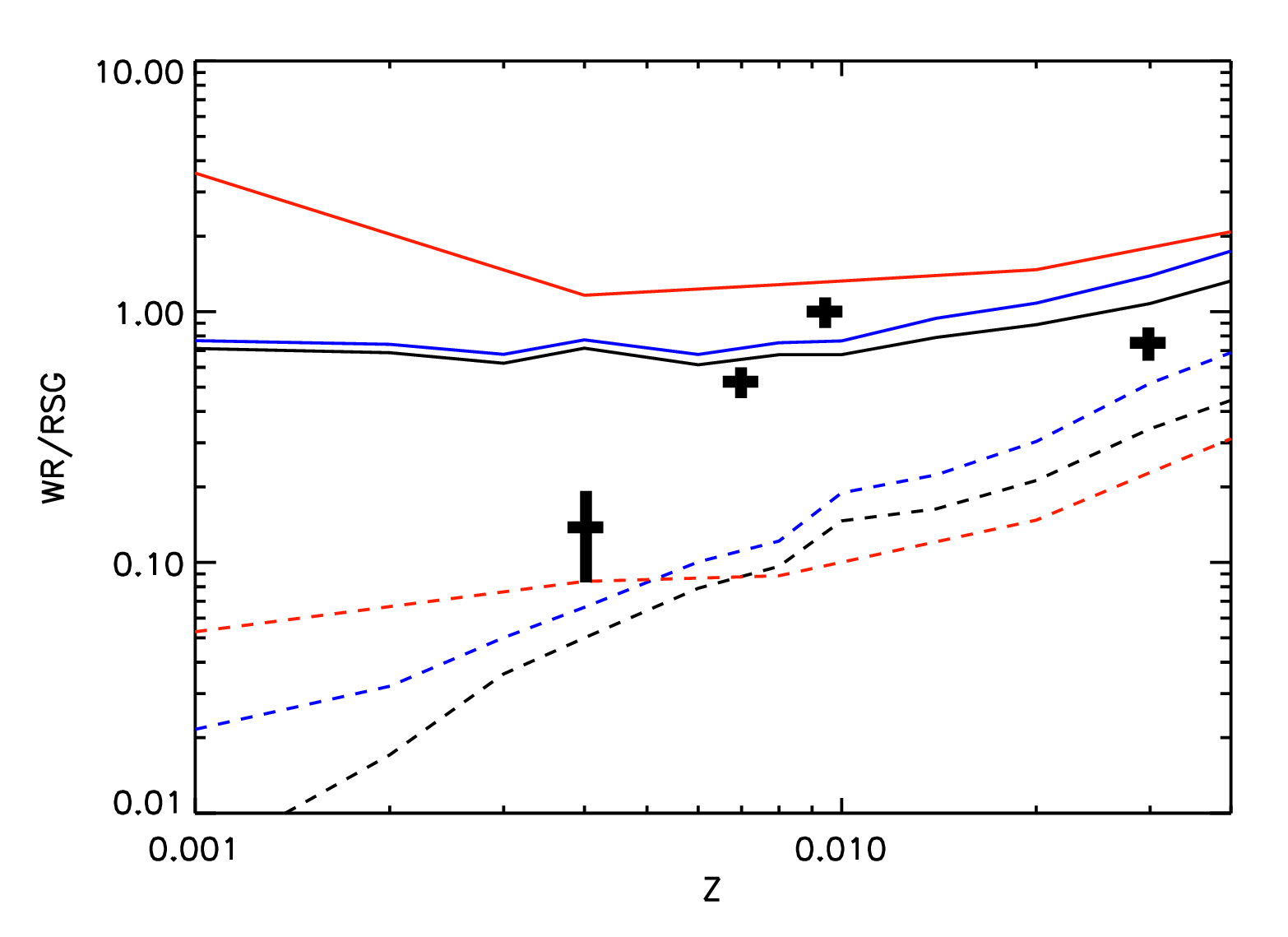}
\includegraphics[width=\columnwidth]{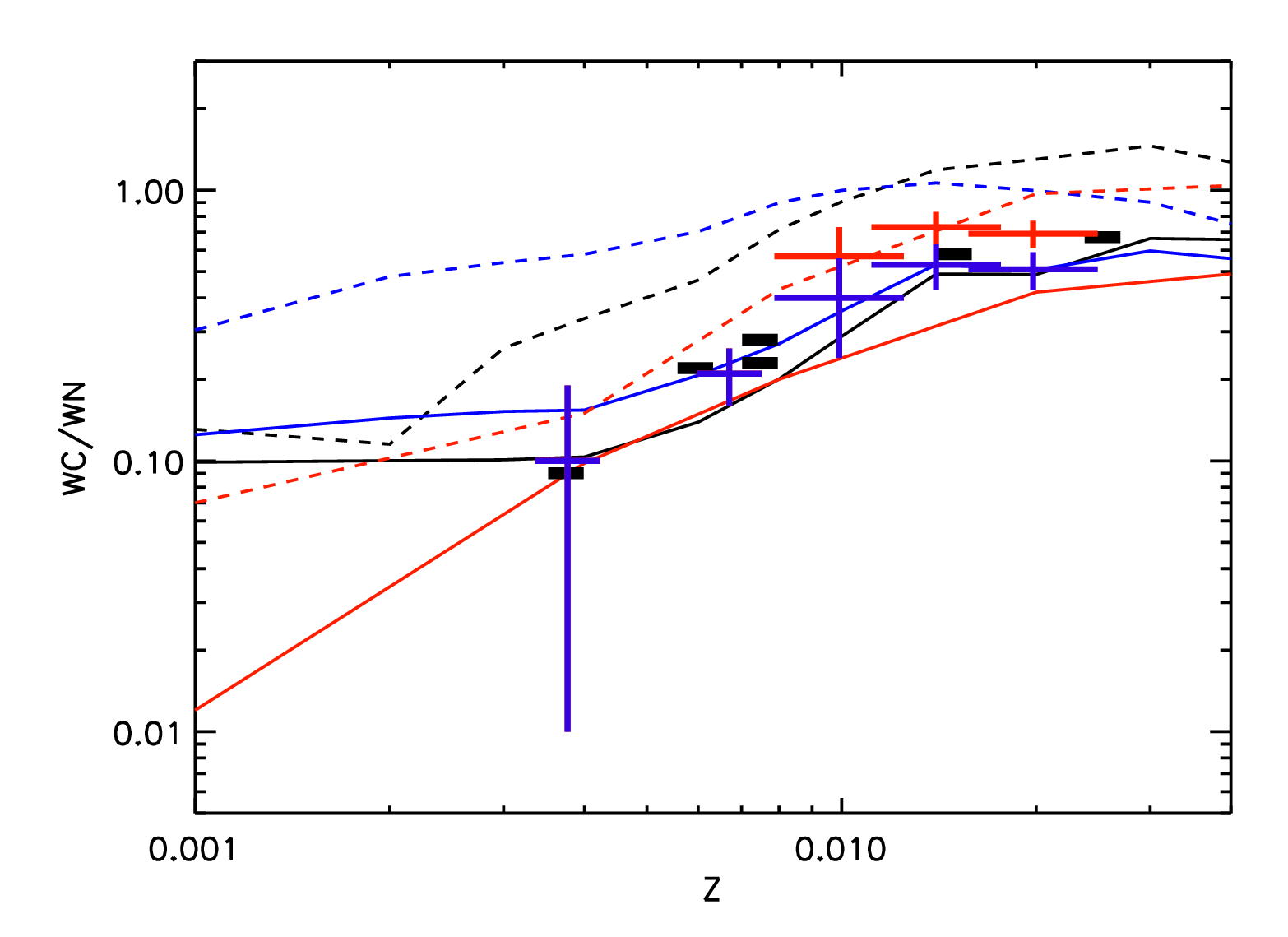}
\caption{Stellar type number ratios for massive star populations. Solid lines show \bpass\ binary models, and dashed lines show single models. Points with error bars show observational constraints taken from the literature. All ratios include only stars with log (L/L$_\odot$)$>4.9$, except the WR/O star ratio where the dotted (binary) and dash-dotted (single) lines include all O-stars. The red lines show results for constant star formation from BPASS v1.0 while the blue and black lines show similar results from BPASS v2.1 with two different IMFs (M$_\mathrm{max}$=300 and M$_\mathrm{max}$=100\,M$_\odot$ respectively). WR/O ratio data is taken from \citet{1994A&A...287..803M}, the BSG/RSG ratio data is from \citet{2003AJ....126.2867M} and the WR/RSG data comes from Phil Massey (private communication). The WC/WN ratio data plotted in blue and red and is taken from \citet{2015MNRAS.447.2322R} with the blue points omitting the dusty WC stars. The WC/WN ratios plotted in black are from \citet{2005A&A...429..581M}.}
\label{FigJJ9b}
\end{center}
\end{figure*}

\subsubsection{Stellar type ratios}

\begin{table*}
\caption{Assumed parameters for identifying different stellar types in the number counts.} \label{tab:stellartypes}
\begin{center}
\begin{tabular}{llcl}
Type & &$\log(T_{\rm}/{\rm K})$  &  Other \\
\hline\hline
O &(BSG)  & $\ge 4.48$   &  --\\
Of &(BSG) & $\ge 4.519$  &  $\log g < 3.676 \log(T_{\rm}/{\rm K})+13.253$ \\
B &(BSG)  & 4.48 to 4.041  &--\\
A &(BSG)  & 4.041 to 3.9  &--\\
FG &(YSG) & 3.9 to 3.660    &--\\
K &(RSG)  & 3.660 to 3.550  &--\\
M &(RSG)  &$<$ 3.550   &--\\
WNH & (WR)   &  $\ge 4.45$  & $X_{\rm surf} \le 0.4$\\
WN  & (WR) & $\ge 4.45 $  & $X_{\rm surf} \le 10^{-3}$, $(n(C)+n(O))/n(He)\le 0.03$. \\
WC  & (WR) & $ \ge 4.45$  & $X_{\rm surf} \le 10^{-3}$, $(n(C)+n(O))/n(He) > 0.03$. \\
\end{tabular}
\end{center}
\end{table*}

An important validation test to consider is not just whether we reproduce the location of these stars in the HR diagram but whether we reproduce the correct relative numbers. We show in Figure \ref{FigJJ9a} the predicted numbers of massive stellar types (log(L/L$_\odot$)$>4.9$), specifically O stars, BSG, YSG, RSG, WR (total), WN and WC stars as a function of metallicity for constant star-formation and also with age assuming an instantaneous starburst. O stars and BSGs are the most populous classes at all metallicities, as expected for the stars that exist when stars burn hydrogen to helium. The numbers of the remaining subtypes vary smoothly with metallicity. In general the WR stars decrease in number at lower metallicities since the formation of these requires mass loss and the action of stellar winds is less efficient in the absence of metals. We note that binary populations, which provide another avenue for mass loss, exhibit a weaker metallicity trend than the single star population.

Considering the evolution of an instantaneous burst, we see that the different stellar types appear at different stellar population ages. Binary interactions extend the lifetime of the WR stars, especially WC stars, to later times. This highlights the difficulty of precisely aging a stellar population from the fact that WR stars exist alone. In our RSG numbers we see a late time peak at around 100Myrs in both metallicities shown in the figure. This is due to a small contribution of asymptotic giant branch stars at these ages which meet the temperature and luminosity constraints for selection as RSGs in our model. Observationally they would also be difficult to distinguish.

Similar predictions to these have been used for many years to constrain stellar models by comparing the model to the number of different stellar types observed in an entire galaxy, ideally at known metallicity \citep[e.g.][]{1994A&A...287..803M}. Here we perform a similar test, comparing \bpass\ predictions of the WR to O, BSG to RSG, WR to RSG and WC to observed WN ratios in nearby galaxies (see figure caption for data sources). One problem with such comparisons is uncertainty in how complete each observational sample is, especially for the WR to O star ratio where both stellar types are hot and difficult to find in optical surveys. The comparison will also be somewhat dependent on the star formation history and here we assume continuous star formation at a constant rate. 

The panels in Figure \ref{FigJJ9b} show the current \bpass\ v2.1 predictions (black and blue lines given two IMF mass limits, $M_\mathrm{max}=100$ or 300\,M$_\odot$) for these ratio in comparison to the observed ratios which exist and also previous \bpass\ v1.0 predictions (red line).  For all these number and stellar type number ratios we only include stars with $\log(L/L_{\odot})>4.9$, since \citet{2003AJ....126.2867M} used this as a luminosity cut-off to make sure that AGB stars are not included in their sample (although we note that a small number of our AGB models do indeed satisfy this criterion). WR stars do not typically have luminosities below this value \citep{2016MNRAS.459.1505M}. It is consistent to also apply this to our O star number predictions for comparison to the data, but we note that this excludes the many O stars that lie just below this limit in our models. In the upper left panel of \ref{FigJJ9b} we include predictions for the WR/O ratio that include all O stars as well as only those above our luminosity limit (dotted and dot-dashed lines for binary and single stars respectively).

At Solar metallicity the effect of binaries, and of omitting the lower luminosity O stars, is small. These effects grow at decreasing metallicity, since stars at low metallicity have hotter effective temperatures for the same mass, and thus more O stars are present in the population. Comparison to the observed ratios suggests that models including all main-sequence O stars provide a good agreement to the data, except in the highest metallicity cases. The reason for this is probably related to the fact that the observed sample is old  and may suffer from significant incompleteness. Modern, more complete, surveys may find quite different results, but it is encouraging that \bpass\ models reproduce the observed trend in the observed ratios.

The BSG/RSG ratio observational sample \citep{2003AJ....126.2867M} only contains ratios for the SMC and LMC. Again binary evolution predicts values within a factor of a few of the observed ratios, but somewhat underpredicts the observed BSG fraction, likely due to the lack of rotational mixing in our models. More rotational mixing would lead to more BSGs \citep{2014A&A...570L..13C} and so we would expect their inclusion to boost our predicted ratio in future and consider the fraction in \bpass\ v2.1 to be a lower limit. We note that the MIST models \citep{2016ApJ...823..102C} predict a higher BSG/RSG ratio than \bpass.

The WR/RSG ratio was presented in our first study with the \bpass\ stellar models \citep{2008MNRAS.384.1109E}. In v1.0 it was our worst fit of the stellar population ratios since the observational data showed a pronounced trend in ratio with metallicity which the models did not reproduce. Here we compare our models to updated observational numbers (P. Massey, private communication). We see that our binary populations are far closer to the new observed ratios which show a significantly weaker trend. The observational data now shows very little variation in population ratio with metallicity above $\sim$0.3\,Z$_\odot$, consistent with the ratios in our models. In the \bpass\ v2.1 model population, the ratio continues to remain flat at lower metallicities. In the observational data set, the SMC is an outlier. However this ratio is based on only 13 WR stars, suggesting a considerable uncertainty in small number statistics, and potential incompleteness. Alternatively, as the observed data point lies between the single star and binary population predictions from \bpass\ it is possible that in the case of the SMC, the effective binary fraction could be smaller.

Finally, the WC/WN ratio is the focus of our final comparison between stellar populations and is primarily determined by the WR mass-loss rates. In this figure we show the latest observations from \citet{2015MNRAS.447.2322R} as well as the older sample from \citet{2005A&A...429..581M}. In \citet{2008MNRAS.384.1109E} we found that \bpass\ binary WC/WN ratio predictions lay below what was then thought to be the observed ratio. \citet{2007ApJ...662L.107V}  discussed in detail how binary populations affect this ratio: binary interactions produce WR stars from low-mass stars which would not otherwise evolve into the WR phase in the single star evolution case. The products of such evolutionary pathways tend to be WN stars rather than WC stars and hence more interacting binaries in a model set causes a drop in the WC/WN ratio. Recent observational surveys presented here show that the ratio is indeed lower than previously thought, in closer agreement to the binary populations predictions. The excellent agreement shown in the bottom right panel of Figure \ref{FigJJ9b} is strong circumstantial evidence that binary interactions are responsible for a good number of WR stars and that single-star stellar winds are not strong enough to create every WR star we see in the sky.

In summary our population models are consistent with the observed location of post-main sequence massive stars on the HR diagram, as well as the relative numbers of stars we see in galactic stellar populations. This is important as RSG and WR stars have unique spectral signatures in the emergent spectrum from a stellar population. RSGs dominate the longer redder wavelengths while WR stars are bright in the UV, produce hard ionizing radiation and also strong broad stellar emission lines. The ability of our models to reproduce the stars in resolved populations is good evidence that we likely predict these spectral features accurately.

\subsection{Stellar Remnants}

Compact stellar remnants are the end states of virtually all stars. Since their formation depends primarily on core mass and composition at the end of nuclear burning, their populations can be affected by binary interactions during and after the main sequence lifetime of their progenitors. Thus the white dwarfs, neutron stars and black holes in a  stellar population present a test of models such as \bpass.

\begin{figure*}
\begin{center}
\includegraphics[width=2\columnwidth]{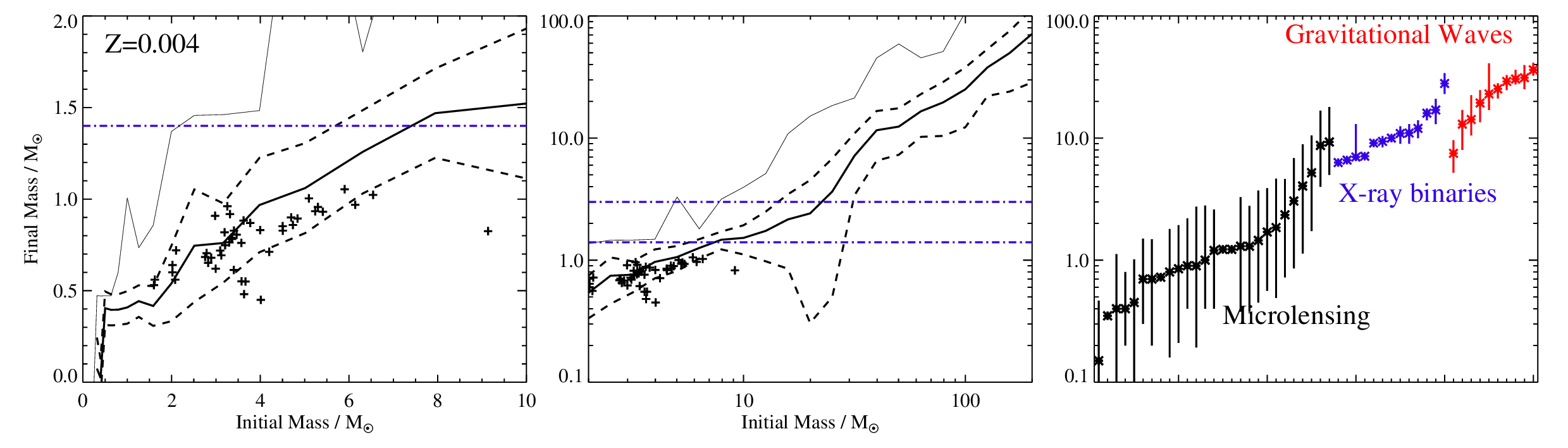}\\
\includegraphics[width=2\columnwidth]{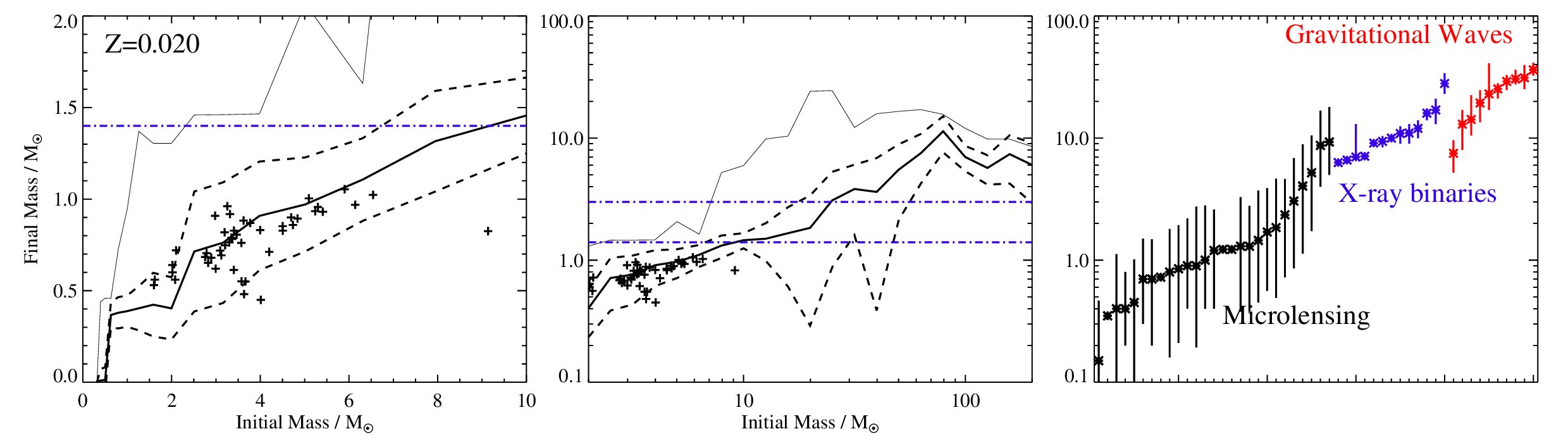}
\caption{Initial to final mass relations for compact remnants at two metallicities in the \bpass\ models. The solid central line is the mean remnant mass for that initial mass, dashed lines are the 1\,$\sigma$ lines and the upper solid line is the maximum remnant mass for that initial mass. This can exceed the initial mass due to binary interaction. The left and central panel show white dwarf data from \citet{2005MNRAS.361.1131F} and \citet{2009ApJ...705..408K} with the two panels scaled to show different mass ranges for clarity. The horizontal lines indicate the 1.4 and 3.0\,M$_\odot$ limiting masses above which we presume neutron stars and black holes to form respectively. The right-hand panels show the cumulative distribution of observed masses for compact remnants, i.e. the known population sorted by mass and plotted against an arbitrary index number. These are derived from gravitational microlensing within the Galaxy \citep[black,][]{2016MNRAS.458.3012W}, the observed mass of black holes in X-ray binaries \citep[blue, compiled by][]{2010MNRAS.403L..41C} and the 10 black holes identified as the sources in candidate gravitational wave transients to date \citep[red,][]{2016PhRvL.116x1102A,2016PhRvX...6d1015A,2017PhRvL.118v1101A,2017arXiv170909660T}. These demonstrate the range of observed final masses, but their initial masses are unconstrained.}\label{FigJJ10a}
\end{center}
\end{figure*}

The best extant observational data on remnants is on white dwarfs where observers have been able to calculate the observed mass of the white dwarf as well as its initial mass. This is only possible when the white dwarf is part of a stellar cluster from which the age can be estimated, together with the elapsed cooling time of the white dwarf since it was formed. This provides an opportunity to test our initial to final mass relation against these observations. We show in Figure \ref{FigJJ10a} the observed relations \citep{2005MNRAS.361.1131F,2009ApJ...705..408K} compared to our model predictions at two metallicities. 

It is important to stress that with the introduction of binary models, the relation between these two parameters is no longer single-valued; there is now a range of possible final masses for any given initial mass. This may naturally explain some of the scatter that is seen in the relation in observed datasets - a result that is difficult to explain in the absence of binary interactions.

\begin{figure*}
\begin{center}
\includegraphics[width=2\columnwidth]{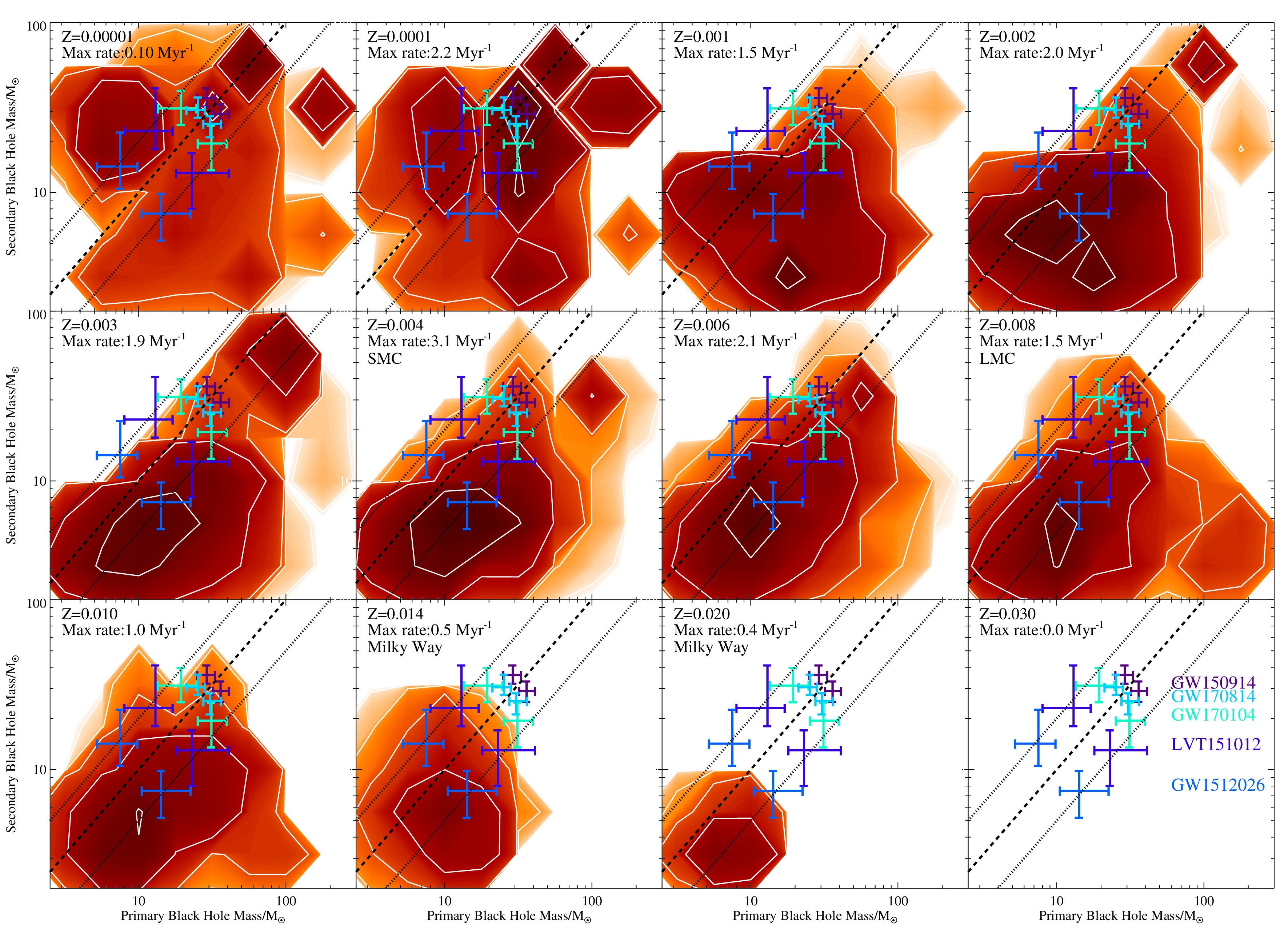}
\caption{The predicted masses and merger rate for black hole binaries as a function of metallicity, assuming continuous star formation at a rate of 3\,M$_\odot$\,yr$^{-1}$ over a 10 Gyr period. The primary is defined as the more massive star at formation (rather than merger). The masses of the black hole progenitors of the five detected gravitational wave transients (four confirmed events, one lower significance candidate) are indicated \citep{2015CQGra..32g4001L,2016PhRvL.116x1102A,2016PhRvX...6d1015A,2017PhRvL.118v1101A,2017arXiv170909660T}  - each appears twice depending on the evolutionary history. Lines indicate mass ratios of unity, 0.5 and 2.0.}\label{FigJJnewcont}
\end{center}
\end{figure*}

The final to initial mass relation for neutron stars and black holes is more difficult to estimate both observationally and in population synthesis models. As yet no black hole has been convincingly associated with a specific initial mass. This may be possible in future if one or more black hole binaries are located in co-eval stellar clusters of a known age, but  has yet to be achieved observationally. In Figure \ref{FigJJnewcont} instead we compare the range of remnant masses to three different observed data sets: the sample of \citet{2016MNRAS.458.3012W} from gravitational microlensing within the Galaxy, the observed mass of black holes in X-ray binaries \citep[compiled by][]{2010MNRAS.403L..41C} and lastly the 10 black holes identified as the sources in gravitational wave transients to date \citep{2016PhRvL.116x1102A,2016PhRvX...6d1015A,2017PhRvL.118v1101A,2017arXiv170909660T}. This sample remains too small to statistically recover the observed mass distribution, but it does provide constraints on the {\it range} of masses for black holes which can compared to the range of final masses in our models. As discussed in \citet{2016MNRAS.462.3302E}, formation of the most massive black holes  requires very low metallicity stellar populations. 

Compact object binaries lose energy through gravitational radiation. As a result, the binary orbit hardens until eventually the compact objects come into contact and merge, emitting a gravitational wave transient in the process. We have looked at the merger times of compact remnant mergers, and made predictions based on our baseline v2.0 models, in \citet{2016MNRAS.462.3302E}. Our rough volume average rate estimate for neutron star-neutron star mergers is 210 to 1600\,Gpc$^{-3}$\,yr$^{-1}$, for black hole-neutron star mergers 0.07 to 62\,Gpc$^{-3}$\,yr$^{-1}$ and for black hole to black hole mergers 8 to 120\,Gpc$^{-3}$\,yr$^{-1}$. These predictions are compared against the gravitational wave transients reported to date in Figure \ref{FigJJnewcont} - all five events are consistent with emerging from environments at $<0.5$\,Z$_\odot$, with the two most massive events most likely arising at significantly lower metallicities, in good agreement with our earlier work. Interestingly all events are consistent with a final mass ratio $<2$, suggesting that the parameter space for detectable mergers may be a little narrower than the population in our models. We note here that a number of uncertainties remain in making such predictions. Key amongst these are the remnant lifetimes and merger timescales. These are dependent on uncertainties such as supernova kicks, the resulting orbital eccentricity and CEE, all of which introduce associated uncertainty into predictions of event rates, even if the age distribution and star formation history of potential sources are known. As discussed in section \ref{sec:kicks}, we are currently investigating a new model for supernova kicks \citep{2016MNRAS.461.3747B} but this is still largely unconstrained by observations. Until we have a stronger constraint on the distribution of such kicks, and a larger observational sample for comparison and verification, we choose not to refine our predictions further here.

\section{SUPERNOVAE AND TRANSIENTS}\label{sec:tests_sne}

Massive stars can be defined as those that end their lives in a core-collapse supernova. Here we outline some of the outputs from our populations that concern  supernovae; namely the delay time distribution, the relative rates and the predicted location of SN progenitors on the HR diagram.

\begin{figure*}
\begin{center}
\includegraphics[width=2\columnwidth]{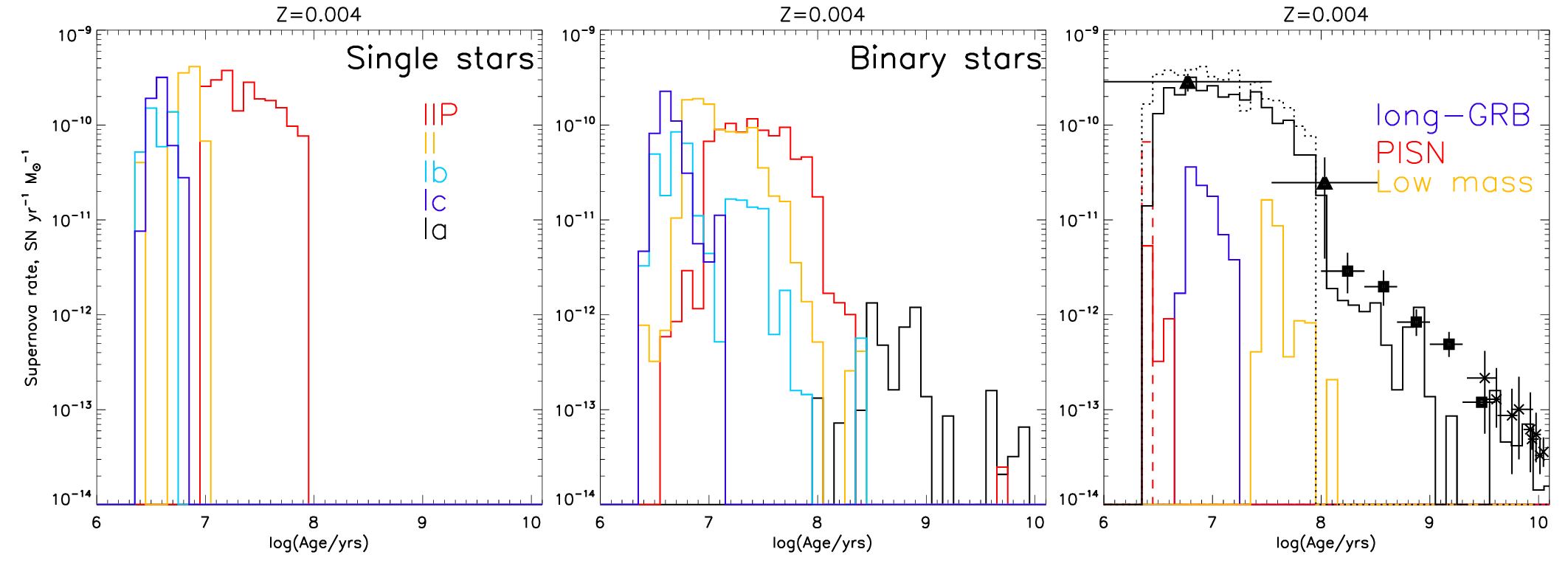}\\
\includegraphics[width=2\columnwidth]{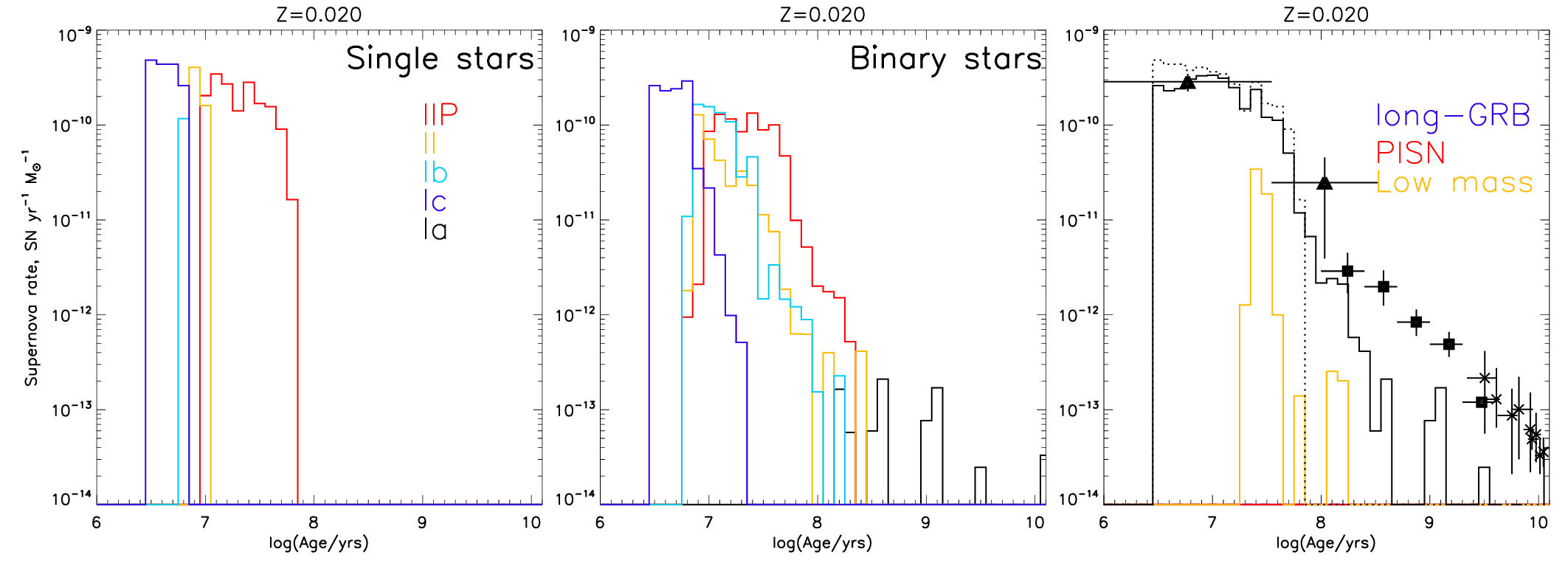}
\caption{Supernova rate distributions with stellar population age. In the left-hand panels we show single star models at two metallicities, while in the central panels we show the equivalent rates for a binary population.  The right-hand panels consider the total supernova rate (including type Ia events, black) with the solid line giving binary models and the dotted single star models.  The triangle points are the observed core-collapse supernova rates reported by \citet{2010MNRAS.407.1314M}. The data points indicate type Ia SN rates from \citet[crosses]{2010ApJ...722.1879M} and \citet[squares]{2008PASJ...60.1327T}}\label{FigJJ13a}
\end{center}
\end{figure*}

In Figure \ref{FigJJ13a} we show the delay time distribution for different core-collapse event types for our single star and binary populations at two different metallicities. We compare our predictions to observed delay-time distributions in a similar method to that performed by \citet{2017A&A...601A..29Z}. The left-hand panels show the predictions for single star populations. We see that, at Solar metallicity, the oldest supernovae occur at stellar population ages no older than 80\,Myrs, and there is a clear progression of supernova types from Ic to Ib, then II and IIP. At lower metallicity this sequence is less clearly defined. 

We consider binary populations in the central panels of Figure \ref{FigJJ13a}. With the presence of binaries the core-collapse SNe extend over a much broader rate of ages, continuing to occur up to a stellar population age of a few times 100\,Myrs. This is a common prediction from binary population synthesis as recently highlighted by \citet{2017A&A...601A..29Z}. There is still a general trend of SN types from Ic, Ib, II to IIP at both metallicities, but there is also now much more overlap. This is due to the existence of a variety of evolutionary pathways with different strengths of interactions, rather than a single pathway for each stellar mass. The binary predictions also show estimates for type Ia SNe, which become possible at late times as core-collapse SNe decline. The rate of type Ia events increases as the metallicity of the stellar population drops below Solar in our models. There may be some observational support for this; \citet{2012AJ....144..177Q} identified a tendency for white dwarfs to be observed more commonly in low luminosity dwarf host galaxies, which tend to be sub-Solar in metallicity in the local Universe. We stress again that our type Ia rates are currently highly approximate and do not include much of the detailed physics required (see section \ref{sec:snrs}). However it is encouraging that our estimated delay times for these event are consistent with when other population synthesis models suggest they should occur \citep[][and references therein]{2014ARA&A..52..107M}.

The final pair of panels in Figure \ref{FigJJ13a} (right) show rarer transient event types. They indicate that PISN and long-GRBs form by pathways considered in \bpass\ only occur at low metallicities and at the younger ages. It should be noted that our long-GRB rates represent a lower limit since there are certainly higher metallicity pathways to forming GRBs, which are seen up to Solar metallicity. These events likely appear in our models as a subset of the type Ic supernovae. We also show where low-mass type explosions may occur. These sources have progenitors between 1.5 and 2.0\,M$_\odot$ - the observational appearance of any transient resulting from these remains uncertain \citep{2016MNRAS.461.2155M}. However we have included them for completeness and we note that their total rate is low relative to core-collapse, or even type Ia events. Future work will attempt to simulate explosions in these stellar models, yielding a better idea of whether they should be included in the population synthesis of SN rates.

\begin{figure*}
\begin{center}
\includegraphics[width=2\columnwidth]{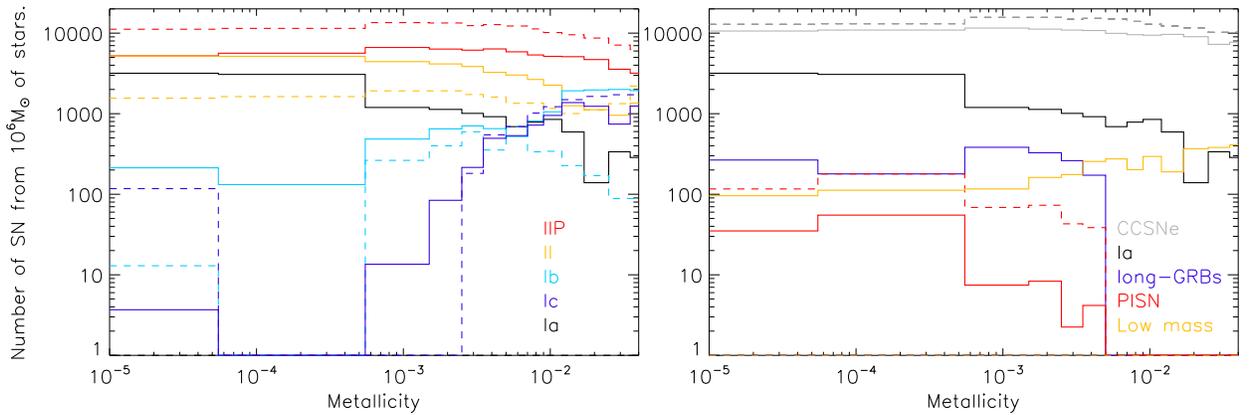}
\caption{Total number of supernovae arising from a 10$^6$ M$_\odot$ stellar population within 10\,Gyr of its formation, shown as a function of metallicity. Solid lines indicate binary populations, while dashed lines give equivalent single star populations}\label{FigJJ13b}
\end{center}
\end{figure*}

We now take the total number of core-collapse SNe arising over the first 10\,Gyr after the formation of $10^6$\,M$_\odot$ of stars at each metallicity, and compare them in Figure \ref{FigJJ13b}. We see the expected trend with metallicity as a consequence of the metal dependence of stellar winds driving mass loss. For example, there are more IIP SN at lower metallicity at the expense of the type Ib and Ic SNe, since fewer stars lose their hydrogen envelope before supernova. We also see the rate of type Ia SNe increasing at lower metallicity for the same reason. Binary interactions weaken this trend by providing alternate mechanisms for mass loss at lowest metallicities, but do not remove it.

Our predicted long-GRB and PISN rates are relatively stable below a metallicity of $Z=0.004$, at which they start to occur. Our predicted rates are, of course, subject to uncertainties, for example whether we include the low-mass events in our sample, or whether we include SNe where a black hole is formed rather than a neutron star remnant since for some of these the explosion may be concealed within the event horizon. \citet{2017ApJ...837..120G} compared our predicted ratios of type Ib/c and II SNe to the observed SN type ratios and found good agreement. However we caution that relative rates determined from observations are still somewhat dependent on sample completeness and observational bias and often have poor metallicity constraints.

\begin{figure*}
\begin{center}
\includegraphics[width=2\columnwidth]{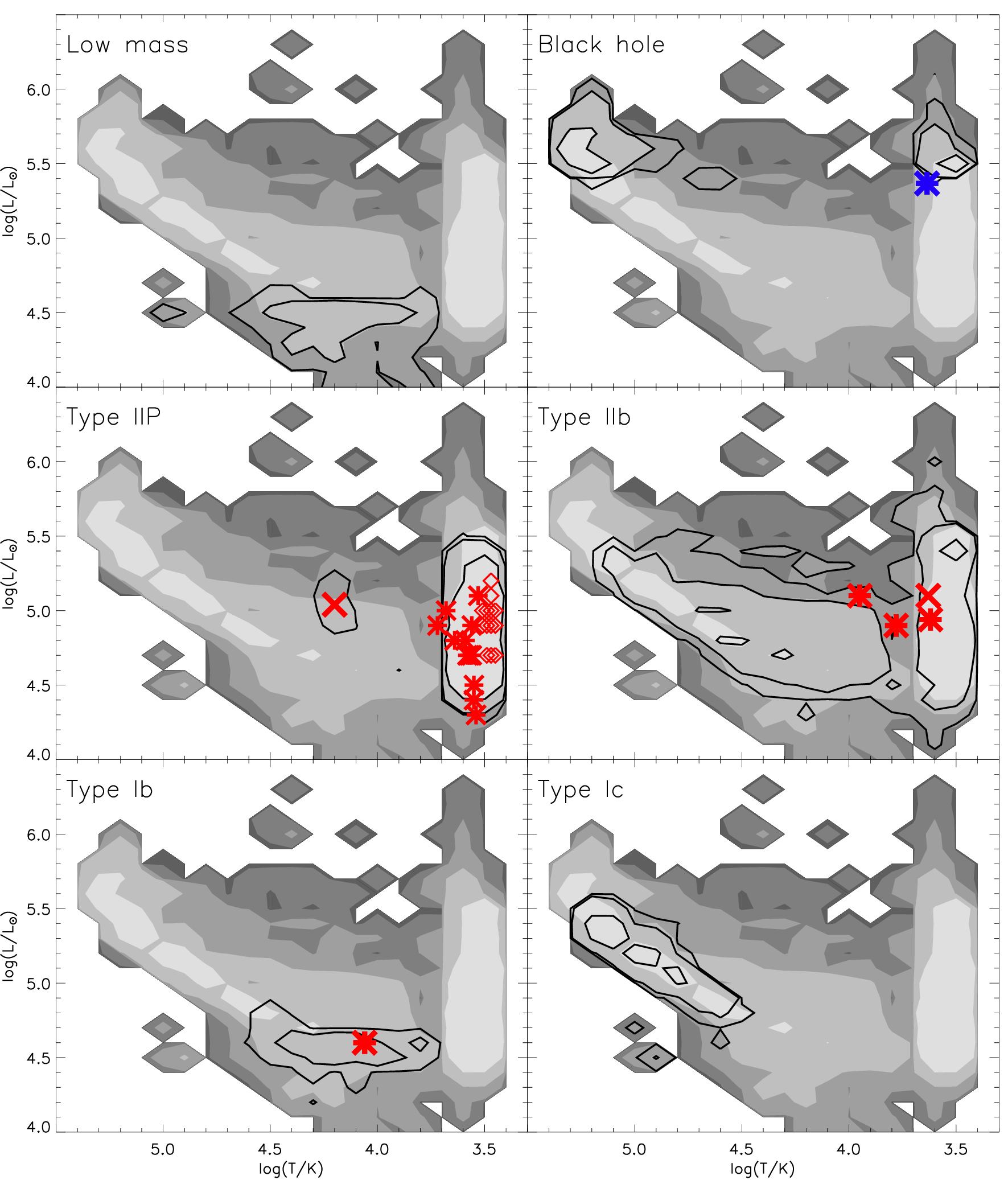}
\caption{Prediction of the final location of SN progenitors on the HR diagram. The blue asterisk is the RSG that vanished \citep[i.e. the candidate black hole formation event][]{2017MNRAS.469.1445A}, diamonds indicate pre-explosion upper limits on luminosity from \citep{2015PASA...32...16S}, the X in the IIP panel is the progenitor of SN1987A \citep{1987ApJ...321L..41W} and in the IIb panel is SN1993J \citep{1994AJ....107..662A}. Underlying greyscale contours give the total core-collapse SN progenitor distribution from a $10^6$\,M$_\odot$ stellar population given our assumed IMF, while overlying line contours indicate expected progenitors for each subclass. We average the populations from $Z=$0.008 to 0.020 to allow for metallicity spread in the observational data.}\label{FigJJ14}
\end{center}
\end{figure*}

We can also test our models by considering the observed progenitors of SNe. If a SN occurs within about 20 to 30\,Mpcs of the Sun, and the host galaxy has been imaged by the Hubble Space Telescope before the transient event, there is a very good chance that the progenitor star could be detected in a pre-explosion image. We show in Figure \ref{FigJJ14} the observed locations in the HR diagram for a sample of well-constrained progenitors, as well as the predictions from our stellar models as to where the SN should be. Here the underlying greyscale contour map represents the locations of all supernova events expected in a stellar population with our IMF regardless of type and when they occur, while the line contours on each panel select only those expected to produce a given SN subtype. For the observed datapoints use the sample of \citet{2015PASA...32...16S} supplemented by SNe 1987A \citep{1987ApJ...321L..41W} and 1993J \citep{1994AJ....107..662A}, as well as the recent `failed supernova' event in NGC\ 6946  \citep{2017MNRAS.469.1445A}. 

As Figure \ref{FigJJ14} shows, type IIP progenitor stars are both predicted to be and observed to be primarily RSGs. The one exception is the progenitor of SN1987A which is a BSG. The type IIb SN progenitors detected to date all appear to be YSGs or RSGs. Here there is overlap with our models but it is interesting that we also predict some progenitors that should be substantially hotter than we have observed to date. When the hydrogen envelope drops below about 1\,M$_\odot$ in mass, the surface temperature in our models can change rapidly for small amounts of further mass loss and so is very sensitive to time-step and exact conditions. The discrepency between models and observations for these stars may thus arise from the difficulty of predicting the correct surface temperature. Our models favour BSG or RSG rather than YSG progenitors. 

For the type Ib/c SNe only one progenitor has been observed, iPTF13bvn \citep{2013ApJ...775L...7C,2016MNRAS.461L.117E}. As described by \citet{2013MNRAS.436..774E}, we also have a large number of non-detections consistent with the prediction that the progenitors are all hot and faint in the optical (although UV-bright). The Ib progenitors while fainter than those of type Ic events, are cooler and thus brighter optically and so were the first to be observed \citep[as predicted by][]{2012A&A...544L..11Y}. 

Finally, we note that the progenitor of the only-known candidate black hole forming event \citep{2017MNRAS.469.1445A} was identified through the disappearance of a luminous star in the well-studied galaxy NGC\ 6946. This star was expected to go supernova and thus it has been suggested that any explosion energy was contained within the event horizon of a forming black hole at the star's core. The progenitor star was the most luminous RSG in the sample which fits with our understanding that more massive RSGs should form black holes and not have visible supernovae. This provides a test of our prescriptions for when black holes are formed as remnants within our model set.

\section{SPECTRAL SYNTHESIS VALIDATION: RESOLVED AND UNRESOLVED STELLAR POPULATIONS}\label{sec:tests_unresolved}

Population synthesis models such as ours  are used to interpret the integrated spectral appearance of unresolved stellar populations. However both stellar binaries and resolved stellar clusters provide an opportunity to explore the colours and spectral properties of individual stars within the context of a co-eval stellar population.

\begin{figure*}
\begin{center}
\includegraphics[width=\columnwidth]{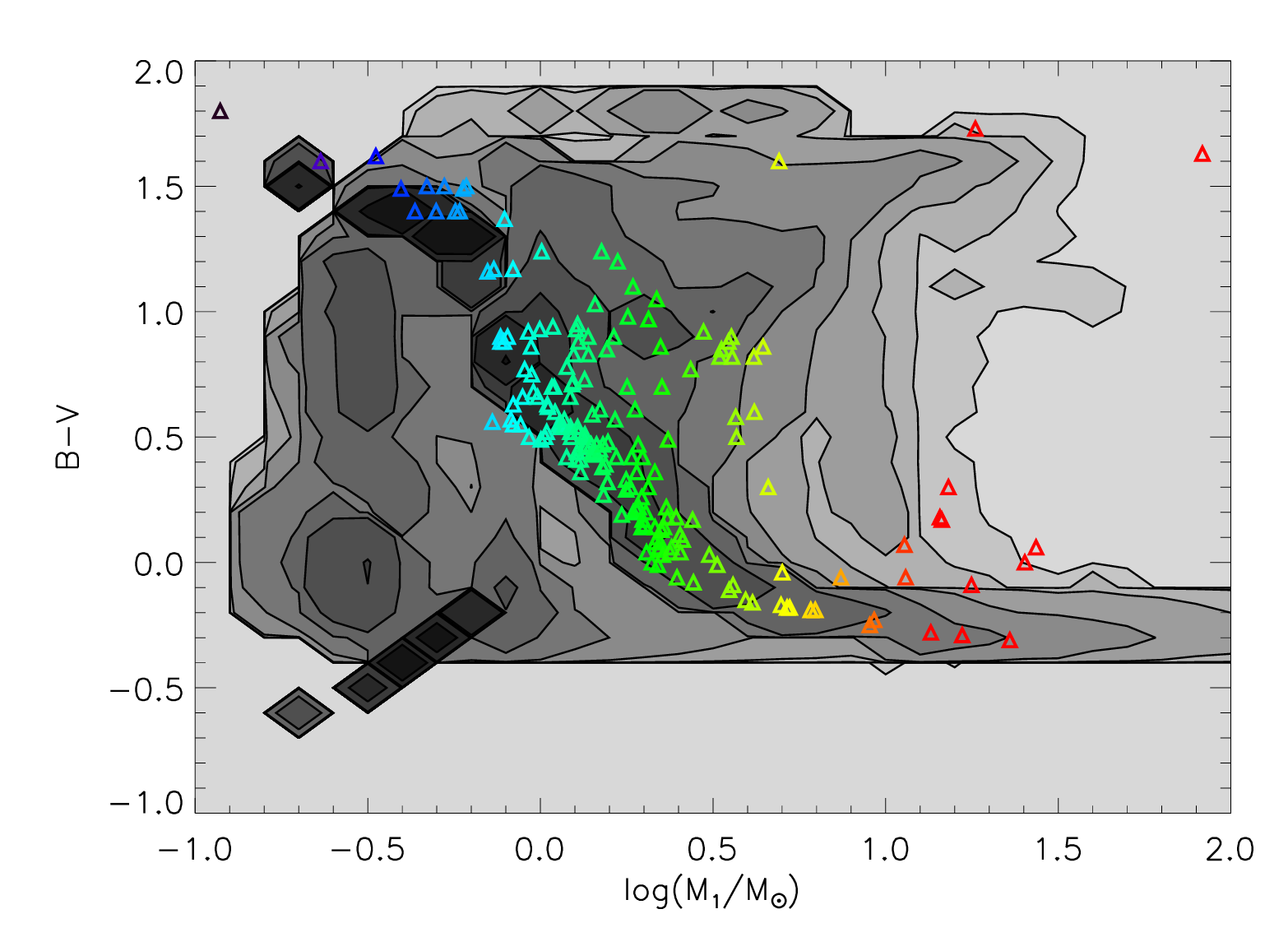}
\includegraphics[width=\columnwidth]{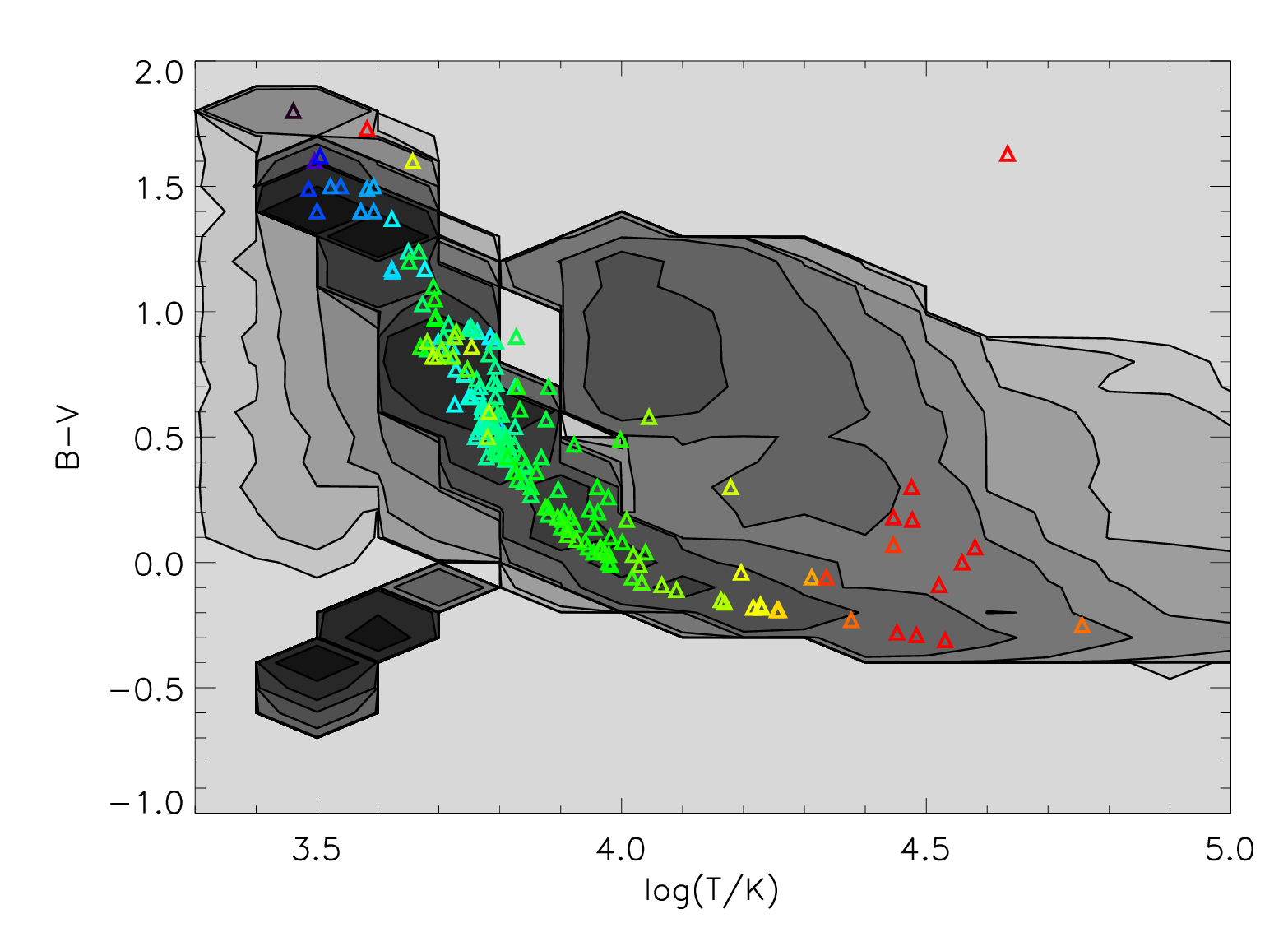}
\caption{The eclipsing binary star sample presented in Figure \ref{Fig2}, but now showing the B-V colour of the binary versus both the mass and surface temperature of the primary star. Observational data points are colour-coded by mass, while the underlying contours indicate the parameter space accessed by the same binary population as in Figure \ref{Fig2}.}\label{Fig3}
\end{center}
\end{figure*}

In Figure \ref{Fig3}, we revisit the sample of eclipsing binaries discussed in section \ref{sec:tests_resolved}. We  compare the $B-V$ colours of these sources to the synthetic colours predicted by our spectral synthesis models.  The observed colours are broadly consistent with the expected behaviour in both temperature and mass for these well-constrained  systems, with the models successfully reproducing the stellar main sequence and also accounting for much of the scatter in colour at intermediate masses.  There are, however, a few notable failures. In particular, the highest mass system in the observed sample (WR20a) shows colours considerably redder than predicted for its temperature and mass, while other massive stars are also a few tenths of a magnitude redder than might be expected. This likely reflects the fact that massive stars are typically found in young star forming regions behind large amounts of dust. In fact  WR20a is at distance of nearly 8kpc towards the Galactic centre and thus the colours are significantly reddened. Lower mass stars are typically in older regions with less dust attenuation, so it is unsurprising that these show better agreement with our models.

\begin{figure*}
\begin{center}
\includegraphics[width=2\columnwidth]{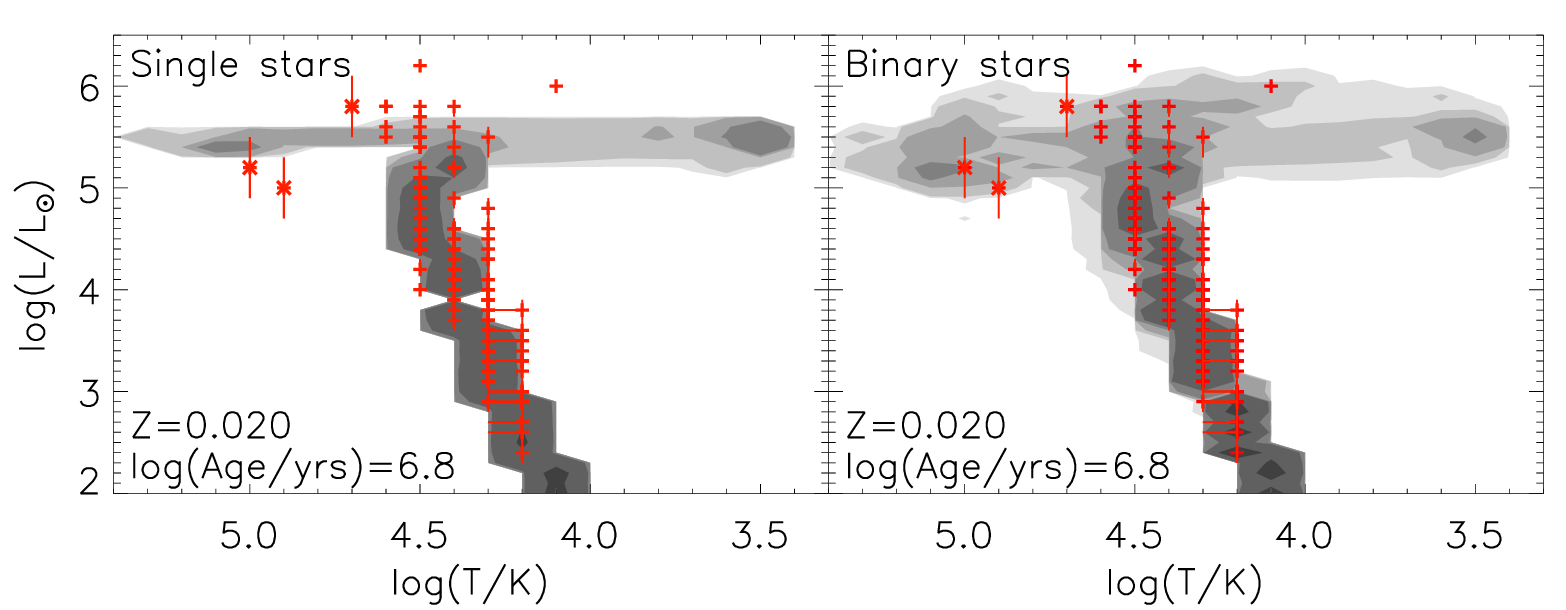}\\
\includegraphics[width=2\columnwidth]{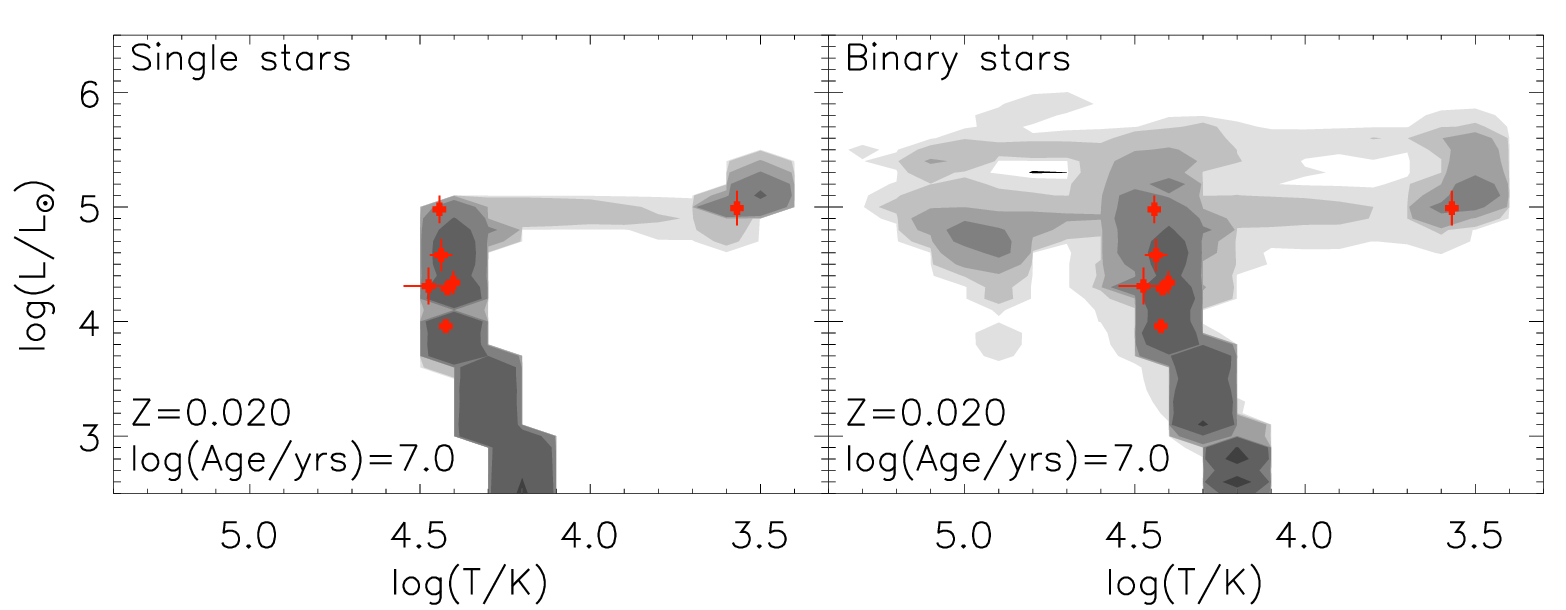}\\
\includegraphics[width=2\columnwidth]{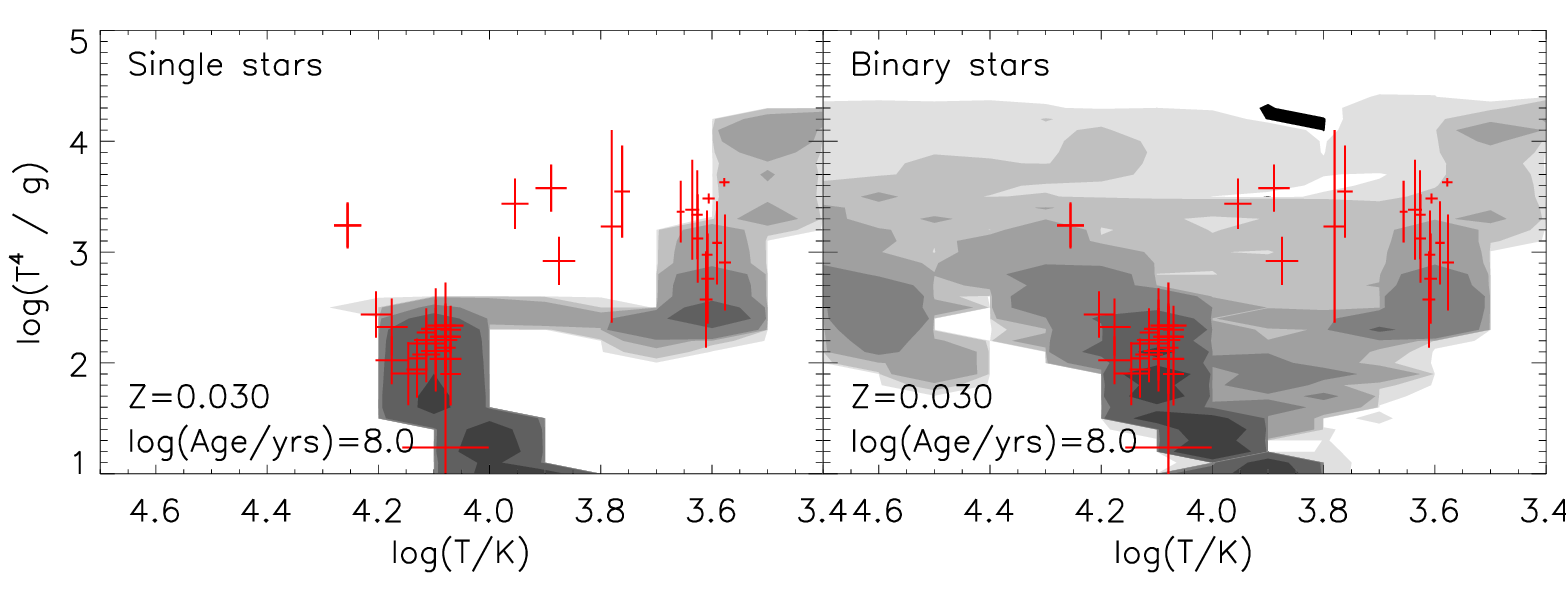}\\
\caption{Comparison of predictions of the HR diagrams for the clusters Cygnus OB2 (top), Upper Scorpius (middle) and NGC6067 (lower, spectroscopic HR diagram) and the location of observed stars in these clusters. The left panels are for single-star population and the right-hand panels include interacting binaries. The ages are chosen such that the binary star population has qualitatively the best fit.}\label{FigJJ8a}
\end{center}
\end{figure*}

\begin{figure*}
\begin{center}
\includegraphics[width=2\columnwidth]{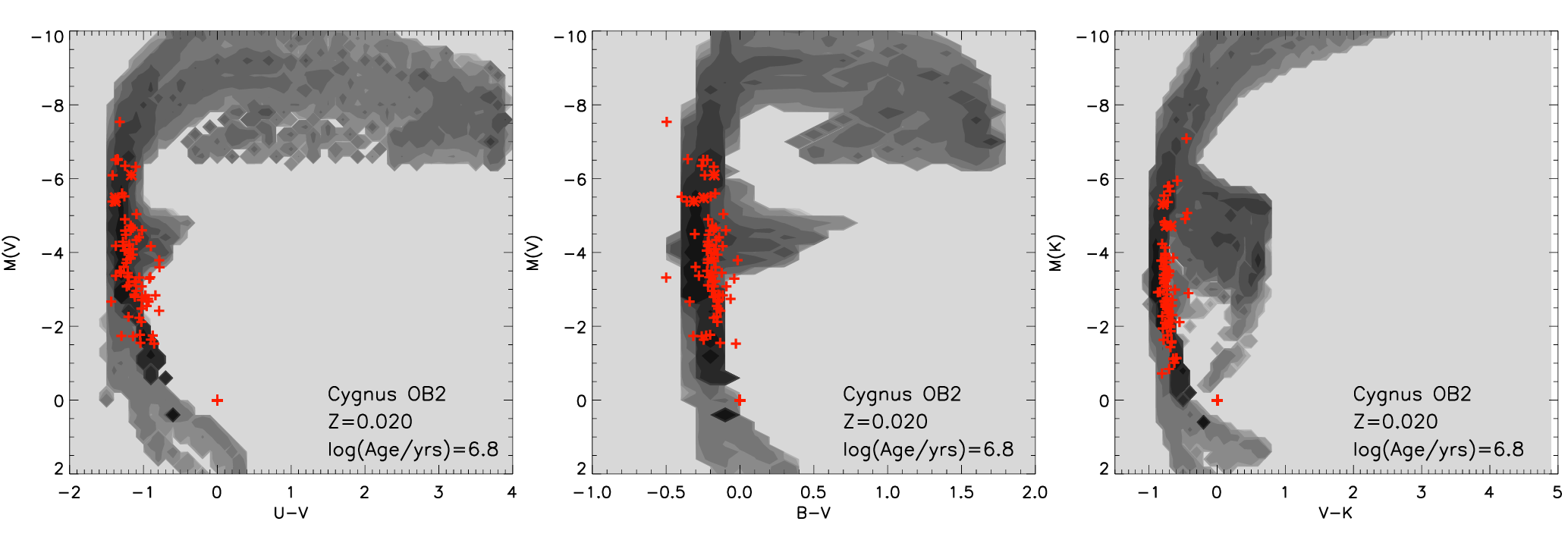}
\includegraphics[width=2\columnwidth]{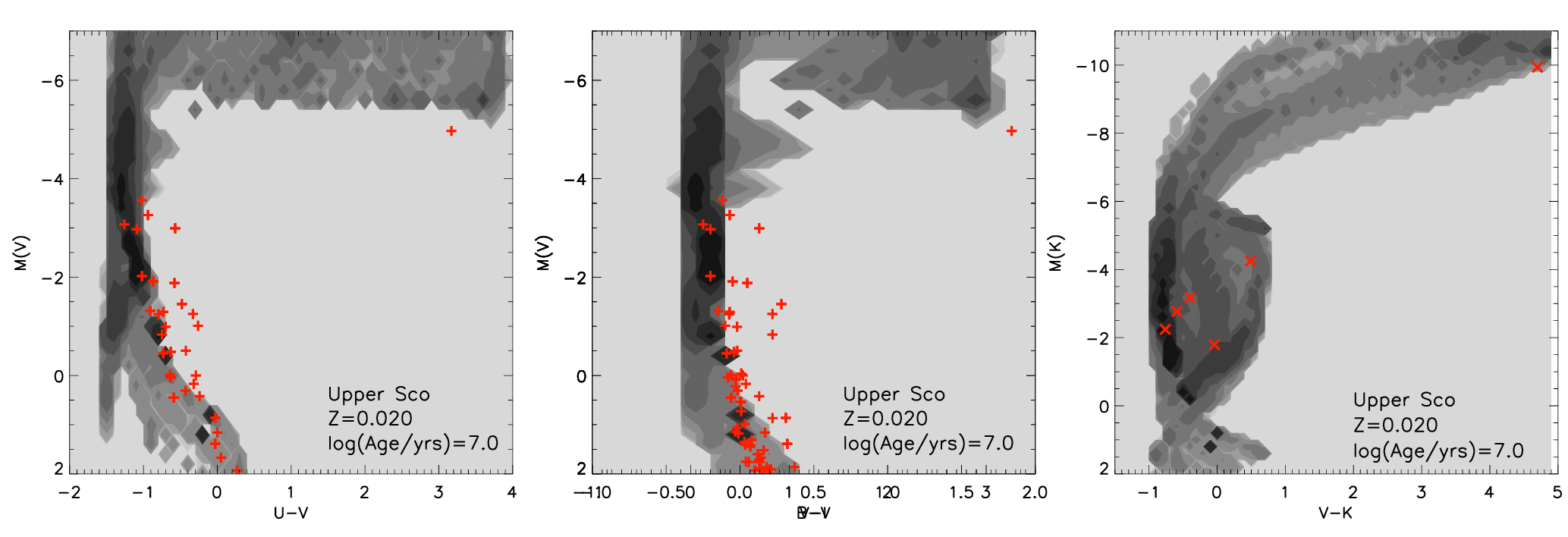}
\includegraphics[width=2\columnwidth]{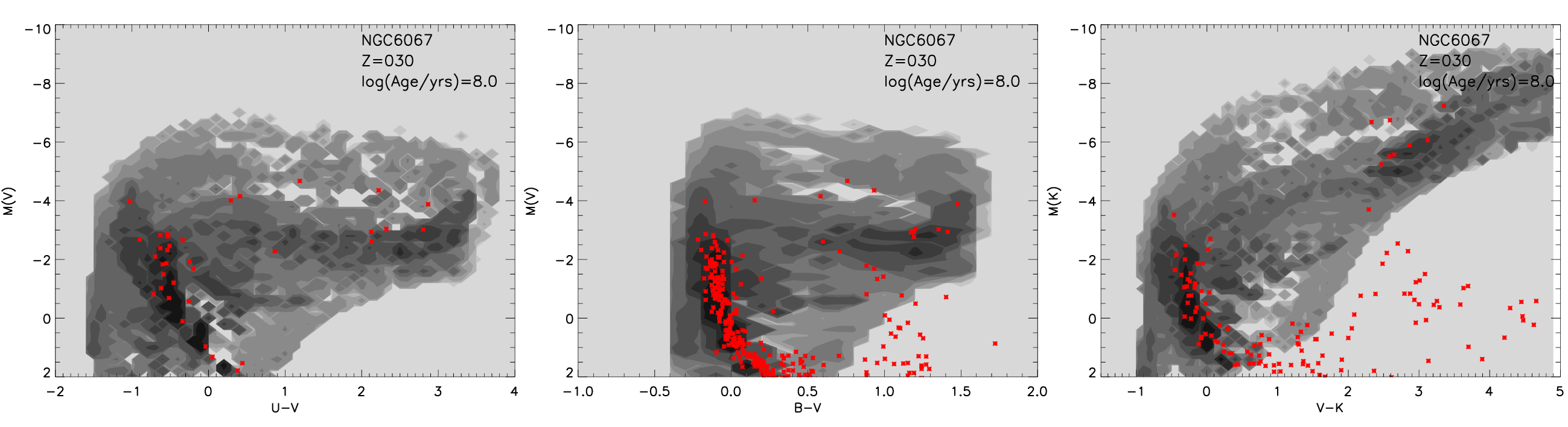}
\caption{Comparison of predictions of stellar locations on the CMDs for the clusters Cygnus OB2, Upper Scorpius and NGC6067.}\label{Fig:colour-mag1}
\end{center}
\end{figure*}

In Figures  \ref{FigJJ8a} and  \ref{Fig:colour-mag1} we compare the theoretical HR diagrams and the observational colour-magnitude diagrams for the same three clusters. Any incompatibility between our population synthesis (based on evolutionary models) and spectral synthesis (which combines these with atmosphere models) should be revealed by the comparison.  Age estimates for each cluster are based on a determination by eye and by comparing to earlier literature of the ages of the clusters. We typically use the binary star models. Fits by single-star populations are also possible if blue-straggler stars are ignored. 

The Cygnus OB2 association \citep{2015MNRAS.449..741W} and Upper Scorpius \citep{2012ApJ...746..154P} are best fit in our models at stellar population ages log(age/years)=6.8 and 7.0 respectively, while NGC\,6067 \citep{2017MNRAS.469.1330A} is fit at log(age/years)=8.0. This age for Cygnus OB2 is at the higher end of the range suggested by previous studies, as we would expect with a binary population. For Upper Sco, the strongest constraint is given by the existence of Antares, and this fixes our models to an approximate age of 10\,Myr, very similar to the 11\,Myr found by  \citet{2012ApJ...746..154P}. Our age for NGC\,6067 agrees with previous estimates. The HR diagrams for single star models are presented on the  left of Figure \ref{FigJJ8a} and binaries on the right.  Interacting binaries  broaden the range of properties (and hence colours) expected for cluster members. If we used single star models for these comparisons we would have to use a much younger population to fit the bulk of stars, but we would not be able to match the luminosity of the WR stars. The binary population spreads out the most luminous stars significantly in temperature at late ages. In older clusters blue stragglers typically form a very clear group above the main sequence turn-off; in younger clusters it is  difficult to separate these out and identify the true, non-interacting, main-sequence stars. Colour-magnitude diagrams for these clusters extending from the optical to the near-infrared are shown in Figure \ref{Fig:colour-mag1}. The stellar main sequences predicted from the HR diagrams are  coincident with the location of stars in their colour-magnitude diagrams albeit subject to some uncertainties in extinction effects (which may be present in the blue bands, particularly for Upper Sco), providing a verification test for the atmosphere models which dominate in young stellar populations. 

\begin{figure*}
\begin{center}
\includegraphics[width=2\columnwidth]{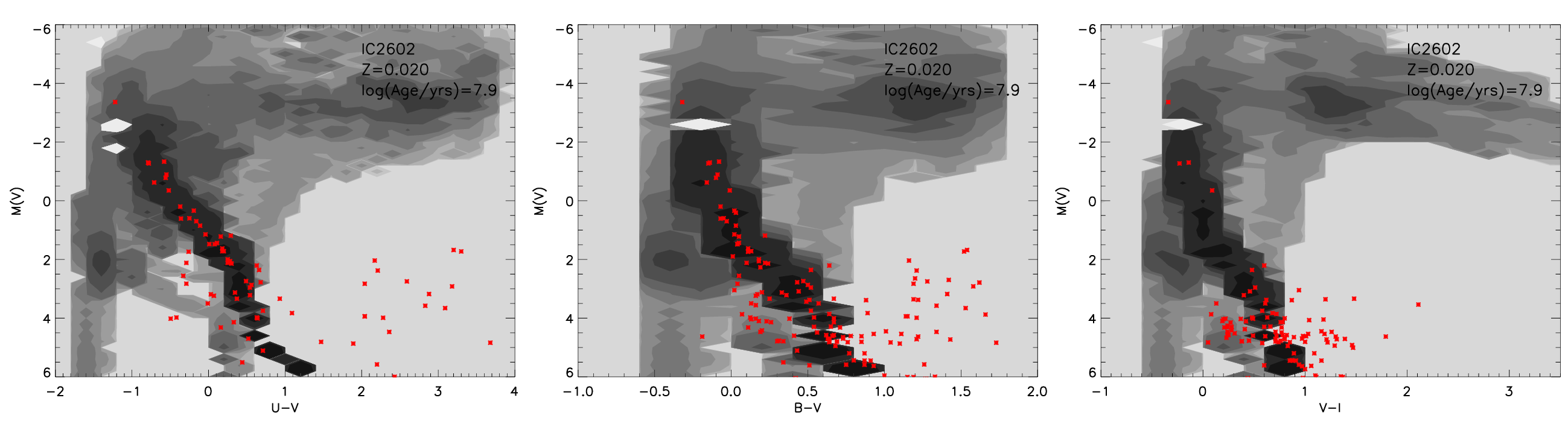}
\includegraphics[width=2\columnwidth]{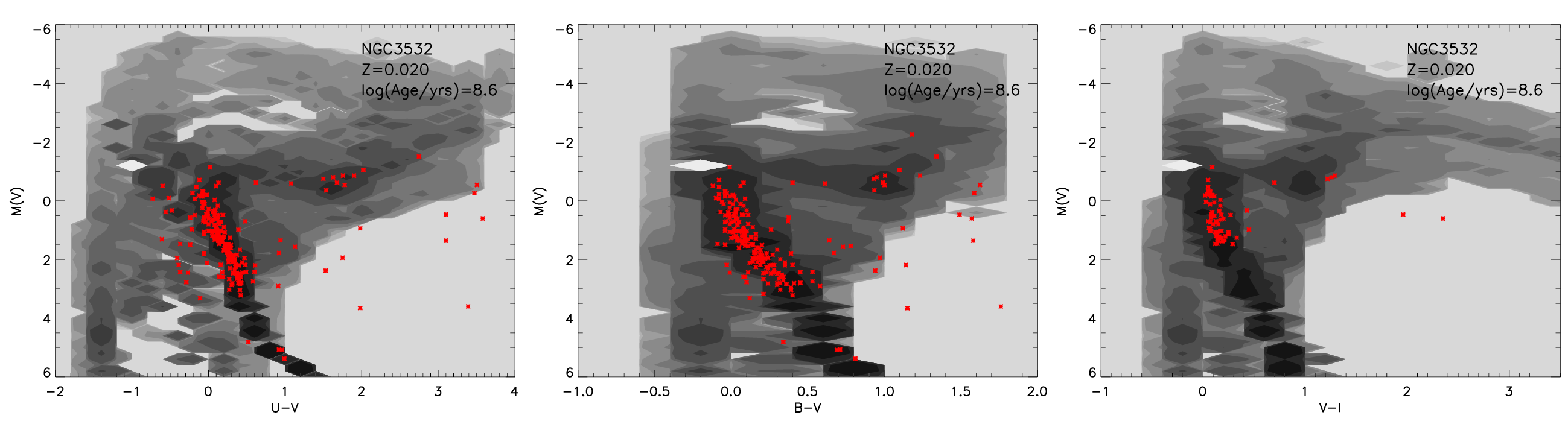}
\includegraphics[width=2\columnwidth]{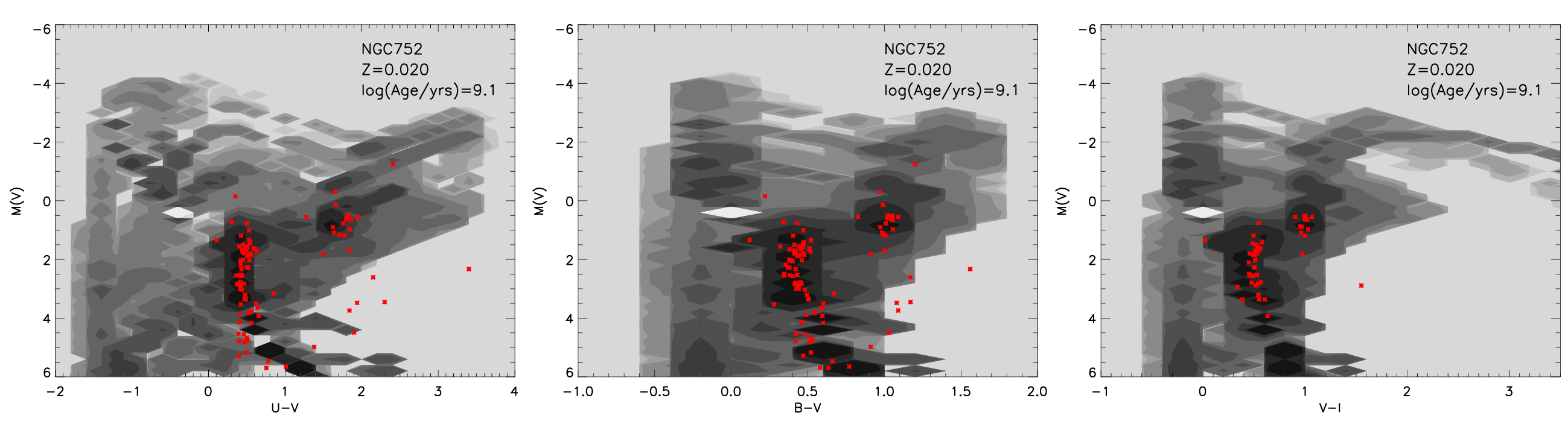}
\caption{Colour-magnitude diagrams for three additional old stellar clusters, IC2602, NGC3532 and NGC752, compared to photometric data drawn from the WEBDA database (see text). In each case, the approximate age has been estimated.}\label{Fig:colour-mag2}
\end{center}
\end{figure*}

In Figure \ref{Fig:colour-mag2} we present colour-magnitude diagrams for three additional old stellar clusters, compared to photometric data for cluster members drawn from the WEBDA database\footnote{http://webda.physics.muni.cz}. In each case the appropriate age has been drawn from our models by eye for comparison. We estimate log(age/years) of 7.9, 8.6 and 9.1 for IC2602, NGC3532 and NGC752 respectively.  These compare to log(age/years) of 7.5, 8.5 and 9.05 from the WEBDA database.   Again a single star (at the top of the main sequence) drives the fit for IC2602, and the existence of this star is possible at an older age in \bpass\ models than in typical single star models since binary models permit the existence of blue stragglers.

As the figure demonstrates, \bpass\ binary models simultaneously reproduce the locations in colour-magnitude space of the main sequence and red giant branch for these clusters. While some scatter remains which cannot be explained by binaries, we cannot rule out the presence of non-cluster interlopers in the available available catalog data particularly at faint absolute magnitudes.

\begin{figure*}
\begin{center}
\includegraphics[width=0.9\columnwidth]{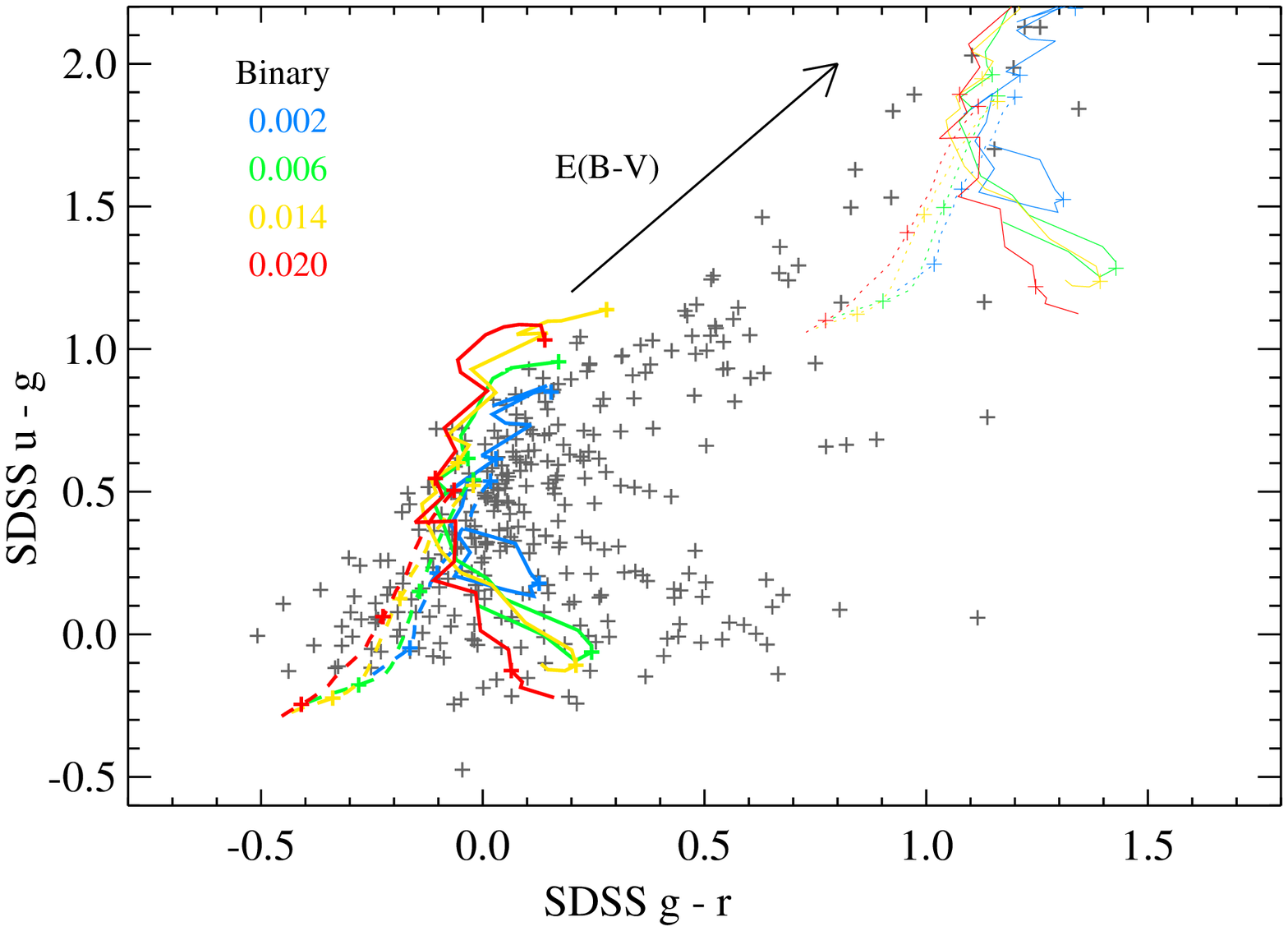}
\includegraphics[width=0.9\columnwidth]{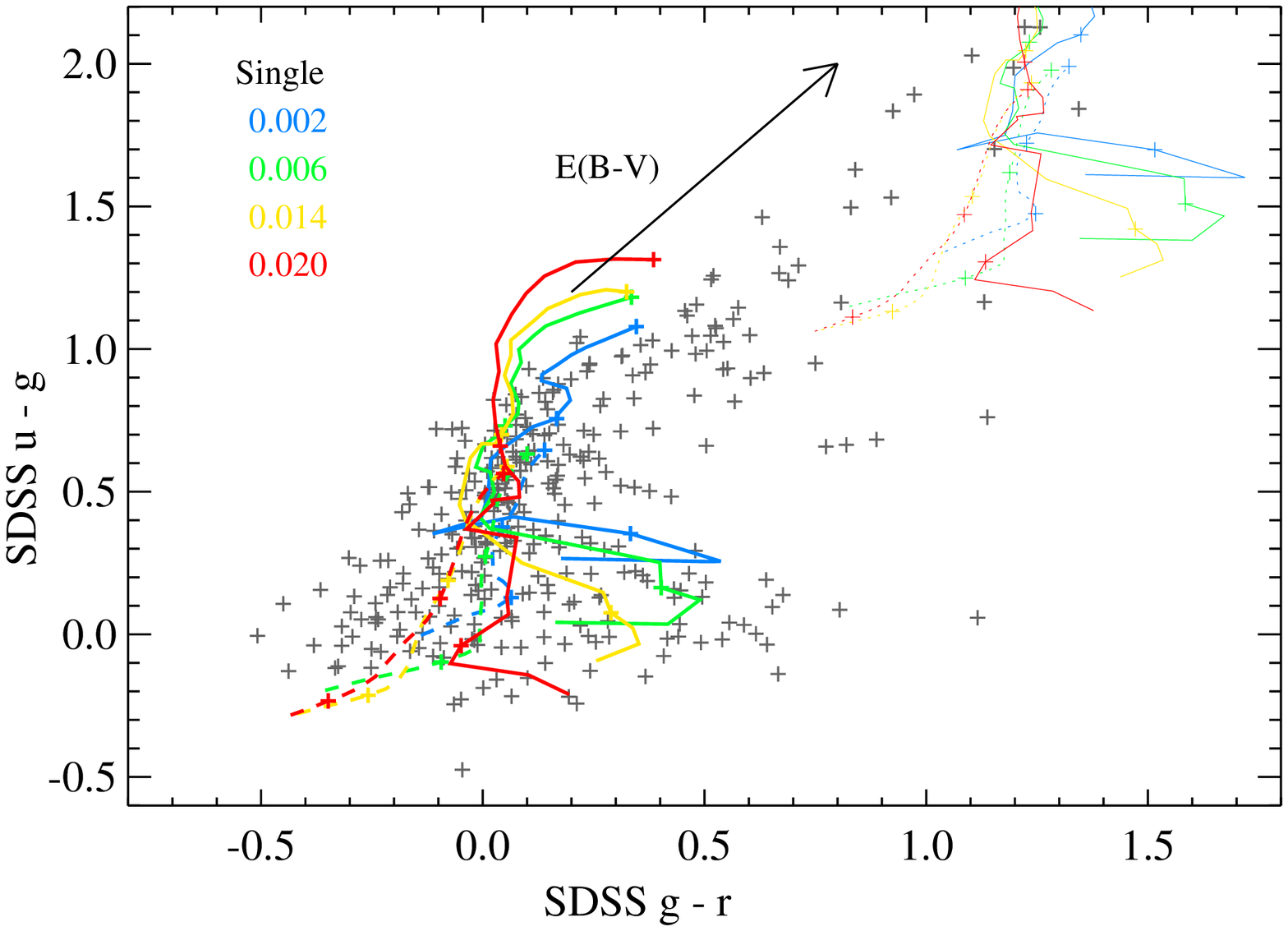}
\includegraphics[width=0.9\columnwidth]{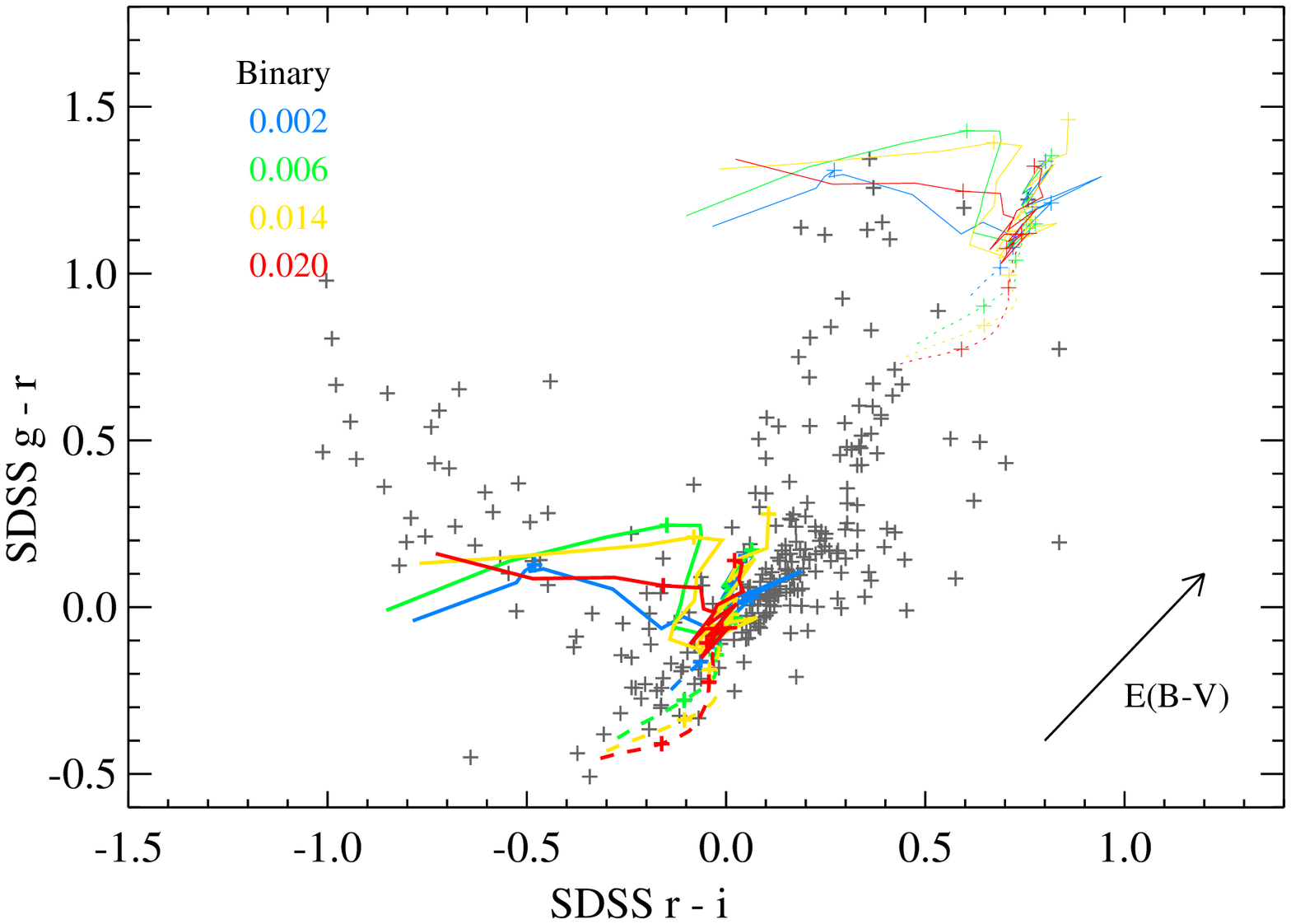}
\includegraphics[width=0.9\columnwidth]{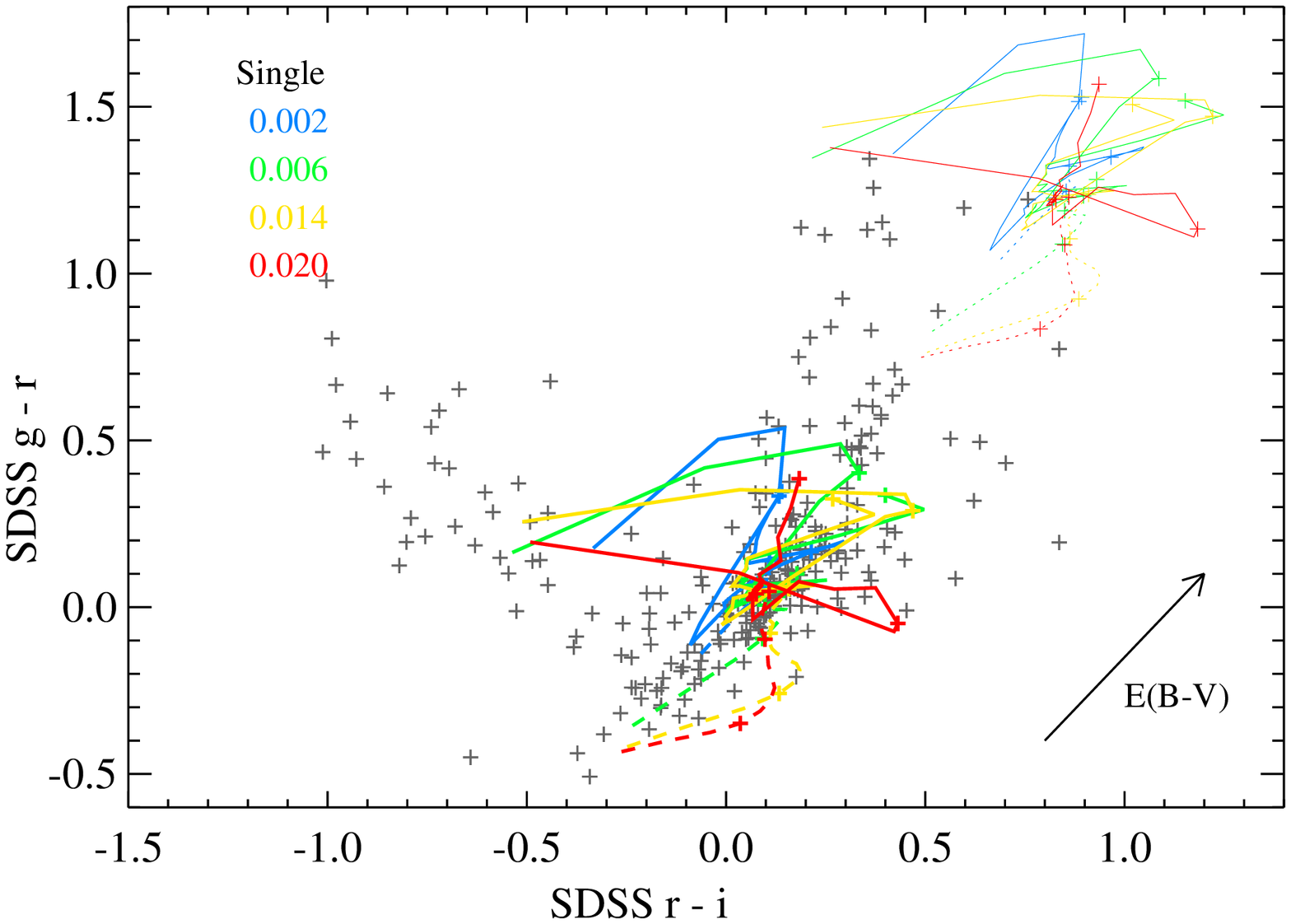}
\includegraphics[width=0.9\columnwidth]{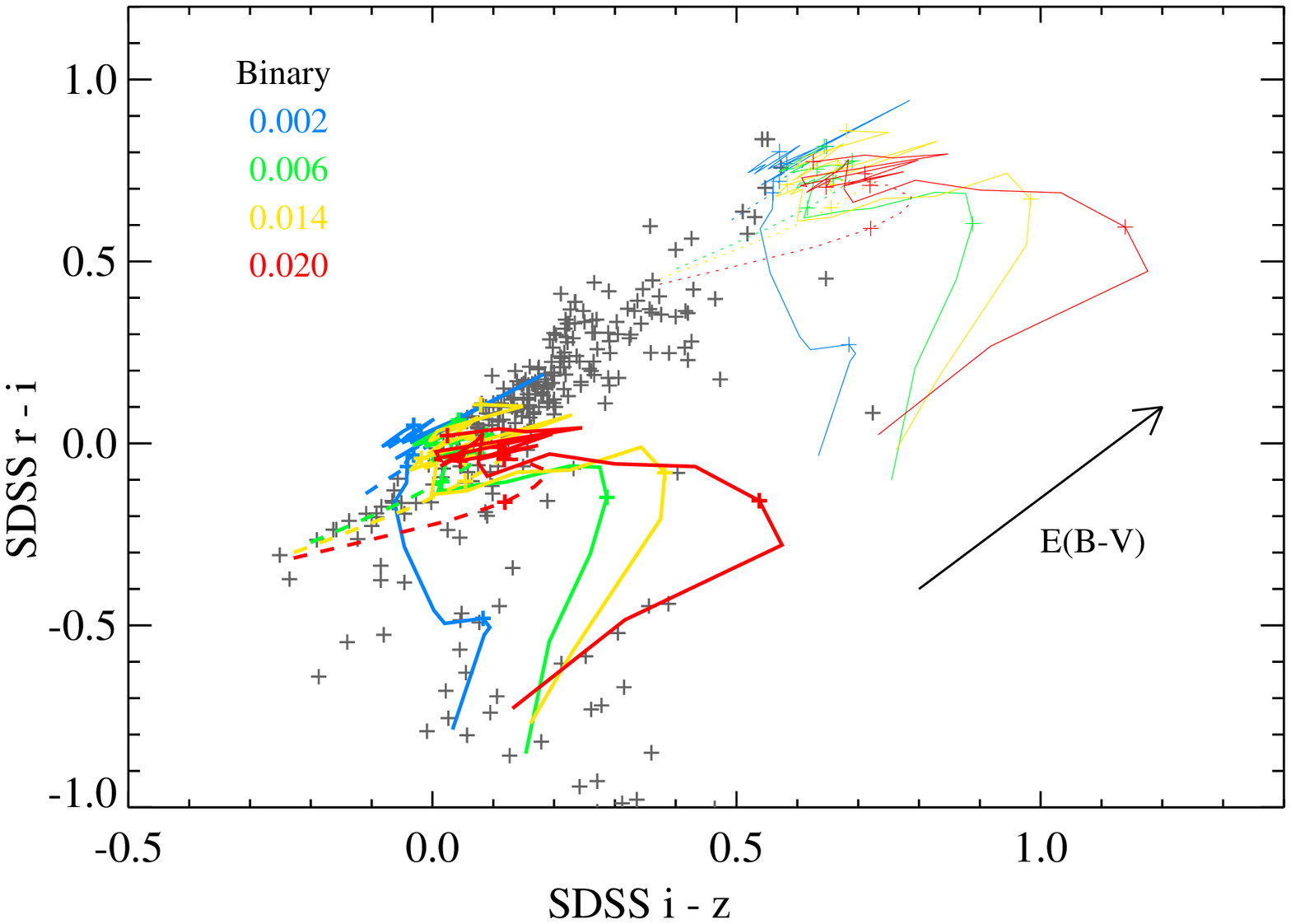}
\includegraphics[width=0.9\columnwidth]{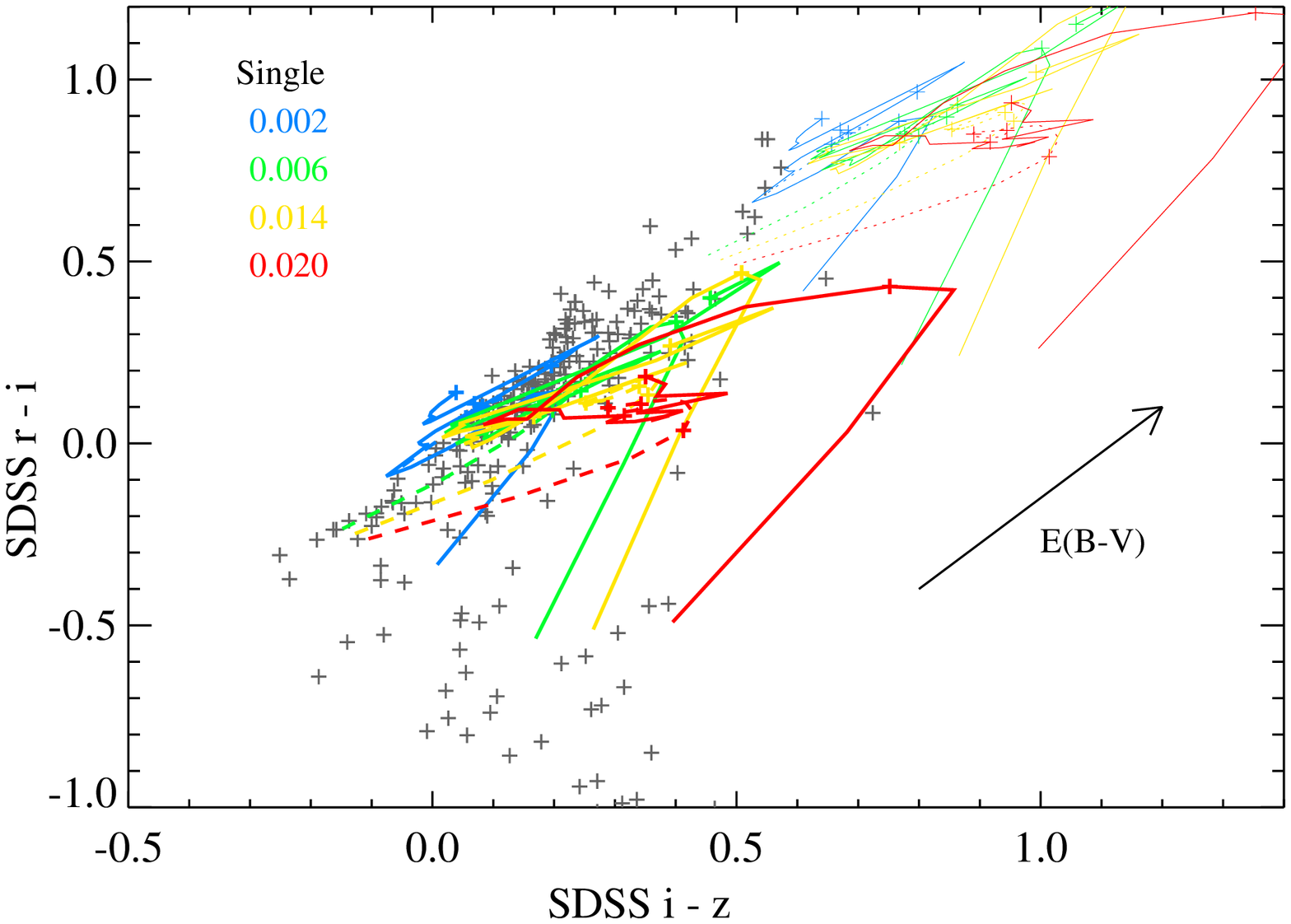}
\caption{The photometric colours of unresolved star clusters in the M33 system, overplotted with evolution tracks at four different metallicities. Solid lines show the evolution with age of \bpass\ models for a single-aged starburst from 5\,Myr (bluest colours) to 1\,Gyr. Dashed lines indicate the same but for a population continuously forming stars at a constant rate of 1\,M$_\odot$\,yr$^{-1}$. The thick lines indicate a population without dust reddening while the offset tracks shown in thin lines have a dust extinction of E(B-V)=1, assuming the Calzetti dust law. Data points are drawn from the photometric catalog of \citet{2010ApJ...720.1674S} and include only their `highly probable' or `confirmed' clusters with m$_g<19.5$.}
\label{fig:m33_clusters}
\end{center}
  \end{figure*}

A final strong test of simple stellar populations can be obtained by comparing the colours derived from our models to those of unresolved but (likely) single-aged stellar clusters in a nearby galaxy. In Figure \ref{fig:m33_clusters} we do so, making use of the M33 star cluster population identified by \citet{2010ApJ...720.1674S}. The colour of an unresolved population evolves rapidly in the first $\sim$10\,Myr after the onset of star formation and thereafter more slowly, and with relatively little dependence on stellar metallicity. In figure \ref{fig:m33_clusters} we do not plot tracks with ages $<5$\,Myr since these are likely to be severely affected by nebular emission and potentially enshrouded by natal dust clouds. With the exception of the youngest sources then, \bpass\ models appear to qualitatively reproduce the range of colours observed in the star clusters of M33, particularly if moderate reddening (with $E(B-V)$ up to 1) is permitted.

In summary by testing our evolution models against observational data on Hertzsprung-Russell diagrams, their respective colour-magnitude diagrams and colour-colour diagrams, we have shown that the evolutionary models and atmosphere models can be used with confidence, especially when studying unresolved clusters.


\section{COMPLEX STELLAR POPULATIONS}\label{sec:complex}

By their nature, the integrated light from most galaxies is challenging to simulate. The stellar component of a galaxy typically consists of a number of low mass individual stellar populations, not necessarily co-eval and each embedded in the interstellar medium. Their light is modified by the interstellar gas and by scattering, absorption and re-emission from dust grains. Nonetheless, the underlying spectral energy distribution of a galaxy, together with the gas excitation and dust temperature, is fundamentally shaped by its stellar population. The nebular continuum and emission lines are primarily driven by Lyman continuum photons and hence by the youngest stars. Thus the interpretation of integrated light from galaxies is also largely dependent on stellar population synthesis models and the range of observed galaxy properties provides a constraint on those models. Here we present observational constraints on the performance of our models in galaxy-scale stellar systems.

\subsection{Wolf-Rayet Galaxies}\label{sec:wr_obs}

A straight forward observational verification test of our \bpass\ models in extragalactic stellar populations is provided by so called `Wolf-Rayet' galaxies, classified through strong emission in regions of the spectrum in which the Wolf-Rayet stars provide the dominant flux contribution, primarily centred on the \heii 4686\AA\ line (the `WR blue bump') and \civ 5808\AA\ (the `WR red bump'). The presence of these massive, stripped-envelope stars is taken to indicate a recent, massive starburst which should thus be possible to match against our single-age starburst models. 

\begin{figure*}
\begin{center}
\includegraphics[width=\columnwidth]{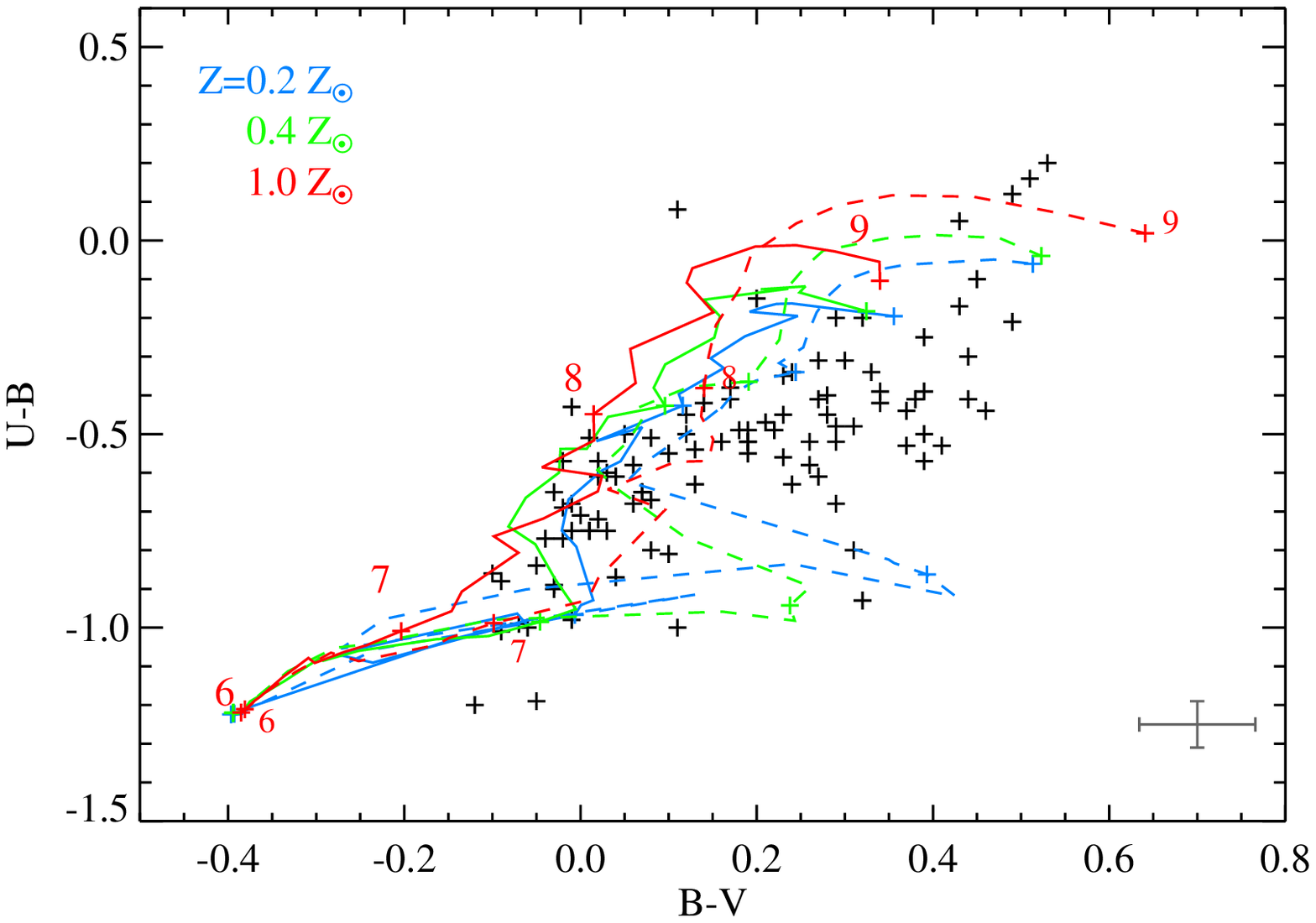}
\includegraphics[width=\columnwidth]{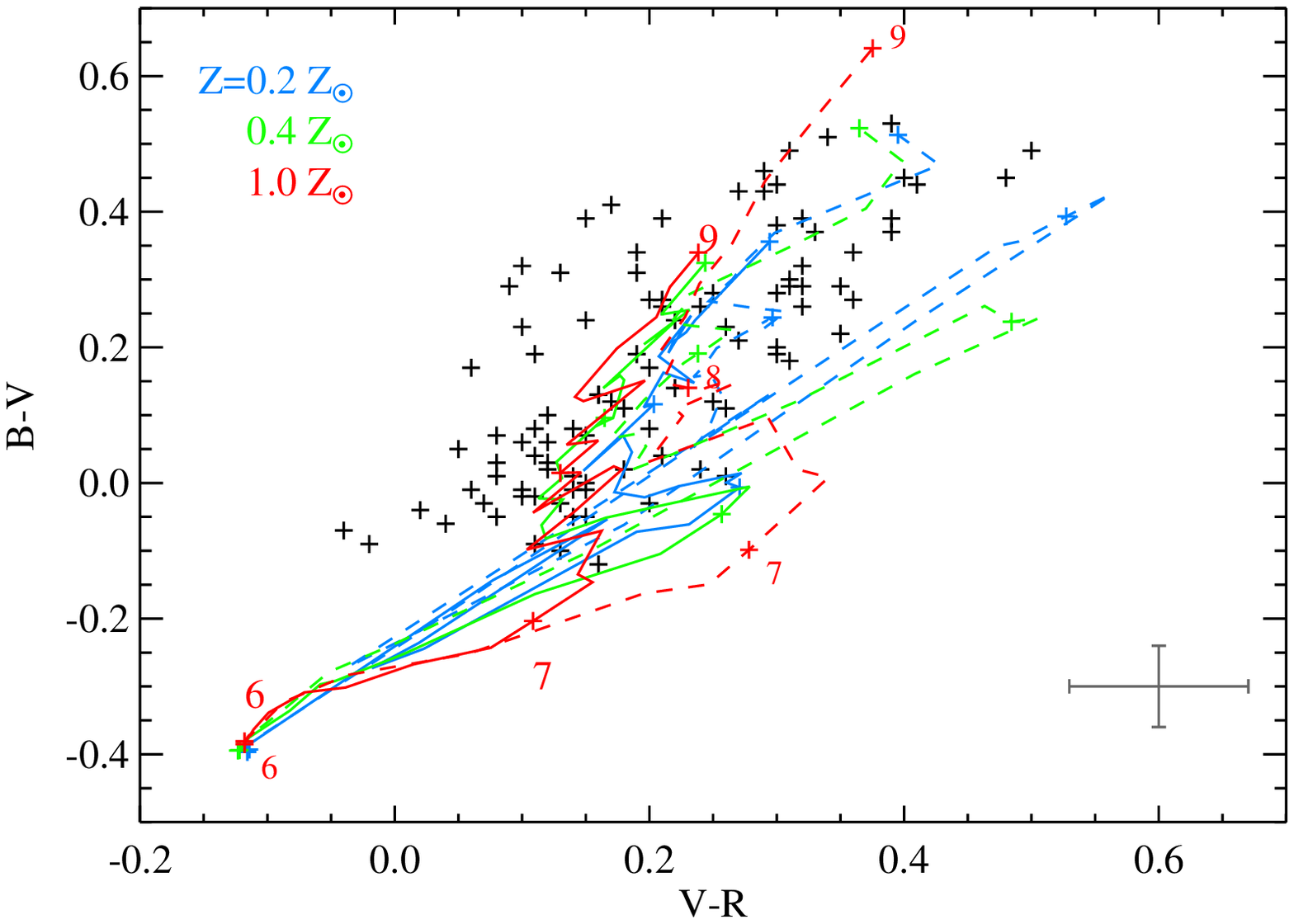}
\caption{The photometric colours (in the Johnson-Vega system) for measured star forming regions in the \citet{2008A&A...491..131L} sample of Wolf-Rayet galaxies. Tracks show the evolution in colour with age for single-age starbursts comprised of single stars (dashed) and binary stars (solid lines), at three different metallicities. The catalog photometry was adjusted by \citeauthor{2008A&A...491..131L} to remove an estimated nebular contribution (typically $\Delta$mag$\sim0.05-0.10$) and is compared to stellar tracks without nebular emission. The median photometric uncertainties of the datapoints in each colour is indicated in the lower right of each panel.}
\label{fig:mag_WR_inst}
\end{center}
\end{figure*}

Observational samples of Wolf-Rayet-dominated stellar populations, particularly those with follow-up spectroscopy, have improved since we presented an initial analysis of these systems with our v1.0 model set \citep{2009MNRAS.400.1019E}.  \citet{2008A&A...491..131L,2010A&A...516A.104L,2010A&A...517A..85L} analysed a sample 20 nearby Wolf-Rayet galaxies, determining the photometric colours of both the integrated galaxy and  distinct star forming regions within it.   In Figure \ref{fig:mag_WR_inst} we compare the predictions of our models to the photometric colours measured for that sample. We omit the photometry for underlying stellar populations (which are typically old) and for companion galaxies and compare against data for which an estimated nebular gas emission component has previously been subtracted. Given the photometric scatter in the data, the observed colours of WR-dominated star forming regions are consistent with both single and binary star models. Although neither set of tracks precisely reproduces the data, the observations are arguably more consistent with our binary models than our single star models, particularly in Johnson $B-V-R$ colour space, although they tend to be $\sim0.2$\,mag redder in $B-V$ than binary tracks, perhaps suggesting that the nebular contribution to the $B$ band has been over-estimated. Unfortunately the broad band colours are only weakly sensitive to metallicity and the uncertainties on the photometric data do not allow a more quantitative study of this sample.

To assess our model predictions for the strength of the Wolf-Rayet stellar features in the spectra of galaxies or their resolved star forming regions, we supplement the \citet{2010A&A...516A.104L} spectroscopic sample with two further studies. The first, presented here for comparison with our own earlier work is the Brinchmann et al (2008) sample of WR-galaxies selected from SDSS. These were selected through use of a virtual narrowband-filter to identify archival spectra with a flux excess at the expected wavelengths, before the broad spectral lines were individually measured. The second is a sample selected for spectroscopic follow up by \citet{2016ApJ...826..194S} as individual WR-hosting star clusters in six nearby galaxies. In common with the \citet{2010A&A...516A.104L} selection, while the galaxies themselves are resolved, the individual star forming regions are not. In each of the samples discussed here, the original authors corrected their spectra for an inferred nebular flux contamination and determined a metallicity (either from the `direct temperature' or through empirical strong line indicator methods), before estimating the flux or equivalent width in the WR red and blue bumps.

\bpass\ stellar population models have evolved since the early results were presented in \citet{2009MNRAS.400.1019E}. We predict comparable blue bump equivalent widths from binary populations to our earlier models at twice Solar metallicity. Our predictions at lower metallicities have risen from $\sim0.5$\AA\ (at 0.05\,Z$_\odot$) to almost 4\,\AA. Our predictions for single star models have also changed, with the new populations generating less emission in the blue bump at late ages than the earlier models (primarily due to an improved treatment of O- and Of-stars). Similarly, our new models predict stronger emission in the red bump than before, particularly at low metallicities and early times. Interestingly, we see a strong metallicity dependence on the line strengths in both Wolf-Rayet features. Both bumps show two peaks in line production with stellar population age, one at $\sim$3\,Myr and attributable primarily to the ionizing flux from short-lived massive WO stars, and the second at $\sim$10\,Myr due to formation of late-WN and WC stars. Evolution into these stellar classes, and the ratio between them, is strongly dependent on metallicity \citep[see][]{2016MNRAS.459.1505M}. At low metallicities, the WC phase is never reached as a result of low mass loss rates; the surviving WN stars remain  luminous in the Helium-dominated blue bump, but do not contribute strongly to the carbon-dominated red bump. At high metallicities, WC stars form and thus a strong peak is seen in the red bump at late times. 

\begin{figure*}
\begin{center}\hspace{-0.5cm}
\includegraphics[width=0.8\columnwidth]{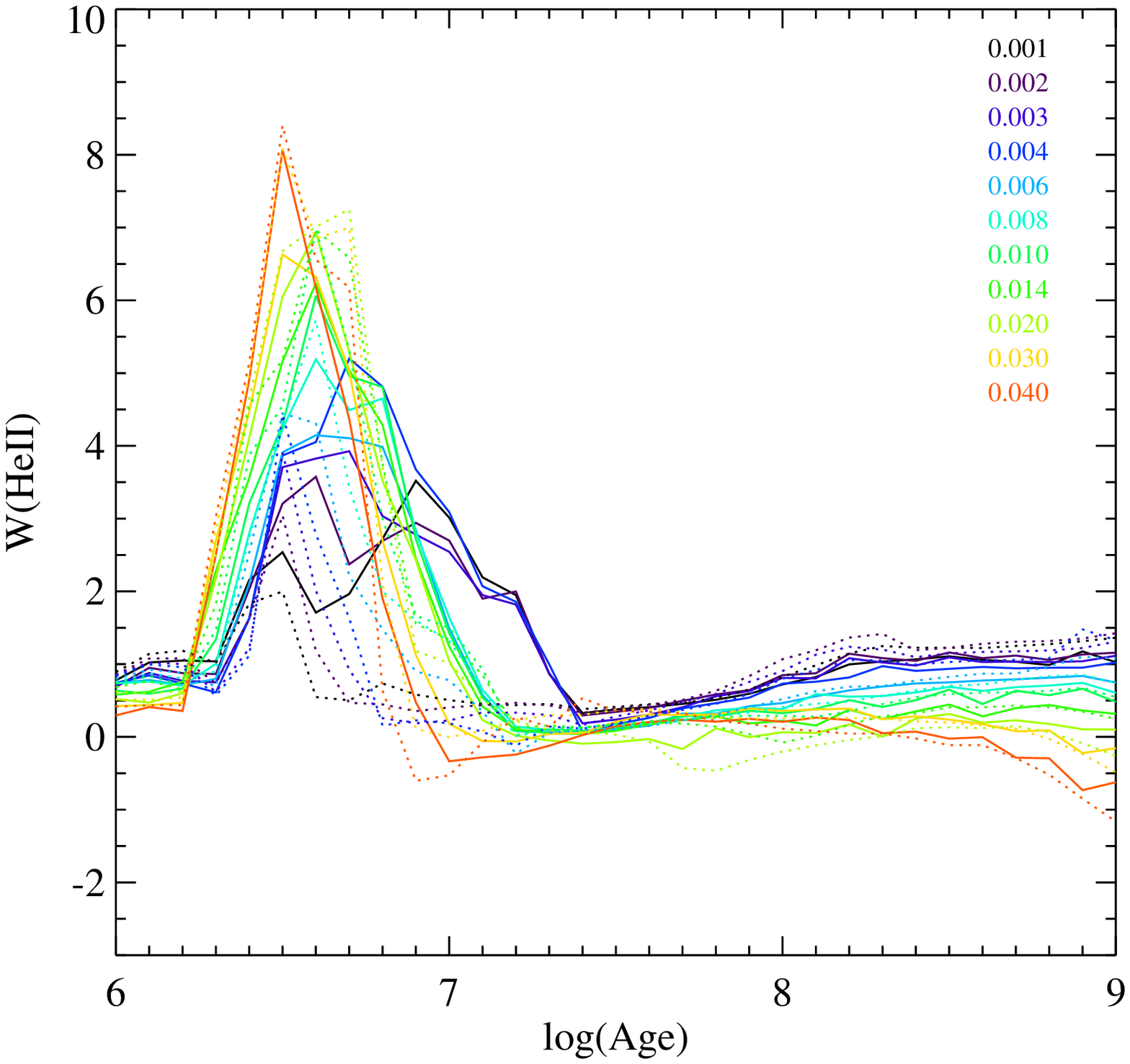}
\includegraphics[width=0.8\columnwidth]{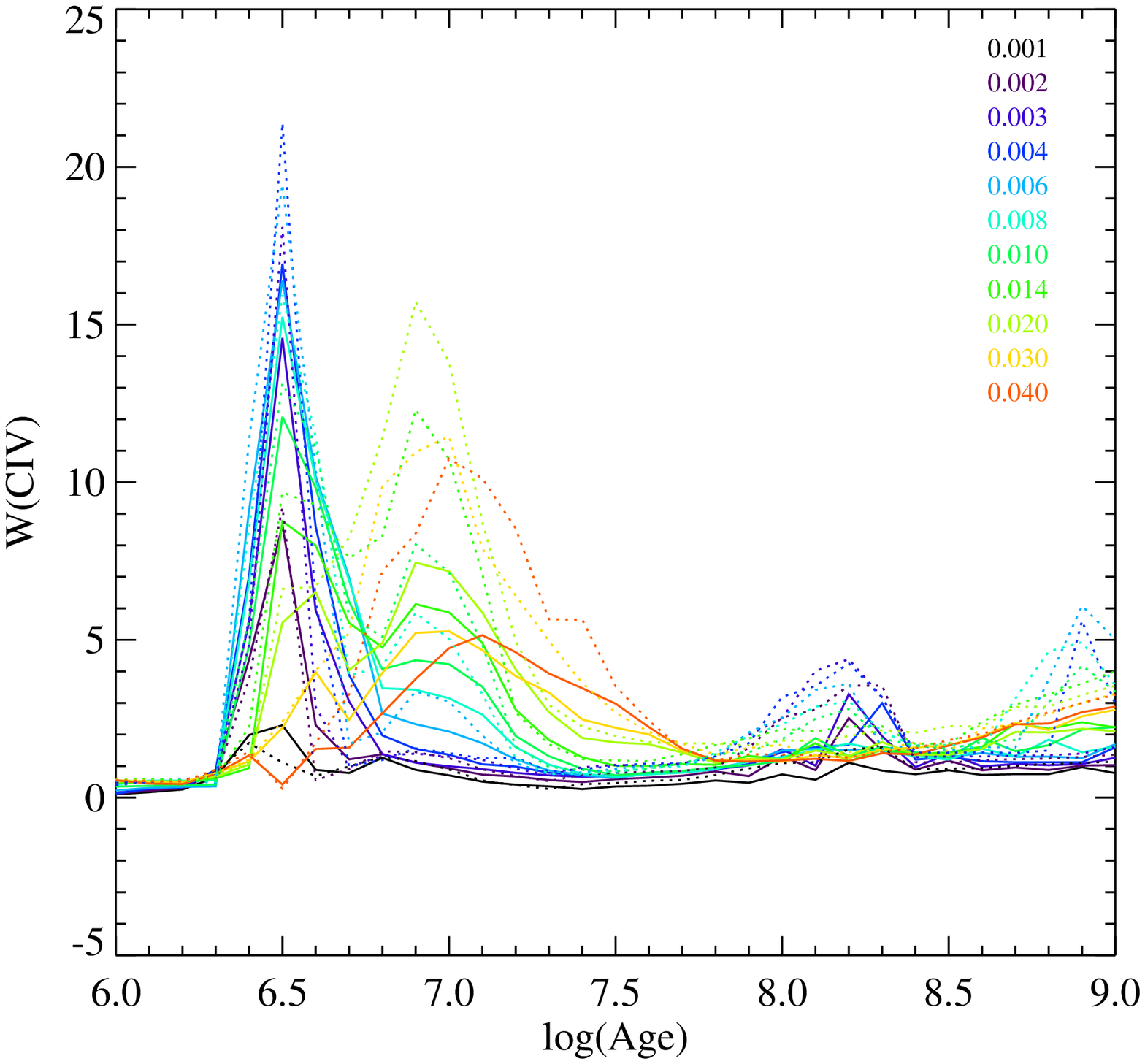}
\includegraphics[width=0.8\columnwidth]{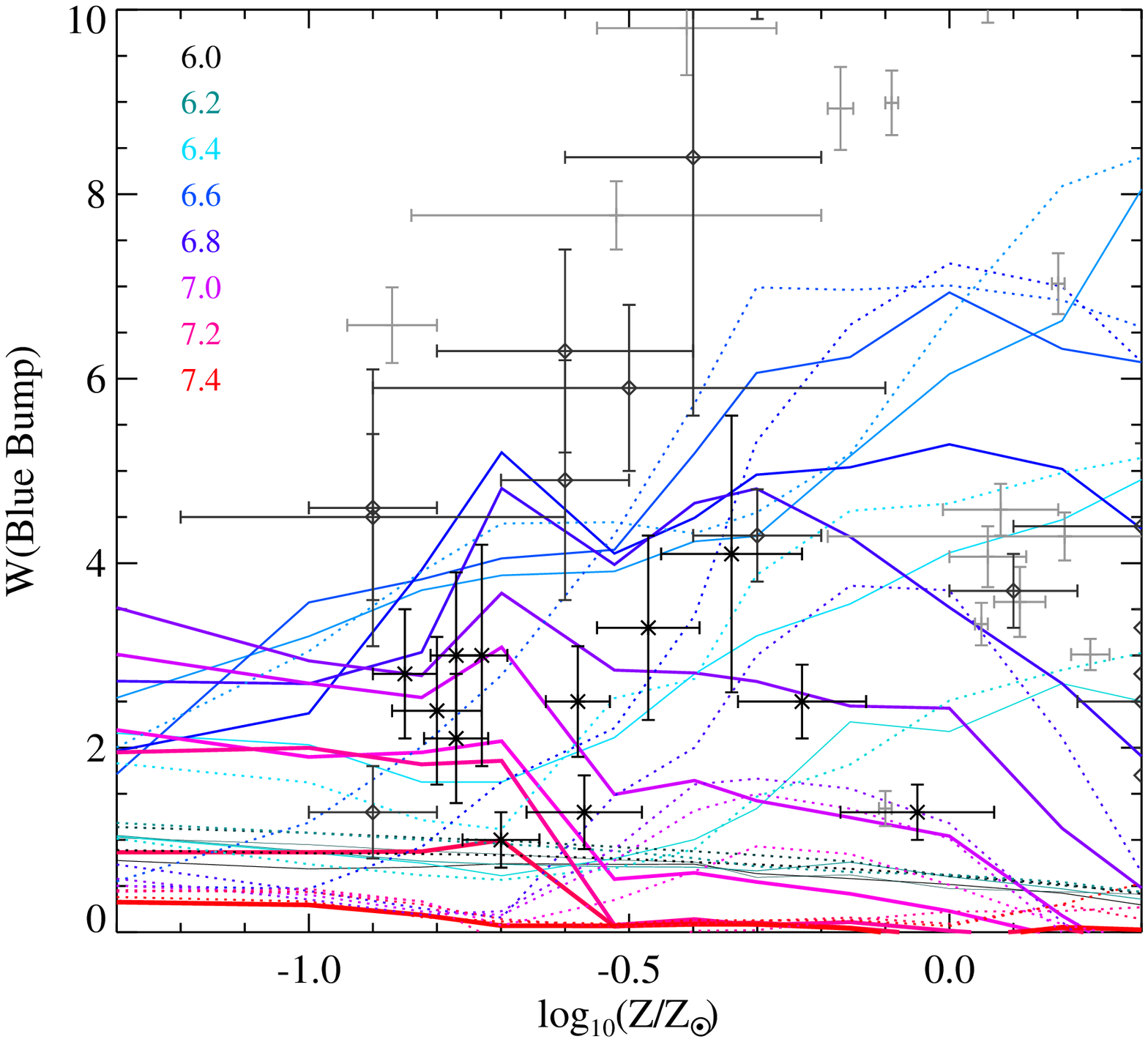}
\includegraphics[width=0.8\columnwidth]{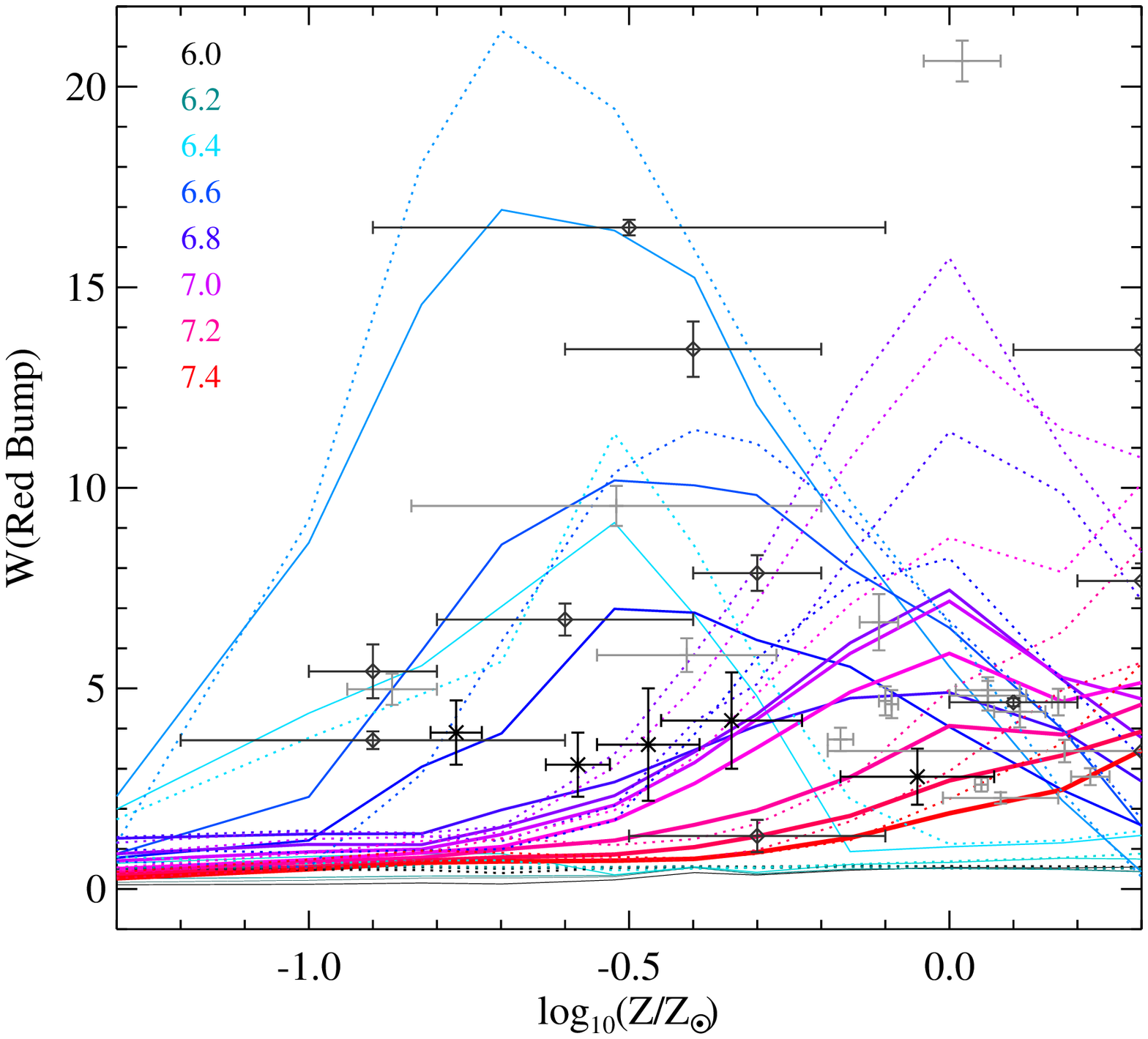}
\includegraphics[width=0.8\columnwidth]{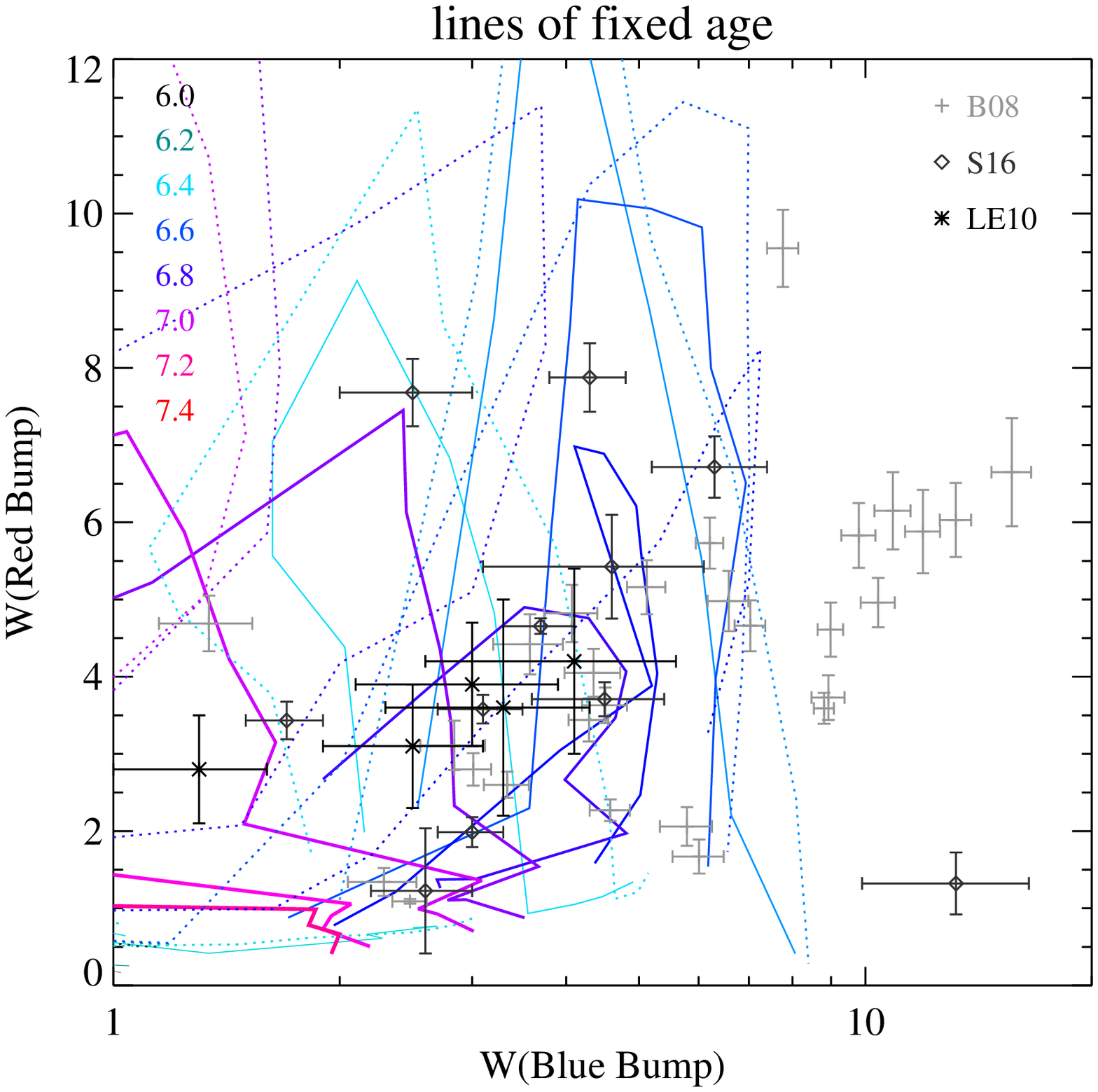}
\includegraphics[width=0.8\columnwidth]{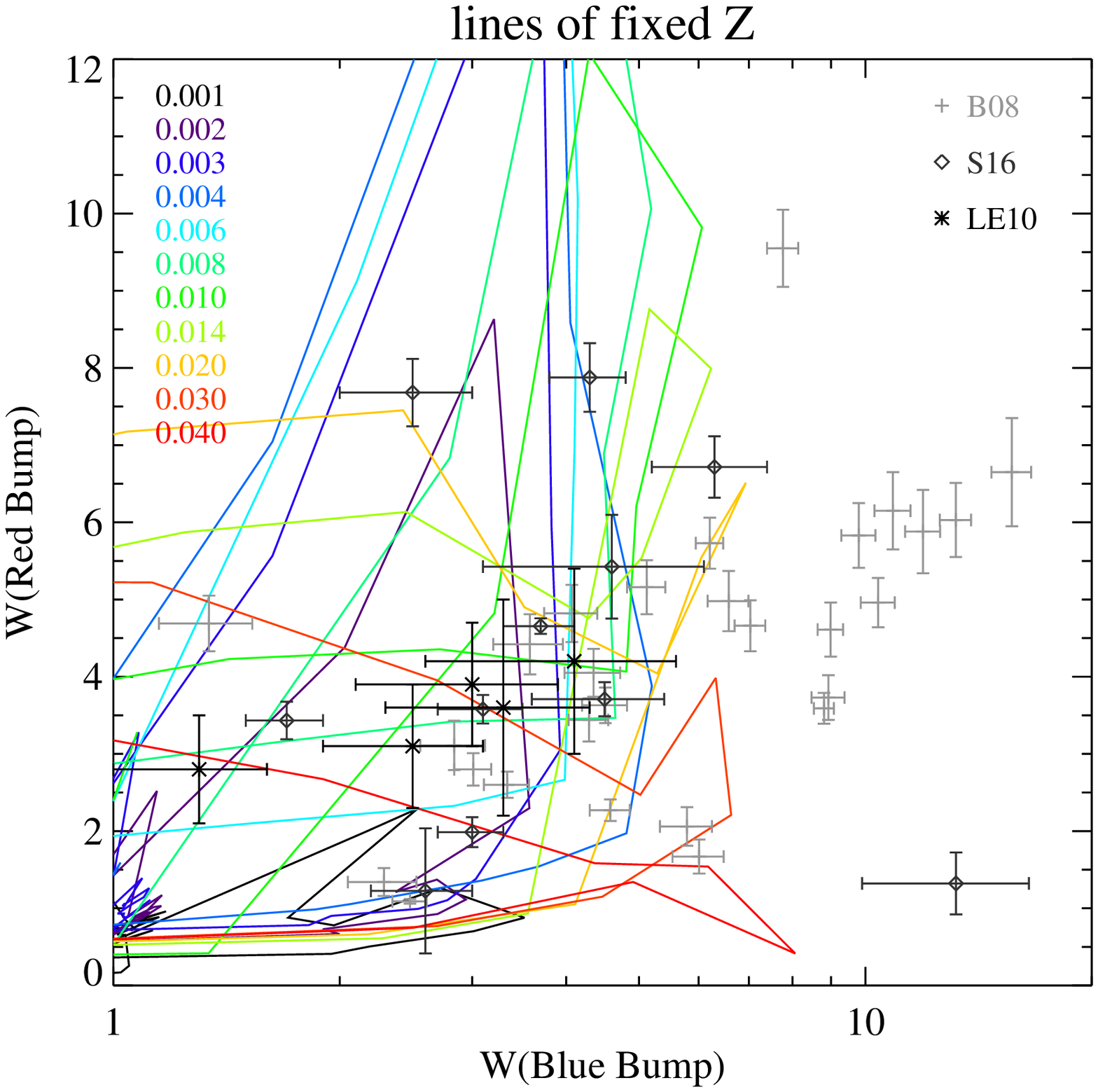}
\caption{The predicted strengths of the Wolf-Rayet `red' and `blue' bumps, centred on \heii 4686\AA\ and \civ 5808\AA\ respectively. All line strengths are given as equivalent widths in angstroms, with positive values indicating emission. Tracks show the evolution in colour with age for single-age starbursts comprised of single stars (dashed) and binary stars (solid lines). The top panels show the time evolution of the line strengths as a function of metallicity. The centre panels compare the strengths of starbursts of equal ages (labelled as log(age/years) in the upper left) at different metallicities. These are compared with the combined catalogue data of \citet[][LE10]{2010A&A...516A.104L}, \citet[][S16]{2016ApJ...826..194S} and Brinchmann et al (2008,B08) in grey scale, with dark regions indicating a higher density of sources. Catalog photometry was adjusted by the original authors to remove estimated nebular contribution and is compared to stellar tracks without nebular emission. In the bottom panels, the strengths of the two bumps are compared simultaneously with tracks at equal age (left) or metallicity (right). Note that model agreement is far better with the LE10 and S16 data than that of B08. }
\label{fig:WR_bumps}
\end{center}
\end{figure*}

The v2.1 models do well at recovering the broad distribution of spectroscopic properties, with models at near-Solar metallicity and ages of $\sim3-10$\,Myr simultaneously reproducing the observed strengths of both Wolf-Rayet bumps in most cases. Having said that, the comparison is not without tension. Younger ages are preferred by fits to the blue bump as opposed to slightly older ages in the red bump. There are also a number of sources for which none of our single-age, single metallicity models reproduce the observed lines; these are found predominantly in the Brinchmann et al (2008) sample. Unlike the other two samples shown in Figure \ref{fig:WR_bumps}, in which individual star forming regions have been identified (if not resolved), this sample comprises entirely unresolved galaxies. The nebular correction for these sources, as well as the underlying old stellar continuum, has been determined by comparison with earlier population synthesis models and so is somewhat dependent both on the stellar models in that code (which did not incorporate binary interactions or rotation) and the presumed star formation history. Given the extremely prominent red and blue bumps in this sample, and their inconsistency with both the semi-resolved samples and our single age starburst models, two possibilities present themselves: the nebular contribution may have been underestimated in the catalogue correction (i.e. approximately one third of the reported line flux has a nebular emission origin), or there is a significant contribution from diffuse interstellar medium between individual star forming regions.  The latter appears unlikely, given the high ionization potentials required to excite the emission lines dominating each bump.

\subsection{Star Formation History}\label{sec:sfh}
The default \bpass\ models produce the stellar output spectrum for a burst of stars formed at a given time before present.  From these it is possible, and often necessary, to construct integrated spectra representing more complex star formation histories such as exponentially rising or falling, or multiple burst histories.  Inevitably this results in the addition of free parameters describing this history, in addition to those associated with the stellar models. Our preferred approach therefore is to explore as limiting cases the evolution of an instantaneous, rapid burst of star formation and of a continuous moderate (1\,M$_\odot$\,yr$^{-1}$) star formation episode as a function of time, while noting that neither case may be appropriate to any given galaxy.

\begin{figure*}
\begin{center}
\includegraphics[width=\columnwidth]{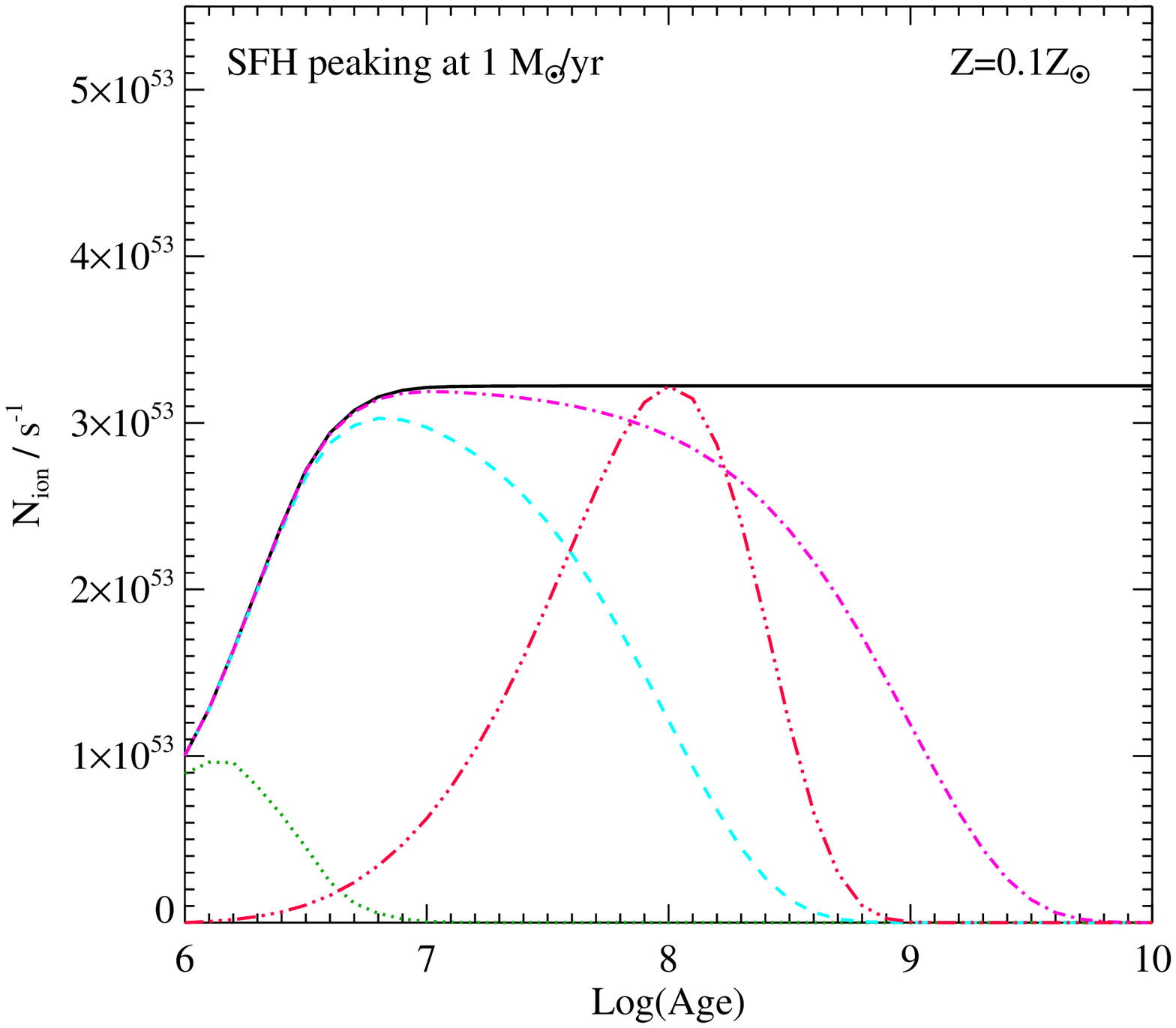}
\includegraphics[width=\columnwidth]{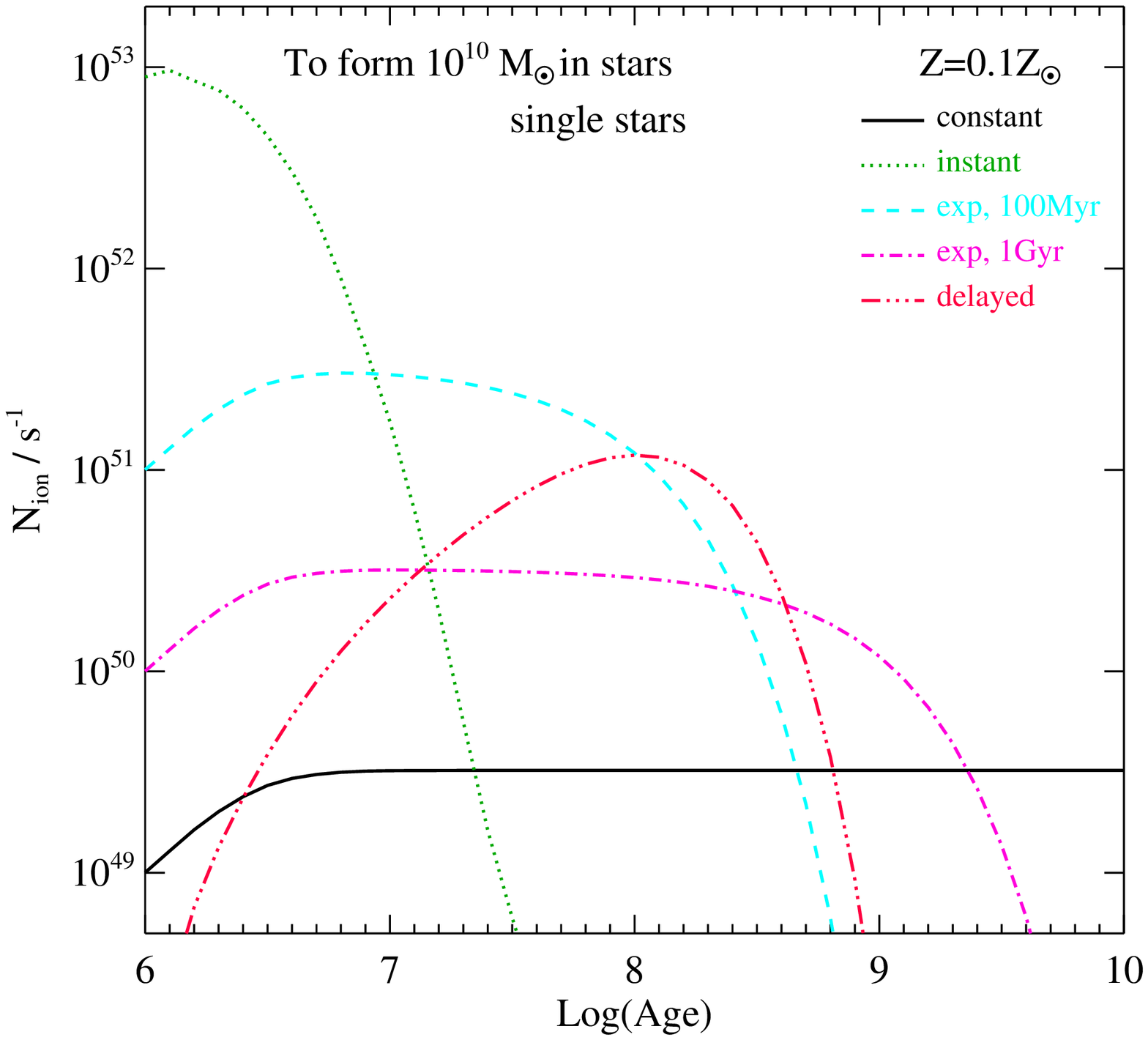}
\includegraphics[width=\columnwidth]{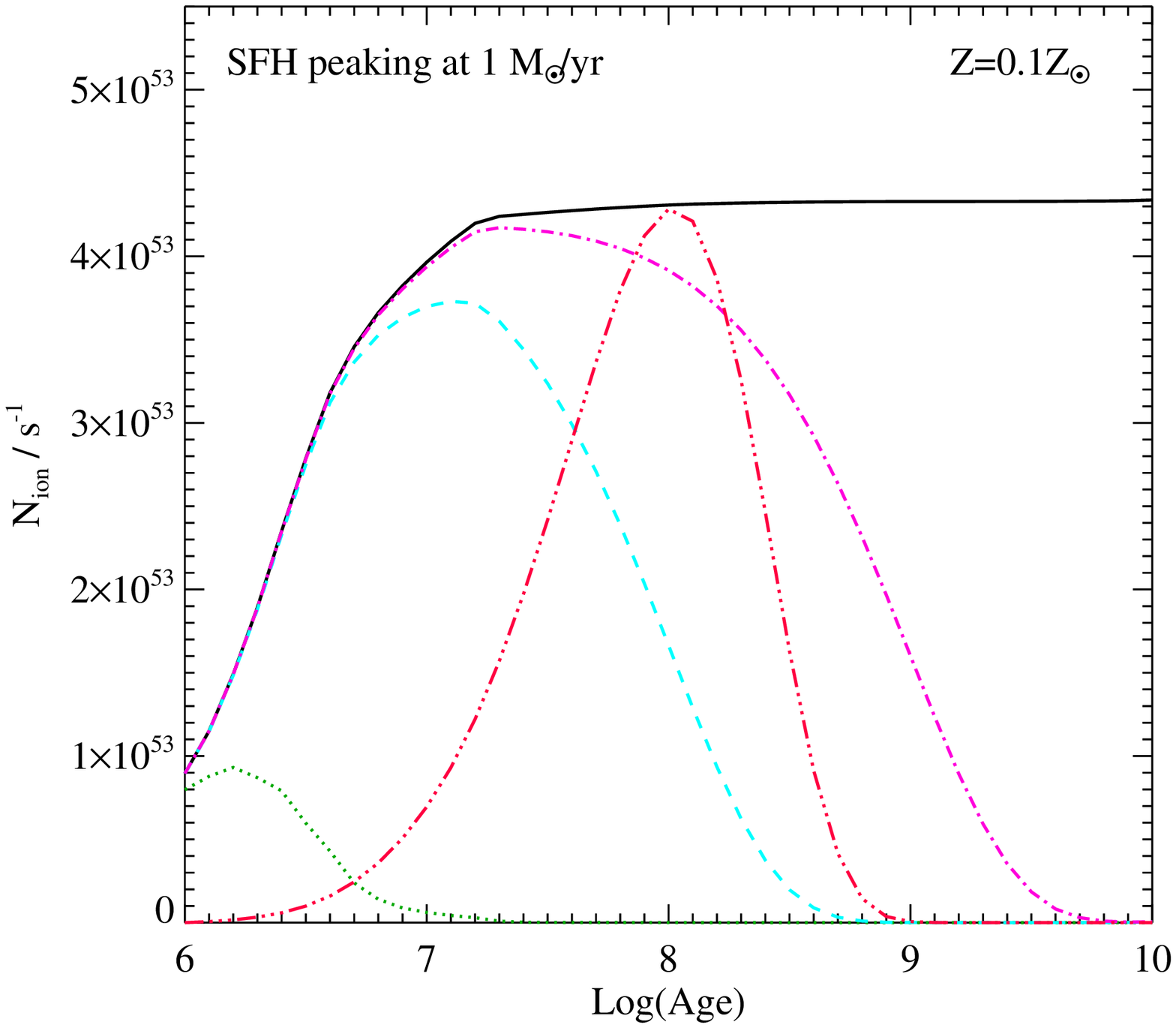}
\includegraphics[width=\columnwidth]{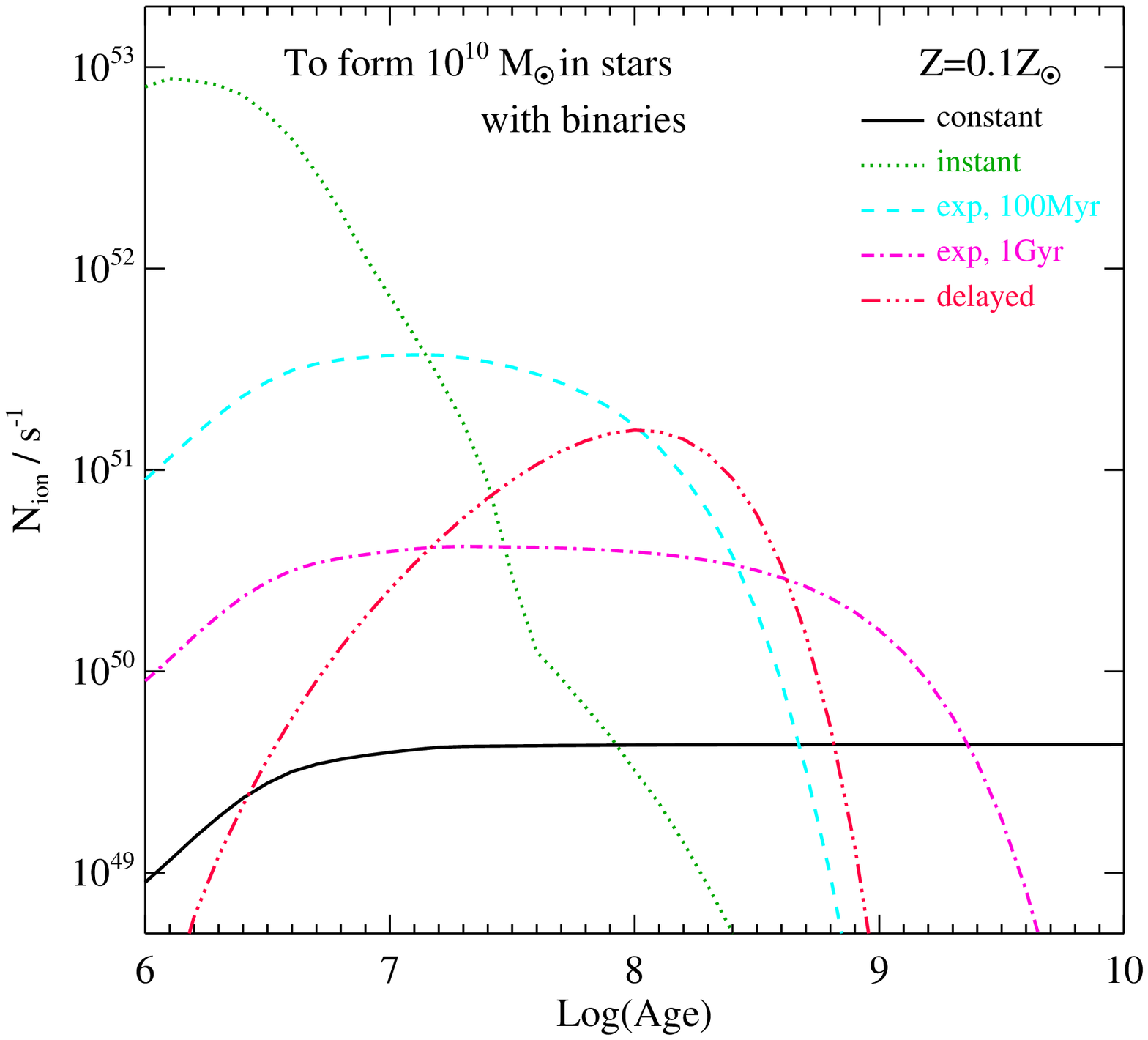}
\caption{The ionizing photon flux (i.e. Lyman continuum emission in photons per second) from a stellar population given five different star forming histories. Upper panels show the results for our single star models. Lower panels illustrate the same but for a population incorporating binaries. The solid line indicates continuous star formation at a constant rate, the dotted line a single, instantaneous starburst at the onset of star formation. The dashed and dot-dashed lines indicate exponentially decaying star formation histories in which the rate varies with time as exp(-t/$\tau$), with $\tau$=100\,Myr and 1\,Gyr respectively. The dot-dot-dashed line indicates a `delayed exponential' star formation history, in which the rate varies as t/$\tau$\,exp(-t/$\tau$). In the left-hand Figure, the maximum star formation rate is scaled to 1\,M$_\odot$\,yr$^{-1}$ ($10^6$\,M$_\odot$ in stars in the first Myr for the instantaneous burst), while in the right-hand Figure, the populations are scaled to produce the same mass, $10^{10}$\,M$_\odot$ by stellar population age 10\,Gyr. }\label{fig:ionizing_sfh}
\end{center}
\end{figure*}

We present an illustration of the resulting uncertainties in Figure \ref{fig:ionizing_sfh}. The figure indicates the time evolution of the ionizing flux from stellar populations which either peak at the same star formation rate (left-hand panels) or produce the same total mass of stars (right-hand panels), but which display a range of star formation histories and either neglect (upper panels) or incorporate (lower panels) the effects of binary evolution. The models shown are at a metallicity of 0.1\,Z$_{\odot}$ since the effects discussed are strongest at lower metallicities and weakest around Solar metallicity. The clearest, most distinctive difference is that the binary population incorporates a higher fraction of hot stars, and so produces higher fluxes at a fixed star formation rate. However, careful inspection reveals a second, also significant factor which is partly obscured by the necessity for logarithmic scaling in the figure. Populations incorporating binaries produce ionizing flux for longer and thus the continuous star formation history does not reach a steady state until later in the binary population starburst, while the more complex star formation histories also peak somewhat later than in the single star evolution case.

As the figure makes clear, any simple prescription which assumes a fixed conversion from ionizing flux, or its proxy H$\alpha$ line emission in the optical, to an `instantaneous' star formation rate neglects rather significant uncertainties on the origin of the ionizing photons and the past star formation history, which are increased in a binary stellar evolution scenario due to the range of possible stellar lifetimes for ionizing sources. Care will also be required in interpreting the colours of composite stellar populations due to contributions from both nebular continuum and emission line reprocessing of this ionizing flux.

A similar stellar lifetime effect is seen in the photometric colours of composite stellar populations. Figure \ref{fig:mag_sfh} compares the colour evolution of the populations (using the SDSS $ugriz$ filter system and neglecting the effects of nebular continuum, see below). For both single and binary evolution models, the instantaneous starburst shows more stochastic variation in colour, as different mass populations reach the end of their main sequence or post-main-sequence lifetimes and a dramatic colour change appears.  In the composite stellar populations, this variation is smoothed out by stars which formed slightly later, reducing the variability in photometric colour of the population. The colour differences between the single and binary models are relatively small ($\Delta$\,mag$\sim$0.1) in the visible bands ($gri$), and at early ages ($\sim10^8$\,years). The behaviour with age in the two populations is similar. The binary population is consistently bluer than its single star counterpart with the colour difference most pronounced in the SDSS $u-g$ colour, and at ages $>1\,$Gyr, where the binary models are as much as 0.8 magnitudes bluer than the single star models. It is notable, if unsurprising, that the shortest wavelength of the standard SDSS colours also most clearly shows the impact of the star formation history, with old instantaneous bursts redder than populations forming stars over a more extended time period. Delayed and exponentially-declining starbursts begin their lives approximating the colour of a continuous starburst, before converging with the colours of an instantaneous starburst at a time comparable to their exponential timescales, $\tau$.

{
\begin{figure*}
\begin{center}
\includegraphics[width=\columnwidth]{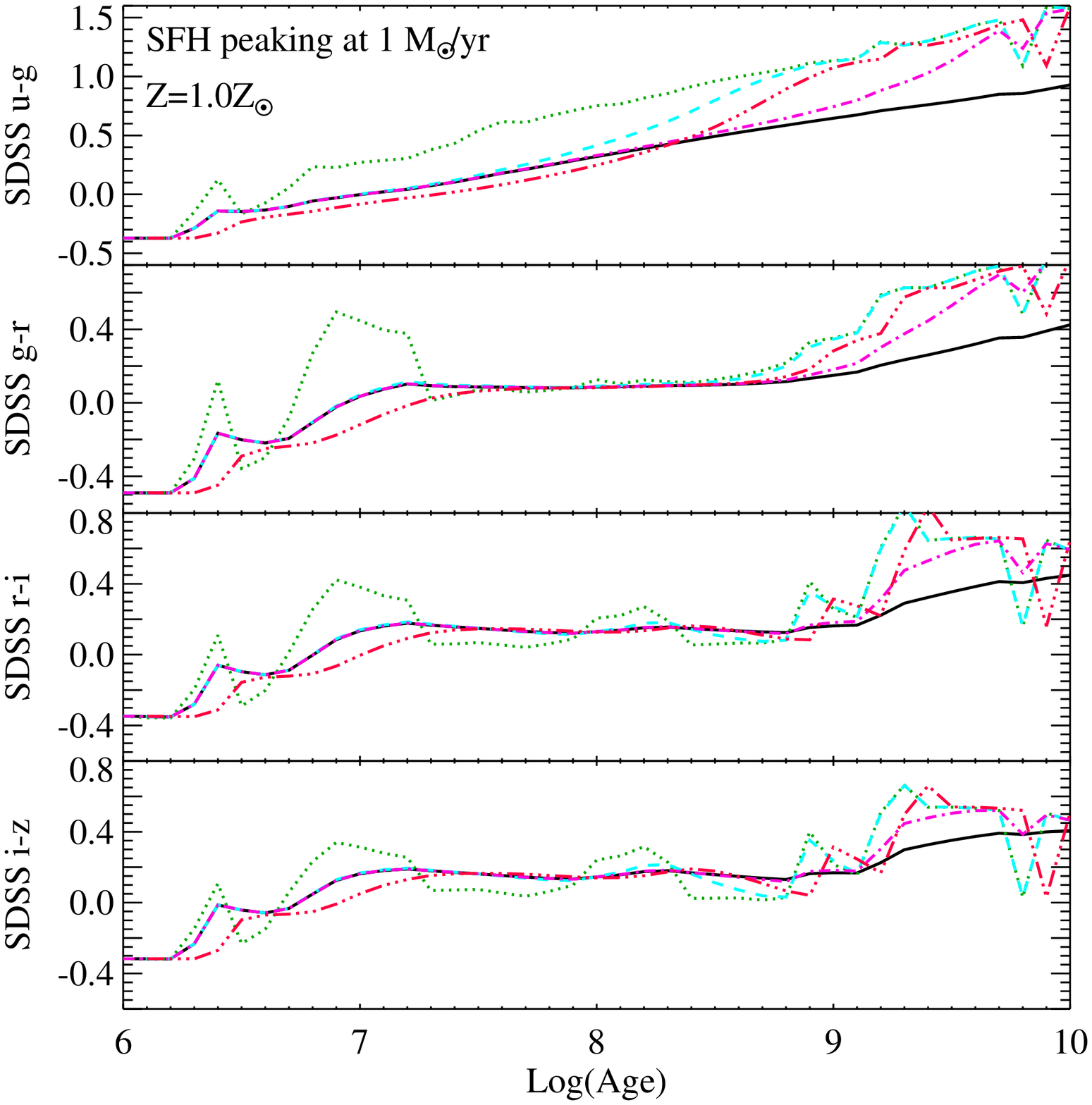}
\includegraphics[width=\columnwidth]{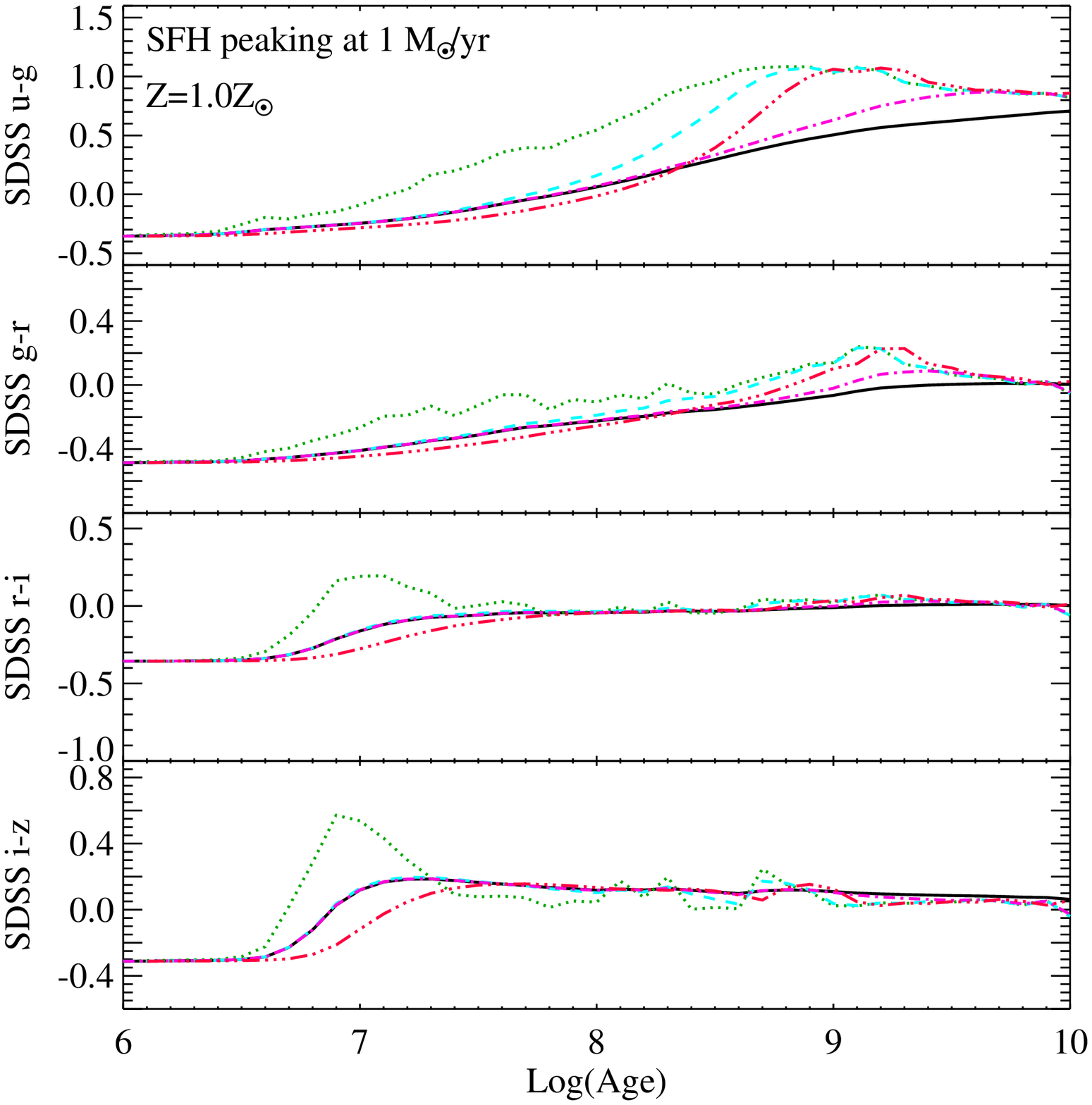}
\caption{The photometric colours (in the SDSS $ugriz$ system) for a stellar population given five different star forming histories. Left panels show the results for our single star models. Right panels illustrate the same but for a population incorporating binaries. Note that nebular emission is not included (see Figure \ref{fig:mag_sfh_cloudy}). Star formation histories are shown defined in Figure \ref{fig:ionizing_sfh}.}\label{fig:mag_sfh}
\end{center}
\end{figure*}

\begin{figure*}
\begin{center}
\includegraphics[width=\columnwidth]{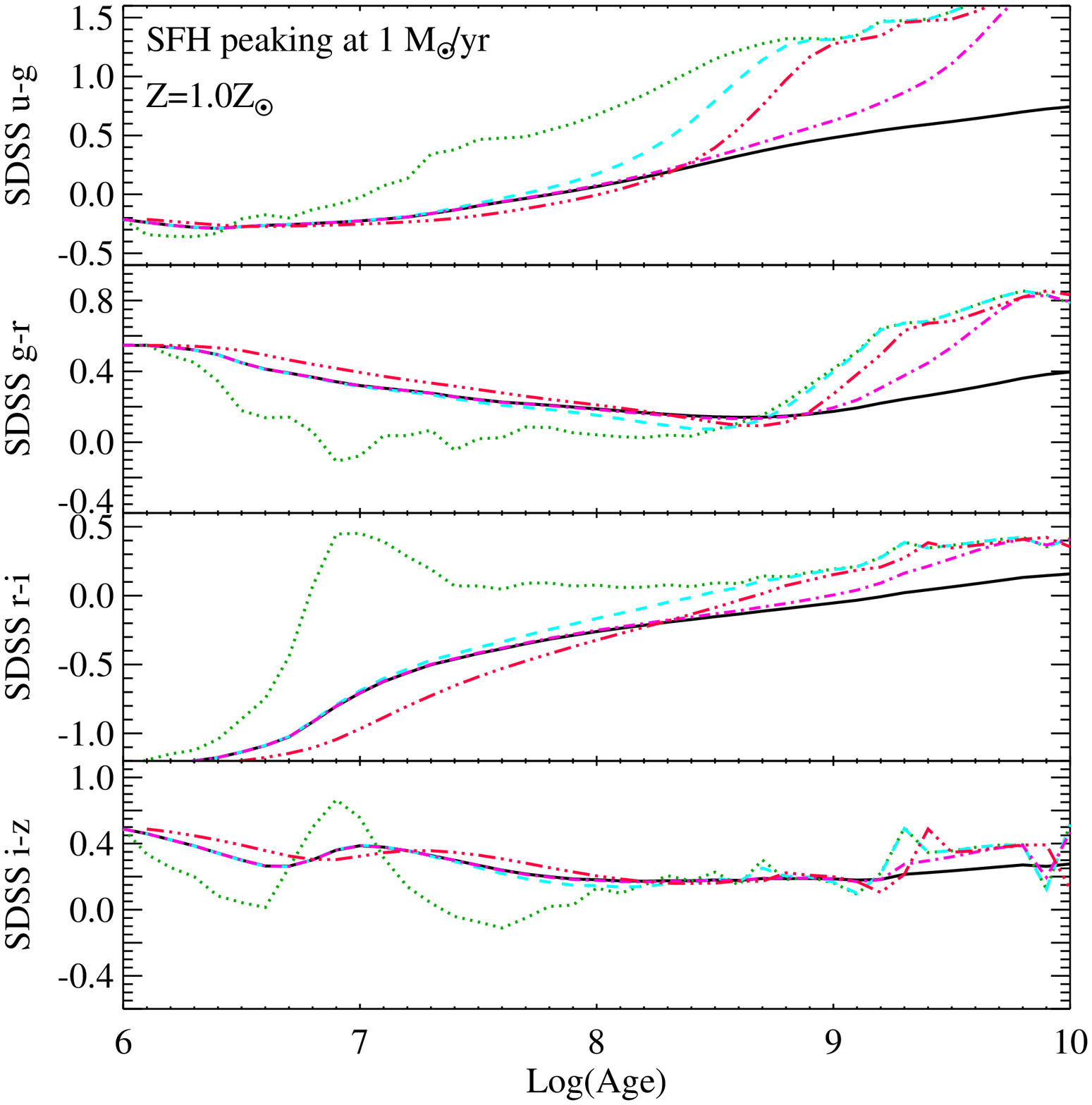}
\includegraphics[width=\columnwidth]{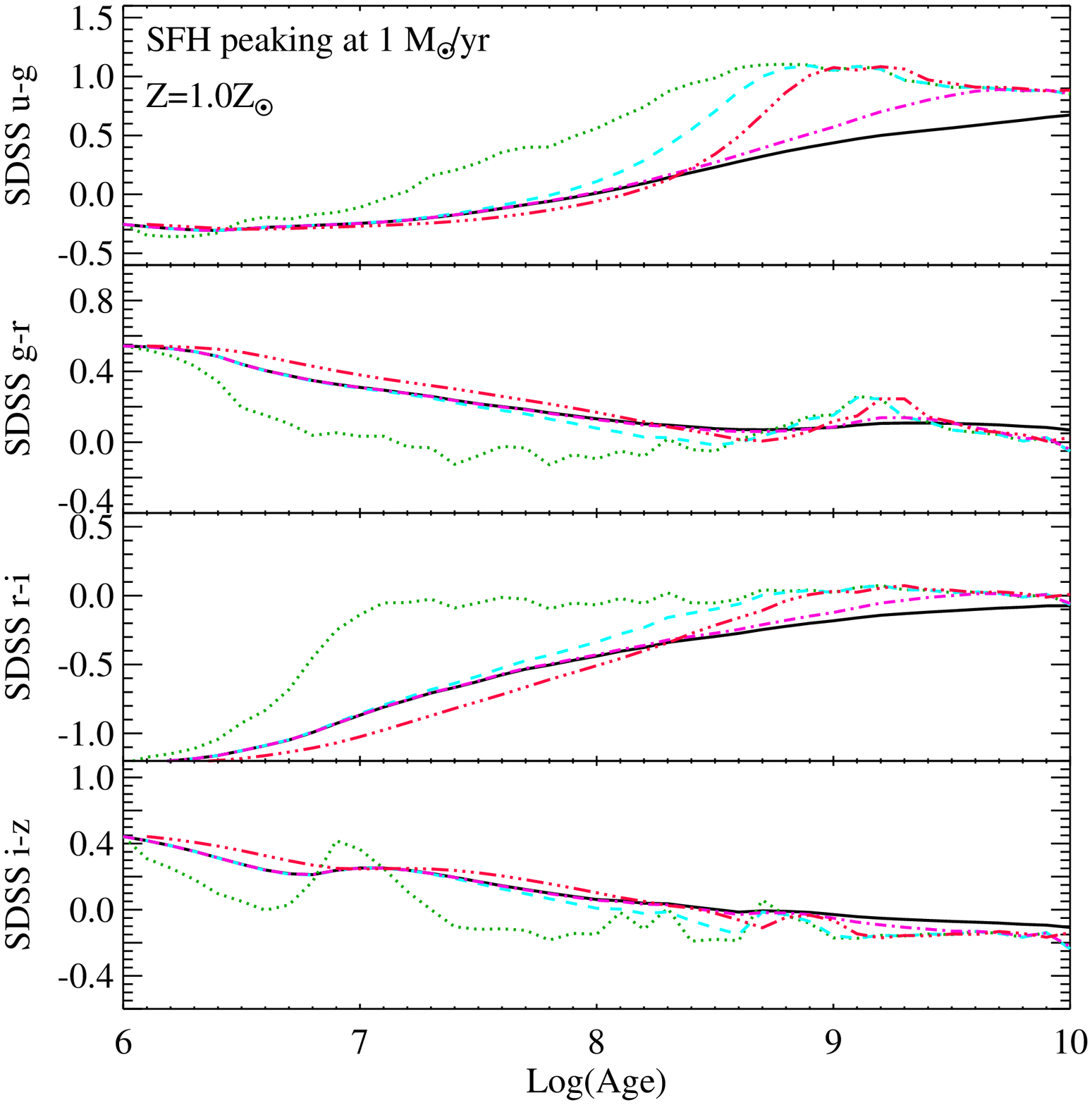}
\caption{The photometric colours (in the SDSS $ugriz$ system) for a stellar population {\bf with} nebular gas effects included at a fixed electron density of $10^2$\,cm$^{-3}$, given five different star forming histories. Left panels show the results for our single star models. Right panels illustrate the same but for a population incorporating binaries. Star formation histories are shown as defined in Figure \ref{fig:ionizing_sfh}.}
\label{fig:mag_sfh_cloudy}
\end{center}
\end{figure*}
}

\subsection{Nebular emission}\label{sec:nebular}

The radiative transfer of this stellar emission through nebular gas screen within the source galaxy is beyond the scope of \bpass\ itself, but can be modelled in detail using a radiative transfer code such as {  CLOUDY} \citep{1998PASP..110..761F} or {  MAPPINGS} \citep[see][and references therein]{2013ApJS..208...10D} to assess the contribution of nebular continuum and line emission. Again, this adds degrees of freedom to any model that attempts to explain observed galaxies, in the form of the interstellar gas density and its geometry, or alternatively in the form of the {\it ionization parameter}. This measures the ratio of the number of ionizing photons to the local gas density, and is thus constructed through the combination of appropriate stellar atmosphere models and a choice of gas distribution \citep[e.g.][]{2017ApJ...840...44B}. As Figure \ref{fig:euv} makes clear, however, the extreme ultraviolet is a spectral region severely affected by the presence of absence of binary evolution. These hard photons are seldom observed, but instead reprocessed by the interstellar medium and thus nebular emission cannot be ignored in any complex stellar population.

Detailed comparison with observed extragalactic spectra requires a grid of models with different gas density and geometry since self-shielding and collisional excitation or de-excitation of ions can alter the transition probabilities in either very sparse or very dense gas. Such a grid is explored in Xiao et al (in prep). Here we present a single model for clarity. Our baseline model for nebular emission is constructed using {  CLOUDY} v13.03. It assumes an electron density of $10^2$\,cm$^{-3}$, distributed in a sphere around the stellar population. In previous work, we have used a gas distribution with an inner radius of 10 parsec, corresponding to an ionization parameter of -2.26 for an established population forming stars with a constant rate of 1\,M$_\odot$\,yr$^{-1}$ at Z$_\odot$ in our binary models, and -2.59 for our single star models, assuming a typical electron fraction of $10^{-3}$. An inner radius of 3\,pc yields equivalent ionization parameters of -1.21 and -1.55. 

We assume that the nebular emission is radiation bounded, allowing for absorption through the diffuse interstellar medium between individual star forming clusters.
Our selected gas density is a typical value for extragalactic star forming H\,{  II} regions in the local Universe, although we note that these range over several orders of magnitude in density \citep[see e.g.][]{2009A&A...507.1327H}, and there are indications that a slightly higher nebular gas density $\sim250$\,cm$^3$ may be more typical of galaxies in the distant Universe \citep[e.g.][]{2016ApJ...816...23S}. Similarly it is possible to identify nebular clouds with inner radii both closer and more distant than in our previous baseline nebular model, and here we opt to show results for an inner gas radius of 3pc, consistent with other work in the field \citep[e.g.][]{2017ApJ...840...44B,2016ApJ...816...23S}.

In Figure \ref{fig:mag_sfh_cloudy}, we calculate the colours for the same complex stellar populations shown in Figure \ref{fig:mag_sfh} but this time with nebular emission included using {  cloudy} for post-processing. A constant gas density of $10^2$\,cm$^{-3}$ is assumed. We caution that the gas density is unlikely to be this high in the vicinity of old stellar populations (which tend to disperse their natal clouds and whose flux is instead intercepted by the cool diffuse ISM), and that this might thus be considered an unphysical model except in the case where an old population and a young one are co-located. The effect of nebular gas processing varies both with age and with wavelength. For simple, single-age starburst populations built from single star models, the effects of nebular gas are only significant at young ages ($<10^7$\,years), while a complex star formation history extends the duration of nebular effects to $\sim10^8$\,years, depending on the details of the star formation timescale. The star formation histories considered here approximate the behaviour of a continuous star formation epoch at early times before converging towards the behaviour of the single age population after the star formation rate has declined, consistent with the behaviour seen in the ionizing photon flux (see Figure \ref{fig:ionizing_sfh}). In the SDSS $u-g$ colour the single star models are redder by $\Delta$mag$\sim0.5$ at early times as the hard blue photons are reprocessed into a cooler nebular continuum, while the reddest optical colour sees a larger shift ($\Delta$mag$\sim0.8$ in $i-z$). The behaviour in the intermediate colours is modified by the presence of strong nebular emission lines, primarily H$\alpha$ which lies in the SDSS $r$-band. As Figure \ref{fig:mag_sfh_cloudy} shows, this leads to very blue colours in $r-i$ at early times, while H$\beta$ and \oiii\ occur in the $g$-band, mitigating variation in the $g-r$ colour. 

The presence of binary evolution models in a population modifies this behaviour. We can see the source of this difference in Figure \ref{FigJJ8a} by comparing the single star to binary star populations. In the latter blue stragglers and stars that have lost their hydrogen envelope to become hot helium stars appear at ages later than 10\,Myrs that are not possible from a single star population. These prolong the period over which nebular emission is significant, smoothing out some of the variation in the single star models with time. Again, populations with nebular emission show redder colours at young population ages than those without, except where nebular emission lines strongly contaminate a photometric band.  However binary populations never reach the reddest colours observed in the single star models.

\subsection{The local star-forming galaxy population}\label{sec:sdss}

To provide an observational validation and test of our \bpass\ spectral synthesis, we consider the Sloan Digital Sky Survey Data Release 7 \citep[SDSS DR7, ][]{2009ApJS..182..543A} emission line galaxy sample first explored by \citet{2004MNRAS.351.1151B,2008MNRAS.385..769B}, and which has been classified by the JHU-MPA collaboration using a combination of SED fitting and strong emission line calibrations for age and metallicity. We select galaxies automatically classified as star forming rather than powered by emission from active galactic nuclei. While the caveats regarding nebular emission uncertainties discussed above apply equally to the strong line metallicity and type classifications, this sample nonetheless provides a useful point of comparison for our models. 

\begin{figure*}
\begin{center}
\includegraphics[width=0.45\columnwidth]{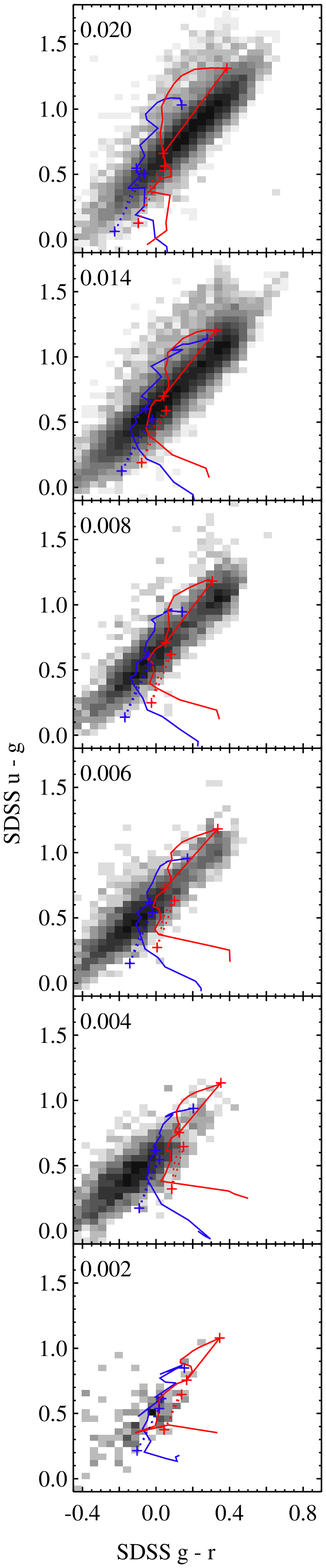}
\includegraphics[width=0.45\columnwidth]{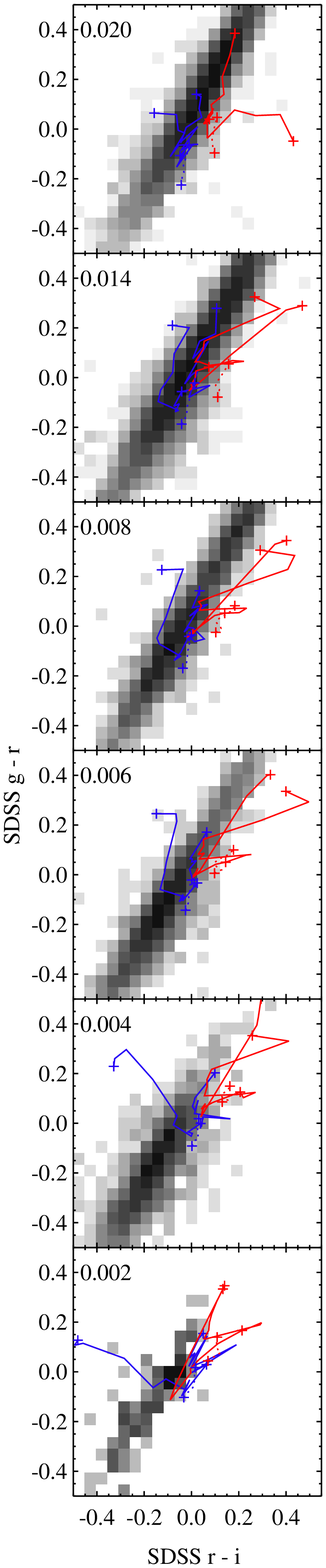}
\includegraphics[width=0.45\columnwidth]{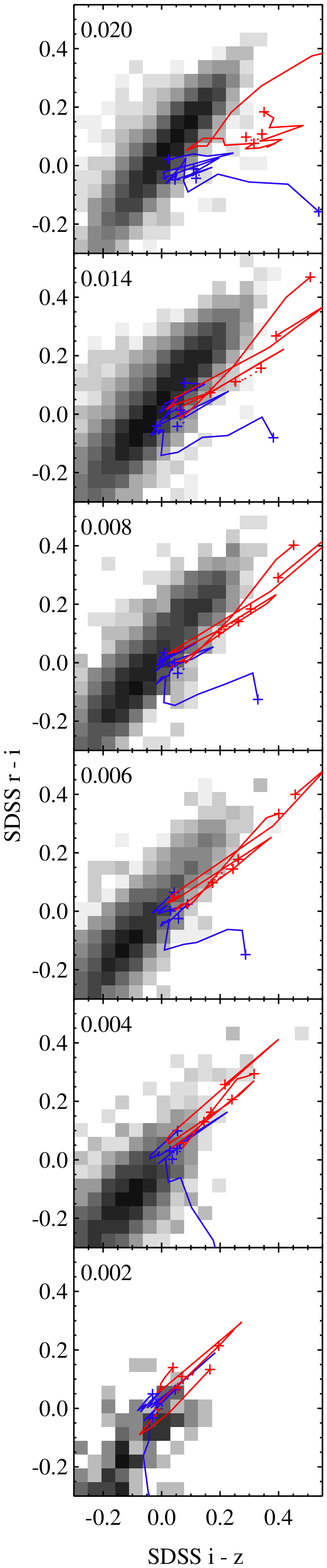}
\caption{The photometric colours of star forming galaxies in the SDSS survey (shown as a density map in greyscale), overplotted with the colour evolution of our models with age, assuming no effect from dust or nebular emission, and plotted as a function of metallicity. Single star tracks are plotted in red, while binary tracks are shown in blue. Tracks are plotted at ages from 1 Myr (bluest colours, i.e. lowest numerical values in $u-g$ on each track) to 1 Gyr with 1 dex increments in age marked by crosses on the tracks.}
\label{fig:mag_sdss1}
\end{center}
\end{figure*}

In Figure  \ref{fig:mag_sdss1} we illustrate the colours of our single-age starburst stellar models, overplotted on this SDSS galaxy data, as a function of stellar population age and metallicity. Given that in this case no correction has been made for dust or nebular emission, the broad agreement, in terms of range of colours probed by star forming galaxies and  their trends,  is good. The models sweep out much of the colour-colour space occupied by the galaxy population during their lifetime. There is evidence, however, that the simple single-age stellar models are not sufficient.  The gradient of the galaxy colour-colour relation in the $u-g-r$ plane is steeper than that of the models, while there appears to be an offset of approximately 0.2 mag in $r-i$, the effects of which become more apparent with increasing metallicity.  There are likely several components adding to an explanation for these offsets:  the role of nebular emission in modifying the continuum flux, the role of dust extinction in individual galaxies and the role of complex star formation histories, as well as the over-representation of AGB stars in our single star models at late times. With increasing present-day metallicity the number of permutations of these factors rises. The well established mass-metallicity relation in the local Universe \citep{2004ApJ...613..898T} means that such galaxies are typically larger, older and thus have been through more interactions, starbursts and stages of galactic evolution than low-metallicity galaxies.

Exploring the full parameter space of complex galaxy models is beyond the scope of this paper; the number of free parameters grows rapidly with the complexity of the model and quickly exceeds the number of constraints unless a full spectral and emission line analysis of each comparison source is undertaken. Nevertheless, to provide a slightly more realistic galaxy model for comparison with observed galaxy populations, we consider a toy stellar population that incorporates a subset of the effects discussed above. In this model, we consider an underlying old stellar population which formed its stars in a short lived exponentially-decaying starburst between 3 and 1\,Gyr before the present. On top of this, we impose a young starburst contributing either 1\% or 50\% of the stellar mass of the galaxy, with an exponentially decaying star formation rate and a decay constant, $\tau=$100\,Myr. Both bursts are at the current gas phase metallicity, although varying this by one bin has little effect on the galaxy colours except at the lowest metallicities. Stars are assumed to be embedded in dense nebular gas (at the same metallicity) for the first 15\,Myr, and thereafter not to be modified by gas absorption or emission. We refer to this as our ``late burst" model. We take a minimal approach to modelling dust extinction, correcting each observed data point for internal extinction as estimated by the JHU-MPA database, and then allowing the models to vary slightly according to an additional internal dust absorption component using the \citet{2000ApJ...533..682C} dust law and an $E(B-V)=0.0, 0.05$ or 0.10, before comparison with the data.  

\begin{figure*}
\begin{center}
\includegraphics[width=0.45\columnwidth]{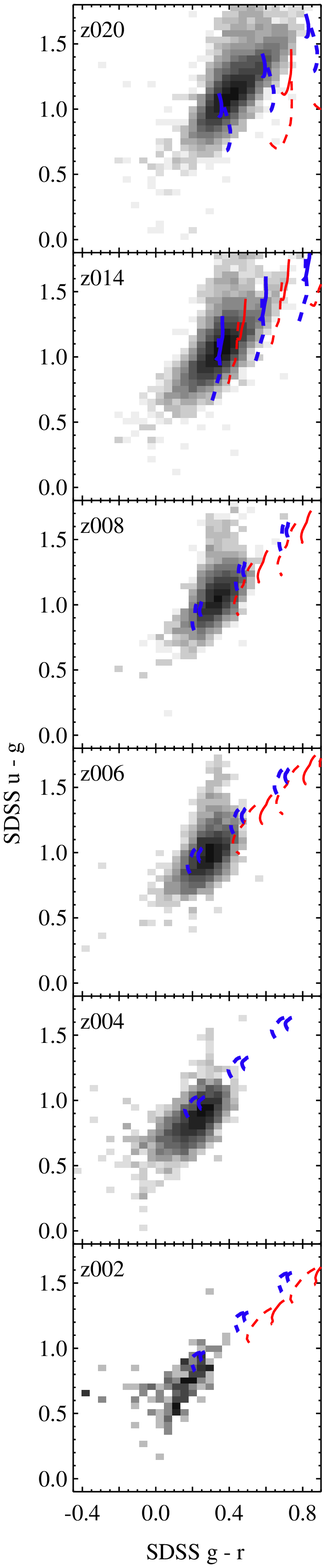}
\includegraphics[width=0.45\columnwidth]{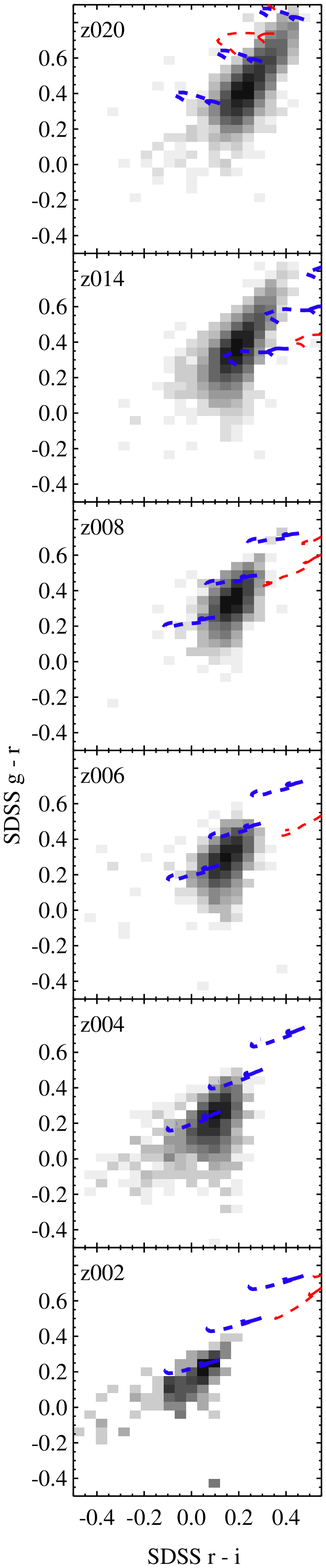}
\includegraphics[width=0.45\columnwidth]{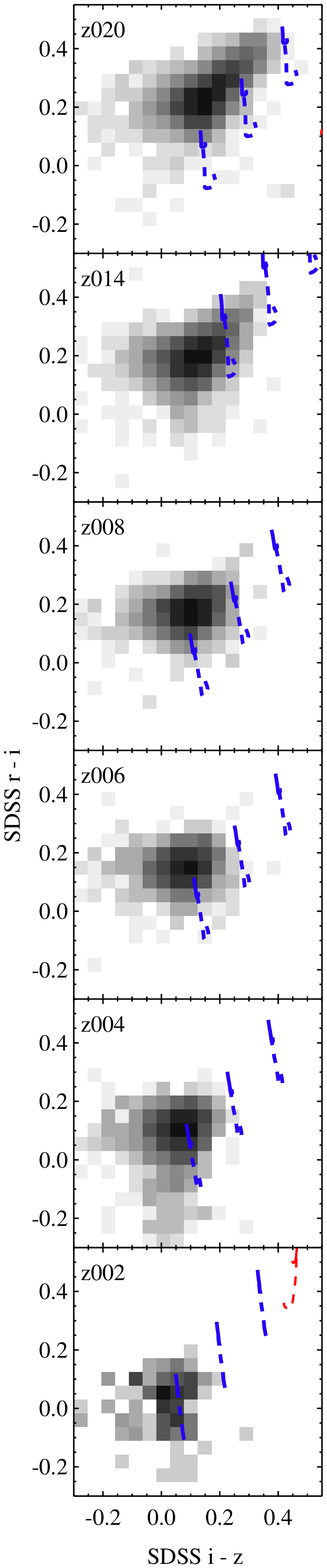}
\caption{The photometric colours of star forming galaxies in the SDSS survey, overplotted with the colour evolution of our complex ``late burst'' star formation model with age and dust extinction. Models shown incorporate an old underlying stellar population, with either 1\% (solid line) or 50\% of mass (dashed line) contributed by a new starburst ranging from 1 to 30\,Myr in age (age along tracks). CLOUDY radiative transfer is used to model the nebular emission in populations up to 15\,Myr in age, but this is neglected in older populations. An initial correction for internal extinction as calculated by the JHU-MPA collaboration is applied to the SDSS photometric data. Additional extinctions of $E(B-V)=0., 0.05$ or 0.1 are applied to the models using the Calzetti dust extinction law and are shown as different tracks with the more extincted tracks moving redwards. Our single star models are shown in red and often lie off the plots, binary models are shown in blue.}
\label{fig:mag_sdss2}
\end{center}
\end{figure*}

As Figure \ref{fig:mag_sdss2} demonstrates, this more complex star formation history improves the agreement between model colours and observed data  for our binary models, but worsens it for our single star models. Most of the variation in colour of the SDSS population is accounted for by fairly minor differences in dust extinction, or perhaps in  appropriate extinction law, with elapsed time since the late starburst producing a spread  of colours away from the dust reddening vector. Notably, when modelling this complex star formation history, our single star models perform significantly worse, and are unable to reproduce the observed colours at most metallicities. Relatively simple star formation histories are favoured by the single star models, but these do show the very red colours associated with overrepresented AGB stars in our single star models at late times.

\subsection{Star Formation Rate Indicators (SFRIs)}\label{sec:sfris}

The presence of massive stars in a population leads to distinctive emission characteristics which scale with the size of the recent star formation event, and which are therefore used as star formation rate indicators for unresolved stellar populations.  Amongst those widely used in both the local and distant Universe are the rest-frame far-ultraviolet continuum luminosity and the luminosity of the H$\alpha$ recombination line \citep[see][for a recent review]{2012ARA&A..50..531K}.

Conventionally, star formation rates are calibrated by comparing observational data to the predicted emission strength from stellar population synthesis models forming stars at a constant rate, with adjustments applied for metallicity, dust extinction and to ensure that different star formation rate indicators produce a consistent star formation rate estimate for a given stellar population. Inevitably, calibration efforts focus on the well-established ongoing star formation episodes in relatively local galaxies, and at near-Solar metallicities, with extrapolations to other environments based on a combination of empirical data and analytic projections drawn from stellar population synthesis and nebular gas modelling \citep[e.g.][]{2004AJ....127.2002K}. 

\begin{figure*}
\begin{center}
\includegraphics[width=0.95\columnwidth]{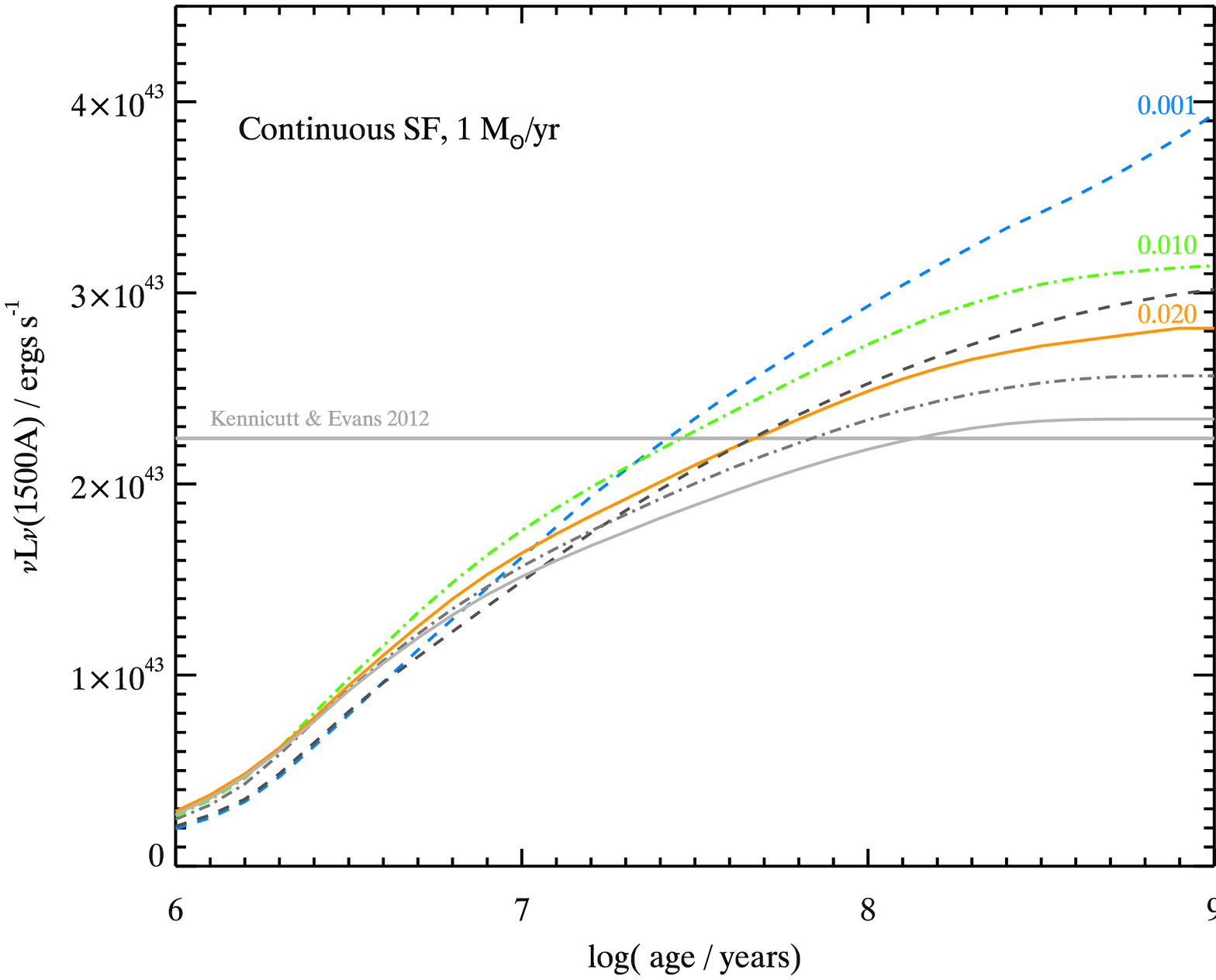}
\includegraphics[width=0.95\columnwidth]{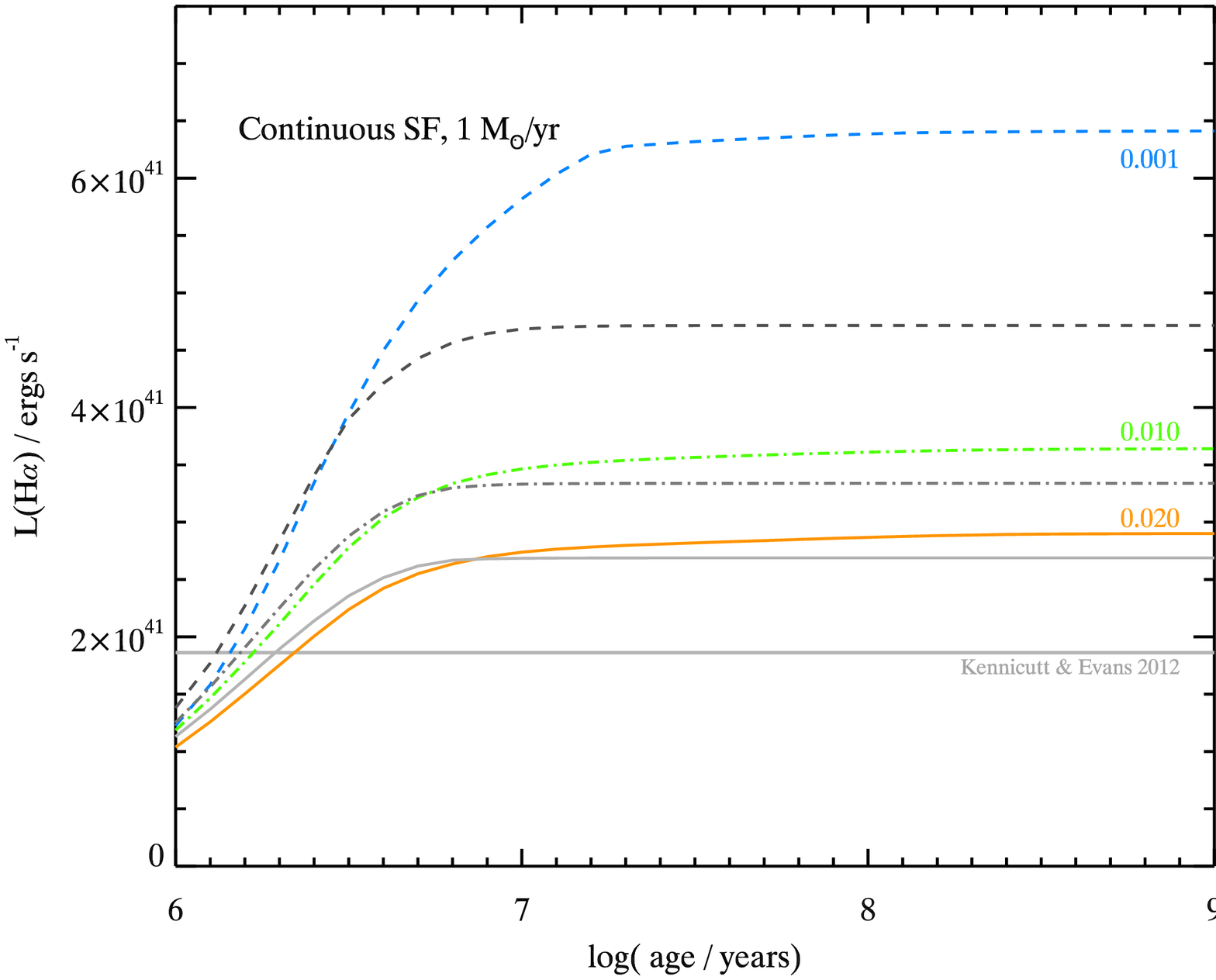}
\caption{The evolution of key star formation rate indicators for young stellar populations.  We show values for three metallicities (corresponding to 0.05, 0.5 and 1\,Z$_\odot$) and for populations incorporating binaries (colour) and without binaries (greyscale with matching linestyle). The solid horizontal line in each case indicates the calibrations recommended by \citet{2012ARA&A..50..531K}.}
\label{fig:sfris}
\end{center}
\end{figure*}

In Figure \ref{fig:sfris} we illustrate the variation in the luminosity per stellar mass of star formation within \bpass\ models with stellar population age, metallicity and the presence or absence of binaries, using our standard model for nebular gas conditions as discussed in section \ref{sec:nebular}. We also compare these to a pair of widely-used SFRI calibrations for the same observational signatures \citep{2012ARA&A..50..531K,2011ApJ...737...67M}. In the case of both the 1500\AA\ rest-frame continuum and the H$\alpha$ line, our calibrations are in good agreement with previous estimates at Solar metallicity. In the case of H$\alpha$ line luminosity, the factor of 1.45 between the Kennicutt \& Evans calibration and that from \bpass\ is entirely accounted for by the adopted IMF.  As discussed in \citet{2016MNRAS.456..485S}, the choice of 300\,M$_\odot$ for an upper mass limit leads to a 45\% increase in the ionizing photon production (and hence Hydrogen recombination line flux) relative to an IMF with a maximum of stellar mass of 100\,M$_\odot$, while the effect of the most massive stars on the rest-frame ultraviolet continuum (which is built up from a broader range of stellar masses) is less dramatic.

When deviating from near-Solar metallicities, \bpass\ typically predicts a higher luminosity for a given star formation rate (or, alternately, fewer stars being formed for a given observed flux) than previous calibrations. This is particularly true when binaries are incorporated into the population synthesis. Quantitative results as a function of metallicity and binary/single star model are given in Table \ref{tab:sfris}, where these values are quoted at an age of 10$^{8.5}$\,years after the onset of constant star formation. Values in this table can be utilised as star formation rate conversions using the prescription:
\begin{equation}
\log(\dot{\mathrm{M}} / \mathrm{M}_\odot\,\mathrm{yr}^{-1}) = \log(L / \mathrm{erg\,s}^{-1}) - C(Z)
\label{eqn:sfri}
\end{equation}
where $L$ is the relevant observed luminosity and $C(Z)$ is the metallicity-dependent value given in the table.

It should be noted that Figure  \ref{fig:sfris} also demonstrates the key importance of understanding star formation timescale when attempting to apply a star formation calibration to an unresolved star forming system, particularly in the distant Universe where stellar populations are typically younger. Single star populations forming stars at a continuous rate reach a stable calibration for H$\alpha$ by about 5\,Myr after the onset of star formation, and in the UV after about 100\,Myr. By contrast, binary populations take longer to establish a constant star formation rate to luminosity relation and in some cases, as for  the rest-frame ultraviolet continuum at the lowest metallicities, it is not clear that this ever occurs. This time-dependent conversion factor needs to be considered when assigning star formation rates to galaxies or unresolved stellar clusters at stellar ages of less than a few hundred Myr \citep[see also][]{2017arXiv170507655G}.

\begin{table}
\caption{The predicted luminosity corresponding to a constant star formation rate of 1\,M$_\odot$\,yr$^{-1}$, i.e. values of $C(Z)$}
\begin{tabular}{lccccc}
Z  & Z/Z$_\odot$ & \multicolumn{2}{c}{ H$\alpha^a$} & \multicolumn{2}{c}{$\nu$L$_\nu$(1500)$^b$}\\
  & & single & binary & single & binary\\
\hline\hline
0.00001 & 0.05\%  &  41.76    & 41.92  &  43.53  & 43.60\\
0.0001  & 0.5\%  &  41.81    & 41.96  &  43.53  & 43.60\\
0.001   & 0.05  &  41.67    & 41.81  &  43.47  & 43.55\\
0.002   & 0.10  &  41.64    & 41.77  &  43.44  & 43.53\\
0.003   & 0.15  &  41.62    & 41.74  &  43.42  & 43.50\\
0.004   & 0.20  &  41.60    & 41.70  &  43.40  & 43.49\\
0.006   & 0.30  &  41.58    & 41.63  &  43.39  & 43.47\\
0.008   & 0.40  &  41.55    & 41.60  &  43.37  & 43.45\\
0.010   & 0.50  &  41.52    & 41.56  &  43.36  & 43.44\\
0.014   & 0.70  &  41.49    & 41.51  &  43.34  & 43.41\\
0.020   & 1.00  &  41.43    & 41.46  &  43.32  & 43.38\\
0.030   & 1.50  &  41.34    & 41.36  &  43.28  & 43.36\\
\hline
0.020$^1$ & & 41.24 & 41.31 & 43.30 & 43.37 \\
0.020$^2$ & & 41.27 & -- & 43.35 & --  \\
\end{tabular}
{\small
$^{a,b}$\,log$_{10}$(luminosity / ergs s$^{-1})$ at log(age/years)=8.5. \\
$^1$ for an IMF truncated at a stellar mass of 100\,M$_\odot$.\\
$^2$ the recommended calibration of \citet{2012ARA&A..50..531K}.} \label{tab:sfris}
\end{table}

\subsection{The distant galaxy population}\label{sec:distant}

A number of recent publications have suggested that star-forming galaxies in the distant ($z>2$) Universe may show signs of a harder typical ultraviolet radiation field than is common in the local Universe \citep[e.g.][]{2016ApJ...826..159S,2015MNRAS.450.1846S,2014MNRAS.444.3466S}. These galaxies are selected primarily on the presence of a Lyman break in their rest-frame ultraviolet continuum \citep[Lyman break galaxies or LBGs,][]{1992AJ....104..941S} or strong Lyman alpha emission \citep[LAEs, e.g.][]{1998AJ....115.1319C}. The galaxies are typically younger and  have simpler stellar populations than in local star forming populations, and they are also likely at significantly sub-Solar ($<0.5$\,Z$_\odot$) metallicities. As Figure \ref{fig:euv} made clear, the role of binaries on extreme ultraviolet emission is a strong function of metallicity.

This is an active and rapidly developing area of research that will likely be revolutionized by the imminent launch of the James Webb Space Telescope (JWST). The NIRSPEC instrument will enable deep spectroscopy of hundreds (if not thousands) of star-forming galaxies at $3<z<8$, from the rest-frame ultraviolet through to the rest-optical \citep[see e.g.][]{2016ASPC..507..305G}. As discussed in \citet{2017arXiv170207303S}, the use of binary evolution models such as \bpass\ may well be necessary to model such galaxies and a number of papers have already been published exploring this area, notably \citet{2016ApJ...826..159S} on applications to galaxies at Cosmic Noon ($z\sim2-3)$, \citet{2016MNRAS.456..485S} and \citet{2016MNRAS.459.3614M} on implications for reionization and \citet{2016MNRAS.458L...6W} for application to semi-analytic galaxy evolution modelling. Here we summarize some key observational tests and applications of \bpass\ in this regime and the key differences between version 2.1 and our earlier v2.0 models.

\subsubsection{Ionizing Photon Production Efficiency}

A key parameter for constraining the reionization history of the Universe, and  one of the most significant  to change behaviour between \bpass\ v2.0 and v2.1, is   the ionizing photon production efficiency $\xi_{\mathrm{ion}} = \dot{N}_\mathrm{ion}/f_\mathrm{esc}\,L_\mathrm{UV}$, i.e. the ratio between the ionizing photon production rate and the rest-frame ultraviolet continuum luminosity for a given fraction of such photons which escape their host galaxy. While the luminosity can be directly observed in the distant Universe, the ionizing photon production rate must be inferred from either models or indirect observations of the strong nebular emission lines generated from the galaxy in question, assuming that a fraction ($1-f_\mathrm{esc}$) of the ionizing continuum is absorbed by the interstellar medium. 

For simplicity, an escape fraction  $f_\mathrm{esc}=0$ is often assumed, yielding $\xi_{0,\mathrm{ion}}$, but in practice the paradigm that the intergalactic medium is ionized by early star formation, and that ionization state is maintained over cosmic time, requires a much higher escape fraction. Observations aimed at constraining that escape fraction compare  measured luminosities at wavelengths shortwards of the ionization edge of hydrogen at $\lambda$(rest)=912\AA\ with  models of what flux the  stellar population is expected to produce, given the observed 1500\AA\ luminosity.  $\xi_{0,\mathrm{ion}}$ is thus a crucial parameter both for estimation of escape fraction and for interpretation of the effects of star forming galaxies on the reionization epoch.

Between v2.0 and v2.1 we have made two important changes to the models that affect this parameter:  we have extended our models to still lower metallicities, and we have improved the stellar atmosphere models in use for the highest gravity O stars (see section \ref{sec:method_atmos_wmbasic}). These models are particularly important at low metallicities (where mass loss is lowest and stars can thus remain more massive), and in the rotationally mixed stars  which dominate the ionizing photon production. 

\begin{figure}
\begin{center}
\includegraphics[width=1.\columnwidth]{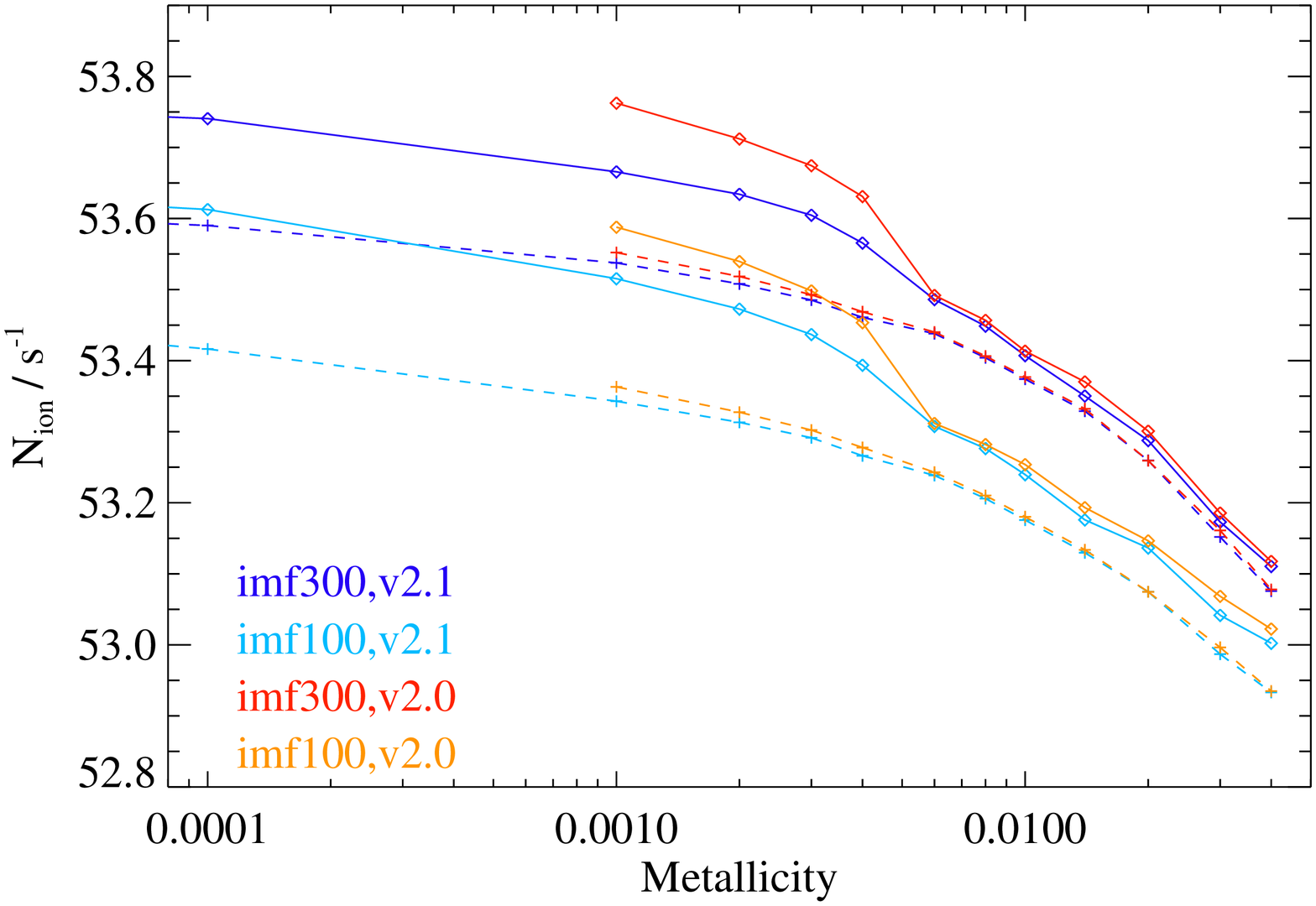}
\includegraphics[width=1.\columnwidth]{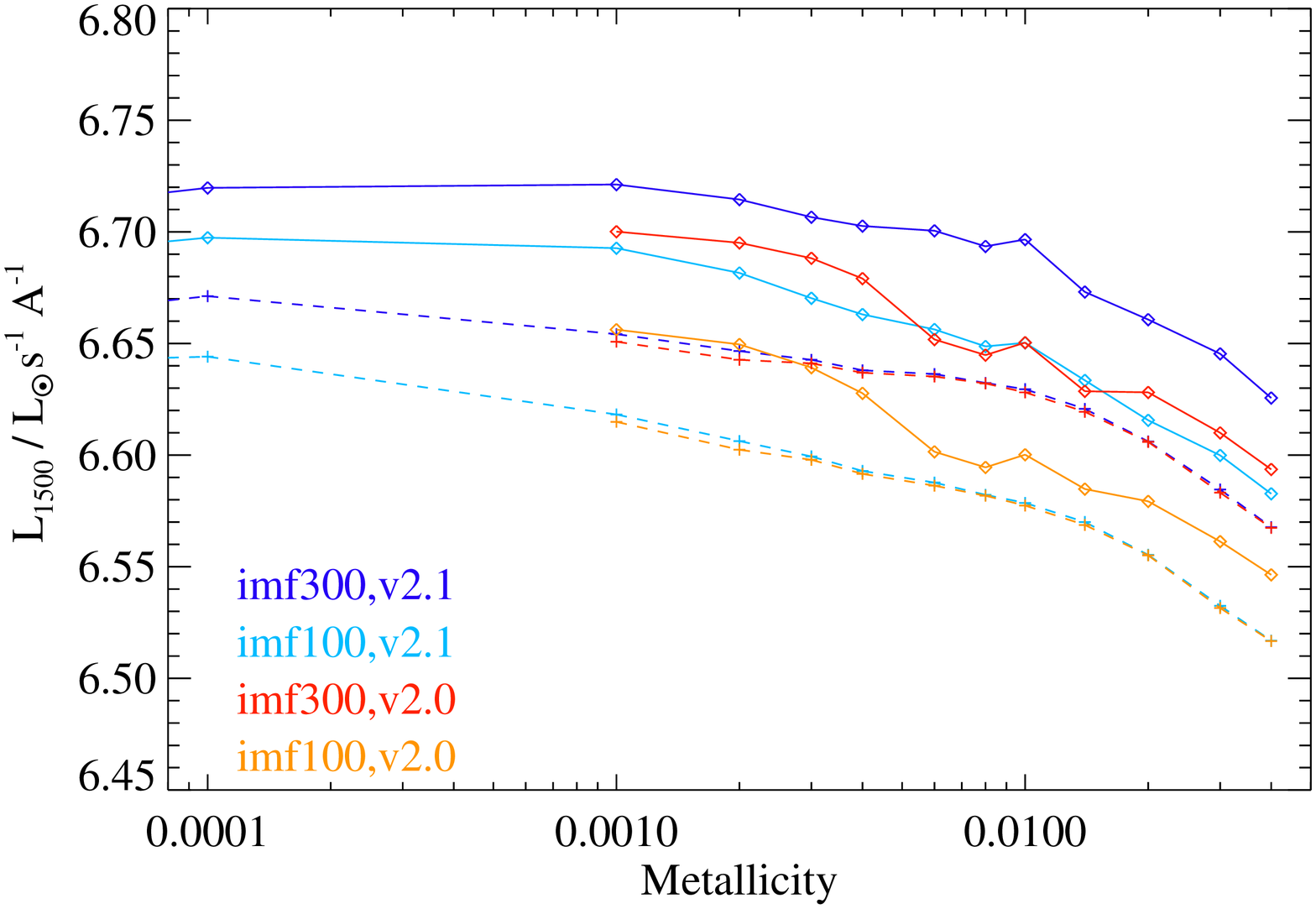}
\includegraphics[width=1.\columnwidth]{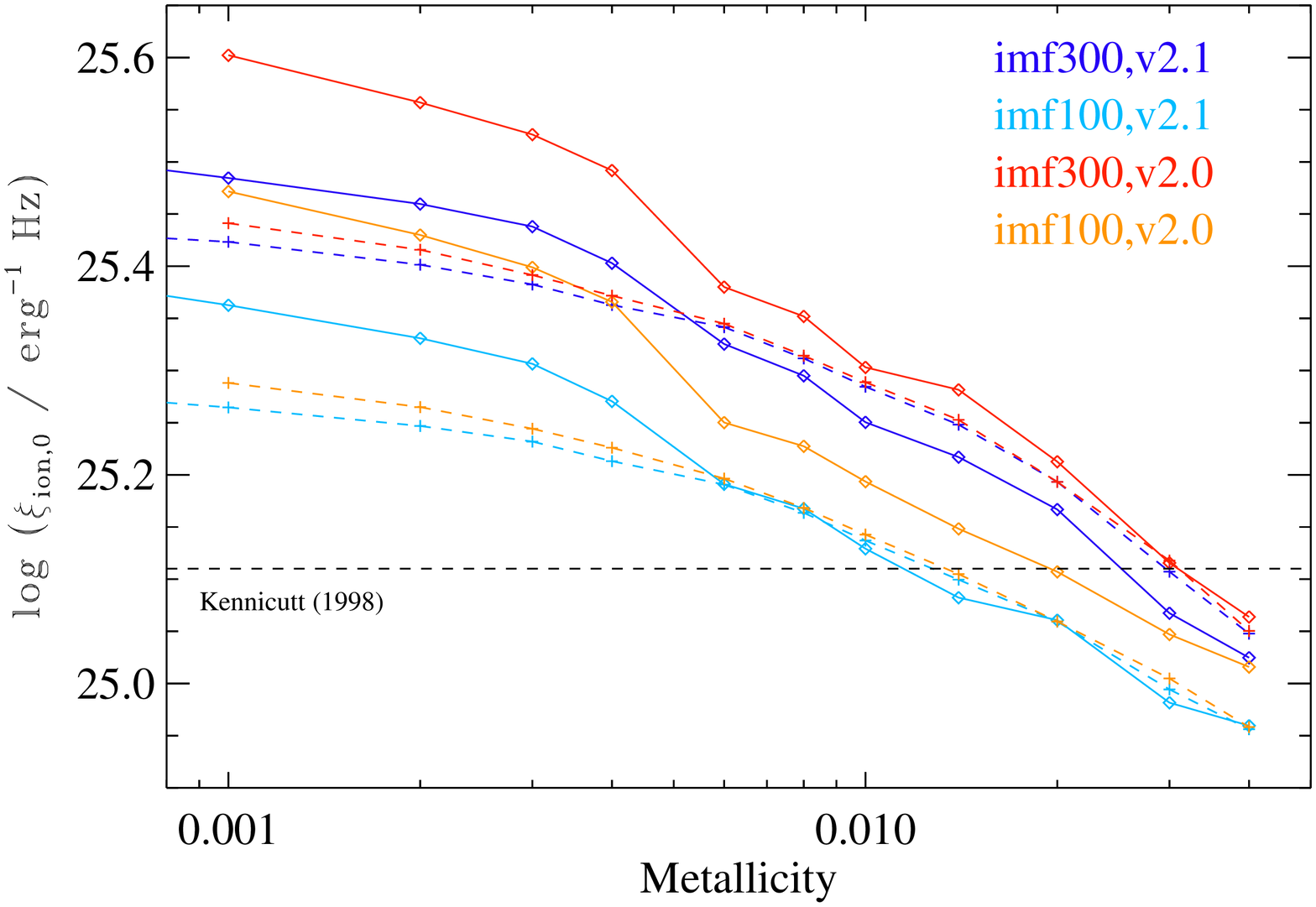}
\caption{A comparison between the ionization characteristics of \bpass\ v2.0 and v2.1 models for a population which has been continuously forming stars at a rate of 1\,M$_\odot$\,yr$^{-1}$ over a 100\,Myr period, shown as a function of metallicity. In the top panel we show the small effect of our improved atmosphere models on the ionizing photon production rate. By contrast the centre panel shows a strong change in rest-ultraviolet luminosity between v2.0 and v2.1. As a result, the lower panel shows a change in the behaviour in $\xi_{0,\mathrm{ion}}$. Single star models are shown with dashed lines, while binary models are shown with solid lines. Our standard IMF is shown (labelled imf300), with an identical IMF with M$_\mathrm{max}$=100\,M$_\odot$ also given for comparison. }
\label{fig:ionize_comp}
\end{center}
\end{figure}

In Figure \ref{fig:ionize_comp} we present a direct comparison between our previous estimates of ionizing photon production rate, ultraviolet luminosity and thus $\xi_{0,\mathrm{ion}}$ and new estimates from v2.1 for the case of continuous star formation observed at 100\,Myr after the onset of the star formation epoch. There are three clear differences in behaviour between our improved v2.1 models and the older v2.0 models.  Firstly, v2.1 does not show the dramatic increase in ionizing photon production rate at metallicities below $Z=0.004$, which was associated with the onset of quasi-homogenous chemical evolution - while this still leads to increased photon production, it is moderated by the new atmosphere models. Secondly, the v2.1 models are  more luminous in the rest-frame ultraviolet than the v2.0 models and the difference between single and binary models has narrowed due to a more rapid build up of continuum flux than in our previous models. Thirdly, as a result of these behaviours, the metallicity evolution of the ionizing photon production efficiency has become less clear cut.  In v2.0, the binary population always exceeded the single star population in $\xi_{0,\mathrm{ion}}$ at a given metallicity. In v2.1, we see {\em lower} photon production efficiencies in the binary population than in single stars at metallicities above  $\sim$0.3\,Z$_\odot$ in our standard IMF, and near-equal efficiencies in IMFs excluding the most massive stars. This behaviour reverses at lower metallicities, where the binary population continues to dominate. 

\begin{figure}
\begin{center}
\includegraphics[width=1.\columnwidth]{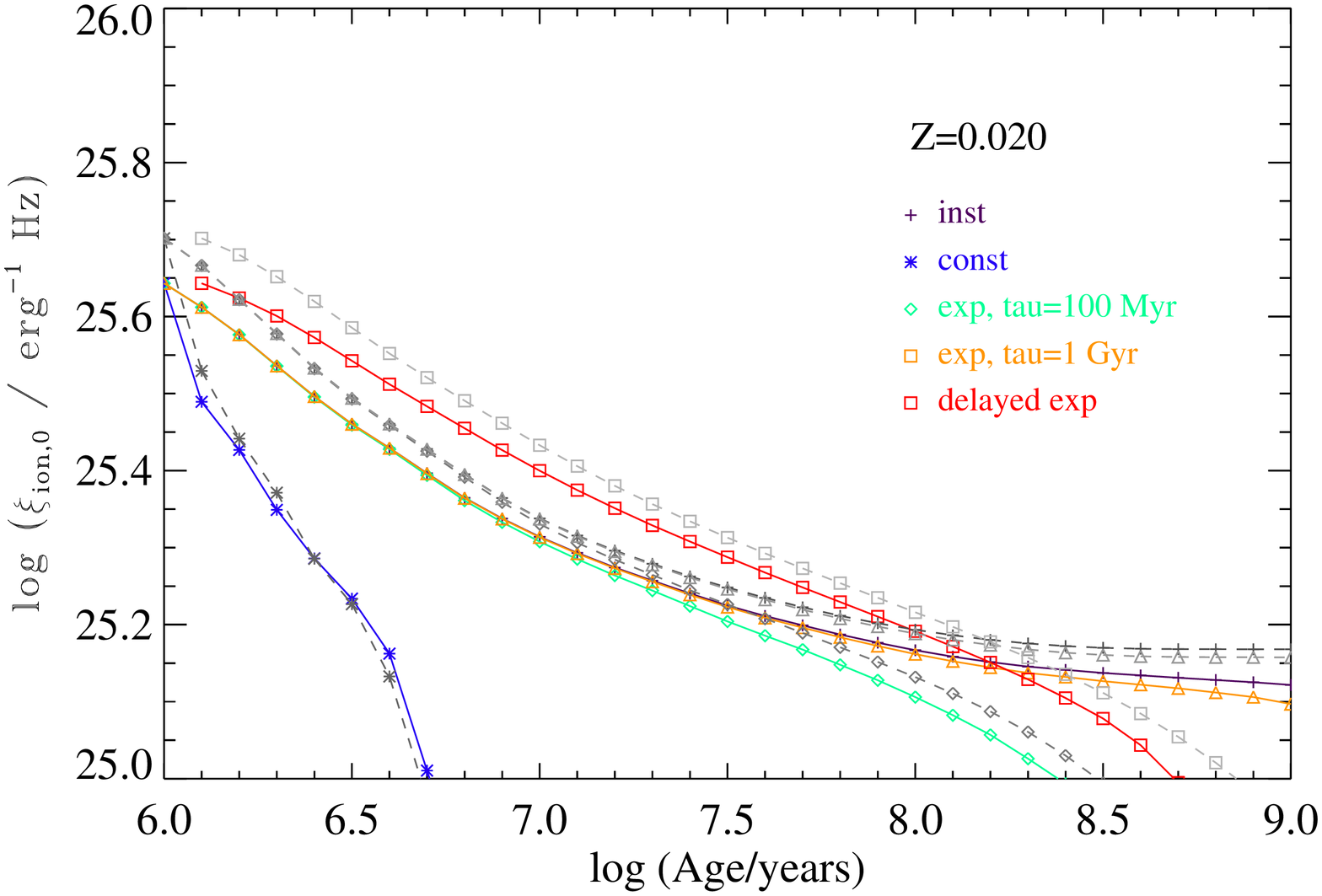}
\includegraphics[width=1.\columnwidth]{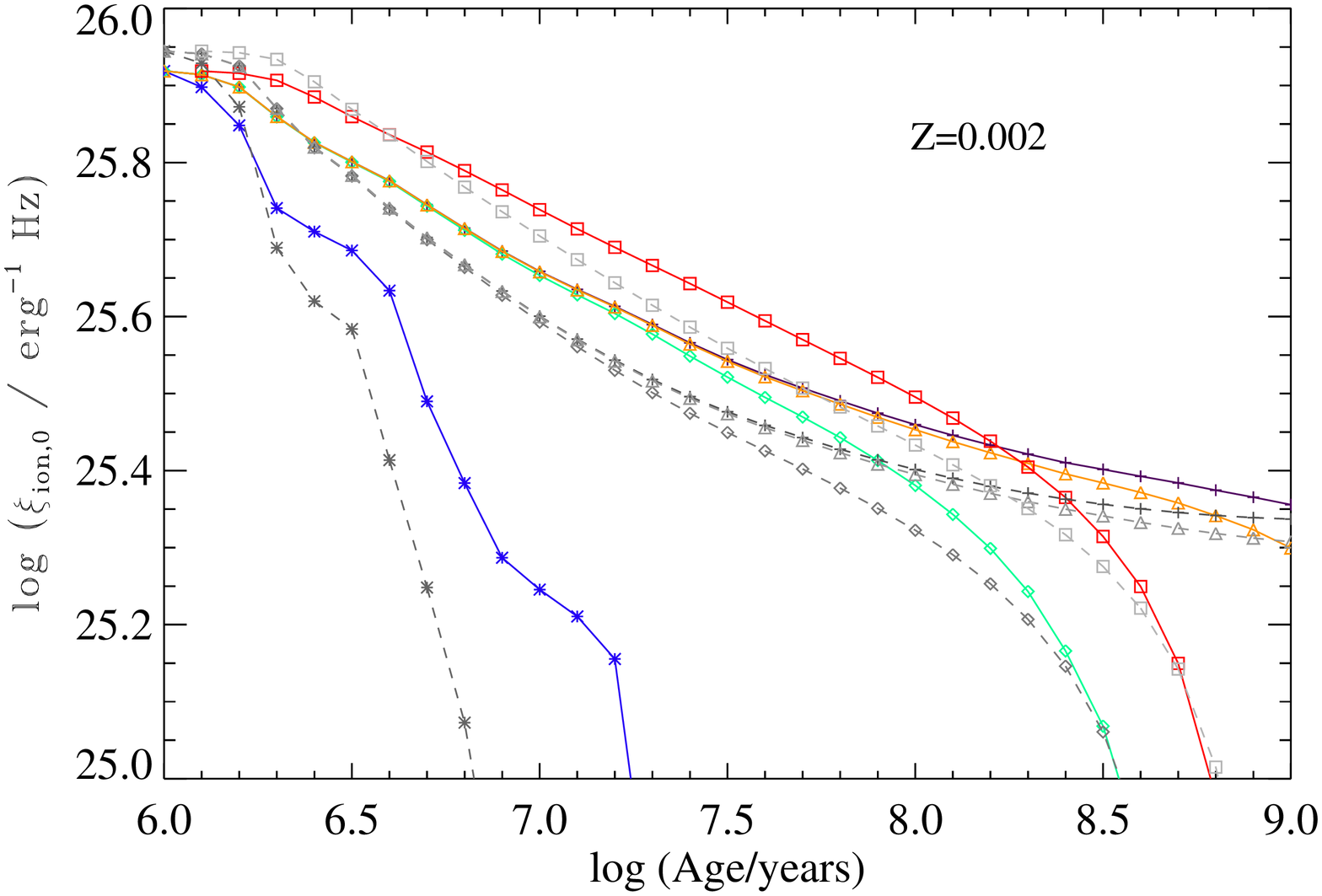}
\caption{The time evolution of ionizing photon production efficiency as a function of star formation history shown for metallicities of $\sim$Z$_\odot$ and 0.1\,Z$_\odot$. Single star models are shown in greyscale, with dashed lines, while binary models are shown in colour with solid lines. The star formation histories shown are those defined in Figure \ref{fig:ionizing_sfh}.}
\label{fig:ionize_age}
\end{center}
\end{figure}

This behaviour is entirely driven by the behaviour of the youngest stars in a given population. As Figure \ref{fig:ionize_age} demonstrates,  the ionizing photon production output of a single-aged, instantaneous-starburst binary population exceeds that of the single star population at all metallicities for ages $>3$\,Myr but the stars below this age produce photons with a far higher efficiency.  In the low metallicity populations, the longer ionizing photon production lifetime of the binary stars leads to the binary population overtaking the single stars at late ages for any ongoing or evolving starburst, while at near-Solar metallicity the rapid mass loss due to stellar winds dominates over binary effects.

The decrease in overall ionizing photon efficiency in \bpass\ v2.1 relative to v2.0  reflects an improvement in stellar  atmosphere modelling, but also emphasizes the significant uncertainties in the stellar models, particularly in the low metallicity regime, and thus the  difficulty of interpreting distant galaxies. We note that an improved treatment of rotation (see section \ref{sec:rotation}) could reverse the shift seen in this version. 

\begin{figure*}
\begin{center}
\includegraphics[width=1.5\columnwidth]{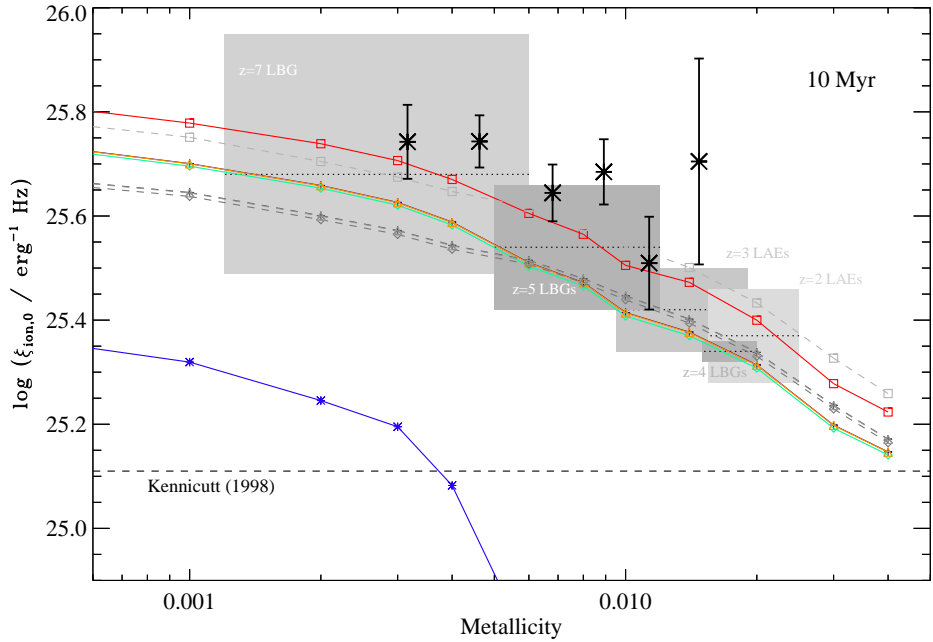}
\caption{The ionizing photon production efficiency as a function of metallicity and star formation history for an ongoing star formation event which has formed stars continuously over the previous 10\,Myr. Grey regions indicate measured values of $\xi_{0,\mathrm{ion}}$ in the distant Universe together with their uncertainties. These samples are unconstrained in metallicity and are shown offset to indicate plausible metallicity ranges given \bpass\ population synthesis models. Data points show results for a sample of local extreme star-forming galaxies (Lyman Break analogues) from \citet{2016MNRAS.459.2591G}, binned by optical emission line metallicity. Star formation histories are colour-coded as in Figure \ref{fig:ionize_age}.}
\label{fig:etaplot}
\end{center}
\end{figure*}

In Figure \ref{fig:etaplot} we compare our current predictions for ionizing photon efficiency with observational constraints. These are drawn from the literature and based on a standard analytic conversion from H$\alpha$ emission line flux to ionizing photon production rate (which itself is somewhat model dependent). Distant galaxies are shown as shaded regions indicating the mean and standard deviation of the (often small) measured sample in $\xi_{0,\mathrm{ion}}$. The metallicity is poorly constrained in these observations and samples are shown offset to indicate a plausible range in metallicity mass fraction. We also show data for a sample of extreme local starburst galaxies identified as analoguous to the $z\sim5$ galaxy population for which metallicity information can be inferred from the optical spectra and allows the data to be binned \citep{2016MNRAS.459.2591G}. In previous versions of this comparison, we have considered a model population continuously forming stars at 1\,M$_\odot$\,yr$^{-1}$ over 100\,Myr; now we compare with  models at 10\,Myr after the onset of star formation although, as Figure \ref{fig:ionizing_sfh} demonstrated, an alternative approach would be to assume a more bursty and less constant star formation history.  

As can be seen, a single age starburst would struggle to reproduce the  high $\xi_{0,\mathrm{ion}}$ values inferred from both distant and local extreme starburst observations, unless captured in the very first few million years. By contrast an ongoing or rising starburst history provides a reasonable match to the observed data at ages of order $\sim$10\,Myr.  

\subsubsection{Optical Recombination Line Ratios}

\begin{figure*}
\begin{center}
\includegraphics[width=\columnwidth]{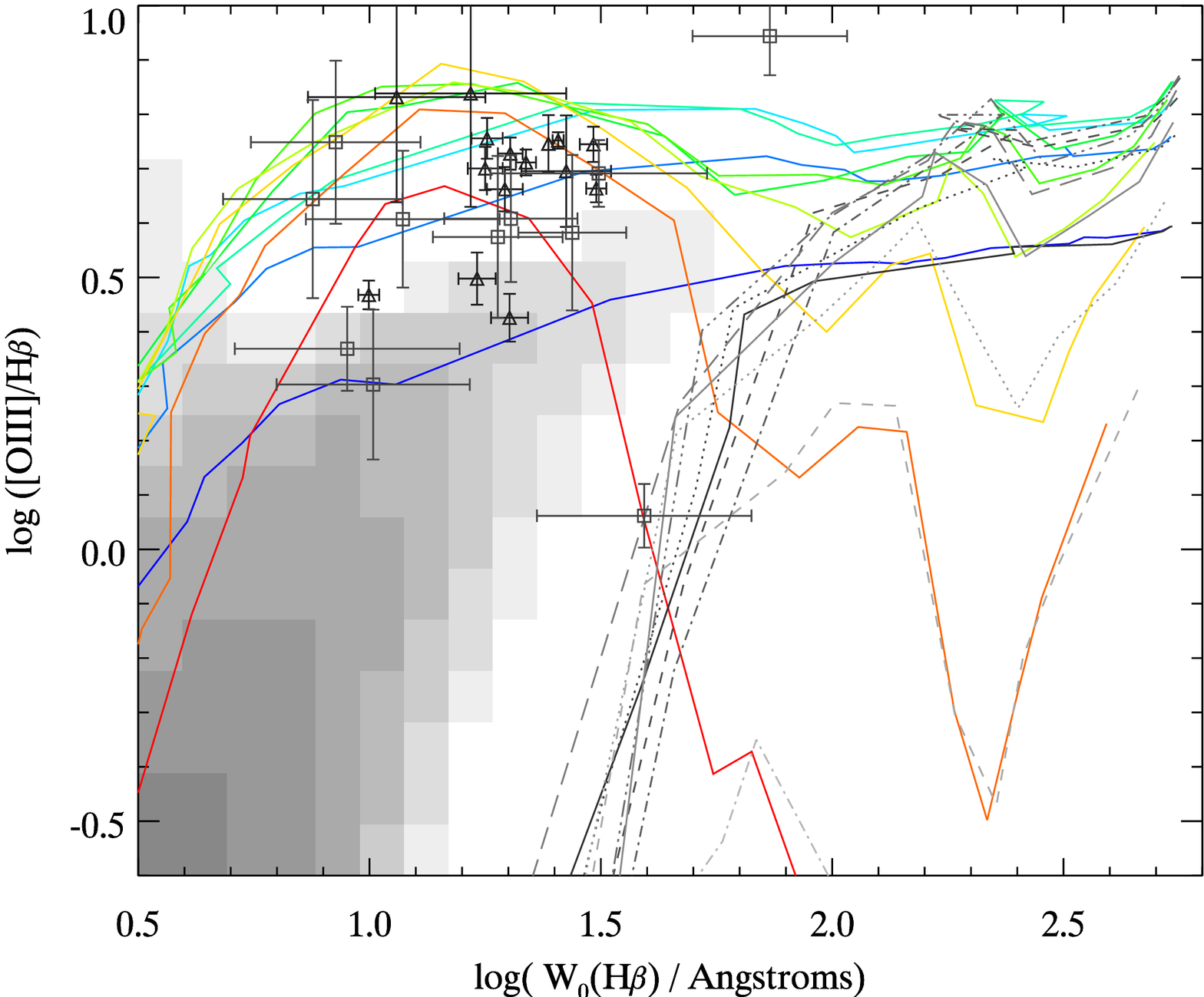}
\includegraphics[width=\columnwidth]{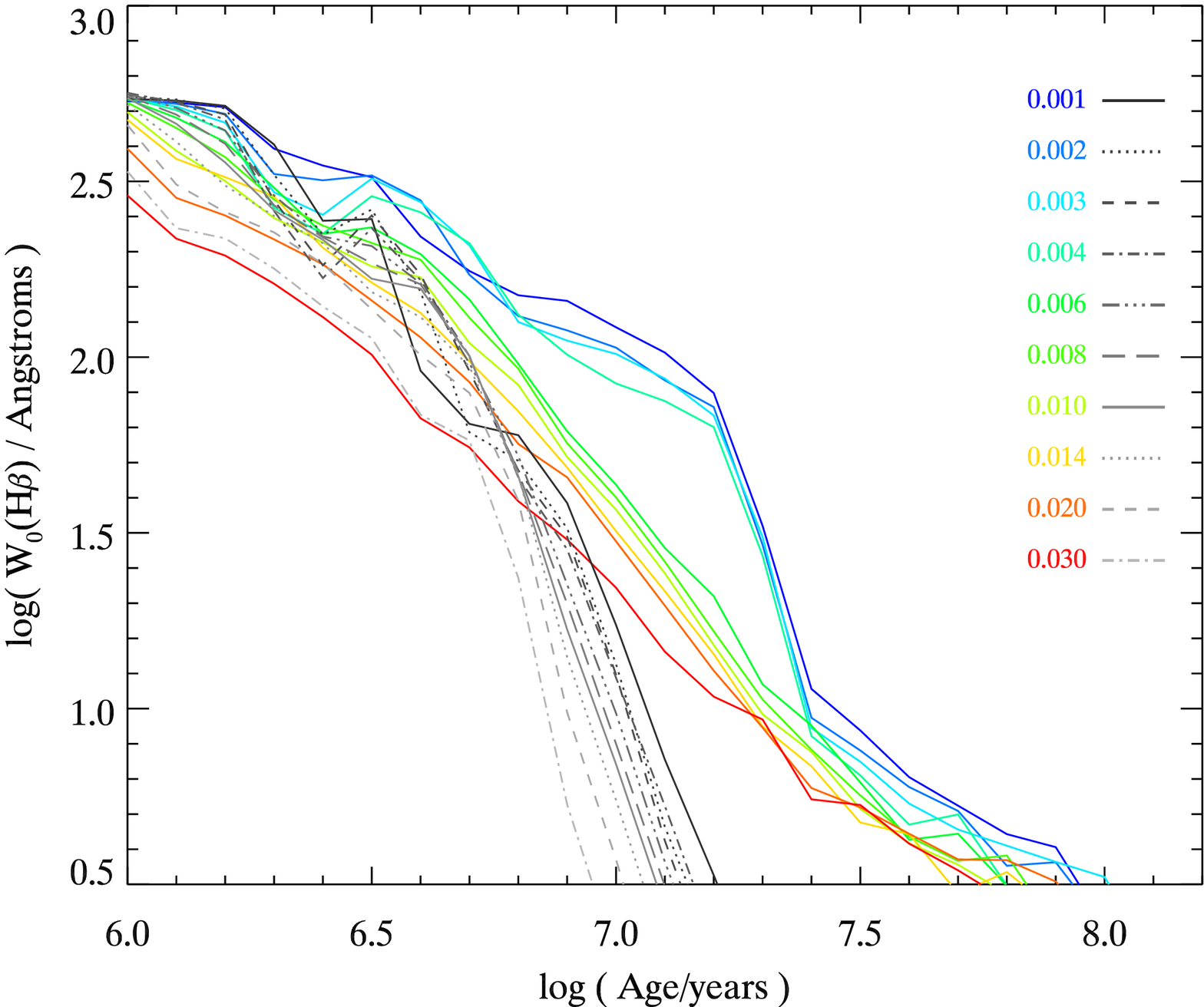}
\caption{The evolution in Balmer emission line strength and \oiii/H$\beta$ line ratio with stellar population age. Age increases along the tracks from high equivalent width to low. Tracks for binary models are shown in colour and those for single star models in greyscale, with different line styles. Overplotted are the data for star-forming galaxies at $z=2-3$ reported by \citet[][triangles]{2016ApJ...820...73H} and \citet[][squares]{2013ApJ...777...67S}. Greyscale indicates the measured properties of the local SDSS star-forming galaxy population.}
\label{fig:linerat1}
\end{center}
\end{figure*}

In \citet{2014MNRAS.444.3466S} we demonstrated that, subject to the usual uncertainties in nebular gas conditions, our v1.1 single age starburst models were capable of reproducing the optical strong line ratios seen in the $z\sim2-3$ galaxy population.  In Figure \ref{fig:linerat1}, we update this comparison incorporating our newer v2.1 models. For consistency we use the same nebular emission model as in the above work, with a fixed electron density (100\,cm$^{-3}$) and cloud geometry (spherical, 10\,pc inner radius), and allow stellar population age to vary along tracks. We confirm our earlier finding that  young stellar populations incorporating binary evolution pathways can reproduce both the strong line ratios and the H$\beta$ equivalent widths seen in the distant population, while single star models struggle to do so.  

We also show the measured properties of the SDSS star forming galaxy population, as defined above. Interestingly these show an offset from both the binary and single star model populations, lying between the two but more consistent with the binary tracks. We note that these have not been corrected for internal dust extinction (which may affect lines more strongly than continuum in some extinction models, and so would push the data towards the single star tracks in equivalent width) but are corrected for fibre losses.  We consider only a simple star formation history in this case, since the line emission spectrum is heavily dominated by the youngest starburst.  A more complex star formation history will tend to slightly reduce the model H$\beta$ equivalent width, again pushing the  data towards the single star models. However as Figure \ref{fig:linerat1} also demonstrates, the evolution of the single star models in line equivalent width is extremely rapid, and so the SDSS population would have to be caught in a very narrow time window some  10\,Myr after the onset of star formation.  By contrast, the binary models provide high H$\beta$ equivalent width over a  more extended time period increasing the probability of populating the parameter space occupied by the SDSS sources. If binary evolution models are preferred there are two likely explanations for the offset of the SDSS galaxies from our models. The first may be use of an inappropriate gas density or cloud geometry for the local population - adjusting these will tend to reduce the model \oiii/H$\beta$ ratio at a given H$\beta$ equivalent width, bringing the binary models better in line with the data \citep[see][]{2014MNRAS.444.3466S}. Alternately, the  binary distribution  parameters assumed (see Table \ref{tab:inputs}), which appear to reproduce those in the distant Universe and simple massive stellar populations, may overestimate the interacting binary fraction in high metallicity galaxies in the local Universe with more complex star formation histories and old underlying stellar populations. The result of assuming a lower binary fraction in low mass stars would be to move the binary models towards the local galaxy data, as further discussed in section \ref{sec:oldpops}.

\begin{figure}
\begin{center}
\includegraphics[width=0.99\columnwidth]{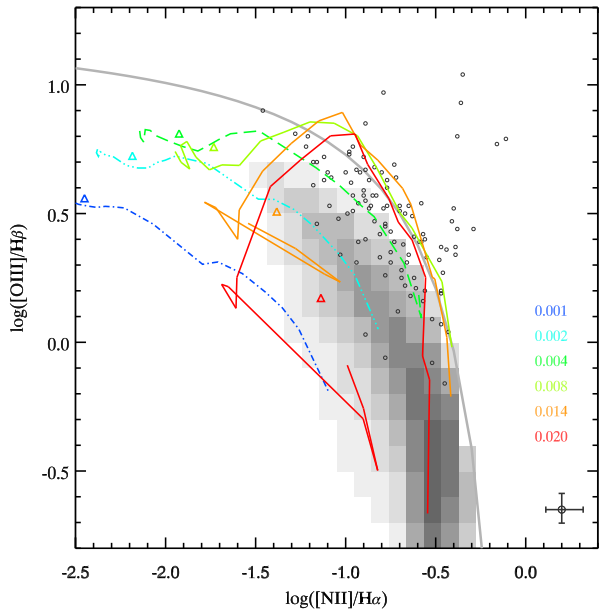}
\caption{The evolution of strong optical emission line ratios with stellar population age. Age increases along the tracks from 1\,Myr to 30\,Myr for instantaneous burst models. Small circles are data for star-forming galaxies at $z=2-3$ reported by \citep{2016ApJ...826..159S}. An indicative typical error bar for the data is shown at the bottom right. Greyscale indicates the measured properties of the local SDSS star-forming galaxy population. The solid grey line shows the Kauffmann et al (2003) condition for distinguishing regions ionized by AGN from star forming galaxies. Triangles indicate the line ratios expected for a binary stellar population with constant star formation over the last 100\,Myr.}
\label{fig:bpt}
\end{center}
\end{figure}

In Figure \ref{fig:bpt} we show an alternate ionization diagnostic - the \citet[][BPT]{1981PASP...93....5B} diagram widely used to distinguish emission powered by star forming regions from that powered by accretion onto an active galactic nucleus (AGN). We show the locus occupied both by local star-forming galaxies (greyscale) and the distant galaxy population reported by \citet[][data points]{2016ApJ...826..159S}. Overlaid are evolutionary tracks for a co-eval simple stellar population with age, at a range of metallicities, assuming the same simple nebular model as above. At ages of $\sim10$\,Myr, the tracks cross the Kauffmann et al (2003) boundary for separating local star forming galaxies, and, as before, more complex star formation histories are required to overlay the SDSS galaxy population, which typically lies somewhere between the ratios expected for a single age starburst and constant, ongoing star formation.   A full exploration of the BPT parameter space occupied by \bpass\ models with different assumed nebular gas conditions will be presented in Xiao et al (in prep).

\begin{figure*}
\begin{center}
\includegraphics[width=\columnwidth]{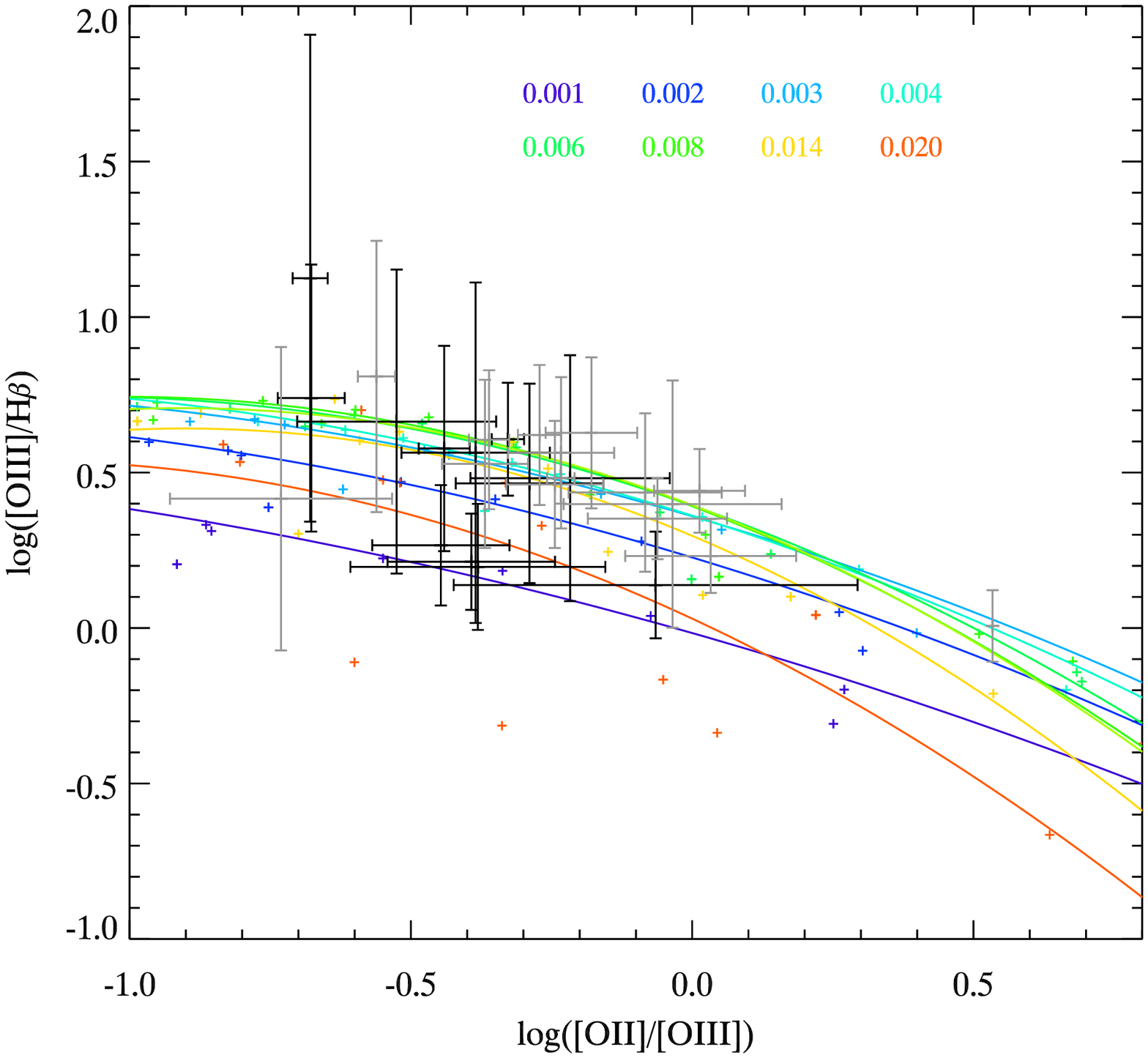}
\includegraphics[width=\columnwidth]{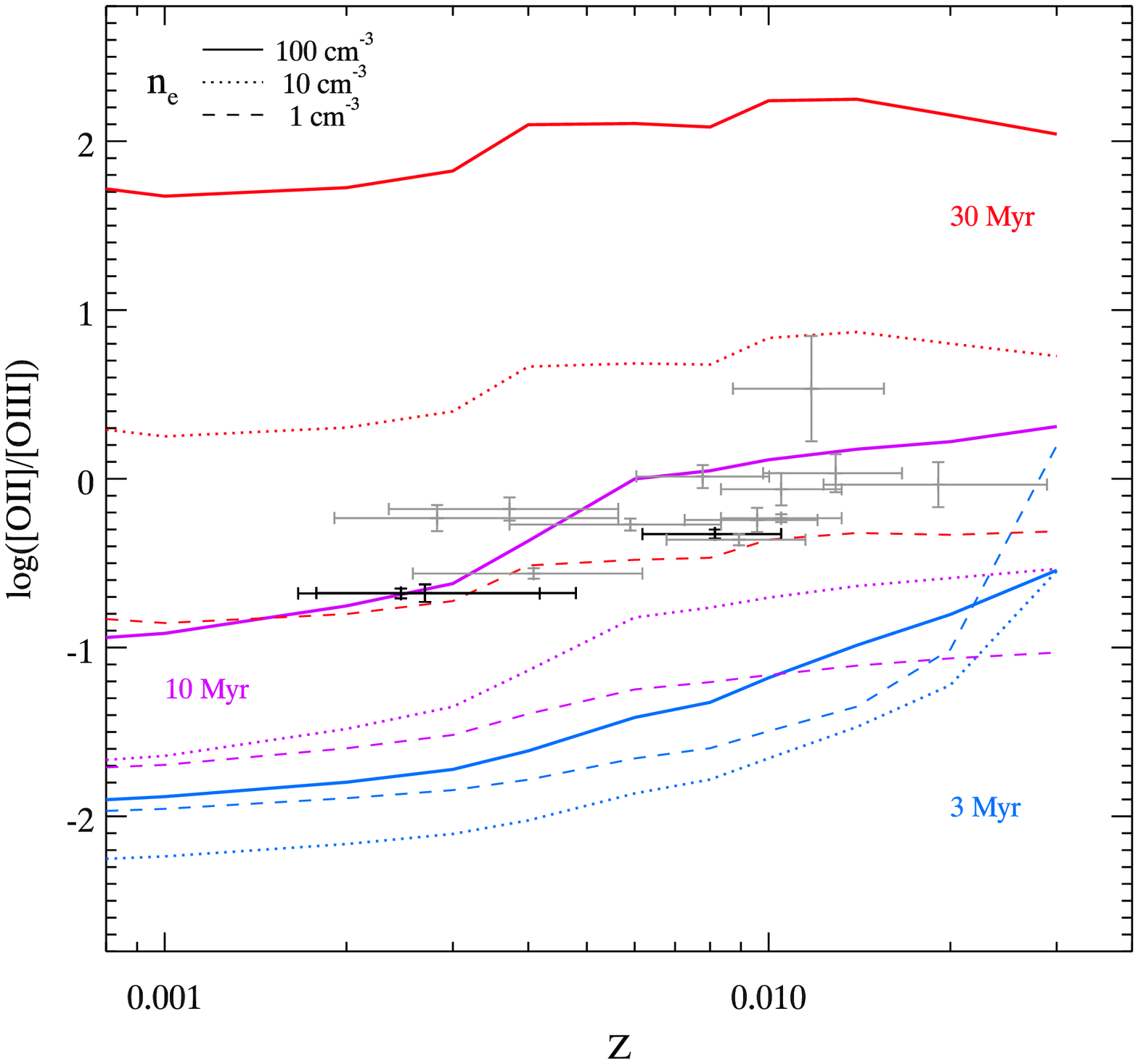}
\caption{The emission line properties of GRB host galaxies compared to \bpass\ v2.1 models. Both panels consider the [OII]/[OIII] excitation diagnostic. In the left-hand panel small coloured points indicate the predicted line fluxes for models with ages from 1 Myr to 100 Myr, and at electron densities ranging from 0.1 to 1000\,cm$^{-3}$. Solid lines are a polynomial fit to all models with $n_e=100$\,cm$^{-3}$ as a function of metallicity and indicate the general trend in the permitted parameter space. The right-hand panel shows the [OII]/[OIII] ratio explicitly as a function of metallicity at three stellar population ages, and three electron densities, showing the ambiguity between an older population and a sparser ISM at any given metallicity. GRB host galaxy data are taken from \citet{2015A&A...581A.125K} and are shown in black for hosts at $z>1$ and in grey for hosts at $z<1$, metallicities for data points are those derived from optical strong line ratios by \citet{2015A&A...581A.125K}. Only objects with measurements in all relevant quantities, and an estimated $E(B-V)<0.2$ are shown. Line ratios have been corrected for internal extinction using the Calzetti dust law.}%
\label{fig:grb_lines}
\end{center}
\end{figure*}

An additional validation test for our models, and an important demonstration of the ambiguity that arises from nebular gas parameters, is their ability to reproduce the line  ratios seen in  gamma-ray burst host galaxies.  The presence of a GRB requires these galaxies to have formed massive stars recently, and thus (as with the WR-galaxies discussed above) their nebular emission tend to be dominated by a young stellar population.  We consider here the TOUGH sample, observed spectroscopically by \citet{2015A&A...581A.125K} with X-SHOOTER on the ESO VLT, for which metallicity and extinction estimates are available.  In Figure \ref{fig:grb_lines} we consider a third diagnostic line ratio, [OII]\,$\lambda$3726,3729\AA/[OIII]$\lambda$5007\AA.  The evolution of this line ratio is not entirely smooth in either stellar population age or nebular electron density, with both playing an important role  in exciting these forbidden lines.  Nonetheless, considering a range of plausible ages (1 Myr to 100 Myr) and  electron densities (0.1 to 1000\,cm$^{-3}$) makes clear the locus of permitted line ratios relative to  [OIII]/H$\beta$ and their good agreement with the observed properties of GRB hosts (upper panel).  Solid lines indicate a polynomial fit to the variation with age at each metallicity and demonstrates the relative insensitivity of this parameter space to metallicity in the range  $Z \sim 0.5-0.7$\,Z$_\odot$, but also the significant scatter around the mean locus given a range of assumptions regarding age and nebular gas conditions.  This point is  perhaps still clearer in Figure \ref{fig:grb_lines} (right-hand panel), which shows the evolution in [OII]/[OIII] as a function of metallicity at three  stellar population ages and three gas densities. There is significant degeneracy between different plausible models.  If interpreted at fixed age and electron densities, the distribution of [OII]/[OIII] line  ratios seen in  the GRB hosts of  \citet{2015A&A...581A.125K} is consistent with a metallicity sequence. However, a variation in stellar population age, or electron density, or both, at fixed metallicity could also provide a good fit to the observations. Given that the  metallicities shown for the data points were themselves derived from the optical recombination line spectrum, and thus ultimately derive from a photoionization model and stellar population fit to a calibration data set, the correct interpretation  is unclear, and likely involves  some evolution in all three parameters.

\subsubsection{Ultraviolet Spectral Features}\label{sec:uvspec}

\begin{figure}
\begin{center}
\includegraphics[width=0.99\columnwidth]{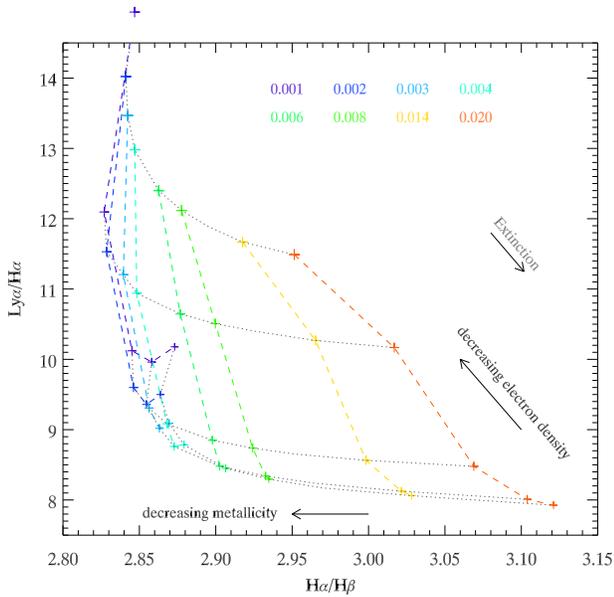}
\caption{The ratio of intrinsic Lyman-$\alpha$ emission to optical Balmer line ratio assuming a range of metallicities and gas parameters. While the Balmer line ratio is only weakly sensitive to the environment, Lyman-$\alpha$ is strongly affected by the metallicity and physical conditions of the nebular gas. Electron densities increase from 1\,cm$^{-3}$ to 10$^4$\,cm$^{-3}$ along coloured (fixed metallicity) lines and are marked at 1\,dex intervals. We assume constant star formation, observed at age $10^8$ years.}
\label{fig:hIlines}
\end{center}
\end{figure}

Ultraviolet emission lines, redshifted into the observed frame optical and near-infrared are crucial evidence for the stellar populations and properties of distant galaxies. A recent study by \citet{2016arXiv160900727B} has already made use of beta versions of our lowest metallicity v2.1 models to explore the interpretation of the anomalous source CR7 \citep{2015ApJ...808..139S} which has been proposed as either a  Population III starburst or, potentially, a direct collapse black hole. While none of the model sets explored in that work provide an unambiguous interpretation for CR7, \bpass\ low metallicity, young starburst models are broadly consistent with the colours of the source in question  (particularly if some degree of enhancement in $\alpha$-elements is assumed) suggesting a stellar origin for the light. The extreme properties of CR7 (specifically its \heii\ line strength) have recently been challenged \citep{2017arXiv170500733S}. Nonetheless, this source and the similar high redshift emission line sources now being identified provide one of the very few constraints on our models at the lowest metallicities. Crucially, they are representative of a regime that may soon be more widely explored by JWST.

As is the case for optical lines, interpretation of the ultraviolet spectrum of a complex stellar population requires an assumption regarding the nebular gas conditions, which are largely unconstrained by observations. \citet{2016ApJ...826..159S} have suggested that the spectra of galaxies at $z\sim2-3$ are best fit by combining very low metallicity stellar spectra (the observed properties of which are most strongly influenced by the iron abundance) with somewhat higher metallicity nebular gas (the properties of which are dominated by oxygen abundance). Again, this may be indicative of strong $\alpha$-element enhancement, as would be expected in the young stellar populations at this redshift.

\begin{figure*}
\begin{center}
\includegraphics[width=1.8\columnwidth]{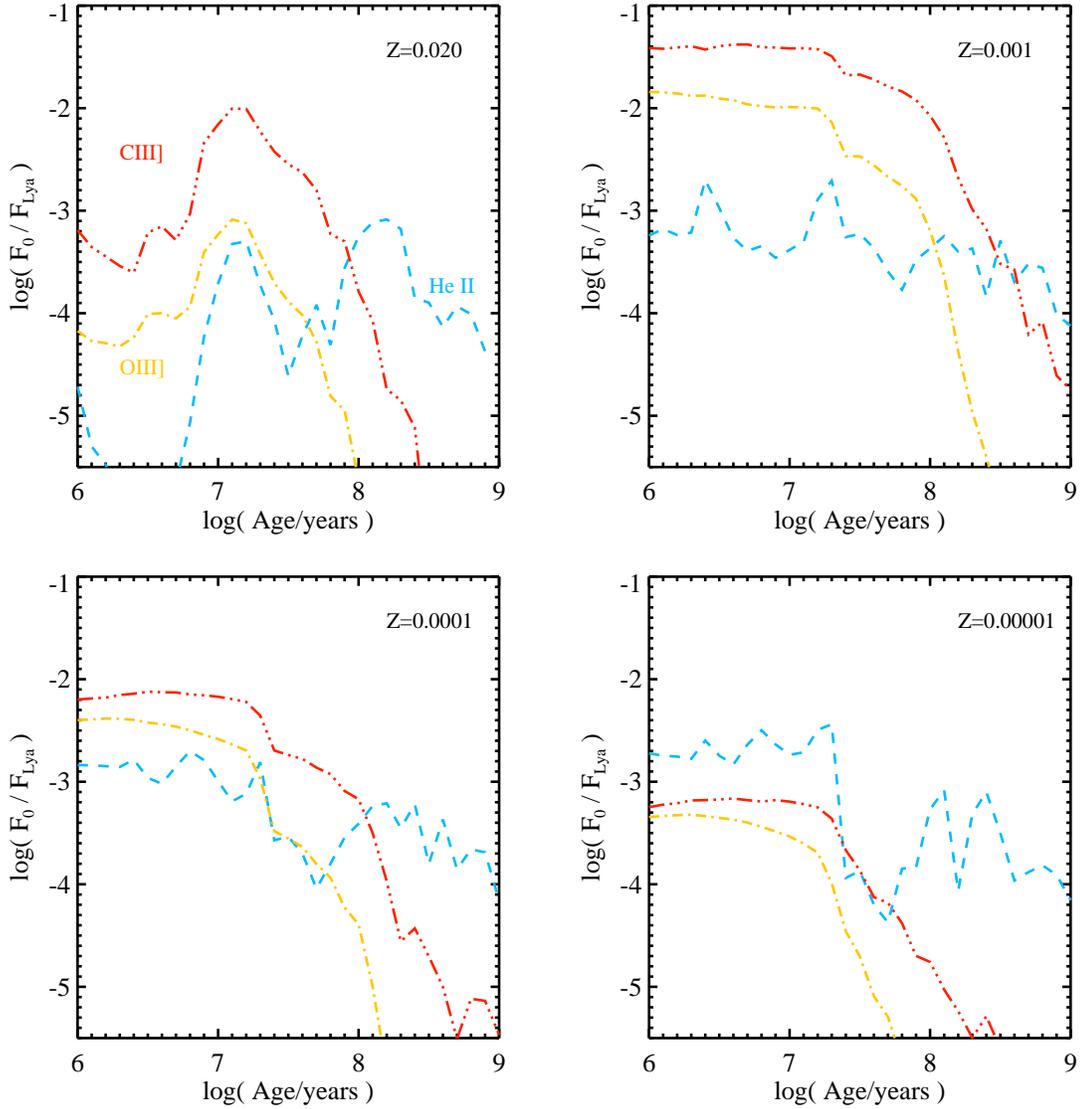}
\caption{The evolution in rest-frame ultraviolet emission line flux ratios for a single-aged simple stellar population at four representative metallicities. Line strengths shown are normalised relative to the Hydrogen Lyman-$\alpha$\ $\lambda$1216\AA\ feature, and are \heii\ $\lambda$1640\AA, \ciii\ $\lambda$1909\AA\ and O\,III] $\lambda$1665\AA. For doublets, the total flux is given.}
\label{fig:uvlines}
\end{center}
\end{figure*}

The difficulty in evaluating nebular effects is particularly true of resonantly scattered lines such as the Lyman-$\alpha$ feature, the ratio of which to the Balmer series is shown in Figure \ref{fig:hIlines} for a continuously star forming population, assuming our usual gas geometry and local electron densities ranging from 1 to 10$^4$\,cm$^{-3}$. In Figure \ref{fig:uvlines}, we simply assume $n_e=10^2$\,cm$^{-3}$ and $Z$(nebular)=$Z$(stellar), in order to demonstrate emission line strengths as a function of age and metallicity for our standard CLOUDY models, normalised relative to the Lyman-$\alpha$ line strength in the same galaxy. Perhaps the most striking effect of decreasing metallicity is the increase in strength of rest-frame ultraviolet emission lines that are seldom seen in normal stellar populations in the local Universe, and which are usually associated with emission from AGN when observed. These lines are not significantly excited in nebular gas irradiated by our equivalent single star models. 

At the lowest metallicity in our models (Z = $1\times10^{-5}$ = 0.05\% Z$_\odot$), \heii\ $\lambda$1640\AA\ is the strongest emission line in the rest-UV longwards of the Lyman-$\alpha$ line. At the low to moderate metallicities more typical of the distant Universe (0.5\% to 10\% Z$_\odot$), the semi-forbidden doublets of \ciii\ $\lambda$1909\AA\ and O\,III] $\lambda$1665\AA\ increasingly exceed the helium line in line flux. All three species decline in strength towards Solar metallicity due to the decline in ionizing photon strength emitted by stellar populations. 

\begin{figure*}
\begin{center}
\includegraphics[width=1.9\columnwidth]{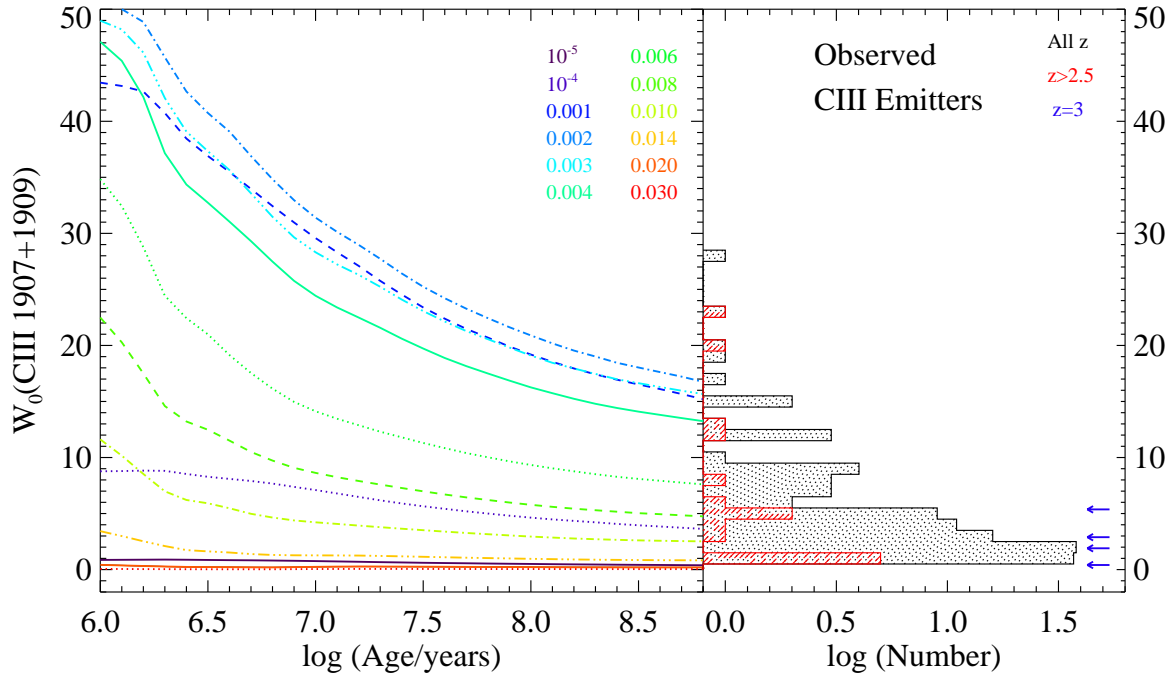}
\caption{The rest-frame equivalent width of the (unresolved) [C\,III] $\lambda$1907\AA, \ciii\ $\lambda$1909\AA\ doublet, as a function of stellar population age and metallicity for a population continuously forming stars at the rate of 1\,M$_\odot$\,yr$^{-1}$. In the right-hand panel, we compare these predictions to the distribution of measured \ciii\ 1909 lines in star-forming galaxies compiled from the literature (see section \ref{sec:uvspec}). Measurements from sources at $z>3$ are highlighted in red, while arrows indicate the measurements obtained from stacks of $z\sim3$ Lyman break galaxies differing in Lyman-$\alpha$ line strength \citep{2003ApJ...588...65S}.}
\label{fig:ciii_data}
\end{center}
\end{figure*}

The [C\,III] $\lambda$1907\AA, \ciii\ $\lambda$1909\AA\ doublet has attracted particular attention of late as the result of its detection in a handful of very distant ($z>6$) star forming galaxies \citep{2015MNRAS.450.1846S,2017MNRAS.464..469S}, and its position in the rest-frame ultraviolet, which means it is observable in the optical in the distant Universe. This line may arise from the atmospheres and winds of massive stars (in which case it will typically be broad) and from ionized nebular gas in H\,{  II} regions (in which it is typically narrow). Both of these are seen locally in the environs of Wolf-Rayet stars and planetary nebulae \citep[e.g.][]{2013A&A...553A.126G}. Observed lines in distant sources tend to be narrow, suggesting a nebular origin, but the constraints on its velocity width are relatively poor in most cases. In Figure \ref{fig:ciii_data} we show the predicted total equivalent width of this doublet as a function of stellar population age and metallicity for a population continuously forming stars at 1\,M$_\odot$\,yr$^{-1}$, after passing through nebular gas. These are compared to the distribution of measured rest-frame equivalent widths in this doublet arising from star forming galaxies at $z=0-8$ \citep[][Maseda et al., in prep]{2014MNRAS.445.3200S,2015ApJ...814L...6R,2016ApJ...821L..27V,2016MNRAS.456.4191P,2017ApJ...838...63D}. We also indicate the rest-frame equivalent widths measured for four stacks of $z\sim3$ LBGs, sorted into quartiles by Lyman-$\alpha$ equivalent width and each containing $\sim$200 galaxies, by \citet{2003ApJ...588...65S}.  Unsurprisingly, the bulk of the observed measurements are consistent with the very low equivalent widths we expect for mature stellar populations at moderate metallicities. It is notable however that, at both $z<3$ and $z>3$, there exists a tail of sources extending to $W_{0,\ciii}>10$\AA. Comparison with our standard CLOUDY models suggest that these are most naturally explained by stellar populations with significantly sub-Solar metallicities, $Z<0.2$\,Z$_\odot$, or very young dominant stellar populations ($<10$\,Myr). In the most extreme observed cases, both of these conditions may be necessary. At the very lowest metallicities in the \bpass\ model set ($Z<0.05$\%\,Z$_\odot$) the hard stellar ionizing spectrum excites very little \ciii\ emission simply because of the scarcity of carbon atoms in the nebular gas. As observations approach this regime in the very distant Universe it is worth noting that, in the presence of a carbon-enriched ISM, the irradiating spectrum will likely excite strong line emission and that such enrichment may result from massive stellar explosions very rapidly after the onset of star formation.  As discussed before, there is scope for further populating this parameter space by varying electron density, local ionization parameter and potentially the $\alpha$-element enhancement of the nebular gas. However the \ciii\ doublet is of particular interest since its line ratio is itself sensitive to electron density and thus potentially allows some such degeneracies to be broken. As samples of deep rest-frame ultraviolet spectra grow, it will likely be possible to break many such degeneracies with observations of both the doublet line ratio and further emission lines in future.

A fuller exploration of the limited observational data on nebular line emission at high redshift is deferred to later work.

\section{FUTURE WORK AND UNCERTAINTIES}\label{sec:future}

Any numerical model, no matter how much effort is expended in its creation, will always be subject to limitations. It is vital to make important caveats clear, as well as their impact on any predictions from that model. Here we outline these for the \bpass\ code and highlight how we aim to resolve them in future versions of the code.

\subsection{Secondary stars}

A key feature of the \bpass\ models that sets them apart from others, except the Brussels code \citep{1998A&ARv...9...63V,2016A&A...589A..64M}, is that all the interacting binary evolution models are evolved in a full detailed stellar evolution code. This is in contrast to the approximation methods employed by rapid population synthesis codes \citep[e.g.][]{2002MNRAS.329..897H}. The rapid method allows for the uncertainties of binary evolution and their impact on predictions to be explored; this would be too computationally intensive for detailed stellar models. The use of detailed models, on the other hand, allow us to accurately follow how the stellar envelope responds to mass loss -- key to determining the eventual mass and fate of the star.

This  impacts on our treatment of the secondary stars. In a binary system after the first supernova the future evolutionary pathways are varied. The system can be unbound so the secondary continues its evolution as a single star, or the binary can remain bound in a range of possible orbits. For rapid population synthesis it is straight forward to simply calculate each possibility as they occur, but in a detailed code this becomes more numerically intensive. Our current solution is to use single star models to represent the evolution when systems are unbound. The range of bound systems is also determined and a range of secondary models calculated in which one of the stars is a compact remnant. 

However a complication arises with how to deal with mass transfer. If the secondary star accretes material (mostly hydrogen) from the primary, we use its final accreted mass at the end of mass transfer as its {\em effective} initial mass when selecting a model to represent its subsequent evolution. If this effective initial mass is greater than 105\% of the actual initial mass then we treat the model as rejuvenated. We use the age of the primary model as an estimate of the rejuvenation age, this is a reasonable approximation for stars that end in a core-collapse supernova but as noted above an overestimate when a white dwarf is formed. Our secondary model contribution to the whole population is split into models that haven't interacted but have simply increased in mass and those that have been rejuvenated, and their contribution is offset in time so that their zero-age is at the post-rejuvenation age. 

The disadvantage of this approach is that we do not model the intricacies of mass accretion onto the secondary stars in detail, and thus do not accurately trace the post-accretion evolution. We simply assume these secondary stars become fully mixed, more massive, main-sequence stars. There is observational evidence that such stars can actually appear very different \citep[e.g.][]{2015MNRAS.447..598S}. The way to improve this is to calculate the accretion onto the secondary in more detail and then calculate its evolution after that accretion. We aim to do this in future but the computational demands are beyond the scope of \bpass\ v2.1 models. The practical outcome of this is that we may underestimate the number of blue supergiants and related objects in our simulations.

\subsection{Rotation and quasi-chemically homogeneous evolution}\label{sec:rotation}

We currently only consider rotation in two contexts. One is to follow the transfer of angular momentum between the stars and orbit during Roche-Lobe overflow and allow for this to instigate common-envelope evolution. The second is in our model of rejuvenation. We assume that if more than 5\% of a star's initial mass is accreted the star rotates rapidly enough to be fully mixed. Furthermore at low metallicites, $Z\le 0.004$, if this occurs for stars more massive than 20\,M${_\odot}$, it is assumed to be fully mixed during its entire main-sequence lifetime, experiencing QHE, and the model is replaced by a model that is fully mixed until it loses all hydrogen. In nature it is more likely there is a continuum of behaviour that we are unable to reproduce without a more sophisticated implementation of rotation in our evolution code. 
Again this is a factor we will return to and improve in future. We also note that QHE may occur due to tidal forces in binary stars at the relatively high metallicity of the LMC \citep{2017A&A...598A..85S} as well as lower metallicities. This is not currently included in any \bpass\ models but is potentially important for gravitational wave sources \citep{2013ApJ...764..166D,2016A&A...588A..50M}.

There is some evidence that such a simple model can work, however. \citet{2013ApJ...764..166D} included a model of rotation in a rapid population synthesis code and found that mass transfer mainly leads to creating rapidly rotating stars during break up, and furthermore that the predicted rotational distribution is similar to that observed in nearby star-forming regions. Our current prescription behaves similarly. Thus while some refinements may be required in specific contexts, we must already reproduce the broad range of behaviour. We note that the primary effect of increasing the number of rotating stars is likely to be boosting the ionizing photon flux from a population.

\subsection{Roche Lobe Overflow, Common envelope evolution and mergers}

Our model of Roche lobe overflow is relatively unsophisticated and we may update it in future with more realistic models \citep[e.g.][]{1988A&A...202...93R,1990A&A...236..385K,2007ApJ...660.1624S}. The complexities of CEE are less uncertain than they used to be, as a result of extensive hydrodynamic simulations in recent years \citep[e.g.][]{2012ApJ...744...52P,2012ApJ...746...74R,2016MNRAS.460.3992N,2016ApJ...816L...9O,2016MNRAS.455.3511S,2017MNRAS.464.4028I}, although this work focuses on low mass ($\sim$1-2\,M$_\odot$) giants. However, while our straight-forward binding/orbital energy prescription for CEE provides results in line with current thinking, how best to model the CEE phase within a 1D hydrostatic evolution code is still unclear \citep{2013A&ARv..21...59I}. 

The only way to improve this model is for more CEE events to be observed and their properties used to refine models: this may include recently observed examples such as those detailed by \citet{2017ApJ...834..107B} and \citet{2017ApJ...835..282M}. Work to unify such observations with simulations is ongoing \citep[e.g.][]{2017ApJS..229...36G}. It may also be necessary in future to model post-CEE systems such as white dwarf binaries \citep[e.g.][]{2013A&A...557A..87T}. The best strategy for model CEE involving a compact remnant remains obscure. This will be resolved by future studies of X-ray binaries as detailed below.

The process of CEE is the one aspect of population synthesis that has the greatest effect on predictions that concern how close two stars reach at the end of the evolution, such as the type Ia supernova rate or the gravitational wave merger rate. By using different models or assumptions very different results can be found. While our CEE model does have the advantage of being able to calculate the binding energy of the star exactly as we have the full detailed stellar structure we have the disadvantage we cannot remove the envelope on timescales similar to those observed in CEE events and hydrodynamic simulations. For CEE in our code we have calculated the approximate values of $\alpha_{\rm CE} \lambda$, the constant reflecting the efficiency of conversion of orbital energy into energy to unbind the envelope. We find that the values range from 100 for the smallest mass ratio systems up to of the order of 2 when the mass ratio is unity before CEE. Most CEE events have values in the range 2 to 30. Thus we find results of a similar magnitude to others despite our quite different implementation.

Our way forward with inclusion of CEE in our stellar models will be comparing our models to known post-CEE binaries as well as compact remnant merger events that give rise to gravitational wave signals. Thus enabling us to study the nature of CEE over the full stellar mass range. 

\subsection{Composition}

\begin{figure}
\includegraphics[width=\columnwidth]{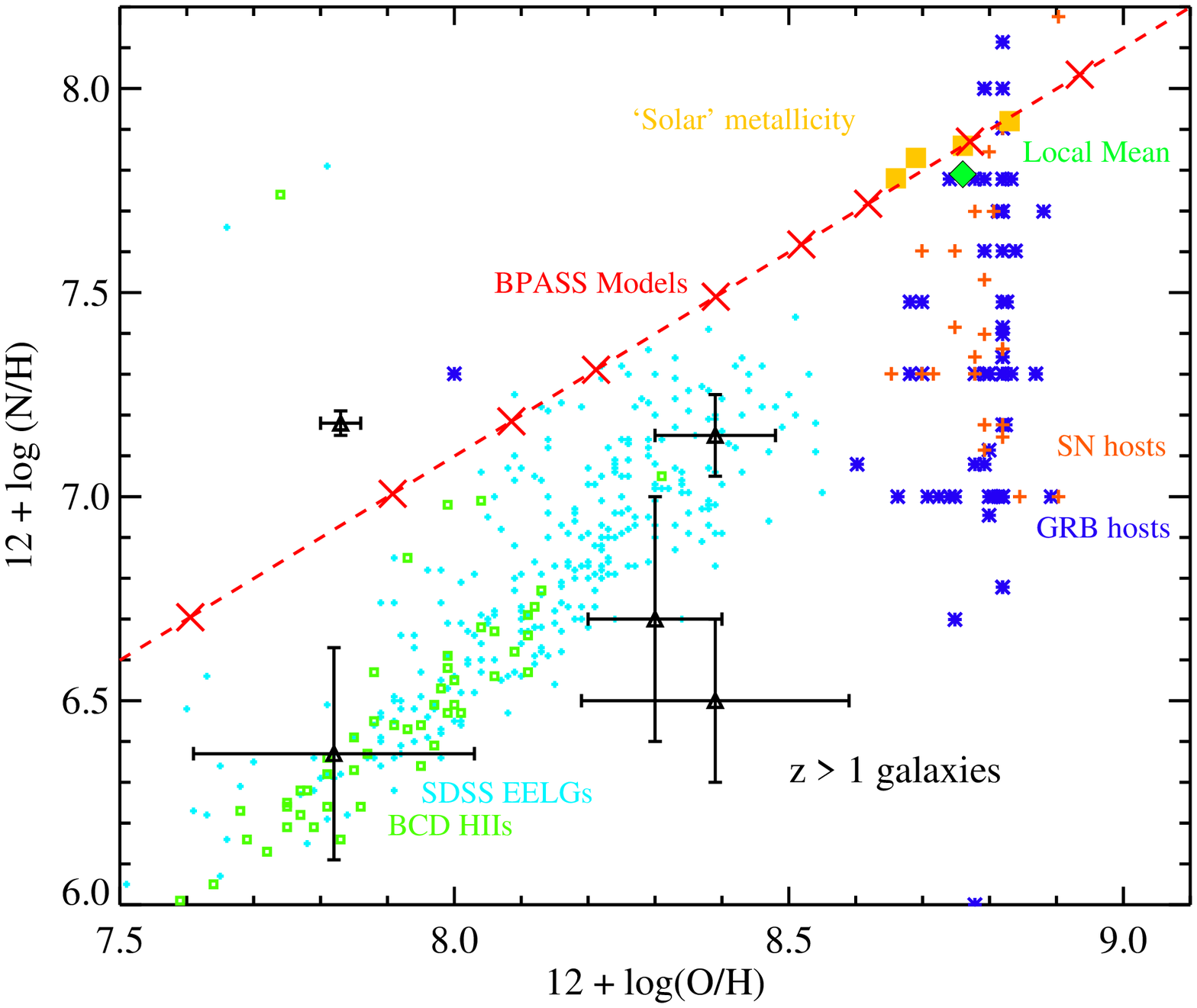} 
\includegraphics[width=\columnwidth]{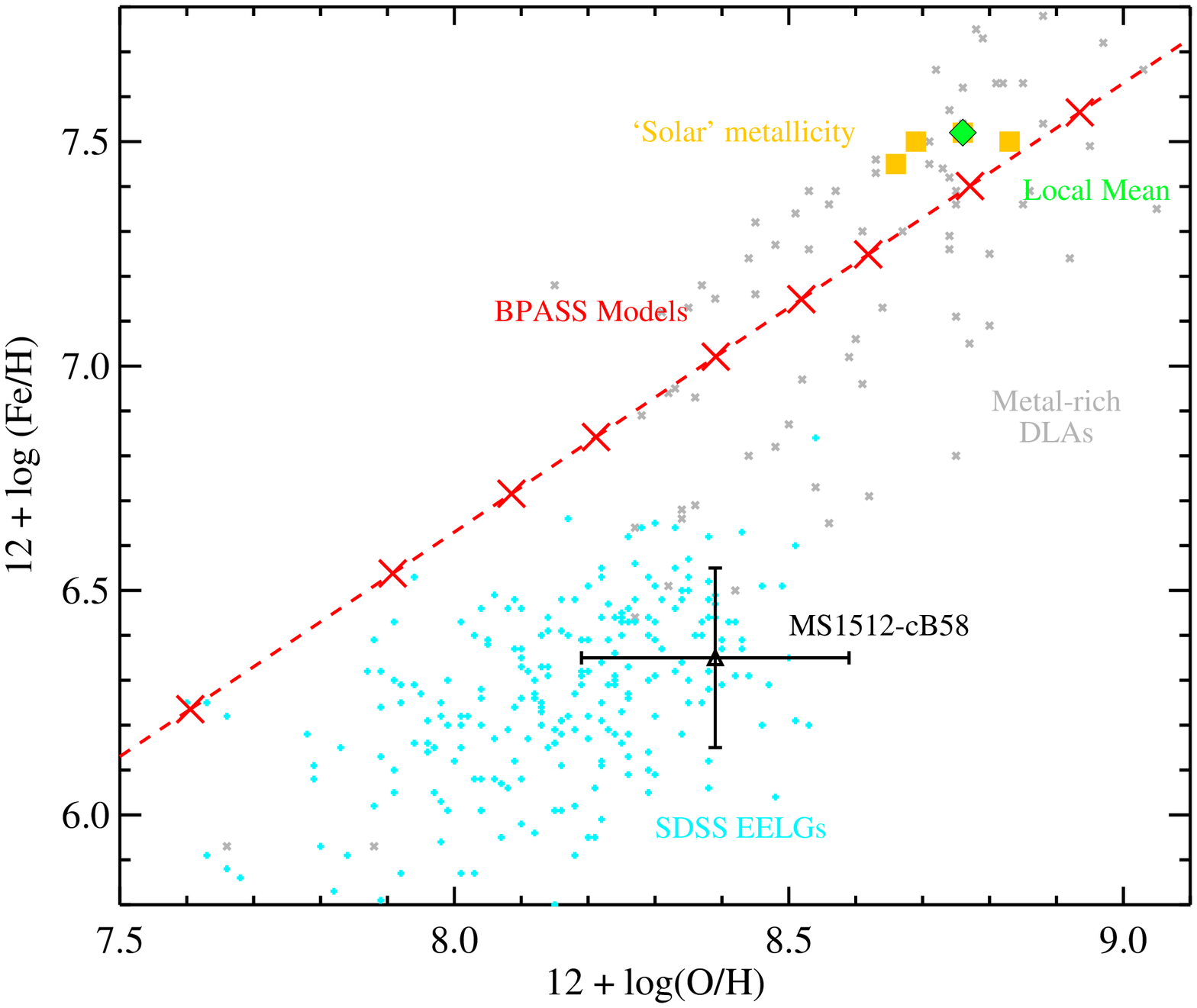} 
\caption{The abundance ratios adopted in our models, shown as large red crosses, together with observational constraints from the literature. Yellow squares indicate the `Solar' abundance ratios compiled in Table \ref{tab:solar}, while the green diamond indicates the mean abundance in the Solar neighborhood estimated by \citet{2012A&A...539A.143N}. Black points indicate high redshift galaxies drawn from the compilation in Table  \ref{tab:abundances}. SDSS metal poor extreme emission line galaxies are shown with small blue symbols \citep{{2006A&A...448..955I}}, while small green squares indicate H\,{  II} regions observed in local blue compact dwarf galaxies \citep{1999ApJ...511..639I}. Nitrogen abundances for GRB (blue asterisk) and SN (orange plus) host galaxies were reported by \citet{2017MNRAS.466.2787C}, while Iron abundances for metal-rich damped lyman alpha absorption systems at $z=0.6-4.8$ (small grey crosses) are drawn from \citet{2015MNRAS.452.4326B}.}
\label{fig:abundances}
\end{figure}

\begin{table*}
\caption{Elemental abundances at `Solar' metallicity}
\hspace*{-0.5cm}
\begin{tabular}{lccccccccccccccc}
  
 &   Z &   X &  Y     & [C/H]  &  [N/H]  &  [O/H] & [Ne/H] & [Mg/H] & [Si/H] & [Fe/H]  \\
 \hline
 \hline
NP12    &  0.014  & 0.710   & 0.276  & 8.33 & 7.79 & 8.76 & 8.09 & 7.56 & 7.50 & 7.52 \\
       &    &    &   & $\pm$0.04 & $\pm$0.04 & $\pm$0.05 & $\pm$0.05 & $\pm$0.05 & $\pm$0.05 & $\pm$0.03 \\

GS98    &  0.017  &  0.735  &  0.248  & 8.52 & 7.92 & 8.83 & 8.08 & 7.58 & 7.55 & 7.50\\
   &    &    &    & $\pm$0.06 & $\pm$0.06 & $\pm$0.06 & $\pm$0.06 & $\pm$0.05 & $\pm$0.05 & $\pm$0.05\\

GAS07   &  0.0122 &  0.7393 & 0.2485 &  8.39 & 7.78 & 8.66 & 7.84 & 7.53 & 7.51 & 7.45$\pm$0.04\\
   &   &   &  &  $\pm$0.05 & $\pm$0.06 & $\pm$0.05 & $\pm$0.06 & $\pm$0.09 & $\pm$0.04 & $\pm$0.04\\

AGSS09  &  0.0134 &  0.7381 &  0.2485 & 8.43 & 7.83 & 8.69 & 7.93 & 7.60 & 7.51 & 7.50\\ 
  &   &   &   & $\pm$0.05 & $\pm$0.05 & $\pm$0.05 & $\pm$0.10 & $\pm$0.04 & $\pm$0.03 & $\pm$0.04\\ 

CLSFB10 & 0.0153 &  0.7321 &  0.2526 & 8.50 & 7.86 & 8.76 & & & & 7.52\\
 &  &   &   & $\pm$0.06 & $\pm$0.12 & $\pm$0.07 & & & & $\pm$0.06\\
\hline
\multicolumn{11}{c}{Values Corrected for Galactic Chemical Evolution for NP12}\\
AGSS09  &   &   &   & 8.53 & 7.95 & 8.77 &  & 7.68 & 7.63 & 7.68\\ 
 &   &   &   & $\pm$0.05 & $\pm$0.05 & $\pm$0.05 &  & $\pm$0.04 & $\pm$0.04 & $\pm$0.04\\ 

CLSFB10 & 0 &   &   & 8.60 & 8.01 & 8.84 & & & & 7.70\\
 &  &   &   & $\pm$0.06 & $\pm$0.12 & $\pm$0.07 & & & & $\pm$0.06\\

\hline
\multicolumn{11}{c}{BPASS model values}\\
WMBasic & 0.02 & 0.7  & 0.28  &   8.52 & 7.92 & 8.83 & 8.08  & 7.59 & 7.55 & 7.52 \\
Stellar models & 0.02  & 0.7  & 0.28  &  8.61  & 8.03 & 8.93 & 8.15  & 7.65  & 7.93  & 7.56 \\
Stellar models & 0.014 & 0.715 & 0.271 &  8.45 & 7.87 & 8.77 & 7.99  & 7.49  & 7.76  & 7.40 \\

\hline\hline
\end{tabular}

{\small Abundances are given in the form of 12+log(X$_n$/H) where X$_n$ is the atomic species in question. NP12 indicates the mean abundances of B type stars in the Solar neighborhood measured by \citet{2012A&A...539A.143N}. The last three lines give the values corresponding to Solar in our stellar models and the corresponding atmosphere models. Sources: GS98 - \citet{1998SSRv...85..161G}; GAS07 - \citet{2007SSRv..130..105G}; AGSS09 - \cite{2009ARA&A..47..481A}; CLSFB10 -  \citet{2011SoPh..268..255C}}\label{tab:solar}
\end{table*}

\begin{table*}
\caption{Abundances of distant galaxies and galaxy composites.}
\hspace*{-0.5cm}
\begin{tabular}{lcccccccccccc}
 Galaxy     & z & [C/H]  &  [N/H]  &  [O/H] & [Ne/H] & [Mg/H] & [Si/H] & [Fe/H]  \\
\hline\hline
 BX418$^1$    & 2.3 & 7.18    &                  & 7.8 \\
    &  & $\pm$0.12    &                  & $\pm$0.1 \\
Lynx$^2$      & 3.36 & 6.90 & 7.18 & 7.83 & 7.08 &  & 6.53-6.71& \\
      &  & $\pm$0.03 & $\pm$0.03 & $\pm$0.03 & $\pm$0.04 &  & $\pm$0.13 & \\
8 O'clock Arc$^3$ & 2.733  & $>$6.28 &  & $>$7.15 &  & & 7.31 & 6.56\\
&  & &  &  &  & & $\pm$0.14 & $\pm0.15$\\
SGAS\,J1051+0017$^4$  & 3.63 &7.5 & 6.7 & 8.3 & 7.13 &  & 6.3 \\
  &  &$\pm$0.4 & $\pm$0.3 & $\pm$0.1 & $\pm$0.23 &  & $\pm$0.5 \\
MS 1512-cB58$^5$  & 2.72 & & 6.50 & 8.39 &  & 7.28 & 7.14 & 6.35\\
 &  & & $\pm$0.20 & $\pm$0.20 &  & $\pm$0.20 & $\pm$0.20 &$\pm$0.20\\
KBSS-LM1 Comp$^6$  & 2.4 & 7.79 & 7.15 & 8.39 & & & 6.58 & \\
  & &$\pm$0.12 & $\pm$0.10 & $\pm$0.09 & & & $\pm$0.12 & \\
CASSOWARY20$^7$  & 1.433 & 6.77 & 6.37 & 7.82 & 7.09 & \\
 &  & $\pm$0.31 & $\pm$0.26 & $\pm$0.21 & $\pm$0.17 & \\
GRB 130606A$^8$ & 5.913 & $>$7.17 & & $>$6.84 &  &  & 6.21 & 5.41\\
 &  &  & &  &  &  & $\pm$0.08 & $\pm$0.08\\
GRB 120327A$^9$ & 2.815 & 8.70 & 6.38 & 6.71 &  & 6.33 & 6.23 & 5.77\\
&  & $\pm$0.17 & $\pm$0.10 & $\pm$0.12 &  & $\pm$0.09 & $\pm$0.09 & $\pm$0.09 \\
\hline
\hline
\end{tabular}

{\small Values are reported in the form 12+log(X$_n$/H) where X$_n$ is the atomic species in question. Sources: $^1$ BX418, \citet{2010ApJ...719.1168E}; $^2$ Lynx arc, \citet{{2004MNRAS.355.1132V}}; $^3$ 8 o'clock arc, \citet{2010A&A...510A..26D}; $^4$ SGAS J105039.6+001730, \citet{2014ApJ...790..144B}; $^5$ MS\,1512-cB58, \citet{2002ApJ...569..742P}; $^6$ the KBSS-LM1 composite galaxy template of \citet{2016ApJ...826..159S} - we use the nebular gas metallicities; $^7$ CASSOWARY20, \citet{2014MNRAS.440.1794J}; $^8$ GRB 130606A afterglow, \citet{2015A&A...580A.139H}; $^9$ GRB 120327A afterglow, \citet{2014A&A...564A..38D}.}\label{tab:abundances}
\end{table*}

Certain aspects of the chemical composition of our models can introduce uncertainty in their interpretation. In particular, we wish to highlight the question of Solar metallicity, the importance of CNO versus iron elements and alpha-enhancement at lower metallicities. As discussed in section \ref{sec:method_evol}, there is little consensus on the numerical value of Solar metallicity. Our standard compositions, determined by the \citet{1996ApJ...464..943I} opacity tables we employ, are listed in Table \ref{tab:solar} and illustrated in Figure \ref{fig:abundances}. These are slightly oxygen-rich at a given total metallicity mass fraction 
compared to some Solar abundance estimates. However this has little impact on the evolution of the stars other than slightly changing the efficiency of the CNO cycle during hydrogen burning. As Figure \ref{fig:abundances} demonstrates the [N/O] ratio in \bpass\ models is comparable to most estimates of Solar metallicity, and to the local mean abundance determined from B stars in the Solar neighborhood \citep{2012A&A...539A.143N}, but is a little high compared to observations of lower metallicity systems in both the local and more distant Universe. However, the figure also demonstrates the very significant scatter seen in this line ratio in observational data, suggesting models with a range of abundance ratios are required. It should also be noted that the abundances by number are also somewhat dependent on the initial hydrogen mass fraction abundance assumed. We use $X=0.7$ as our initial hydrogen mass fraction abundance. This is consistent with the abundances used to calculate our OB star mass-loss rates by \citet{2001A&A...369..574V}, but even if we vary our hydrogen abundance by allowable amounts our [O/Fe] value will be slightly too rich. 

Of more significance is the role of iron, as this is the element which is responsible for most opacity in a stellar interior and for driving the surface stellar winds. In our models the true Solar metallicity can be considered to be represented by either our $Z=0.014$ and 0.020 models, which we define as having 12+log[O/H] values of 8.7 to 8.8, with the higher metallicity model having a Solar iron abundance, while the lower metallicity models have a composition more similar to the local Cosmic abundances defined by \citet{2012A&A...539A.143N} but are slightly less iron rich. Simultaneous observations of oxygen and iron abundances for galaxies are difficult to obtain in the distant Universe, since iron measurements are typically obtained from weak absorption features in the rest-ultraviolet, while oxygen is typically measured from the strong rest-frame optical gas phase emission lines. Nonetheless, measurements from damped lyman-alpha absorption systems suggest that our [Fe/O] abundances track those in the intergalactic medium at high redshifts over a range of metallicities. Agreement with extreme low metallicity star forming galaxies in the local Universe (derived from SDSS) is less clear, and suggests that we may slightly overestimate iron abundance relative to oxygen at low metallicities.

The composition employed by \bpass\ is consistent with all the atmosphere model spectra we use during the hydrogen-rich phase of evolution. For Wolf-Rayet stars this issue is not quite as important since the light element composition reflects the progress of nuclear burning rather the initial composition. We choose to scale mass-loss rates by metallicity from $Z=0.020$ again to be consistent with previous work. 

A final uncertainty regarding the role of stellar composition extends to whether the alpha-element to iron-group elements ratio is constant with total metallicity mass fraction. Alpha elements (which are released by core-collapse supernovae) are formed earlier in the lifetime of a stellar population than iron and other metals (which are formed in type Ia supernovae and massive star winds). A higher abundance of alpha-elements might be expected in lower metallicity systems due to their younger typical stellar populations. At the current time we do not adjust for this as the relative abundance evolution with metallicity is poorly constrained, and is likely heavily dependent on the star formation history in individual cases, thus showing significant scatter in the baseline relation \citep[e.g.][]{2010PASA...27..379W}. Although metallicities of extragalactic systems are typically measured and quoted in terms of [O/H], this does not always relate directly to [Fe/H] which is what drives the greatest changes in stellar evolution. Therefore an approach similar to that of \citet{2016ApJ...826..159S} and \citet{2016arXiv160900727B}, whereby diagnostic emission lines are modelled with an alpha-enriched nebular gas emission component, while the stellar population is modelled at the associated iron abundance, should be considered.

We aim to calculate future model sets with more varied compositions, including $\alpha$-element enhancement, both to more accurately reproduce Galactic composition and those in the high-redshift Universe, but our v2.1 model set provides a single standard abundance model that can be applied across cosmic time and against which variation can be compared.

\subsection{Nebula emission, accretion physics and X-ray binaries}

It should be clearly stated that the baseline \bpass\ models do not include nebula emission from surrounding gas as a default. The response of diffuse interstellar, or even intergalactic, gas to stellar radiation, together with its reprocessing and emission as nebular lines, is a complex topic which has been explored in detail by other authors \citep[e.g.][]{1998PASP..110..761F,2004AJ....127.2002K} and in section \ref{sec:nebular}. The introduction of nebular emission is accompanied by a wealth of free parameters defining the gas geometry, density, abundance ratios, temperatures and thus ionization state, and depletion onto dust grains.  These must be fixed on a case by case basis by analysis of resolved data or emission line ratios, or alternatively a full grid of models, exploring all these free parameters must be compared to the data. We recommend that our stellar population models be reprocessed by a radiative transfer code such as CLOUDY or MAPPINGS before comparison with observed data in contexts where the nebular effects are expected to be significant \citep[see section \ref{sec:nebular} above, ][ and Xiao et al, in prep]{2015MNRAS.452.2597X}. \citet{2014MNRAS.444.3466S} also explores this issue in more detail.

We also currently do not include the accretion luminosity from  X-ray binaries in our model populations since accretion physics is beyond the scope of our models. This can be significant at lower metallicities as the typical black hole mass in our populations increases; therefore our predictions for ionizing fluxes should be considered as a lower bound without this included. While accretion events are short lived, they are also extremely luminous and so may have an appreciable effect on the integrated spectrum of a stellar population in any given time bin. We would thus expect that once an accurate model of accretion onto compact remnants is included there will be an extra source of hard ionizing radiation within our models. This will lead to models with recently formed black-holes possibly taking on some of the observational characteristics of very-low luminosity AGN. We note that our omission of accretion-powered emission includes that of cataclysmic variables and similar sources from our integrated spectra, however CV accretion power would be important only in older population where white dwarfs have had time to form. In those older population there could be other, possibly more pressing, issues if the lower mass stars, such as post-RGB and post-AGB stars (including binaries in these categories), are not well modelled. 

\subsection{Wolf-Rayet star inflation}

It has been known for some time that matching the hydrostatic stellar models of Wolf-Rayet stars to the observed parameters is problematic \citep[see][and references therein]{2012A&A...538A..40G}. This is mainly due to the star having optically thick stellar winds so it is impossible to see the stellar surface, from where the wind is launched in most cases. Here we use the recommended method outlined by the PoWR team to match our WR models to their atmosphere models \citep{2012A&A...538A..40G}. In addition to this, there is growing evidence that the envelopes of WR stars may be still more inflated than predicted by evolutionary calculations \citep{2012A&A...540A.144S,2016MNRAS.459.1505M}. It appears this is only important at near-Solar metallicities but can lead to rather lower stellar temperatures by 0.2 dex. This will impact on predicted line emission from our WR populations, specifically the blue and the red WR bumps, in our spectra and require the use of lower surface gravity models than those currently selected. It is impossible at the current time to evaluate the importance of this effect but future work on modelling resolved and unresolved WR populations may provide the necessary constraints and lead to its implementation in future \bpass\ versions.

\subsection{Neutron star and black hole kicks}\label{sec:kicks}

In these v2.1 models we have continued to use our kick model based on the \citet{2005MNRAS.360..974H} observations of Galactic neutron-star velocities. We determine a velocity at random from Maxwell-Boltzmann distribution with $\sigma=265$\,km\,s$^{-1}$. For black holes we assume the same method but reduce the strength of the kick by a factor of 1.4\,M$_{\odot}$/M$_{\rm rem}$, assuming the distribution represents a momentum distribution for black holes. While there is some evidence to suggest this is the case \citep[e.g.][]{2016MNRAS.456..578M} the true scaling is still uncertain.

We have recently been investigating a new model for neutron-star kicks where the kick velocity is related to the ejecta mass of the exploding star \citep{2016MNRAS.461.3747B}. There is some evidence from simulations of core-collapse that this model may be the correct one to apply \citep{2012ARNPS..62..407J}. Once we obtain better constraints on whether the kick is a match to nature or not, we may switch to this new kick model in future versions of \bpass. We are still considering how to extend this kick model to black holes.

An important uncertainty in neutron star and black hole formation is how we estimate the remnant masses. Ours is a relatively simple method but does produce similar island of stability to that found by other authors \citep{2012ApJ...757...69U,2014ApJ...783...10S,2016ApJ...818..124E}. This can be seen in Figure 6 of \citet{2004MNRAS.348..201E} where contours of remnant masses for given initial mass and metallicity were predicted. The only way to gain insight into the accuracy of our method will be to study future events where the massive stars are observed to disappear without supernovae as found by \citet{2017MNRAS.469.1445A}. Further insights will come from modelling the  black hole mass distribution arising from future gravitational wave sources \citep[see][]{2016MNRAS.462.3302E}, while also using results from gravitational microlensing to determine the black hole mass distribution of single free-floating black holes in the Galaxy \citep{2016MNRAS.458.3012W}.

\subsection{Un- and under-represented stellar types}

Certain stars may be missing within our populations. One important class are luminous-blue variable (LBV) stars.  There are two reasons we do not have these in our populations. One is that we do not know which models will be observed as LBVs. The evolutionary state of these stars is still somewhat uncertain \citep{2011MNRAS.415..773S,2006ApJ...645L..45S,1994PASP..106.1025H,2017ApJ...836...64H}. Additionally if they are the result of transient and irregular mass transfer in a binary system, they may not exist within our set of models as discussed above. This is of course something that will be fruitful to study in future.

For lower mass stars there are limitations in our modelling of AGB stars and the helium flash. In both cases within the evolution code used it is difficult to evolve through these phases. Therefore in our single star models AGB models never form white dwarfs. The models end once the core has grown to the Chandrasekhar mass. In future we will replace these models with synthetic AGB models \citep{2004MNRAS.350..407I} or recalculate the models with more realistic mass-loss rates after 2nd dredge-up. We note however that in our binary populations many stars do not evolve to the AGB phase due to binary interactions before 2nd dredge-up.

To get around the core helium flash we have had to make pseudo-evolution models to circumvent this for stars below about 2\,M$_{\odot}$ for our single-star populations. Because of the large number of binary models we have not yet searched in detail for failed models due to the core helium flash. Most of our efforts have been concentrated on massive stars within our model populations. However in future we will be able to extend the evolution of these low mass stars.

\subsection{Low Mass Clusters and IMF Sampling}

Motivated by recent observations, particularly those of the Tarantula Nebula \citep{2012Sci...337..444S,2013A&A...558A.134D}, we have favoured models of a stellar population drawn from an IMF that extends to an upper mass limit of 300\,M$_\odot$. However as discussed by \citet{2012MNRAS.422..794E} and references therein, a fully populated IMF requires a very large initial molecular cloud; a gas cloud with a total mass of only 1000\,M$_\odot$, for example, is unlikely to generate even a single 300\,M$_\odot$ star, although evidence for a relation between the most massive star in a cluster and its total mass remains unclear \citep[e.g.][]{2014ApJ...793....4A,2014ApJ...780...27P}. The result is potentially a strong stochasticity in the high mass stellar population of low mass star forming regions (i.e. a star of given mass either forms or does not in any given molecular cloud). Since these most massive stars dominate the light at early ages, this can alter the observable properties of the integrated light measured from the population \citep[e.g.][]{2010A&A...512A..79H,2013A&A...558A.134D}. In common with other stellar population synthesis models, \bpass\ assumes a sufficiently large population of stars is formed that the IMF is fully sampled, and outputs are normalised to a total mass of 10$^{6}$\,M$_\odot$. While these should comfortably scale up to larger mass starbursts, we note that caution should be used when studying star forming regions of significantly lower total mass.  For such systems, our equivalent set of data products including an IMF which only samples up to 100\,M$_\odot$ may be more appropriate. Note that we do not provide IMFs generated with stochastic sampling since these are heavily dependent on initial presumed total mass.

\begin{figure*}
\begin{center}
\begin{tabular}{cc}
\includegraphics[width=0.8\columnwidth]{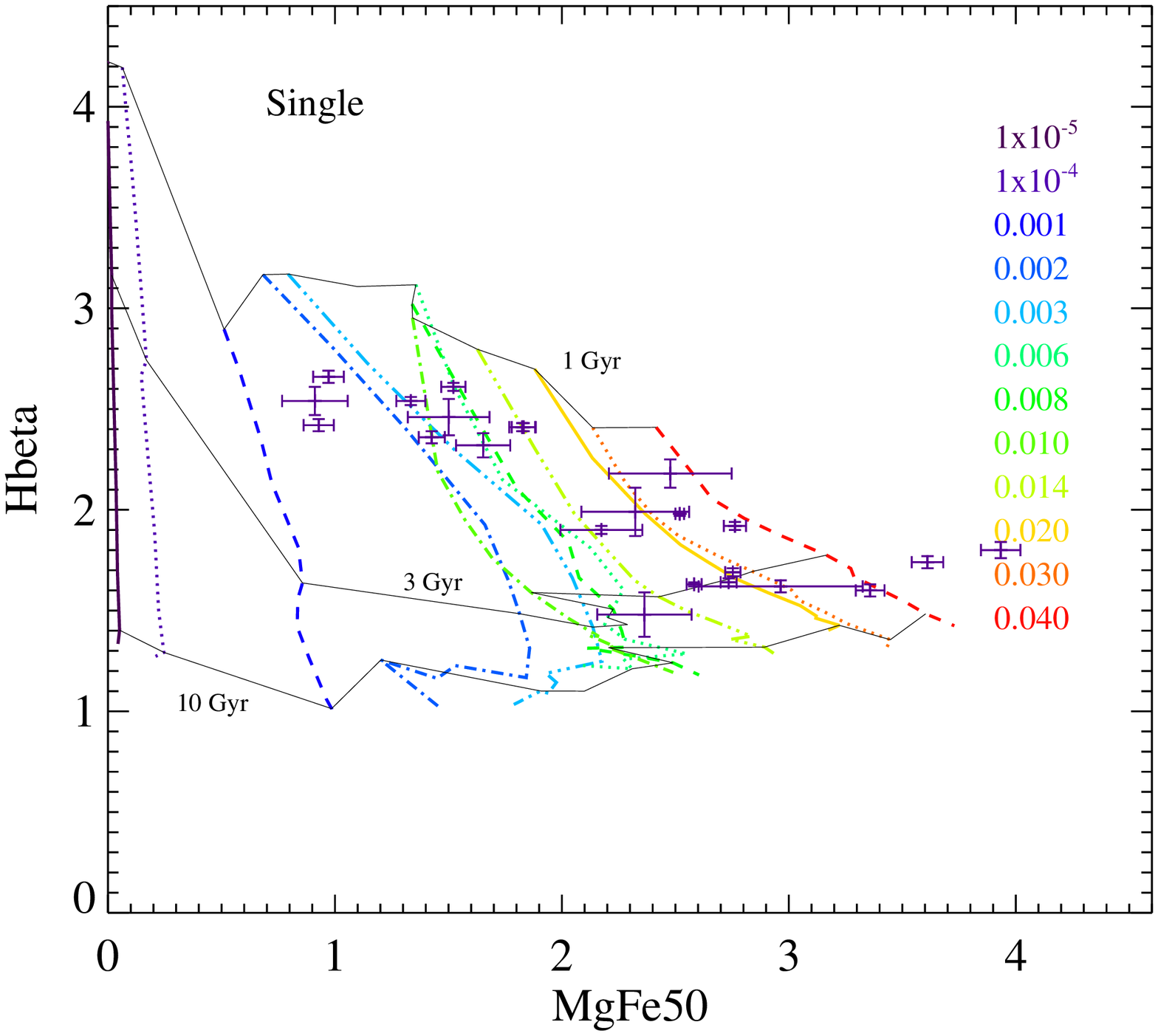} &
\includegraphics[width=0.8\columnwidth]{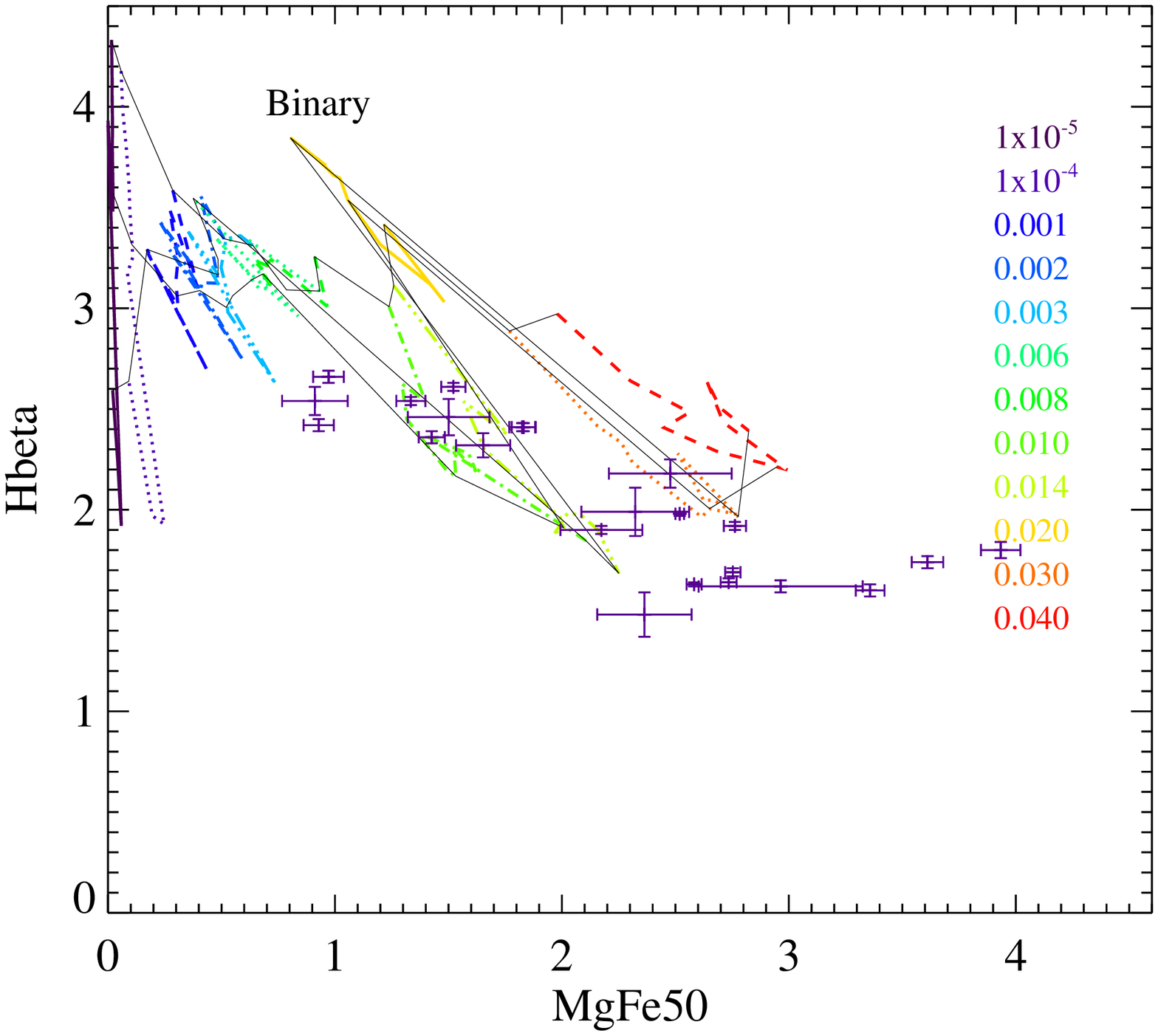} \\
\includegraphics[width=0.8\columnwidth]{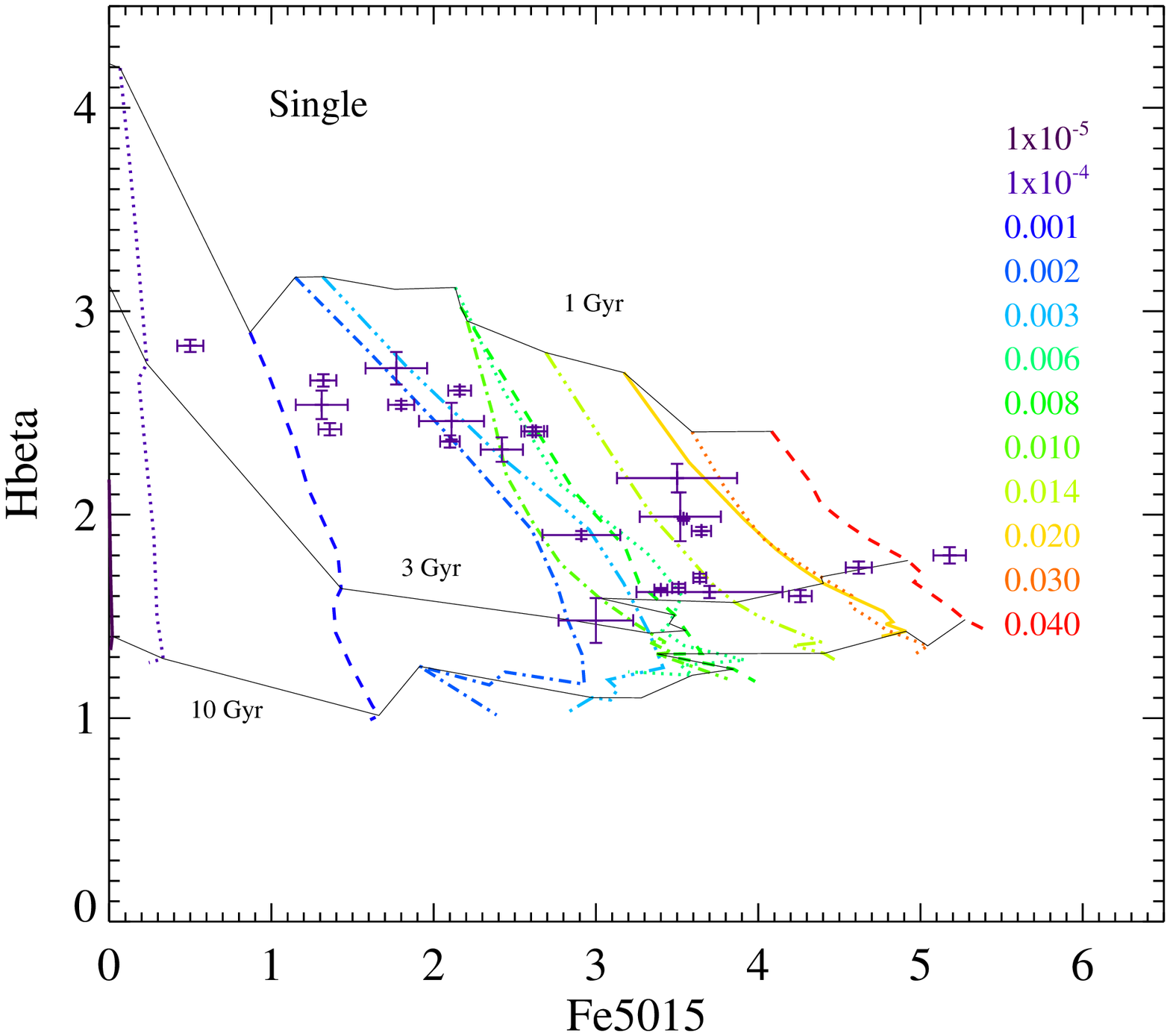} &
\includegraphics[width=0.8\columnwidth]{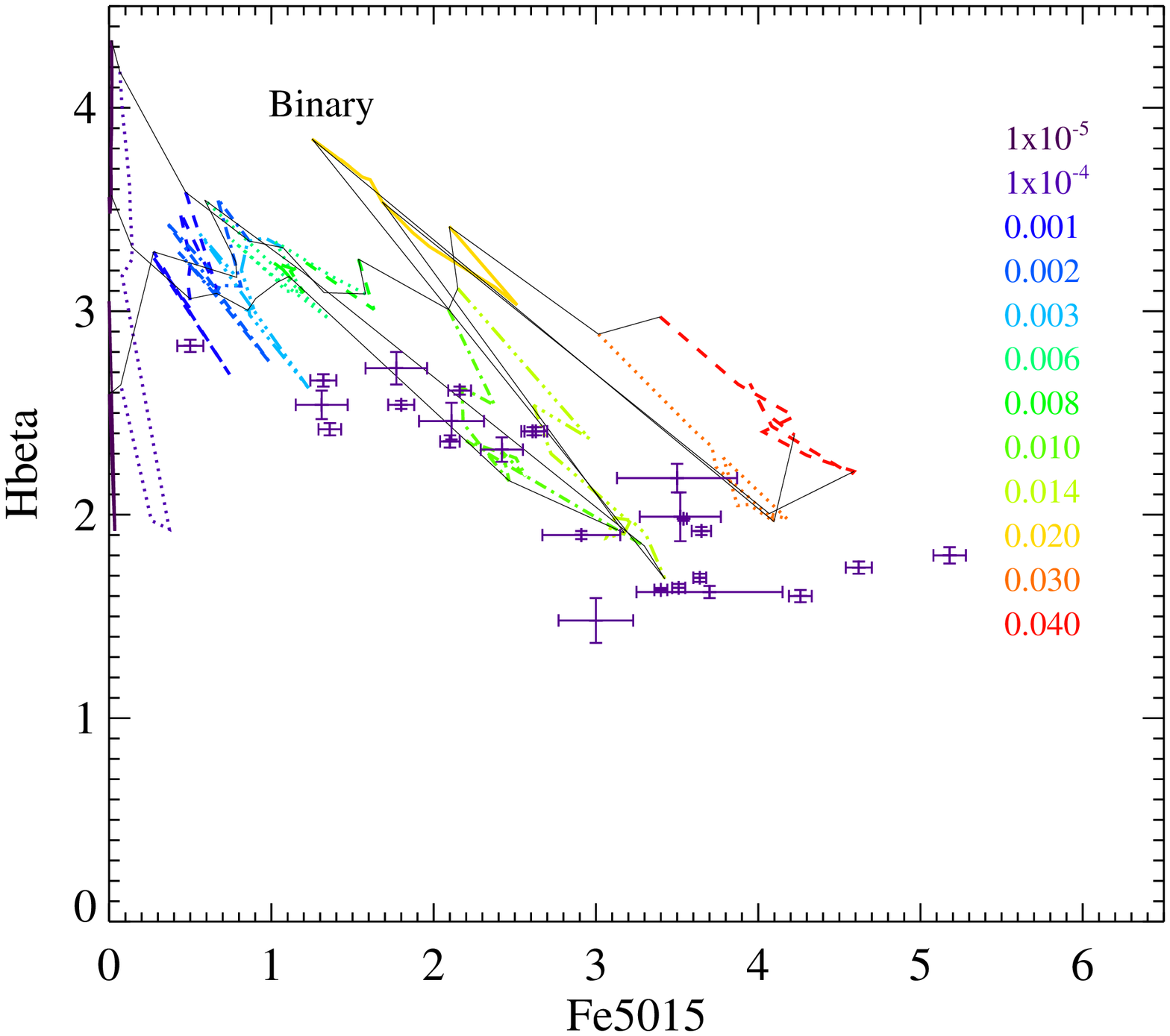} \\
\includegraphics[width=0.8\columnwidth]{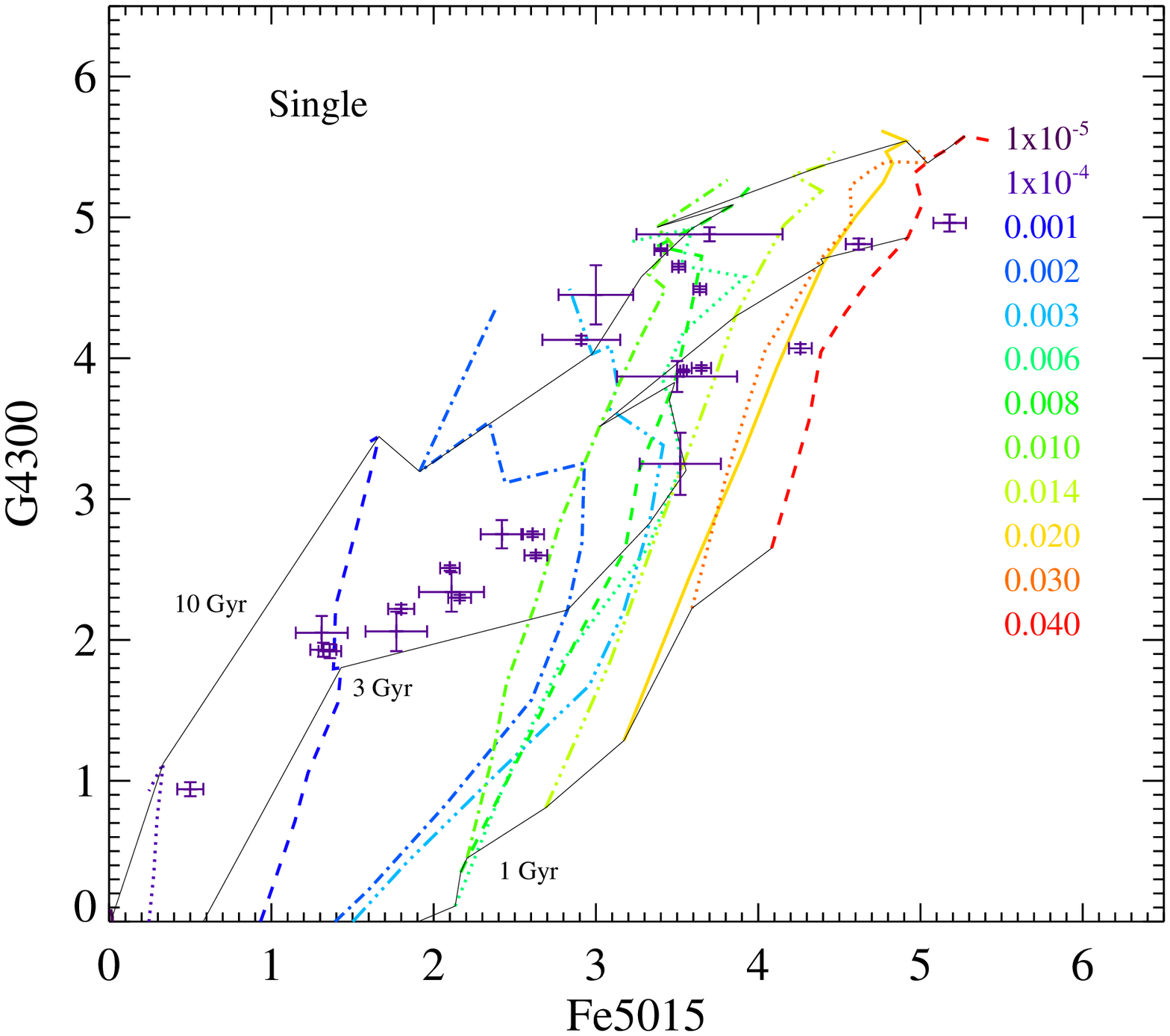} &
\includegraphics[width=0.8\columnwidth]{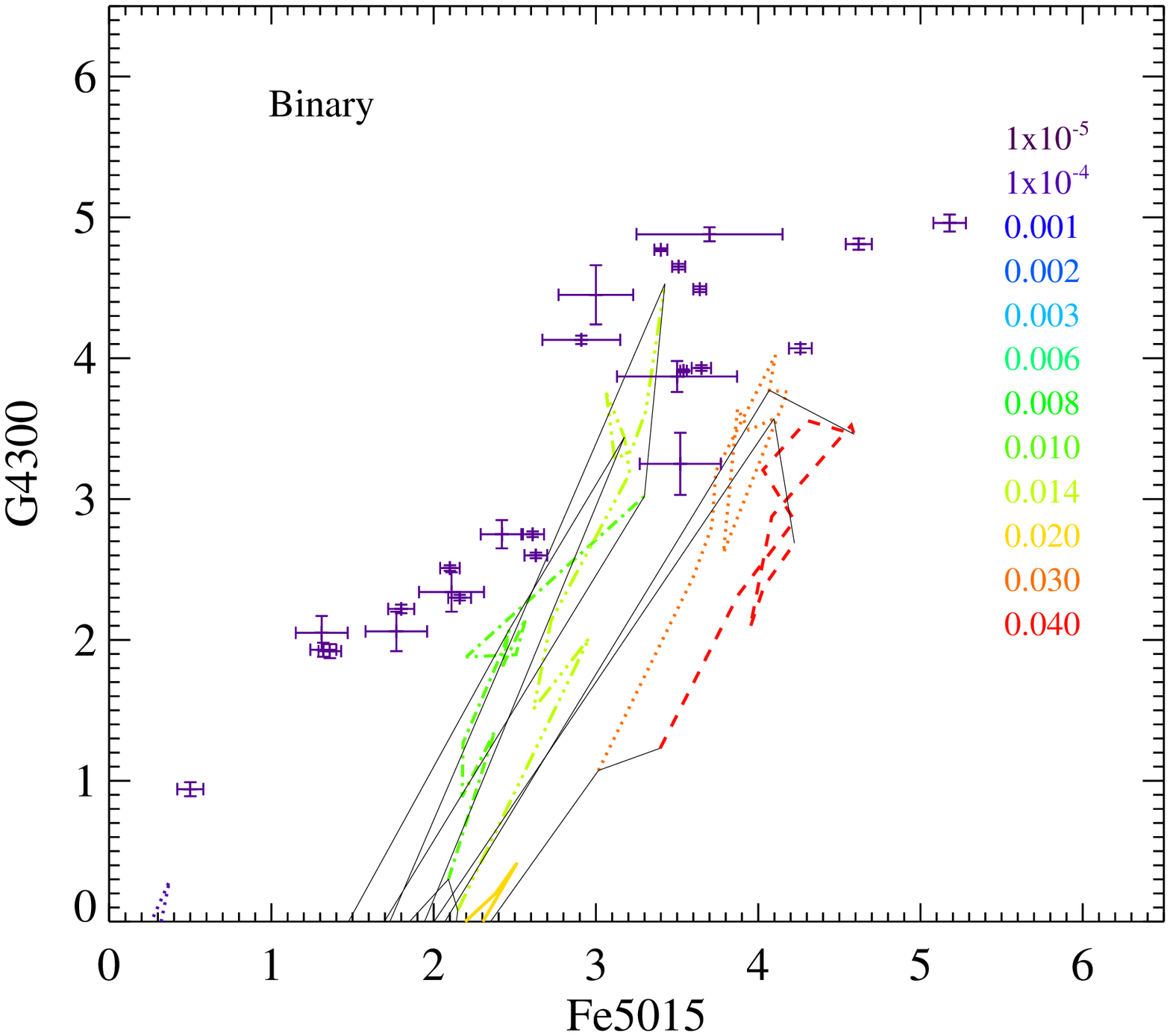} \\
\end{tabular}
\end{center}
\caption{Lick Indices calculated from \bpass\ models at stellar population ages between 1\,Gyr and 12\,Gyr in age, at a variety of metallicities. Data points indicate the indices measured for Galactic globular clusters by \citet{2016ApJS..227...24K}, adjusted to the original \citep{1994ApJS...94..687W} index scale using the zeropoints given by \citet{2007ApJS..171..146S}. The MgFe50 index was defined by \citet{2010MNRAS.408...97K} as insensitive to abundance ratios. In all cases, the single star models perform better than the binary models for these very old stellar populations, as discussed in section \ref{sec:oldpops}}\label{fig:lick}
\end{figure*}

\subsection{Old stellar populations}\label{sec:oldpops}
The majority of observational tests presented in this paper have been for relatively young stellar populations ($< 1$\,Gyr).  At older ages, \bpass\ binary models remain relatively blue compared to single star populations, and do not reach full quiescence until later (see Figure \ref{fig:seds}). 
The reasons why \bpass\ stays blue are two-fold:

The first effect is due to the width of the time bins at late ages and relatively sparse sampling of stellar masses - since each bin is very broad, working with a single age starburst is going to result in all stars in a certain mass range undergoing the bulk of their dramatic evolution in one bin, and not necessarily in adjacent bins. Since there is not an infinite mass resolution in the stellar models, and since stars of different masses die in different ways, that leads to instability in the models at late times - i.e. the behaviour at log(age)=9.3 might be very different from log(age)=9.2 or 9.4 - this is inherent in the stellar population but also something of a modelling resolution issue and is one we plan to address in future by addition of further stellar models. Some of these fluctuations are also due to our poor estimate of rejuvenation ages for secondary stars where the primary did not explode in a core-collapse supernova. An interim approach is to average several age bins at late ages (effectively degrading the age resolution further, but averaging over some of this stochastic variation). We could also interpolate between initial stellar masses but stellar evolution is non-linear and binary evolution even more so. The best solution to this problem is to increase the number of low-mass stellar models in our grid.

The second issue that needs to be considered is the binary fraction as a function of mass.  Since our models were originally tuned to young populations, our focus was on the massive stars, and their binary interaction fraction. This is fixed using a binary period distribution matched to observations (i.e. all the stars in our binary models are technically in binaries, but in many of them the binary is too wide to have any effect on their evolution so they behave like single stars, with the right interacting fraction).  However one thing we have not accounted for systematically (because until recently it has been very poorly constrained by observations, especially as a function of metallicity) is the binary fraction in lower mass stars. The blue colours in our binary models at late ages result to a large extent from interactions between fairly low ($\sim$ a few Solar) mass stars and their companions rejuvenating the population.  Again, the large time bins are going to make this rather stochastic with some bins more affected than their neighbours. It appears likely our binary models are overestimating the interacting binary fraction in these low mass stars. One reason we also provide the single star models is to provide some capacity to vary the binary fraction by combining the two in proportion, reducing the interaction fraction overall and it might be appropriate to do this incrementally as a function of age (i.e. mass range of dominant stars). Recent work by \citet{2017ApJS..230...15M} has provided much tighter constraints on these parameters, although still at Solar metallicity and we are likely to modify our initial binary parameter distributions in future work.

Our single star models perform generally better at late ages, particularly in the ultraviolet and optical but the excess of AGB stars identified in section \ref{sec:method_seds} and Figure \ref{fig:sed_comparison} at late times remains an issue to be explored in future work. Treatment of the pulsating phases of AGB stars is currently neglected, and may have a substantial impact on the survival of such stars and their radii (and hence interactions). While these results are features of the models, rather than necessarily problems in them, we recognise that this can cause significant issues for SED fitting and for measurement of absorption line features such as the Lick indices \citep{1994ApJS...94..687W,1997ApJS..111..377W}. As figure \ref{fig:lick} demonstrates, our binary models fail to reproduce the absorption line indexes observed in old globular clusters within the Milky Way, while these are well fit by our single star populations - as might be expected if the binary fraction in low mass stars is overestimated. This is not an unambiguous interpretation: the binary fraction in globular clusters may differ from that in the field or extragalactic environments, and more work is needed to clarify the situation. Improved treatment of white dwarf atmospheres is also likely to improve performance here in future versions, since these can have a pronounced effect on the Balmer line indices. 

Nonetheless, binary models do offer some interesting prospects for understanding old stellar populations that currently challenge interpretations. An example is the UV-upturn phenomenon in elliptical galaxies with $>10$\,Gyr-old stellar populations \citep[e.g.][]{1995ApJ...442..105D}. The origin of the ultraviolet emission in these sources is unclear, but is likely associated with a substantial population of helium-rich stars. While it is possible that these systems were formed in regions with helium enhancement \citep[e.g.][]{2011ApJ...740L..45C}, that raises questions regarding the origin of such enhancement. An alternative is that they host stars whose atmospheres have been stripped through binary interactions, and a binary model such as \bpass\ is essential for exploring this possibility. 

In future it may be possible to use Lick indices, observations of UV-upturn galaxies and similar data to further constrain the binary model and explore $\alpha$-enhancement in these models, but at present we do not recommend using our v2.1 binary models to fit populations more than a few gigayears old without due caution.

\subsection{Dust and radio components}\label{sec:dustandradio}

In common with most other spectral synthesis codes we do not account for re-emission of extincted light processed into thermal black body radiation by dust grains (or indeed polyaromatic hydrocarbons). We recommend the use of an energy-balance formalism such as that applied by MAGPHYS \citep{2008MNRAS.388.1595D} to predict stellar population effects on emission in the infrared and submillimetre \citep[see][for an example of this approach]{2015MNRAS.446.3911S}. The detailed physics of how binary interactions affect dust must consider supernova and stellar wind dust yields, as well as the destruction of dust grains by energetic events such as supernovae or intense local radiation fields.  

Given that \bpass\ is a full stellar population synthesis code, able to make predictions for all phases of stellar evolution and rates of stellar death, it should be possible to implement a dust yield model at a later date, but this is beyond the scope of the current version, and will rely on future resourcing.

Similarly, emission in the radio continuum arises from a combination of thermal and non-thermal components, with the latter believed to track the supernova rate in a stellar population \citep[see e.g.][]{1992ARA&A..30..575C,2017ApJ...836..185T}. While a crude estimate of the time evolution in radio emission can therefore be constructed from the rate of supernovae in current models \citep[see][for preliminary work on this]{{2017arXiv170507655G}} a detailed treatment again awaits future work.


\section{CONCLUSION}\label{sec:conclusions}

In this paper we have presented the numerical methodology, input models and primary outputs of version 2.1 of the Binary Population and Spectral Synthesis (\bpass) stellar population modelling code. This code builds composite stellar populations from a set of detailed stellar evolution models spanning masses from 0.1-300\,M$_\odot$ and at metallicities from 0.05\% to 200\% of Solar. It combines these with publically available stellar atmosphere models to predict the spectral properties of simple stellar populations at ages from 1\,Myr to 100\,Gyr. We have addressed its performance in different regimes and presented observational validation tests demonstrating its efficacy, and its limitations, in  a wide variety of contexts.

These include:
\begin{enumerate}
\item  The physical characteristics and evolutionary state of well-constrained eclipsing binary stars, which are well matched by model predictions.

\item The population ratios of different massive stellar types, including massive stars, as a function of metallicity, which are generally consistent with observed ratios across a range of metallicities. 

\item The Hertzsprung-Russell diagrams and photometric colours of single-aged stellar clusters, which are generally well fit in our models, but suggest caution is needed in fitting old stellar populations, or those dominated by certain rare stellar categories.

\item The distribution and properties of compact stellar remnants, which show good agreement between observations and models.

\item  The rates, progenitors and characteristics of astrophysical transients, which agree well with observational constraints, but hint that those sources collapsing to black holes may be underrepresented in supernova surveys.

\item  The colours and diagnostic line strengths of Wolf-Rayet galaxies, which can be simultaneously fit at moderate stellar population ages. These are somewhat older in our binary models than single star models would suggest.

\item The colours of the local star forming galaxy population, which show the difficulty and degeneracy in modelling systems with complex star formation histories and substantial gas nebular or dust components. Models can be found which provide a good fit to the data, but these introduce a large number of tunable free parameters.

\item The properties of distant galaxies, including their ionizing photon production, optical line ratios and ultraviolet emission lines, which are generally significantly better fit in our binary models than single star models, and which may be older than single star models suggest, but in which nebular gas again plays an important role. 

\end{enumerate}

In the interests of transparency and clarity, we have also discussed key caveats and uncertainties in the models. Some of these, particularly those relating to under-represented rare populations, composition, inflation and non-stellar components, are present in all stellar population synthesis codes. Others, including period distributions, mass transfer, kicks and rotational mixing, arise from the increased complexity that is inevitable in the modelling of stellar binary interactions. In particular, we identify several aspects of old stellar populations ($>$1\,Gyr) as problematic in our current model set and a target for future development.

Despite the presence of these uncertainties, the tests described above indicate that our physically-motivated minimal assumptions result in a spectral synthesis model set which provides a good description of the extant stellar populations in a broad range of environments, to an extent comparable to, or exceeding, existing single star spectral synthesis models. Given the ever-growing number of observational constraints on extreme stellar populations, and the expected impact of JWST on the observation of low metallicity systems in the distant Universe, the need for such models is likely to grow rather than  diminish over coming years, and we aim to continue to develop the \bpass\ model infrastructure in response to improvements both in stellar modelling and observational constraints.

All outputs of the current \bpass\ v2.1 data release can be found at {\tt http://bpass.auckland.ac.nz}, and are mirrored at {\tt http://warwick.ac.uk/bpass}.


\begin{acknowledgements}
This work would not have been possible without use of the NeSI Pan Cluster, part of the NeSI high-performance computing facilities. New Zealand's national facilities are provided by the NZ eScience Infrastructure and funded jointly by NeSI's collaborator institutions and through the Ministry of Business, Innovation \& Employment's Research Infrastructure programme. URL: https://www.nesi.org.nz.

JJE acknowledges travel funding and support from the University of Auckland.
ERS acknowledges travel funding and support from the University of Warwick.
We thank Robert Izzard for helping us to obtain atmosphere models.
We acknowledge very many useful and interesting discussions with \bpass\ users and collaborators past, present and future, too numerous to name. We also thank the referee of this paper for their hard work and constructive input. We note
that \bpass\ development began when the PIs were post-doctoral researchers and we thank our previous employers (University of Bristol, University of Cambridge, Queen's University Belfast and the Institute d'Astrophysique de Paris) for supporting our use of time on this work. \bpass\ is human-resource limited and  further iterations will be dependent on the PIs securing research funding in future years.\end{acknowledgements}


\end{document}